\pdfoutput=1
\documentclass[12pt,a4paper]{article}
\usepackage{ifthen} % for conditional statements
\newboolean{pdflatex}
\setboolean{pdflatex}{true} % False for eps figures 

\newboolean{articletitles}
\setboolean{articletitles}{true} % False removes titles in references

\newboolean{uprightparticles}
\setboolean{uprightparticles}{false} %True for upright particle symbols

\newboolean{inbibliography}
\setboolean{inbibliography}{false} %True once you enter the bibliography

\def\paperauthors{LHCb collaboration} % Leave as is for PAPER, CONF and FIGURE
\def\paperasciititle{Measurement of CP-averaged observables in the B0->K*0mu+mu- decay} % Set ASCII title here !! MAKE sure it's only ASCII characters !! 
\def\papertitle{Measurement \\ of \CP-averaged observables \\ in the $\decay{\Bd}{\Kstarz\mup\mun}$ decay} % Latex formatted title
\def\paperkeywords{{High Energy Physics}, {LHCb}} 
\def\papercopyright{\the\year\ CERN for the benefit of the LHCb collaboration} 
\def\paperlicence{CC BY 4.0 licence}
\def\paperlicenceurl{https://creativecommons.org/licenses/by/4.0/}

\usepackage[top=1in, bottom=1.25in, left=1in, right=1in]{geometry}

\usepackage{microtype}
\usepackage{lineno}  % for line numbering during review
\usepackage{xspace} % To avoid problems with missing or double spaces after
\usepackage{caption} %these three command get the figure and table captions automatically small

\usepackage{graphicx}  % to include figures (can also use other packages)
\usepackage{color}
\usepackage{colortbl}
\graphicspath{{./figs/}} % Make Latex search fig subdir for figures
\usepackage{amsmath} % Adds a large collection of math symbols
\usepackage{amssymb}
\usepackage{amsfonts}
\usepackage{upgreek} % Adds in support for greek letters in roman typeset

\newcommand*\patchAmsMathEnvironmentForLineno[1]{%
\expandafter\let\csname old#1\expandafter\endcsname\csname #1\endcsname
\expandafter\let\csname oldend#1\expandafter\endcsname\csname
end#1\endcsname
 \renewenvironment{#1}%
   {\linenomath\csname old#1\endcsname}%
   {\csname oldend#1\endcsname\endlinenomath}%
}
\newcommand*\patchBothAmsMathEnvironmentsForLineno[1]{%
  \patchAmsMathEnvironmentForLineno{#1}%
  \patchAmsMathEnvironmentForLineno{#1*}%
}
\AtBeginDocument{%
\patchBothAmsMathEnvironmentsForLineno{equation}%
\patchBothAmsMathEnvironmentsForLineno{align}%
\patchBothAmsMathEnvironmentsForLineno{flalign}%
\patchBothAmsMathEnvironmentsForLineno{alignat}%
\patchBothAmsMathEnvironmentsForLineno{gather}%
\patchBothAmsMathEnvironmentsForLineno{multline}%
\patchBothAmsMathEnvironmentsForLineno{eqnarray}%
}

\usepackage{hyperxmp}

\usepackage[pdftex,
            pdfauthor={\paperauthors},
            pdftitle={\paperasciititle},
            pdfkeywords={\paperkeywords},
            pdfcopyright={Copyright (C) \papercopyright},
            pdflicenseurl={\paperlicenceurl}]{hyperref}
\usepackage[colorinlistoftodos,textsize=scriptsize]{todonotes}

\usepackage[bottom,flushmargin,hang,multiple]{footmisc}

\usepackage[all]{hypcap} % Internal hyperlinks to floats.

\usepackage{xspace} 
\usepackage{upgreek}

\def\lhcb   {\mbox{LHCb}\xspace}

\def\babar  {\mbox{BaBar}\xspace}
\def\belle  {\mbox{Belle}\xspace}

\def\rich   {RICH\xspace}

\def\MagUp {\mbox{\em Mag\kern -0.05em Up}\xspace}

\ifthenelse{\boolean{uprightparticles}}%
{

 \def\Pmu         {\ensuremath{\upmu}\xspace}

 \def\Ppi         {\ensuremath{\uppi}\xspace}

 \def\Pphi        {\ensuremath{\upphi}\xspace}

 \def\Ppsi        {\ensuremath{\uppsi}\xspace}

 \def\PDelta      {\ensuremath{\Delta}\xspace}                 
 \def\PXi         {\ensuremath{\Xi}\xspace}                 
 \def\PLambda     {\ensuremath{\Lambda}\xspace}                 
 \def\PSigma      {\ensuremath{\Sigma}\xspace}                 
 \def\POmega      {\ensuremath{\Omega}\xspace}                 
 \def\PUpsilon    {\ensuremath{\Upsilon}\xspace}

 \def\PB      {\ensuremath{\mathrm{B}}\xspace}                 
                  
 \def\PD      {\ensuremath{\mathrm{D}}\xspace}

 \def\PJ      {\ensuremath{\mathrm{J}}\xspace}                 
 \def\PK      {\ensuremath{\mathrm{K}}\xspace}

 \def\Pb      {\ensuremath{\mathrm{b}}\xspace}                 
 \def\Pc      {\ensuremath{\mathrm{c}}\xspace}

 \def\Pi      {\ensuremath{\mathrm{i}}\xspace}

 \def\Pp      {\ensuremath{\mathrm{p}}\xspace}

 \def\Ps      {\ensuremath{\mathrm{s}}\xspace}

 \def\thebaroffset{0.0em}
}
{

 \def\Pmu         {\ensuremath{\mu}\xspace}

 \def\Ppi         {\ensuremath{\pi}\xspace}

 \def\Pphi        {\ensuremath{\phi}\xspace}

 \def\Ppsi        {\ensuremath{\psi}\xspace}                 
                  
 \mathchardef\PDelta="7101
 \mathchardef\PXi="7104
 \mathchardef\PLambda="7103
 \mathchardef\PSigma="7106
 \mathchardef\POmega="710A
 \mathchardef\PUpsilon="7107
                  
 \def\PB      {\ensuremath{B}\xspace}                 
                  
 \def\PD      {\ensuremath{D}\xspace}

 \def\PJ      {\ensuremath{J}\xspace}                 
 \def\PK      {\ensuremath{K}\xspace}

 \def\Pb      {\ensuremath{b}\xspace}                 
 \def\Pc      {\ensuremath{c}\xspace}

 \def\Pi      {\ensuremath{i}\xspace}

 \def\Pp      {\ensuremath{p}\xspace}

 \def\Ps      {\ensuremath{s}\xspace}

 \def\thebaroffset{0.18em}
}
\newcommand{\offsetoverline}[2][\thebaroffset]{\kern #1\overline{\kern -#1 #2}}%

\makeatletter
\ifcase \@ptsize \relax% 10pt
  \newcommand{\miniscule}{\@setfontsize\miniscule{4}{5}}% \tiny: 5/6
\or% 11pt
  \newcommand{\miniscule}{\@setfontsize\miniscule{5}{6}}% \tiny: 6/7
\or% 12pt
  \newcommand{\miniscule}{\@setfontsize\miniscule{5}{6}}% \tiny: 6/7
\fi
\makeatother

\DeclareRobustCommand{\optbar}[1]{\shortstack{{\miniscule (\rule[.5ex]{1.25em}{.18mm})}
  \\ [-.7ex] $#1$}}

\def\mup        {{\ensuremath{\Pmu^+}}\xspace}
\def\mun        {{\ensuremath{\Pmu^-}}\xspace} % muon negative (\mum is taken)

\def\mumu       {{\ensuremath{\Pmu^+\Pmu^-}}\xspace}

\def\squark    {{\ensuremath{\Ps}}\xspace}

\def\cquark    {{\ensuremath{\Pc}}\xspace}
\def\cquarkbar {{\ensuremath{\overline \cquark}}\xspace}
\def\ccbar     {{\ensuremath{\cquark\cquarkbar}}\xspace}
\def\bquark    {{\ensuremath{\Pb}}\xspace}
\def\bquarkbar {{\ensuremath{\overline \bquark}}\xspace}
\def\bbbar     {{\ensuremath{\bquark\bquarkbar}}\xspace}

\def\pion   {{\ensuremath{\Ppi}}\xspace}

\def\pim    {{\ensuremath{\pion^-}}\xspace}

\def\kaon    {{\ensuremath{\PK}}\xspace}
\def\Kbar    {{\ensuremath{\offsetoverline{\PK}}}\xspace}

\def\KorKbar {\kern \thebaroffset\optbar{\kern -\thebaroffset \PK}{}\xspace}

\def\Kp      {{\ensuremath{\kaon^+}}\xspace}
\def\Km      {{\ensuremath{\kaon^-}}\xspace}

\def\Kstarz  {{\ensuremath{\kaon^{*0}}}\xspace}
\def\Kstarzb {{\ensuremath{\Kbar{}^{*0}}}\xspace}
\def\Kstar   {{\ensuremath{\kaon^*}}\xspace}

\def\D       {{\ensuremath{\PD}}\xspace}

\def\DorDbar {\kern \thebaroffset\optbar{\kern -\thebaroffset \PD}\xspace}

\def\Dp      {{\ensuremath{\D^+}}\xspace}
\def\Dm      {{\ensuremath{\D^-}}\xspace}

\def\DpDm    {\ensuremath{\Dp {\kern -0.16em \Dm}}\xspace}

\def\B       {{\ensuremath{\PB}}\xspace}
\def\Bbar    {{\ensuremath{\offsetoverline{\PB}}}\xspace}

\def\BorBbar {\kern \thebaroffset\optbar{\kern -\thebaroffset \PB}\xspace}
\def\Bz      {{\ensuremath{\B^0}}\xspace}

\def\Bd      {{\ensuremath{\B^0}}\xspace}
\def\Bdb     {{\ensuremath{\Bbar{}^0}}\xspace}
\def\BdorBdbar {\kern \thebaroffset\optbar{\kern -\thebaroffset \Bd}\xspace}
\def\Bu      {{\ensuremath{\B^+}}\xspace}

\def\Bp      {{\ensuremath{\Bu}}\xspace}

\def\Bs      {{\ensuremath{\B^0_\squark}}\xspace}
\def\Bsb     {{\ensuremath{\Bbar{}^0_\squark}}\xspace}
\def\BsorBsbar {\kern \thebaroffset\optbar{\kern -\thebaroffset \Bs}\xspace}

\def\jpsi     {{\ensuremath{{\PJ\mskip -3mu/\mskip -2mu\Ppsi}}}\xspace}
\def\psitwos  {{\ensuremath{\Ppsi{(2S)}}}\xspace}

\def\Y#1S{\ensuremath{\PUpsilon{(#1S)}}\xspace}

\def\proton      {{\ensuremath{\Pp}}\xspace}

\def\Lz          {{\ensuremath{\PLambda}}\xspace}

\def\LorLbar     {\kern \thebaroffset\optbar{\kern -\thebaroffset \PLambda}\xspace}

\def\Lb           {{\ensuremath{\Lz^0_\bquark}}\xspace}

\newcommand{\decay}[2]{\ensuremath{#1\!\to #2}\xspace} 

\def\to                 {\ensuremath{\rightarrow}\xspace}

\def\qsq       {{\ensuremath{q^2}}\xspace}

\def\eps   {{\ensuremath{\varepsilon}}\xspace}

\def\CP                {{\ensuremath{C\!P}}\xspace}

\def\BdToKstmm    {\decay{\Bd}{\Kstarz\mup\mun}}

\def\BdToJPsiKst  {\decay{\Bd}{\jpsi\Kstarz}}

\def\bsll     {\decay{\bquark}{\squark \ell^+ \ell^-}}

\def\AT#1     {\ensuremath{A_{\mathrm{T}}^{#1}}\xspace}           % 2

\def\C#1      {\ensuremath{\mathcal{C}_{#1}}\xspace}                       % 9
\def\Cp#1     {\ensuremath{\mathcal{C}_{#1}^{'}}\xspace}                    % 7
\def\Ceff#1   {\ensuremath{\mathcal{C}_{#1}^{\mathrm{(eff)}}}\xspace}        % 9  
\def\Cpeff#1  {\ensuremath{\mathcal{C}_{#1}^{'\mathrm{(eff)}}}\xspace}       % 7
\def\Ope#1    {\ensuremath{\mathcal{O}_{#1}}\xspace}                       % 2
\def\Opep#1   {\ensuremath{\mathcal{O}_{#1}^{'}}\xspace}                    % 7

\newcommand{\nospaceunit}[1]{\ensuremath{\text{#1}}}       
\newcommand{\aunit}[1]{\ensuremath{\text{\,#1}}}       
\newcommand{\tev}{\aunit{Te\kern -0.1em V}\xspace}
\newcommand{\gev}{\aunit{Ge\kern -0.1em V}\xspace}
\newcommand{\mev}{\aunit{Me\kern -0.1em V}\xspace}
\newcommand{\kev}{\aunit{ke\kern -0.1em V}\xspace}
\newcommand{\ev}{\aunit{e\kern -0.1em V}\xspace}
 
\newcommand{\mevc}{\ensuremath{\aunit{Me\kern -0.1em V\!/}c}\xspace}
\newcommand{\gevc}{\ensuremath{\aunit{Ge\kern -0.1em V\!/}c}\xspace}
\newcommand{\mevcc}{\ensuremath{\aunit{Me\kern -0.1em V\!/}c^2}\xspace}
\newcommand{\gevcc}{\ensuremath{\aunit{Ge\kern -0.1em V\!/}c^2}\xspace}
\newcommand{\gevgevcccc}{\ensuremath{\gev^2\!/c^4}\xspace} % for q^2

\newcommand{\nospacegev}{\nospaceunit{Ge\kern -0.1em V}\xspace}
\def\fb   {\ensuremath{\aunit{fb}}\xspace}
\def\invfb   {\ensuremath{\fb^{-1}}\xspace}

\def\deriv {\ensuremath{\mathrm{d}}}

\def\gsim{{~\raise.15em\hbox{$>$}\kern-.85em
          \lower.35em\hbox{$\sim$}~}\xspace}
\def\lsim{{~\raise.15em\hbox{$<$}\kern-.85em
          \lower.35em\hbox{$\sim$}~}\xspace}

\newcommand{\Real}{\ensuremath{\mathcal{R}e}\xspace}

\def\PDF {PDF\xspace}

\def\mrad{\aunit{mrad}}

\def\tell1  {TELL1\xspace}
\def\ukl1   {UKL1\xspace}

\usepackage{cite} % Allows for ranges in citations
\usepackage{mciteplus}
\def\eos        {\mbox{\textsc{Eos}}\xspace}

\def\flavio     {\mbox{\textsc{Flavio}}\xspace}

\def\Mkpi  {\ensuremath{m(\Kp\pim)}\xspace}
\def\Mkpimm{\ensuremath{m(\Kp\pim\mumu)}\xspace}

\def\BdPsitwosKst  {\decay{\Bd}{\psitwos\Kstarz}}

\def\BzToKstphimm{\decay{\Bz}{\Pphi(1020)(\to\mumu)\Kstarz}}

\def\BsTophiKKmm  {\decay{\Bs}{\Pphi(1020)(\to\Kp\Km)\mumu}}
\def\LambdapKmm    {\decay{\Lb}{\proton\Km\mumu}}

\def\JPsimm  {\decay{\jpsi}{\mumu}}

\def\thetal {\ensuremath{\theta_{l}}\xspace}
\def\thetak {\ensuremath{\theta_{K}}\xspace}

\def\BuKMuMu     {\decay{\Bp}{ \Kp \mup\mun}}

\newcommand{\comment}[1]{}
\def\eos {{\tt EOS}}

\definecolor{darkred}{rgb}{0.6,0.0,0.0}
\definecolor{darkgreen}{rgb}{0.0,0.5,0.0}
\definecolor{lightgreen}{rgb}{0.75,1.0,0.75}
\definecolor{lightred}{rgb}{1.0,0.75,0.75}
\definecolor{lightblue}{rgb}{0.75,0.75,1.0}
\definecolor{darkblue}{RGB}{100,100,200}
\definecolor{verylightblue}{rgb}{0.9,0.9,1.0}
\definecolor{verylightred}{rgb}{1.0,0.9,0.9}
\definecolor{lightgray}{rgb}{0.9,0.9,0.9}
\definecolor{verylightgray}{rgb}{0.95,0.95,0.95}
\definecolor{darkgray}{rgb}{0.75,0.75,0.75}
\definecolor{orange}{rgb}{1.0,0.75,0.0}

\usepackage{longtable} % only for template; not usually to be used in PAPERs
\usepackage{multirow}
\usepackage{afterpage}

\begin{document}

%%%%%%%%%%%%%%%%%%%%%%%%%
%%%%% Title     %%%%%%%%%
%%%%%%%%%%%%%%%%%%%%%%%%%
\renewcommand{\thefootnote}{\fnsymbol{footnote}}
\setcounter{footnote}{1}

% %%%%%%% CHOOSE TITLE PAGE--------
%\onecolumn
%\input{title-LHCb-INT}
%\input{title-LHCb-ANA}
%\input{title-LHCb-CONF}
%\input{title-LHCb-FIGURE}
% $Id: title-LHCb-PAPER.tex 88824 2016-03-04 19:52:15Z clangenb $
% ===============================================================================
% Purpose: LHCb-PAPER journal paper title page template
% Author: 
% Created on: 2010-09-25
% ===============================================================================

%%%%%%%%%%%%%%%%%%%%%%%%%
%%%%%  TITLE PAGE  %%%%%%
%%%%%%%%%%%%%%%%%%%%%%%%%
\begin{titlepage}
\pagenumbering{roman}

% Header ---------------------------------------------------
\vspace*{-1.5cm}
\centerline{\large EUROPEAN ORGANIZATION FOR NUCLEAR RESEARCH (CERN)}
\vspace*{1.5cm}
\hspace*{-0.5cm}
\begin{tabular*}{\linewidth}{lc@{\extracolsep{\fill}}r}
\ifthenelse{\boolean{pdflatex}}% Logo format choice
{\vspace*{-2.7cm}\mbox{\!\!\!\includegraphics[width=.14\textwidth]{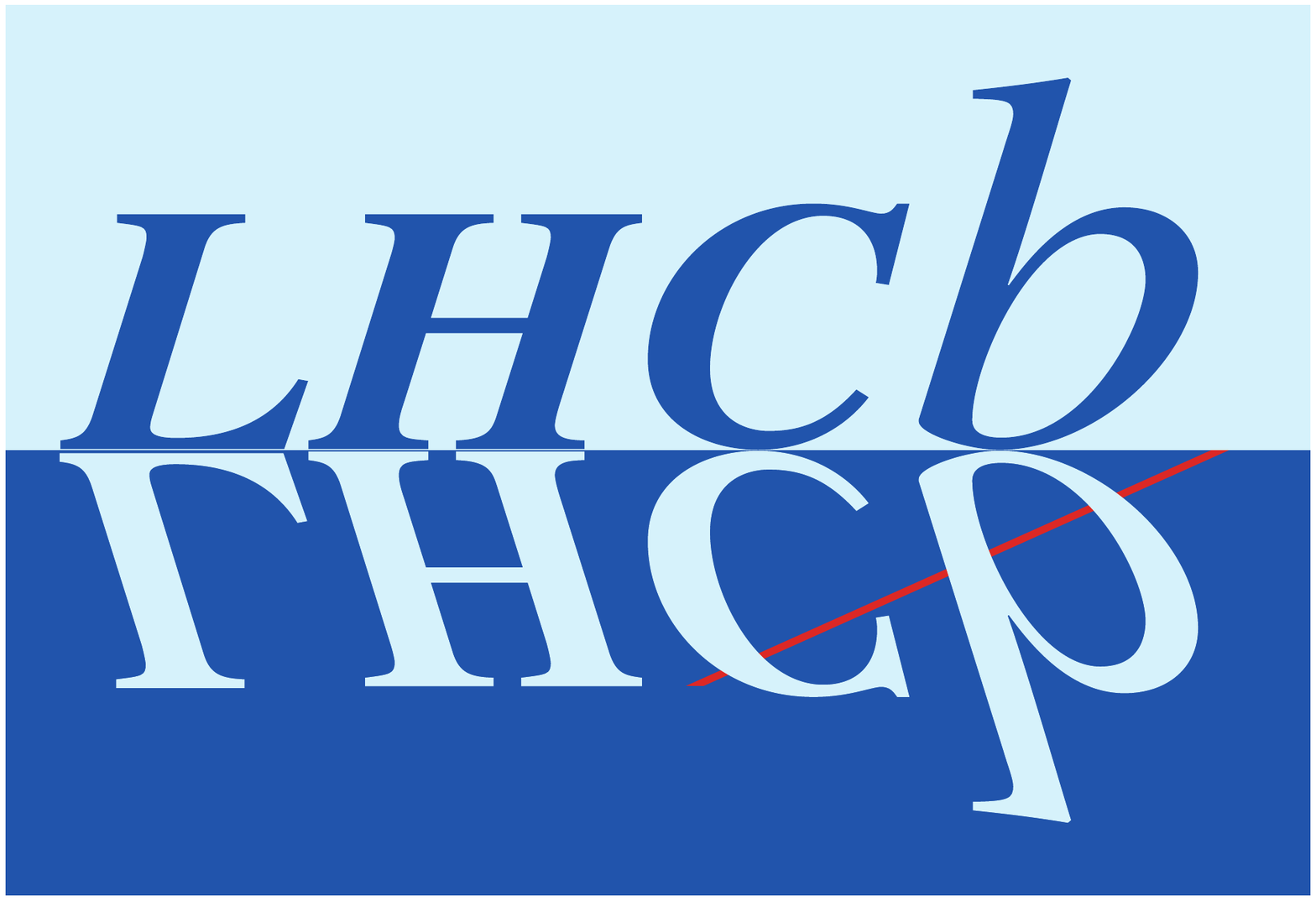}} & &}%
{\vspace*{-1.2cm}\mbox{\!\!\!\includegraphics[width=.12\textwidth]{lhcb-logo.eps}} & &}%
\\
 & & CERN-EP-2020-027\\  % ID 
 & & LHCb-PAPER-2020-002 \\  % ID 
 & & \today \\ 
 & & \\
\end{tabular*}

%\vspace*{3.0cm}
\vspace*{1.2cm}

{\normalfont\bfseries\boldmath\huge
\begin{center}
% DO NOT EDIT HERE. Instead edit macro in main.tex to keep metadata correct
  \papertitle 
\end{center}
}

\vspace*{2.0cm}

% Authors -------------------------------------------------
\begin{center}
%In the footnote, replace 'paper' by 'Letter' in case of submission to PRL or PLB 
% Edit macro in main.tex to keep metadata correct
\paperauthors\footnote{Authors are listed at the end of this paper.}
\end{center}

\vspace{\fill}

\begin{abstract}
  \noindent
  An angular analysis of the $B^{0}\rightarrow K^{*0}(\to\Kp\pim)\mu^{+}\mu^{-}$ decay is presented using a data set corresponding to an integrated luminosity of $4.7{\mbox{\,fb}^{-1}}$ of $pp$ collision data collected with the LHCb experiment. The full set of  $C\!P$-averaged observables are determined in bins of the invariant mass squared of the dimuon system. Contamination from decays with the $\Kp\pim$ system in an S-wave configuration is taken into account. The tension seen between the previous LHCb results and the Standard Model predictions persists with the new data. The precise value of the significance of this tension depends on the choice of theory nuisance parameters.

\end{abstract}

\vspace*{2.0cm}

\begin{center}
  Submitted to Phys. Rev. Lett.
\end{center}

\vspace{\fill}

{\footnotesize 
% Edit macro in main.tex to keep metadata correct
\centerline{\copyright~\papercopyright, license \href{\paperlicenceurl}{\paperlicence}.}}
\vspace*{2mm}

\end{titlepage}

\newpage
\setcounter{page}{2}
\mbox{~}

\cleardoublepage

%\twocolumn
% %%%%%%%%%%%%% ---------

\renewcommand{\thefootnote}{\arabic{footnote}}
\setcounter{footnote}{0}

%%%%%%%%%%%%%%%%%%%%%%%%%%%%%%%%
%%%%%  Table of Content   %%%%%%
%%%%%%%%%%%%%%%%%%%%%%%%%%%%%%%%
%%%% Uncomment if desired
%\tableofcontents
\cleardoublepage

%%%%%%%%%%%%%%%%%%%%%%%%%
%%%%% Main text %%%%%%%%%
%%%%%%%%%%%%%%%%%%%%%%%%%

\pagestyle{plain} % restore page numbers for the main text
\setcounter{page}{1}
\pagenumbering{arabic}

%% Uncomment during review phase. 
%% Comment before a final submission.
% \linenumbers
Decays mediated by the quark-level transition \bsll, where $\ell$ represents a lepton, have been the subject of intense recent study, as angular observables~\cite{LHCb-PAPER-2015-051, Sirunyan:2017dhj, Aaboud:2018krd, Wehle:2016yoi, Aubert:2006vb, Aaltonen:2011ja, LHCb-PAPER-2018-029, LHCb-PAPER-2015-023}, branching fractions~\cite{LHCb-PAPER-2016-012, LHCb-PAPER-2015-023, LHCb-PAPER-2015-009, LHCb-PAPER-2014-006} and ratios of branching fractions between decays with different flavours of leptons~\cite{LHCb-PAPER-2019-009, Abdesselam:2019lab, Lees:2012tva, LHCb-PAPER-2017-013,Abdesselam:2019wac} have been measured to be in tension with Standard Model (SM) predictions. 
Such decays are suppressed in the SM, as they proceed only through amplitudes that involve electroweak loop diagrams. The decays are sensitive to virtual contributions from new particles, which could have masses that are inaccessible to direct searches. 
The observed anomalies with respect to SM predictions can be explained consistently in New Physics models that introduce an additional vector or axial-vector contribution~\cite{Alguero:2019ptt,Aebischer:2019mlg,Arbey:2019duh,Ciuchini:2019usw,Kowalska:2019ley,Alok:2019ufo,Altmannshofer:2014cfa,Crivellin:2015mga,Celis:2015ara,Falkowski:2015zwa,Hiller:2014yaa,Gripaios:2014tna,Varzielas:2015iva,Barbieri:2016las, Sheng:2018vvm,Hiller:2018wbv, Crivellin:2017zlb,Sala:2017ihs, Ko:2017lzd}. 
However, there is still considerable debate about whether some of the observations might instead be explained by hadronic uncertainties associated with the transition form factors, or by other long-distance effects~\cite{Jager:2014rwa,Lyon:2014hpa,Ciuchini:2015qxb, Bobeth:2017vxj}. 

The LHCb collaboration presented a measurement of the angular observables of the  \decay{\Bz}{\Kstarz\mumu} decay in Ref.~\cite{LHCb-PAPER-2015-051} and found 
that the data could be explained by modifying the real part of the vector coupling strength of the decays, conventionally denoted $\Real(C_{9})$. The analysis used the nuisance parameters from Ref.~\cite{Beaujean:2013soa}, implemented in the \eos~software package described in Ref.~\cite{Bobeth:2010wg}, and found a 3.4 standard deviation ($\sigma$) tension with the SM value of $\Real(C_{9})$. The tension observed depends on the values of various SM nuisance parameters, including form-factor parameters and subleading corrections used to account for long-distance QCD interference effects with the charmonium modes. 
Using the \flavio~software package~\cite{Straub:2018kue}, with its default SM nuisance parameters, gives a tension of $3.0\,\sigma$ with respect to the SM value of $\Real(C_{9})$ when fitting the angular observables from Ref.~\cite{LHCb-PAPER-2015-051}. The nuisance parameters include a recent treatment of the subleading corrections~\cite{Straub:2015ica,Altmannshofer:2014rta} that was not available at the time of the previous analysis. 

This letter presents the most precise measurements of the complete set of \CP-averaged angular observables in the decay \BdToKstmm. The data set corresponds to an integrated luminosity of $4.7\invfb$ of \proton\proton collisions collected with the \lhcb experiment. The data were taken in the years 2011, 2012 and 2016, 
at centre-of-mass energies of $7$, $8$ and $13\tev$, respectively.
The analysis uses the same technique as the analysis described in Ref.~\cite{LHCb-PAPER-2015-051} but the data sample contains approximately twice as many \Bd decays, owing to the addition of the 2016 data. The \bbbar production cross-section increases by roughly a factor of two between the Run~1 and 2016 datasets~\cite{LHCB-PAPER-2017-037}. 
The same 2011 and 2012 (Run~1) data as in Ref.~\cite{LHCb-PAPER-2015-051} are used in the present analysis. The results presented in this letter supersede the previous LHCb publication.
The combination of the Run~1 data set with the 2016 data set requires a simultaneous angular fit to account for efficiency and reconstruction differences between years.
 Throughout this letter, \Kstarz is used to refer to the $\Kstar(892)^0$ resonance and the inclusion of charge-conjugate processes is implied. 
The \Kstarz meson is reconstructed through the decay $\Kstarz\to\Kp\pim$. 

The final state of the \BdToKstmm decay can be described by
the invariant mass squared of the dimuon system, \qsq, the invariant mass of the $\Kp\pim$ system, and the three decay angles,  $\vec{\Omega}=(\cos\thetal,\cos\thetak,\phi)$. 
The angle between the $\mup$ ($\mun$) and the direction opposite to that of the $\Bd$ ($\Bdb$) in the rest frame of the dimuon system is denoted $\thetal$.
The angle between the direction of the $\Kp$~($\Km$) and the $\Bd$~($\Bdb$) in the rest frame of the $\Kstarz$~($\Kstarzb$) system is denoted $\thetak$. 
The angle between the plane defined by the dimuon pair and the plane defined by the kaon and pion in the $\Bd$~($\Bdb$) rest frame is denoted $\phi$. 
A full description of the angular basis is provided in Ref.~\cite{LHCb-PAPER-2013-019}. 

Following the definitions given in Refs.~\cite{Altmannshofer:2008dz,LHCb-PAPER-2015-051}, the \CP-averaged angular distribution of the \BdToKstmm decay with the $\Kp\pim$ system in a P-wave configuration can be written as 
\begin{equation}
\begin{split}
\left.\frac{1}{\deriv(\Gamma+\bar{\Gamma})/\deriv q^2}\,\frac{\deriv^4(\Gamma+\bar{\Gamma})}{\deriv\qsq\,\deriv\vec{\Omega}}\right|_{\rm P} =
\frac{9}{32\pi} \Big[
 & \tfrac{3}{4} (1-{F_{\rm L}})\sin^2\thetak + {F_{\rm L}}\cos^2\thetak \\
\phantom{\Big[} +& \tfrac{1}{4}(1-{F_{\rm L}})\sin^2\thetak\cos 2\thetal\\
\phantom{\Big[}-& {F_{\rm L}} \cos^2\thetak\cos 2\thetal + {S_3}\sin^2\thetak \sin^2\thetal \cos 2\phi\\
\phantom{\Big[}+& {S_4} \sin 2\thetak \sin 2\thetal \cos\phi + {S_5}\sin 2\thetak \sin \thetal \cos \phi\\
\phantom{\Big[}+& \tfrac{4}{3} {A_{\rm FB}} \sin^2\thetak \cos\thetal + {S_7} \sin 2\thetak \sin\thetal \sin\phi\\
+& {S_8} \sin 2\thetak \sin 2\thetal \sin\phi + {S_9}\sin^2\thetak \sin^2\thetal \sin 2\phi  \Big]\,,
\end{split}
\label{eq:pdfpwave}
\end{equation}

\noindent where ${F_{\rm L}}$ is the fraction of the longitudinal polarisation of the \Kstarz meson, ${A_{\rm FB}}$ is the forward-backward asymmetry of the dimuon system and $S_i$ are other \CP-averaged observables~\cite{LHCb-PAPER-2015-051}. 
The $\Kp\pim$ system can also be in an S-wave configuration, which modifies the angular distribution to
\begin{equation}
\begin{split}
\left.\frac{1}{\deriv(\Gamma+\bar{\Gamma})/\deriv q^2}\,\frac{\deriv^4(\Gamma+\bar{\Gamma})}{\deriv\qsq\,\deriv\vec{\Omega}}\right|_{{\rm S}+{\rm P}} ~=~ &
(1-F_{\rm S}) \left.\frac{1}{\deriv(\Gamma+\bar{\Gamma})/\deriv q^2}\,\frac{\deriv^4(\Gamma+\bar{\Gamma})}{\deriv\qsq\,\deriv\vec{\Omega}}\right|_{\rm P}\\
& + \frac{3}{16\pi} F_{\rm S}\sin^{2}\theta_{l}  \\
& + \frac{9}{32\pi} ( S_{11}  + S_{13} \cos 2\theta_{l} ) \cos\theta_{K}  \\
& + \frac{9}{32\pi} ( S_{14} \sin 2\theta_{l} + S_{15} \sin\theta_{l} ) \sin\theta_{K}\cos\phi  \\
& + \frac{9}{32\pi} ( S_{16} \sin\theta_{l} + S_{17} \sin 2\theta_{l} ) \sin\theta_{K}\sin\phi \, ,
\end{split}
\label{eq:pdfswave}
\end{equation}
\noindent where $F_{\rm S}$ denotes the S-wave fraction and the coefficients $S_{11}$, $S_{13}$--$S_{17}$ arise from interference between the S- and P-wave amplitudes. 
Throughout this letter, $F_{\rm S}$ and the interference terms between the S- and P-wave are treated as nuisance parameters.

Additional sets of observables, for which the leading $\Bz \to \Kstarz$ form-factor uncertainties cancel, can be built from $F_{\rm L}$, $A_{\rm FB}$ and $S_{3}$--$S_{9}$. Examples of such {\it optimised} observables include the $P^{(\prime)}_{i}$ series of observables~\cite{DescotesGenon:2012zf}. The notation used in this letter again follows Ref.~\cite{LHCb-PAPER-2015-051}, for example $P_{5}' = S_{5}/\sqrt{F_{\rm L}(1-F_{\rm L})}$.

The LHCb detector is a single-arm forward spectrometer covering the pseudorapidity range $2 < \eta < 5$, described in detail in Refs.~\cite{Alves:2008zz,LHCb-DP-2014-002}. The detector includes a vertex detector surrounding the proton-proton interaction region, tracking stations on either side of a dipole magnet, ring-imaging Cherenkov (\rich) detectors, electromagnetic and hadronic calorimeters and muon chambers.

Simulated signal events are used in this analysis to determine the impact of the detector geometry, trigger, reconstruction and candidate selection on the angular distribution of the signal.
The simulation is produced using the software described in Refs.~\cite{Sjostrand:2006za,*Sjostrand:2007gs,LHCb-PROC-2010-056,Lange:2001uf,Golonka:2005pn,Allison:2006ve, *Agostinelli:2002hh,LHCb-PROC-2011-006}.
 Corrections derived from the data are applied to the simulation to account for mismodelling of the charge multiplicity of the event, \Bd momentum spectrum and \Bd vertex quality.
Similarly, the simulated particle identification (PID) performance is corrected to match that determined from control samples selected from the data~\cite{Anderlini:2202412, Aaij:2018vrk}.

The online event selection is performed by a trigger, which comprises a hardware stage, based on information from the calorimeter and muon
systems, followed by a software stage that applies a full event
reconstruction~\cite{LHCb-DP-2012-004}. 
Offline, signal candidates are formed from a pair of oppositely charged tracks that are identified as muons, combined with a \Kstarz candidate. 

% Simulated events are used to determine the acceptance function, which models the effect of the reconstruction and the selection on the angular and \qsq distributions. 
The distribution of the reconstructed \Kp\pim\mumu invariant mass, \Mkpimm, is used to discriminate signal from background. This distribution is fitted simultaneously with the three decay angles. 
The distribution of the reconstructed $\Kp\pim$ mass, \Mkpi, depends on the $\Kp\pim$ angular-momentum configuration and is used to constrain the S-wave fraction.
The analysis procedure is cross-checked by performing a fit of the $\bquark \to \ccbar \squark$ tree-level decay \BdToJPsiKst, with  \JPsimm, which results in the same final-state particles. Hereafter, the $\decay{\Bd}{\jpsi(\to\mumu)\Kstarz}$
decay and the equivalent decay via the \psitwos resonance are denoted by \BdToJPsiKst and \BdPsitwosKst, respectively.

Two types of backgrounds are considered: combinatorial background, where the selected particles do not originate from a single $\bquark$-hadron decay; and peaking backgrounds, where a single decay is selected but with some of the final-state particles misidentified. The combinatorial background is distributed smoothly in \Mkpimm, whereas the peaking backgrounds can accumulate in specific regions of the reconstructed mass.
In addition, the decays \BdToJPsiKst, \BdPsitwosKst and \BzToKstphimm are removed by rejecting events with \qsq in the ranges \mbox{$8.0 < \qsq < 11.0 \gevgevcccc$}, \mbox{$12.5 < \qsq < 15.0 \gevgevcccc$} or \mbox{$0.98 < \qsq < 1.10 \gevgevcccc$}. 

The criteria used to select candidates from the Run~1 data are the same as those described in Ref.~\cite{LHCb-PAPER-2015-051}. 
The selection of the 2016 data follows closely that of the Run~1 data. 
Candidates are required to have \mbox{$5170<\Mkpimm<5700\mevcc$} and \mbox{$795.9<\Mkpi<995.9\mevcc$}. 
The four tracks of the final-state particles are required to have significant impact parameter~(IP) with respect to all primary vertices (PVs) in the event. The tracks are fitted to a common vertex, which is required to be of good quality. The IP of the \Bz candidate is required to be small with respect to one of the PVs. The vertex of the \Bz candidate is required to be significantly displaced from the same PV. The angle between the reconstructed \Bz momentum and the vector connecting this PV to the reconstructed \Bz decay vertex, $\theta_{\rm DIRA}$, is also required to be small.  To avoid the same track being used to construct more than one of the final state particles, the opening angle between every pair of tracks is required to be larger than 1\mrad. 

Combinatorial background is reduced further using a boosted decision tree~(BDT) algorithm~\cite{Breiman,AdaBoost}. The BDT is trained entirely on data with \BdToJPsiKst candidates used as a proxy for the signal and candidates from the upper-mass sideband \mbox{$5350<\Mkpimm<7000\mevcc$} used as a proxy for the background. The training uses a cross-validation technique~\cite{Blum:1999:BHB:307400.307439} and is performed separately for the Run 1 and 2016 data sets. 
The input variables used are the reconstructed \Bz decay-time and vertex-fit quality, the momentum and transverse momentum of the \Bz candidate, $\theta_{\rm DIRA}$,  PID information from the \rich detectors and the muon system, and variables describing the isolation of the final-state tracks~\cite{LHCb-PAPER-2011-004}.
Variables are only used in the BDT if they do not have a strong correlation with the decay angles or $q^2$. 
A requirement is placed on the BDT output to maximise the signal significance. 
This requirement rejects more than $97\%$ of the remaining combinatorial background, while retaining more than $85\%$ of the signal. The signal efficiency of the BDT is uniform in the \Mkpimm and \Mkpi distributions. 

Peaking backgrounds from \BsTophiKKmm, \LambdapKmm, \mbox{\BdToJPsiKst}, \BdPsitwosKst and \decay{\Bdb}{\Kstarzb\mup\mun} decays are considered, 
where the latter constitutes a background if the kaon from the \Kstarzb decay is misidentified as the pion and vice versa.
In each case, at least one particle needs to be misidentified for the decay to be reconstructed as a signal candidate. 
Vetoes to reduce these peaking backgrounds are formed by placing requirements on the invariant mass of the candidates, recomputed with the relevant change in the particle mass hypotheses, and by using PID information. In addition, in order to avoid having a strongly peaking contribution to the $\cos\thetak$ angular distribution in the upper mass sideband, \BuKMuMu candidates with  $\Kp\mumu$ invariant mass within $60\mevcc$ of the \Bp mass are removed. 
The background from \bquark-hadron decays with two hadrons misidentified as muons is negligible. 
 The signal efficiency and residual peaking backgrounds are estimated using simulated events. The vetoes remove a negligible amount of signal. 
 The largest residual backgrounds are from \mbox{\decay{\Bdb}{\Kstarzb\mup\mun}}, \mbox{\decay{\Lb}{p\Km\mumu}} and \mbox{\BsTophiKKmm} decays, at the level of 1\% or less of the expected signal yield.
 This is sufficiently small such that these backgrounds are neglected in the angular analysis and are considered only as sources of systematic uncertainty.

For every $q^2$ bin, a fit is performed in both the standard and the optimised basis.
For each basis, four data sets are fit simultaneously: the \Mkpimm and angular distributions of candidates in the Run~1 data; the equivalent distributions for the 2016 data;  and the \Mkpi distributions of candidates in the Run~1 and the 2016 data sets. 
The signal fraction is shared between the two data sets from each data-taking period. The \CP-averaged angular observables and the S-wave fraction are shared between all data sets.
% A simultaneous fit to the Run~1 data and the 2016 data is performed, with the angular observables as common fit parameters. 
% For each data set, an unbinned maximum-likelihood fit to the distributions of \Mkpimm
% and the three decay angles is used to determine the \CP-averaged angular observables in bins of \qsq, in either the standard or optimised basis; and 
% a simultaneous fit of the \Mkpi invariant mass distribution is used to constrain the S-wave fraction. 
The fitted probability density functions ({\PDF}s) are of an identical form to those of Ref.~\cite{LHCb-PAPER-2015-051}, as are the \qsq bins used.
In addition to the narrow 
$\qsq$ bins, results are obtained for the wider bins \mbox{$1.1 < \qsq < 6.0\gevgevcccc$} and \mbox{$15.0 < \qsq < 19.0\gevgevcccc$}.

The angular distribution of the signal is described using Eq.~(\ref{eq:pdfpwave}). The $P_i^{(\prime)}$ observables are determined by reparameterising
Eq.~(\ref{eq:pdfpwave}) using a basis comprising $F^{}_{\rm L}$, $P^{}_{1,2,3}$ and $P_{4,5,6,8}^{\prime}$.
The angular distribution is multiplied by an acceptance model used to account for the effect of the reconstruction and candidate selection. 
The acceptance function is parameterised in four dimensions, according to
\begin{align}
  \eps(\cos\thetal,\cos\thetak,\phi,q^2) &= \sum_{ijmn} c_{ijmn} L_i(\cos\thetal) L_j(\cos\thetak) L_m(\phi) L_n(q^2)\,,
\label{eq:acceptance}
\end{align}
where the terms $L_h(x)$ denote Legendre polynomials of order $h$ and the values of the angles and \qsq are rescaled to the range $-1 < x < +1$ when evaluating the polynomials.  
For the $\cos\thetal$, $\cos\thetak$ and $\phi$ angles, the sum in Eq.~(\ref{eq:acceptance}) encompasses $L_h(x)$ up to fourth, fifth and sixth order, respectively.
The $q^2$ parameterisation comprises $L_h(x)$ up to fifth order. Simulation  indicates that the acceptance function can be assumed to be flat across \Mkpi.
The coefficients $c_{ijmn}$ are determined using a principal moment analysis of simulated \BdToKstmm decays.
As all of the relevant kinematic variables needed to describe the decay are used in this parameterisation, the acceptance function does not depend on the decay model used in the simulation.

In the narrow \qsq bins, the acceptance is taken to be constant across each bin and is included in the fit by multiplying Eq.~(\ref{eq:pdfswave}) by the acceptance function evaluated with the value of \qsq fixed at the bin centre.  
In the wider \qsq bins, the shape of the acceptance can vary significantly across the bin. In the likelihood, candidates are therefore weighted by the inverse of the acceptance function   
and parameter uncertainties are obtained using a bootstrapping technique~\cite{Efron:1979}. 

The background angular distribution is modelled with second-order polynomials in $\cos\thetal$, $\cos\thetak$ and $\phi$, with the angular coefficients allowed to vary in the fit. 
This angular distribution is assumed to factorise in the three decay angles, which is confirmed to be the case for candidates in the upper mass sideband of the data.

The \Mkpimm distribution of the signal candidates is modelled using the sum of two Gaussian functions with a common mean, each with a power-law tail on the low mass side.
The parameters describing the signal mass shape are determined from a fit to the \mbox{\BdToJPsiKst} decay in the data and are subsequently fixed when fitting the \mbox{\BdToKstmm} candidates.
For each of the \qsq~bins, a scale factor that is determined from simulation is included to account for the difference in resolution between the \mbox{\BdToJPsiKst} and \mbox{\BdToKstmm} decay modes.
A component is included in the \mbox{\BdToJPsiKst} fit to account for \mbox{\decay{\Bsb}{\jpsi\Kstarz}} decays, which are at the level of $\sim 1 \%$ of the \mbox{\BdToJPsiKst} signal yield.
The background from the equivalent Cabibbo-suppressed penguin decay, \mbox{\decay{\Bsb}{\Kstarz\mumu}}~\cite{LHCb-PAPER-2018-004}, is negligible and is ignored in the fit of the signal decay.
The combinatorial background is described well by an exponential distribution in \Mkpimm. 

The $\Kstarz$ signal component in the \Mkpi distribution is modelled using a relativistic Breit-Wigner function for the P-wave component and the LASS parameterisation~\cite{Aston:1987ir} for the S-wave component. 
The combinatorial background is described by a linear function in \Mkpi. 

The decay \BdToJPsiKst is used to cross-check the analysis procedure in the region \mbox{$8.0<\qsq <11.0 \gevgevcccc$}.
This decay is selected in the data with negligible background contamination. The angular structure has been determined by measurements made by the \babar, \belle and \lhcb collaborations~\cite{Aubert:2007hz, Chilikin:2014bkk, Aaij:2013cma}. 
The \BdToJPsiKst angular observables obtained from the Run~1 and 2016 LHCb data, using the acceptance correction derived as described above, are in good agreement with these previous measurements. 

\begin{figure}
  \centering
  \includegraphics[width=0.48\linewidth]{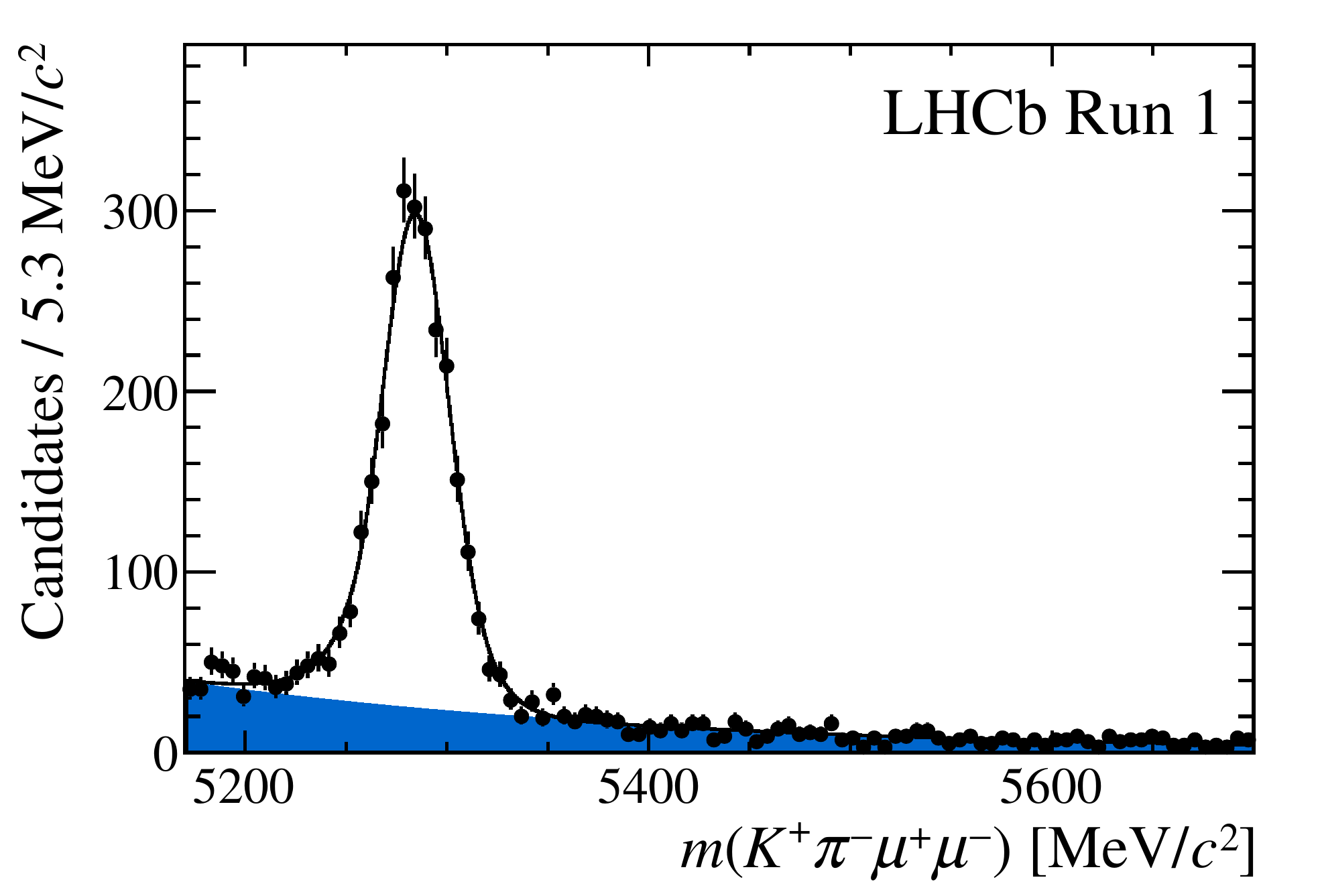}
  \includegraphics[width=0.48\linewidth]{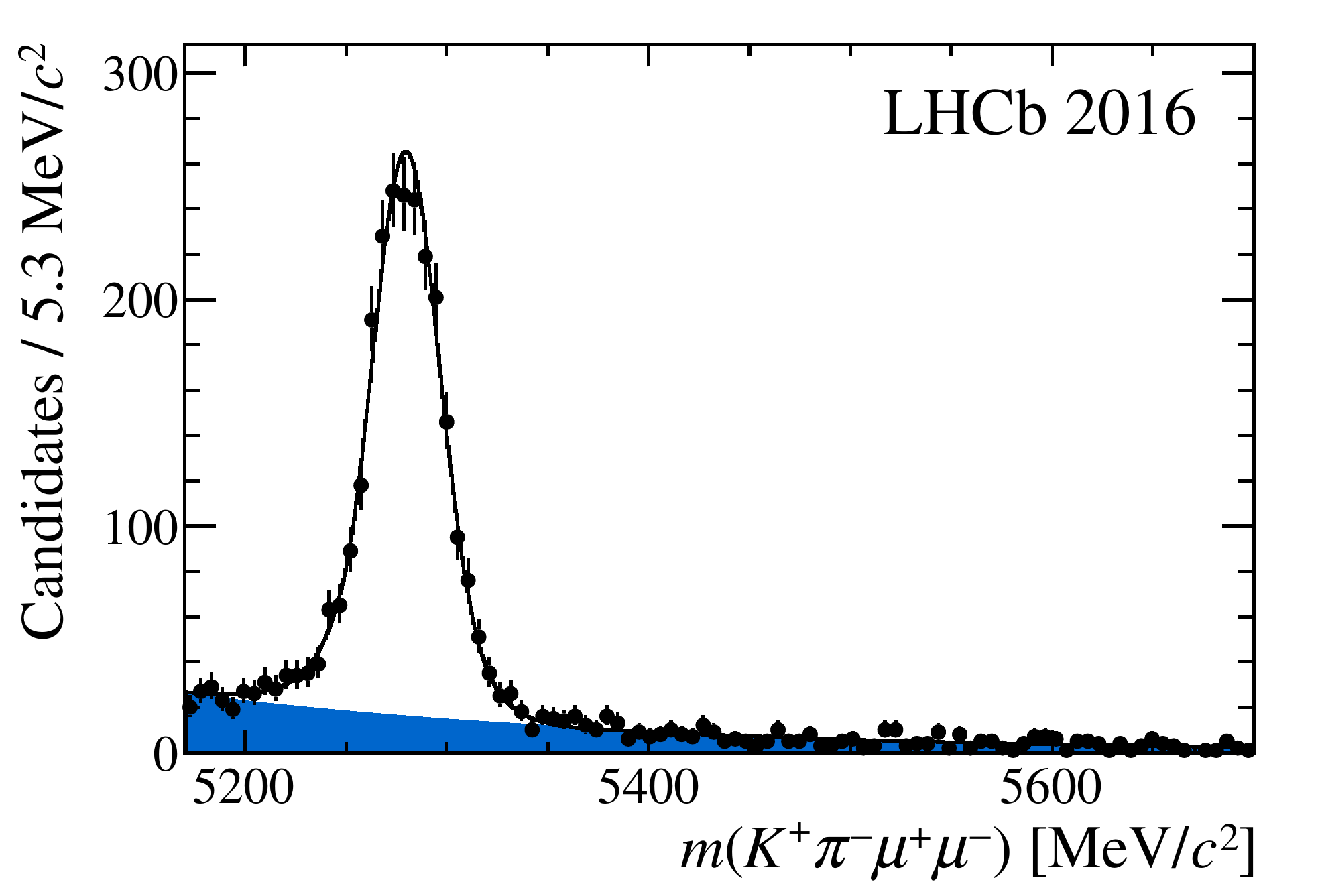}
  \caption{The $\Kp\pim\mumu$ mass  distribution of candidates with $0.1<\qsq<19.0\gev^{2}/c^{4}$, excluding the $\phi(1020)$ and charmonium regions, for the (left) Run~1 data and (right) 2016 data. 
   The background is indicated by the shaded region. 
\label{fig:massfit}}
\end{figure}

Figure~\ref{fig:massfit} shows the projection of the fitted \PDF on the \Kp\pim\mumu mass distribution.
The \BdToKstmm yield, integrated over the $q^2$ ranges \mbox{$0.10 < \qsq < 0.98\gevgevcccc$}, \mbox{$1.1 < \qsq < 8.0\gevgevcccc$}, \mbox{$11.0 < \qsq < 12.5\gevgevcccc$} and \mbox{$15.0 < \qsq < 19.0\gevgevcccc$}, is determined to be $2398 \pm 57$ for the Run~1 data, and $2187 \pm 53$ for the 2016 data.

Pseudoexperiments, generated using the results of the best fit to data, are used to assess the bias and coverage of the fit. 
The majority of observables have a bias of less than 10\% of their statistical uncertainty, with the largest bias being 17\%, and all observables have an uncertainty estimate within 10\% of the true uncertainty. 
The biases are driven by boundary effects in the observables. 
The largest effect comes from requiring that $F_{\rm S}\geq 0$, which can bias $F_{\rm S}$ to larger values. This can then result in a bias in the P-wave observables (see Eq.~\ref{eq:pdfswave}).
 The statistical uncertainty is corrected to account for any under- or over-coverage and a  systematic uncertainty equal to the size of the observed bias is assigned.

The size of other sources of systematic uncertainty is estimated using pseudoexperiments, in which one or more parameters are varied and the angular observables are
determined with and without this variation. The systematic uncertainty is then taken as the difference between the two models. 
The pseudoexperiments are generated with signal yields many times larger than the data, in order to render statistical fluctuations negligible.

The size of the total systematic uncertainty varies depending on the angular observable and the \qsq bin. The majority of observables in both the $S_{i}$ and $P^{(\prime)}_{i}$ basis have a total systematic uncertainty between 5\% and 25\% of the statistical uncertainty. For $F_{\rm L}$, the systematic uncertainty tends to be larger, typically between 20\% and 50\%. The systematic uncertainties are given in Table~\ref{tab:systematics} of the Supplemental Material. 

The dominant systematic uncertainties arise from the peaking backgrounds that are neglected in the analysis, the bias correction, and, for the narrow \qsq bins, from
the uncertainty associated with evaluating the acceptance at a fixed point in \qsq. 
For the peaking  backgrounds,  the  systematic  uncertainty  is  evaluated  by  injecting additional candidates, drawn from the angular distributions of the background modes,  into the pseudoexperiment data.
 The systematic uncertainty for the bias correction is determined directly from the pseudoexperiments used to validate the fit. The systematic uncertainty from the  variation of the acceptance with \qsq is determined by moving the point in \qsq at which the acceptance is evaluated to halfway between the bin centre and the upper or the lower edge. The largest deviation is taken as the systematic uncertainty.
Examples of further sources of systematic uncertainty investigated include the \Mkpi lineshape for the S-wave contribution, the assumption that the acceptance function is flat across the $m(\Kp\pim)$ mass, the effect of the \BuKMuMu veto on the angular distribution of the background and the order of polynomial used for the background parameterisation. 
These sources make a negligible contribution to the total uncertainty.
With respect to the analysis of Ref.~\cite{LHCb-PAPER-2015-051}, 
the systematic uncertainty from residual differences between data and simulation is significantly reduced, owing to an improved decay model for \BdToJPsiKst decays~\cite{Chilikin:2014bkk}.

\begin{figure}[!htb]
    \centering
    \includegraphics[width=0.48\linewidth]{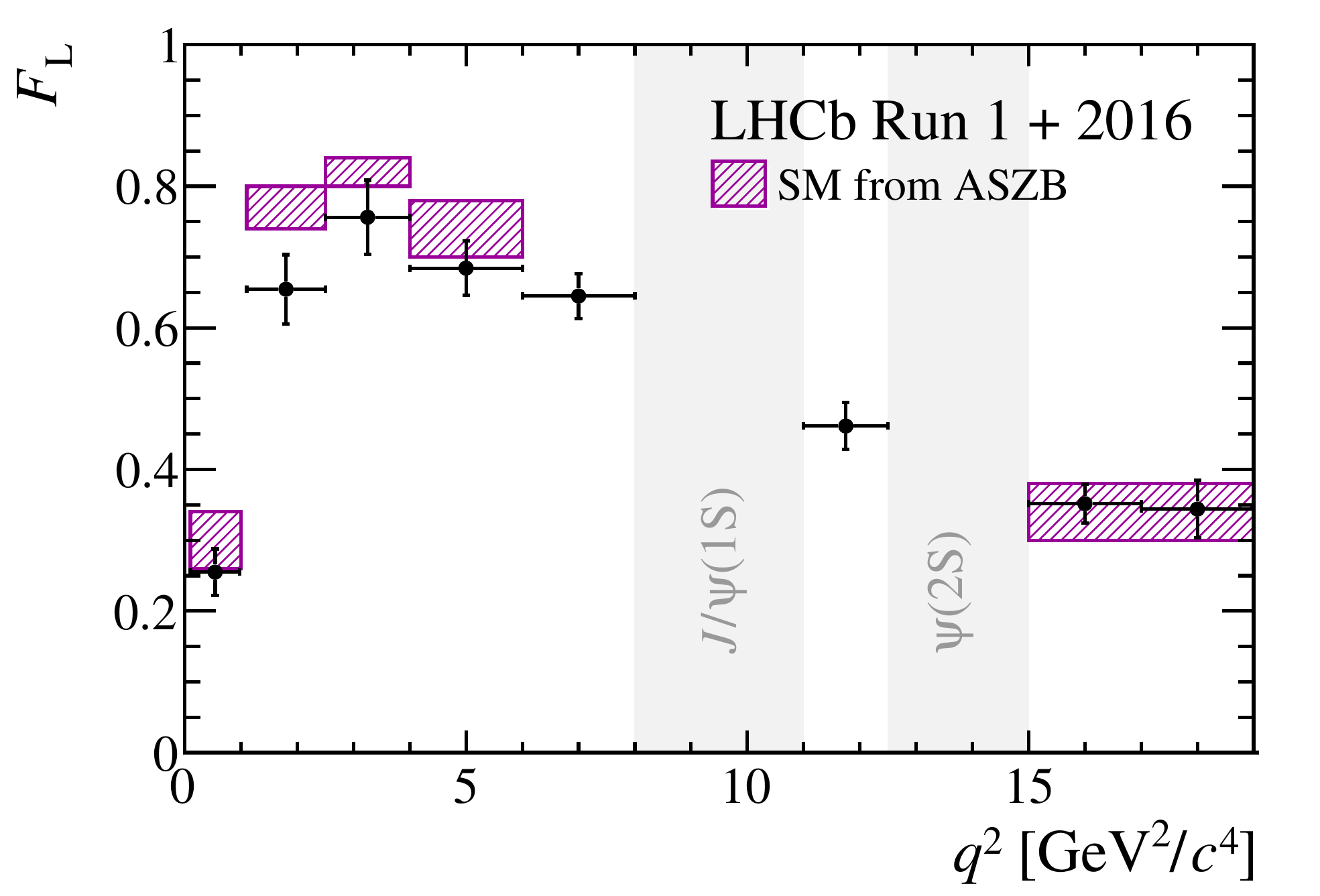} 
    \includegraphics[width=0.48\linewidth]{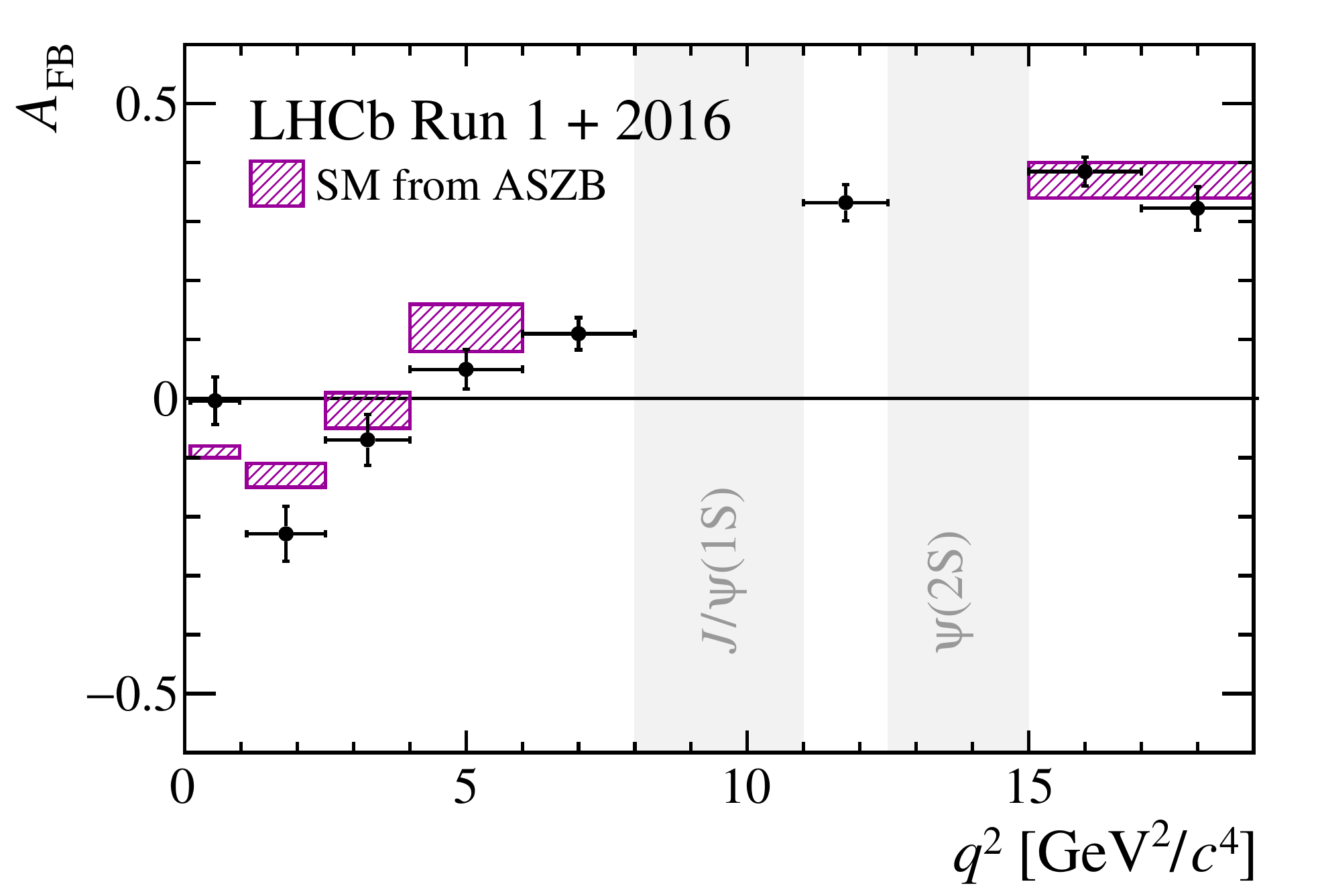} \\
    \includegraphics[width=0.48\linewidth]{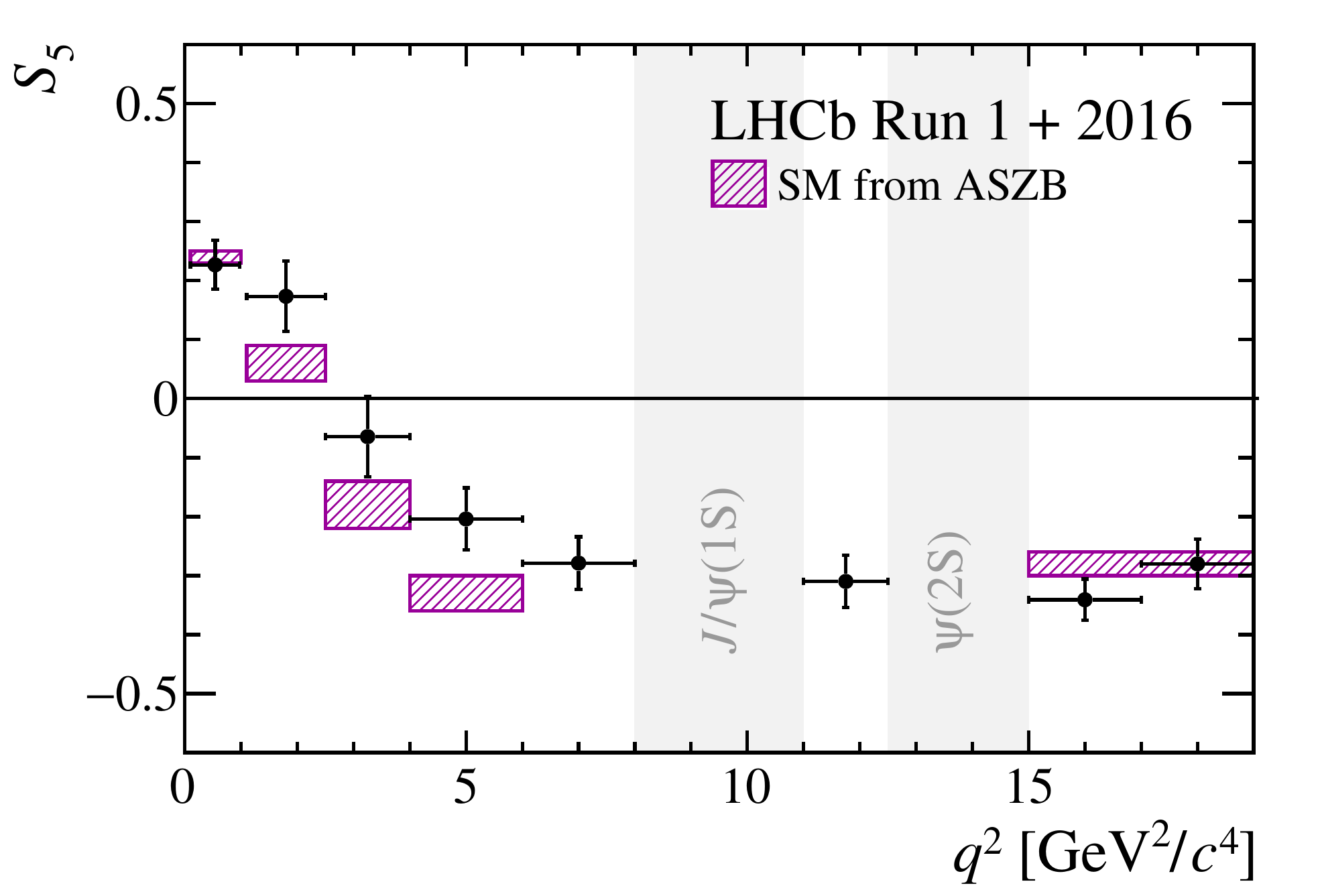} 
    \includegraphics[width=0.48\linewidth]{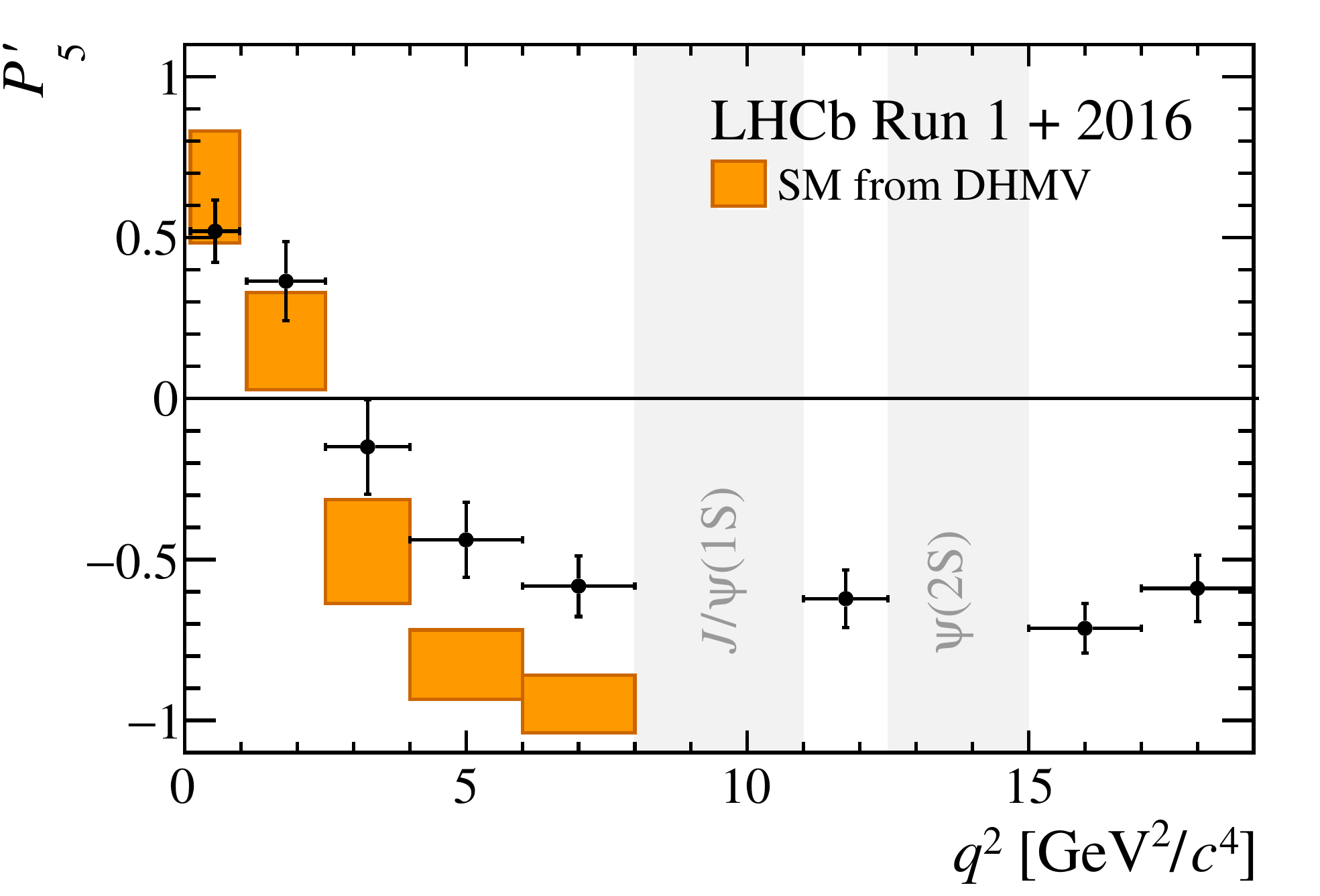} \\
    \caption{
    Results for the \CP-averaged angular observables $F_{\rm L}$, $A_{\rm FB}$, $S_{5}$ and $P_{5}'$ in bins of \qsq.
    The data are compared to SM predictions based on the prescription of Refs.~\cite{Altmannshofer:2014rta, Straub:2015ica}, with the exception of the $P_5'$ distribution, which is compared to SM predictions based on Refs.~\cite{Descotes-Genon:2014uoa, Khodjamirian:2010vf}.}
    \label{fig:results:angobs}
\end{figure}

The \CP-averaged observables $F_{\rm L}$, $A_{\rm FB}$, $S_{5}$ and $P'_5$ that are obtained from the $S_{i}$ and $P_{i}^{(\prime)}$ fits are shown together with their respective SM predictions in Fig.~\ref{fig:results:angobs}. The results for all observables are given in Figs.~\ref{fig:sub:Si} and~\ref{fig:sub:Pi} and Tables~\ref{tab:sub:siresults} and~\ref{tab:sub:piresults} of the Supplemental Material. In addition, the statistical correlation between the observables is provided in Tables~\ref{appendix:likelihood:correlation:average:1}--\ref{appendix:likelihood:correlation:optimised:10}.
The SM predictions are based on the prescription of Ref.~\cite{Altmannshofer:2014rta}, which combines light-cone sum rule calculations~\cite{Straub:2015ica}, valid in the low-\qsq region, with lattice determinations at high \qsq~\cite{Horgan:2013hoa,Horgan:2015vla} to yield more precise determinations of the form factors over the full \qsq range.
For the $P^{(\prime)}_{i}$ observables, predictions from Ref.~\cite{Descotes-Genon:2014uoa} are shown using form factors from Ref.~\cite{Khodjamirian:2010vf}. These predictions are restricted to the region \mbox{$\qsq < 8.0\gevgevcccc$}.
The results from Run~1 and the 2016 data are in excellent agreement. 
A stand-alone fit to the Run~1 data  reproduces exactly the central values of the observables obtained in Ref.~\cite{LHCb-PAPER-2015-051}.

Considering the observables individually, the results are largely in agreement with the SM predictions. 
The local discrepancy in the $P'_5$ observable in the \mbox{$4.0<q^2<6.0\gevgevcccc$} and \mbox{$6.0<q^2<8.0\gevgevcccc$} bins reduces from the 2.8 and $3.0\,\sigma$ observed in Ref.~\cite{LHCb-PAPER-2015-051} to 2.5 and $2.9\,\sigma$. However, as discussed below, the overall tension with the SM is observed to increase mildly.

Using the \flavio~software package~\cite{Straub:2018kue}, a fit of the angular observables is performed varying the parameter $\Real(C_{9})$. 
The default \flavio SM nuisance parameters are used, 
including form-factor parameters and subleading corrections to account for long-distance QCD interference effects with the charmonium decay modes~\cite{Straub:2015ica,Altmannshofer:2014rta}. 
The same \qsq bins as in Ref.~\cite{LHCb-PAPER-2015-051} are included. The $3.0\,\sigma$ discrepancy with respect to the SM value of $\Real(C_{9})$ obtained with the Ref.~\cite{LHCb-PAPER-2015-051} data set changes to $3.3\,\sigma$ with the data set used here. 
The best fit to the angular distribution is obtained with a shift in the SM value of $\Real(C_{9})$ by $-0.99^{+0.25}_{-0.21}$. 
The tension observed in any such fit will depend on the effective coupling(s) varied, the handling of the SM nuisance parameters and the \qsq bins that are included in the fit. For example, the \mbox{$6.0 < \qsq < 8.0\gevgevcccc$} bin is known to be associated with larger theoretical uncertainties~\cite{Altmannshofer:2008dz}. Neglecting this bin, a \flavio fit gives a tension of $2.4\,\sigma$ using the observables from Ref.~\cite{LHCb-PAPER-2015-051} and $2.7\,\sigma$ tension with the measurements reported here.

In summary, using $4.7{\mbox{\,fb}^{-1}}$ of $pp$ collision data collected with the LHCb
  experiment during the years 2011, 2012 and 2016, a complete set of \CP-averaged angular observables has been measured for the \BdToKstmm decay. These are the most precise measurements of these quantities to date.

% Do not include this in analysis note and conference reports
\section*{Acknowledgements}

\noindent We express our gratitude to our colleagues in the CERN
accelerator departments for the excellent performance of the LHC. We
thank the technical and administrative staff at the LHCb
institutes.
We acknowledge support from CERN and from the national agencies:
CAPES, CNPq, FAPERJ and FINEP (Brazil); 
MOST and NSFC (China); 
CNRS/IN2P3 (France); 
BMBF, DFG and MPG (Germany); 
INFN (Italy); 
NWO (Netherlands); 
MNiSW and NCN (Poland); 
MEN/IFA (Romania); 
MSHE (Russia); 
MinECo (Spain); 
SNSF and SER (Switzerland); 
NASU (Ukraine); 
STFC (United Kingdom); 
DOE NP and NSF (USA).
We acknowledge the computing resources that are provided by CERN, IN2P3
(France), KIT and DESY (Germany), INFN (Italy), SURF (Netherlands),
PIC (Spain), GridPP (United Kingdom), RRCKI and Yandex
LLC (Russia), CSCS (Switzerland), IFIN-HH (Romania), CBPF (Brazil),
PL-GRID (Poland) and OSC (USA).
We are indebted to the communities behind the multiple open-source
software packages on which we depend.
Individual groups or members have received support from
AvH Foundation (Germany);
EPLANET, Marie Sk\l{}odowska-Curie Actions and ERC (European Union);
ANR, Labex P2IO and OCEVU, and R\'{e}gion Auvergne-Rh\^{o}ne-Alpes (France);
Key Research Program of Frontier Sciences of CAS, CAS PIFI, and the Thousand Talents Program
 (China);
RFBR, RSF and Yandex LLC (Russia);
GVA, XuntaGal and GENCAT (Spain);
the Royal Society
and the Leverhulme Trust (United Kingdom).

\clearpage
\pagenumbering{roman}
\setcounter{section}{0}

{\noindent\bf\Large Supplemental Material} \\

\noindent This supplemental material includes additional information to that already provided in the main letter. A full set of results for the nominal analysis is presented in both graphical and tabular form in Sec.~\ref{sec:supp:tables}. A complete description of the corresponding systematic uncertainties is given in Sec.~\ref{sec:supp:sys}. The correlations between the angular observables are presented for the $S_{i}$ observables in Sec.~\ref{sec:appendix:likelihood:correlation} and for the $P^{(\prime)}_{i}$ observables in Sec.~\ref{sec:appendix:correlation:optimised}. 
The angular and mass distributions of the selected candidates in the different \qsq bins are shown in Sec.~\ref{sec:sub:signalchannelprojections}.

\section{Results}
\label{sec:supp:tables}

\begin{figure}[!htb]
    \centering
    \includegraphics[width=0.48\linewidth]{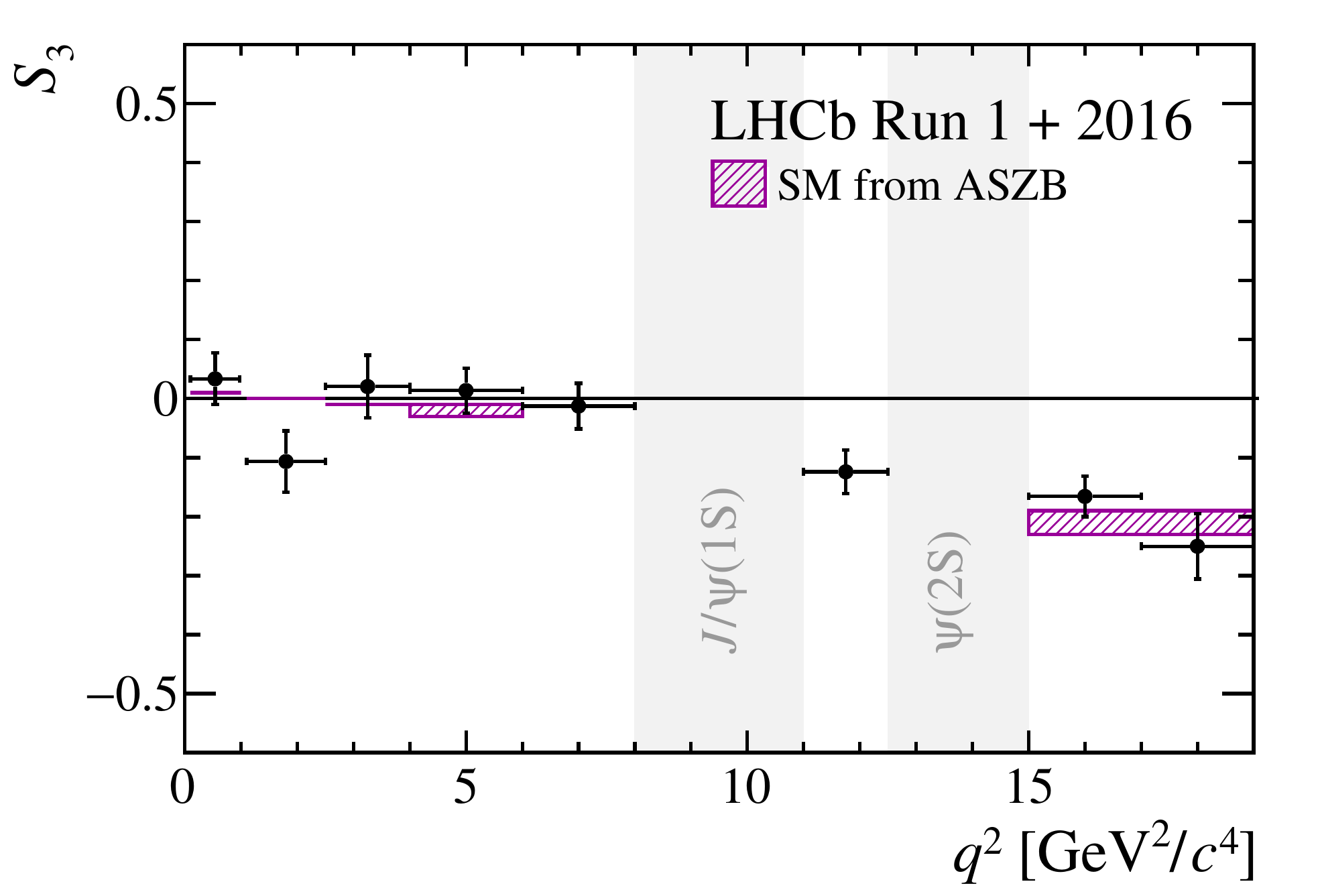} 
    \includegraphics[width=0.48\linewidth]{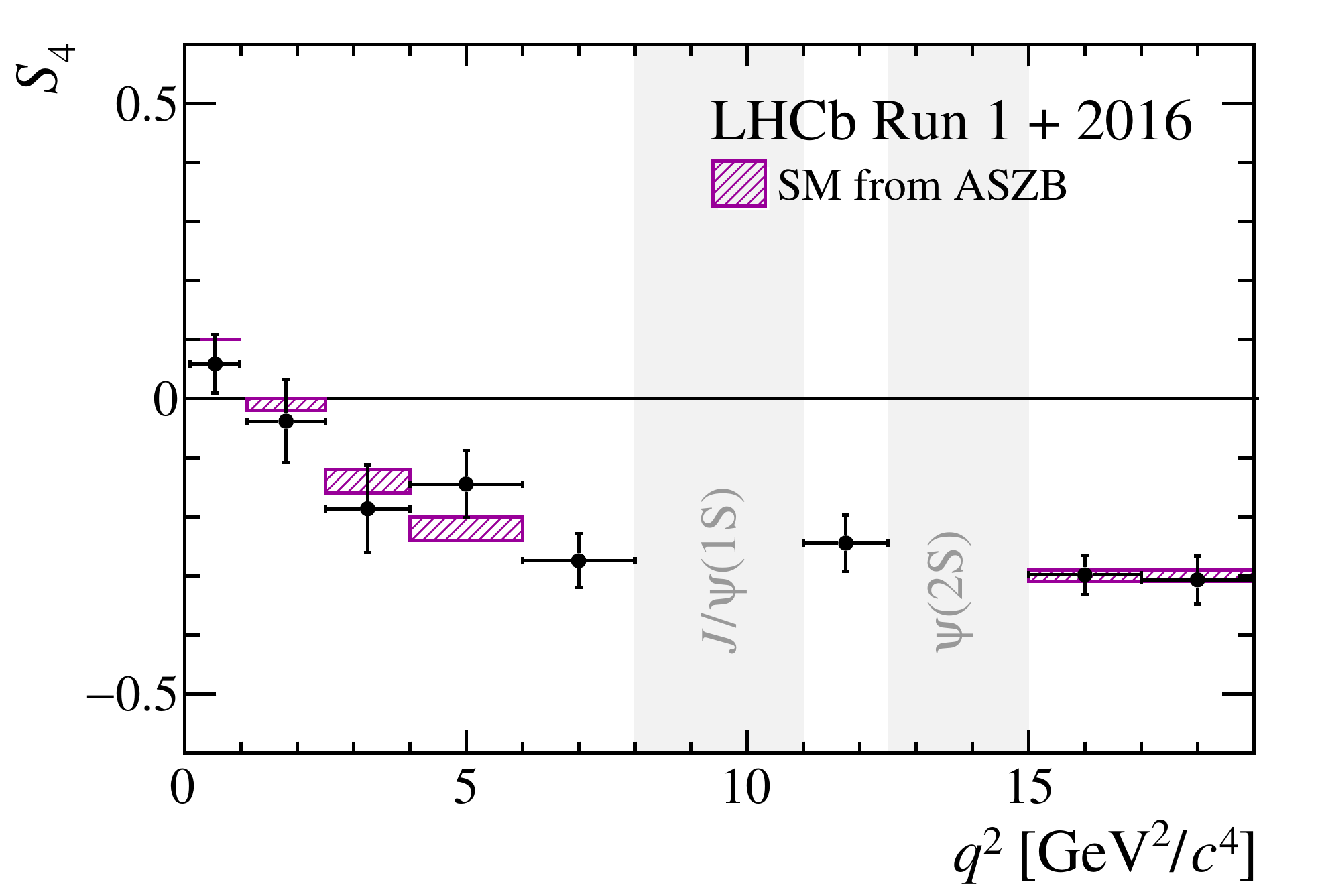} \\
    \includegraphics[width=0.48\linewidth]{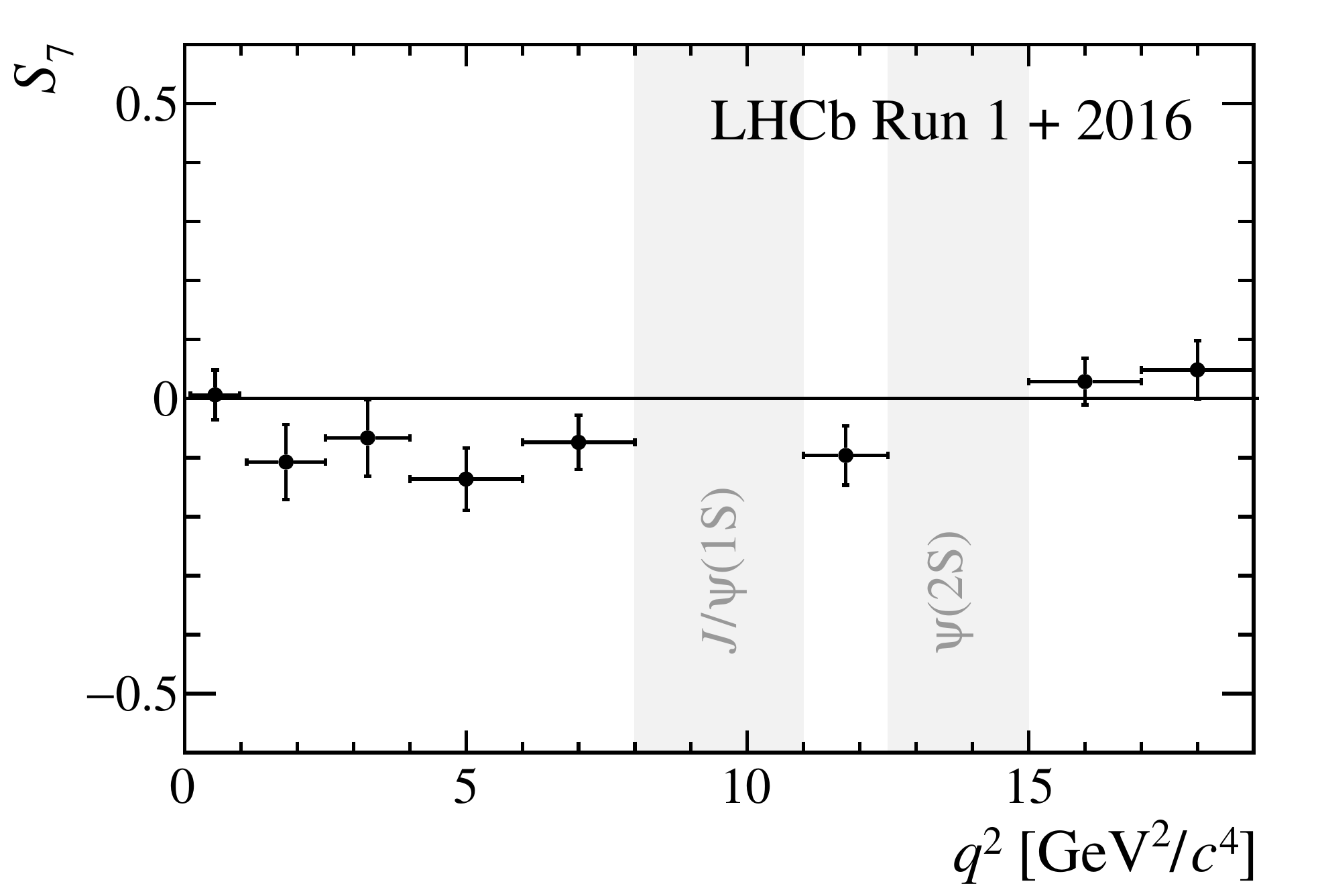} 
    \includegraphics[width=0.48\linewidth]{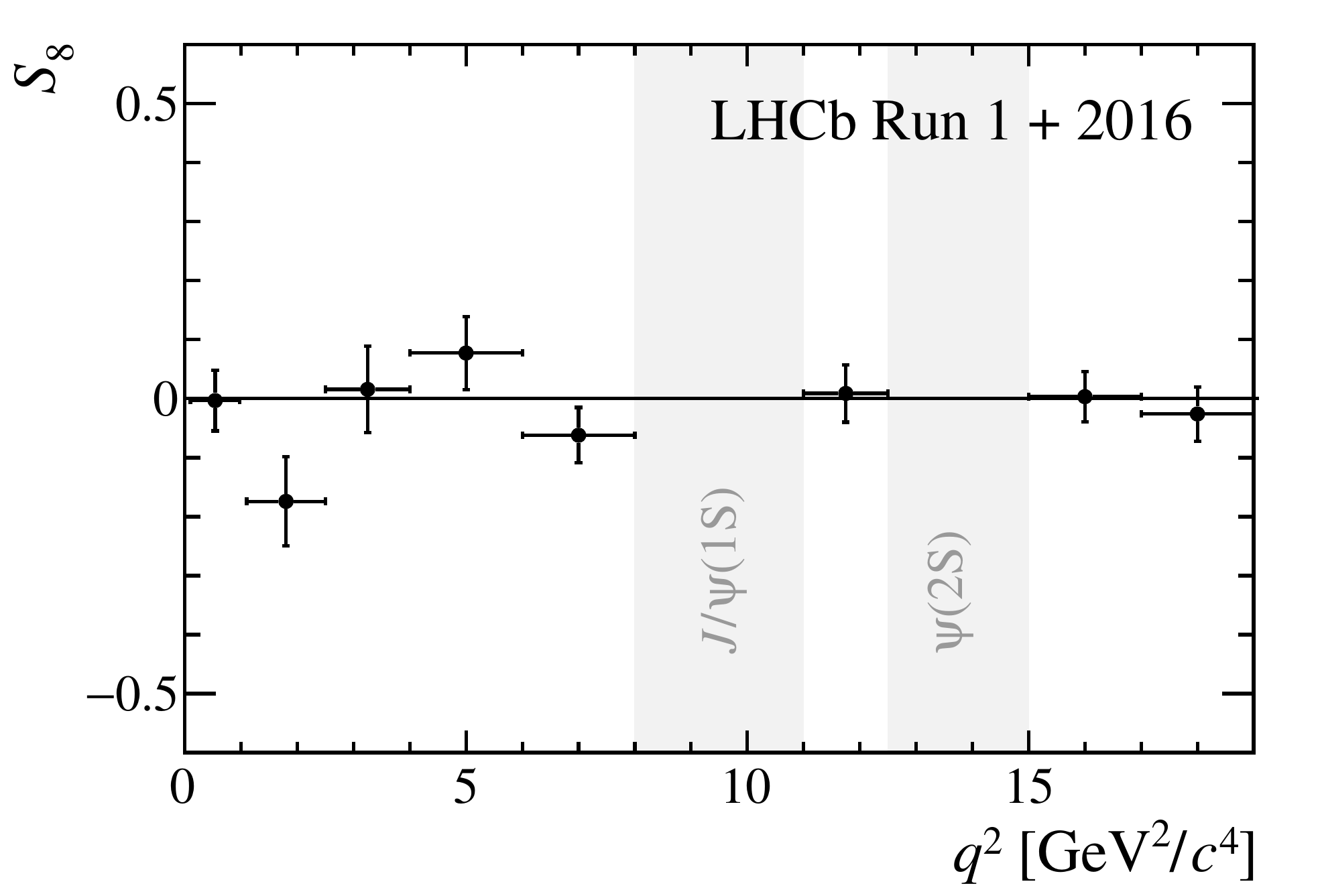} \\
    \includegraphics[width=0.48\linewidth]{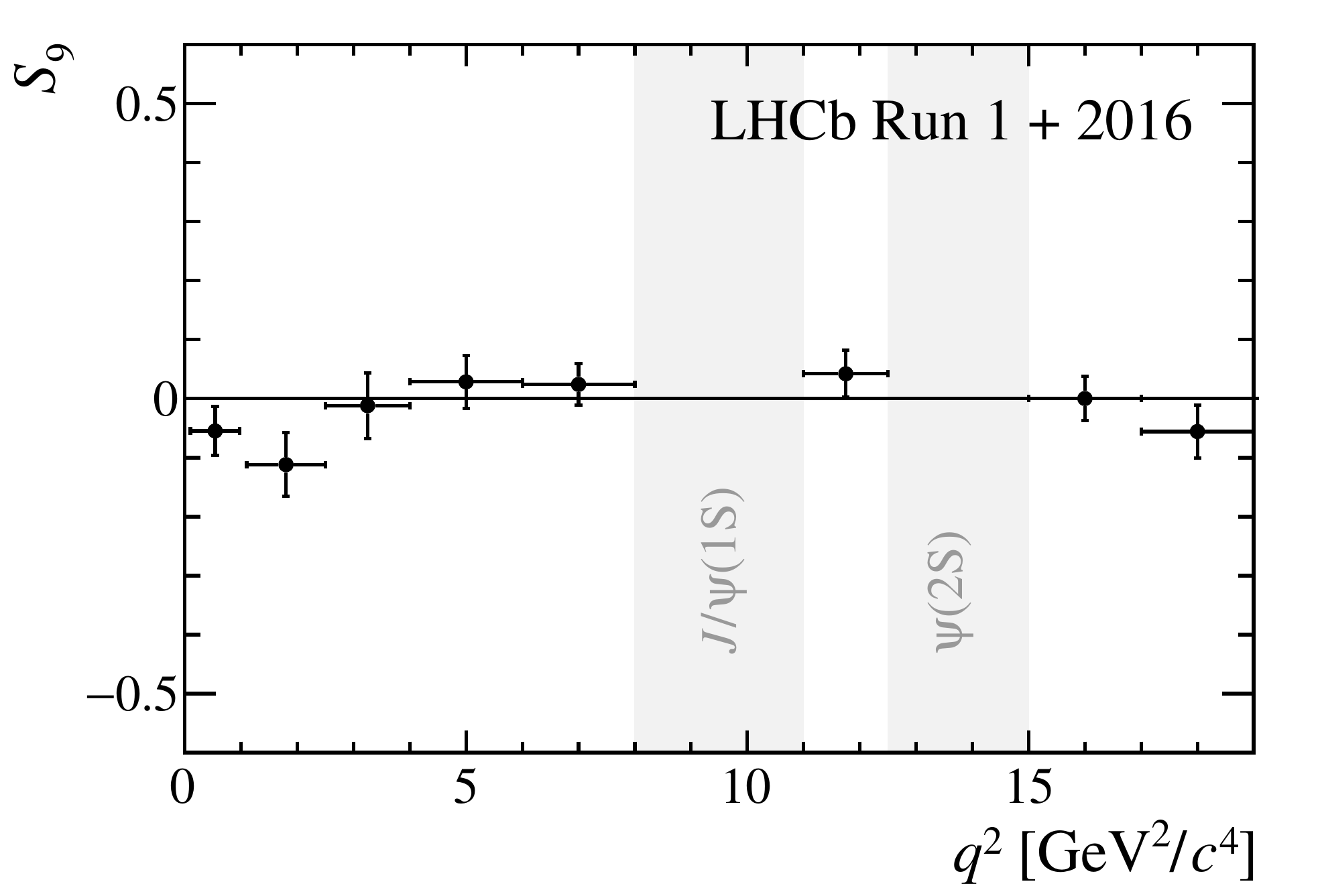}
    \caption{
    Results for the \CP-averaged angular observables $S_3$, $S_4$ and $S_7$--$S_9$ in bins of \qsq.
    The data are compared to SM predictions based on the prescription of Refs.~\cite{Altmannshofer:2014rta, Straub:2015ica}.}
    \label{fig:sub:Si}
\end{figure}

\begin{figure}[!htb]
    \centering
    \includegraphics[width=0.48\linewidth]{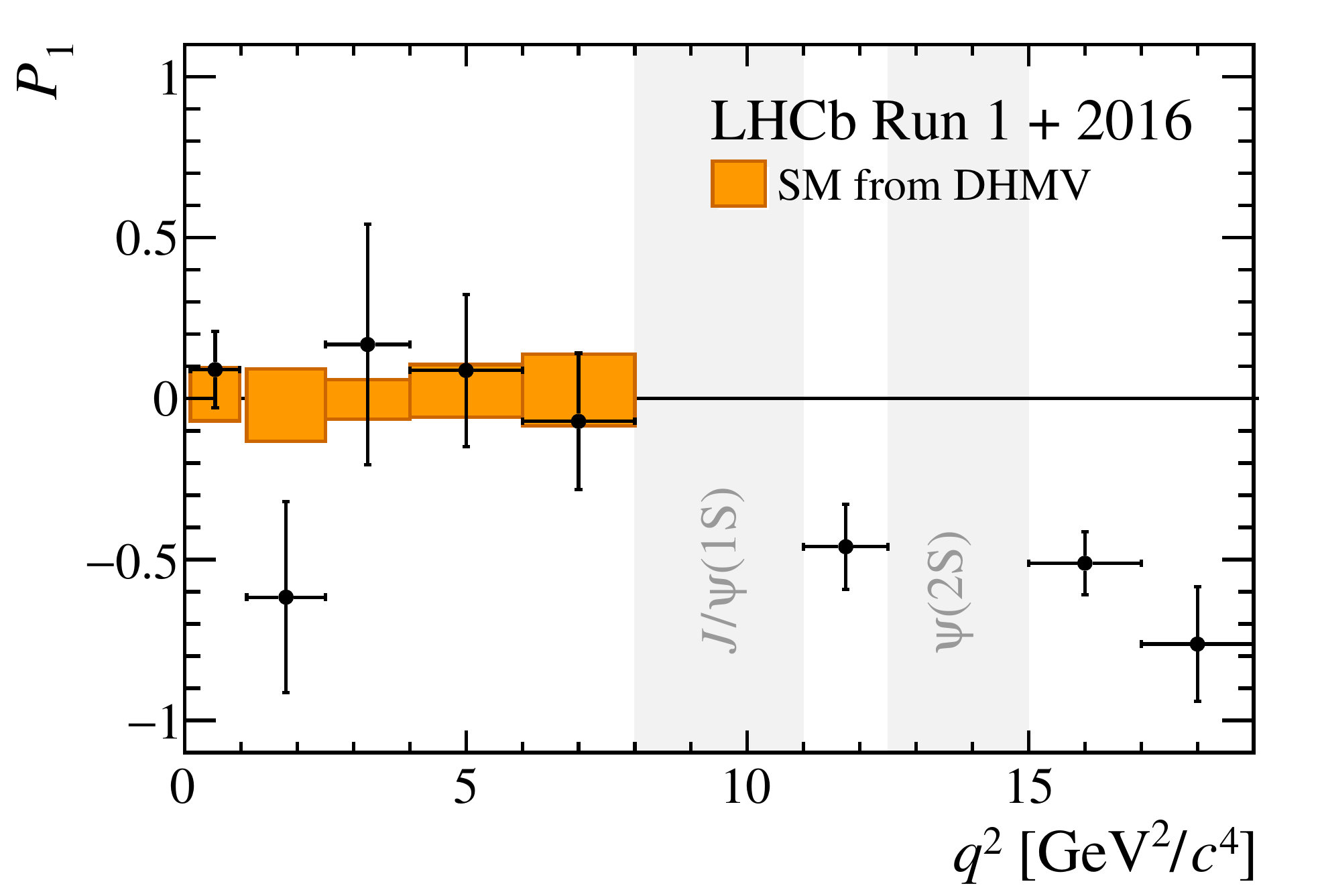} 
    \includegraphics[width=0.48\linewidth]{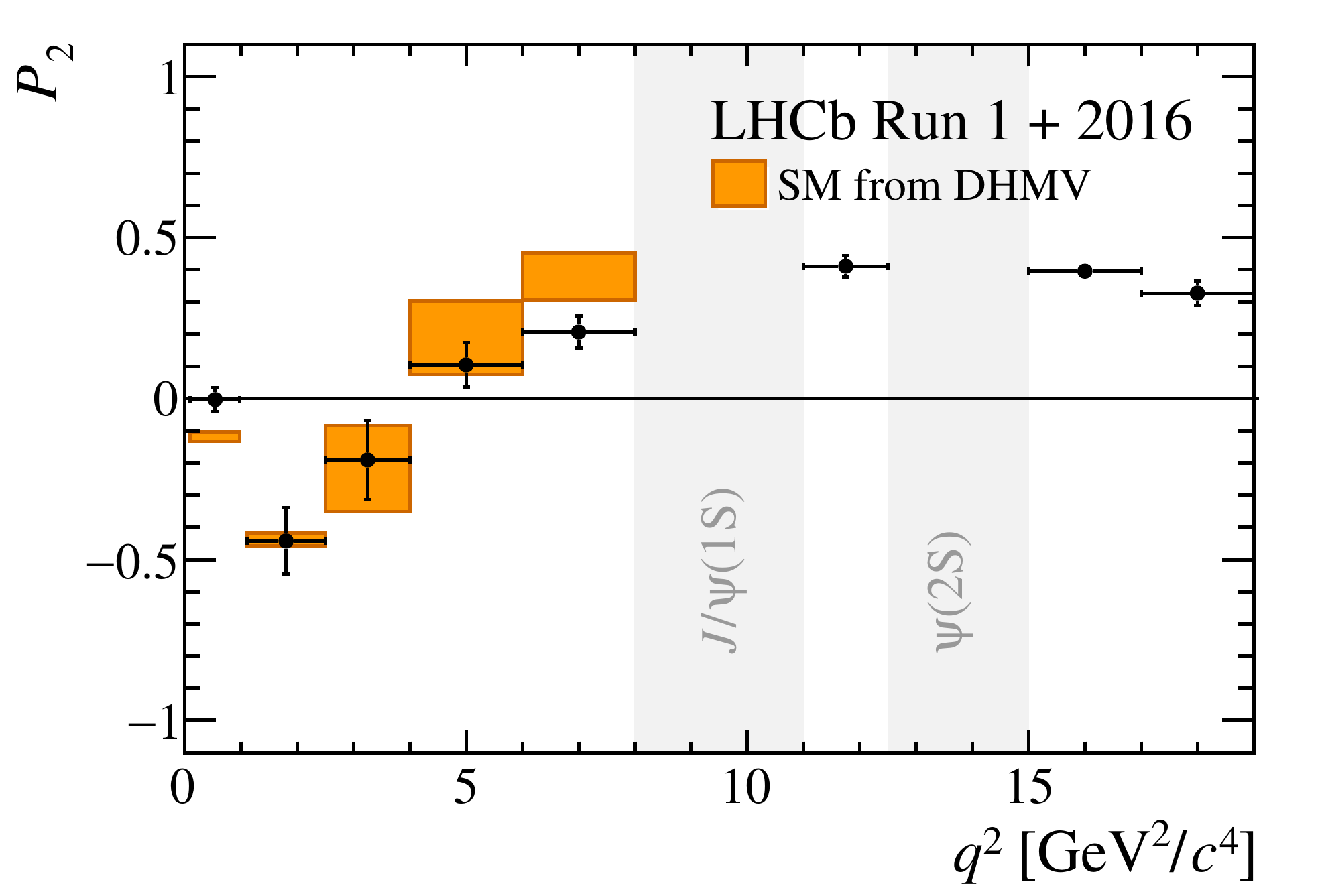} %\\
    \includegraphics[width=0.48\linewidth]{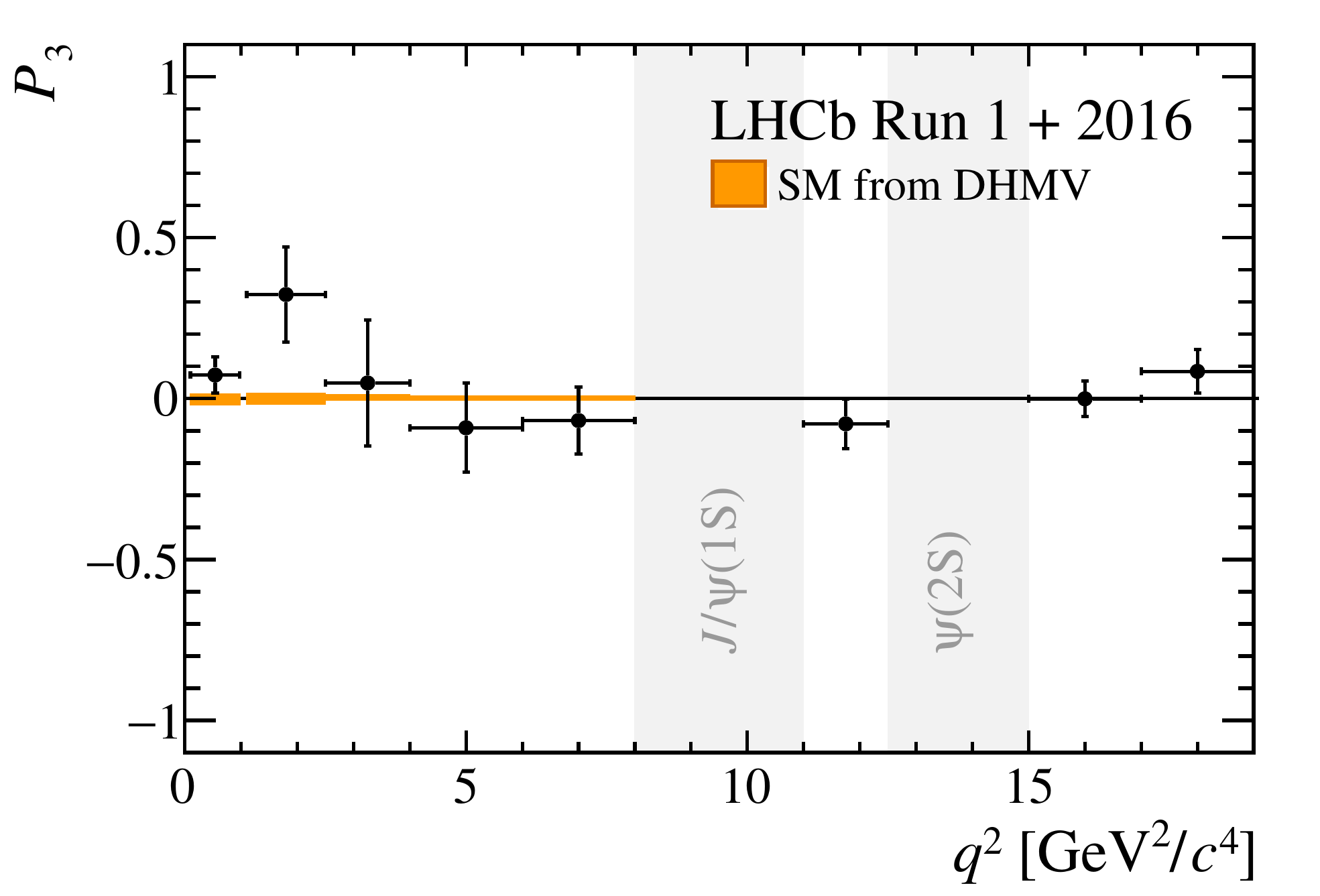} 
    \includegraphics[width=0.48\linewidth]{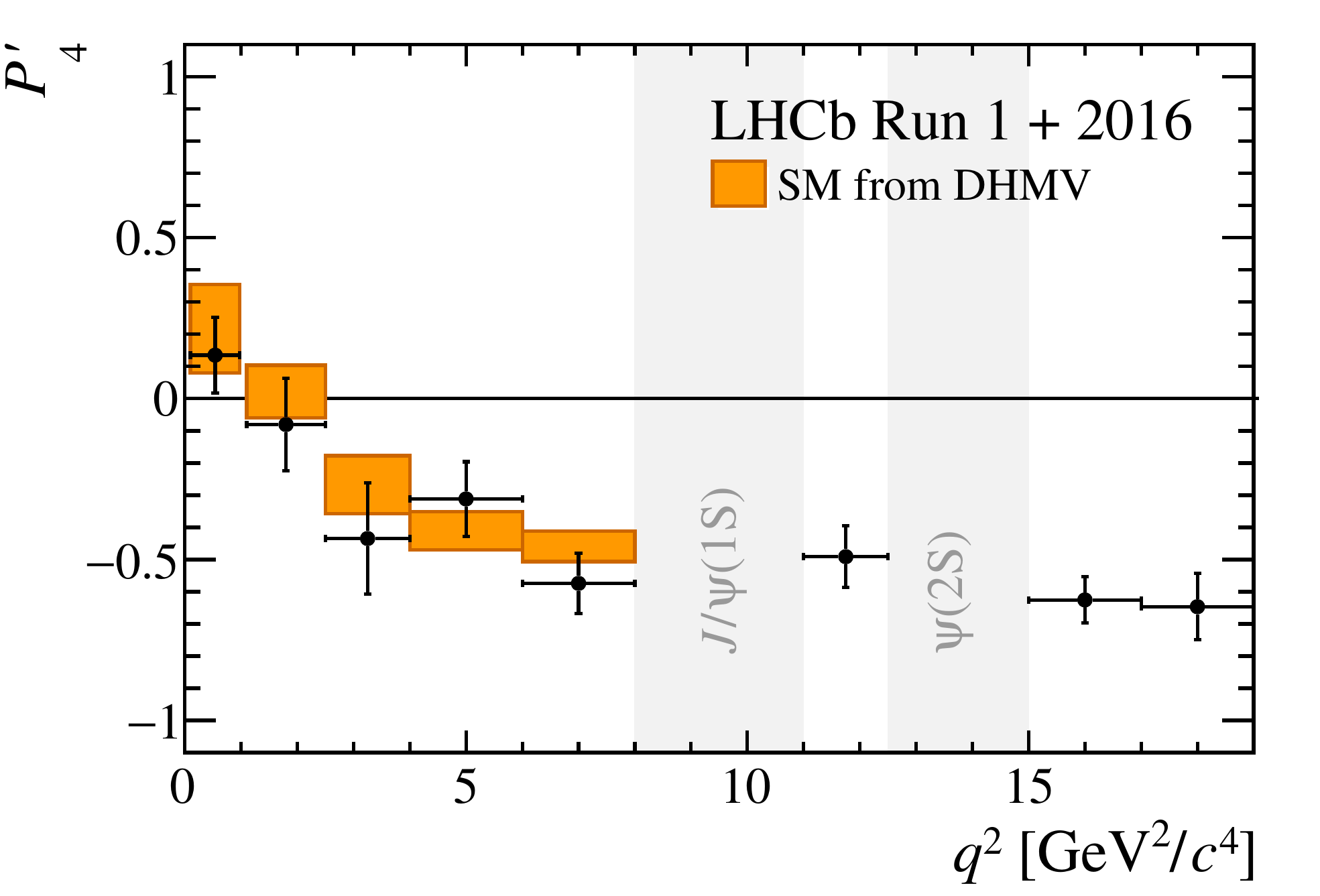} \\
    \includegraphics[width=0.48\linewidth]{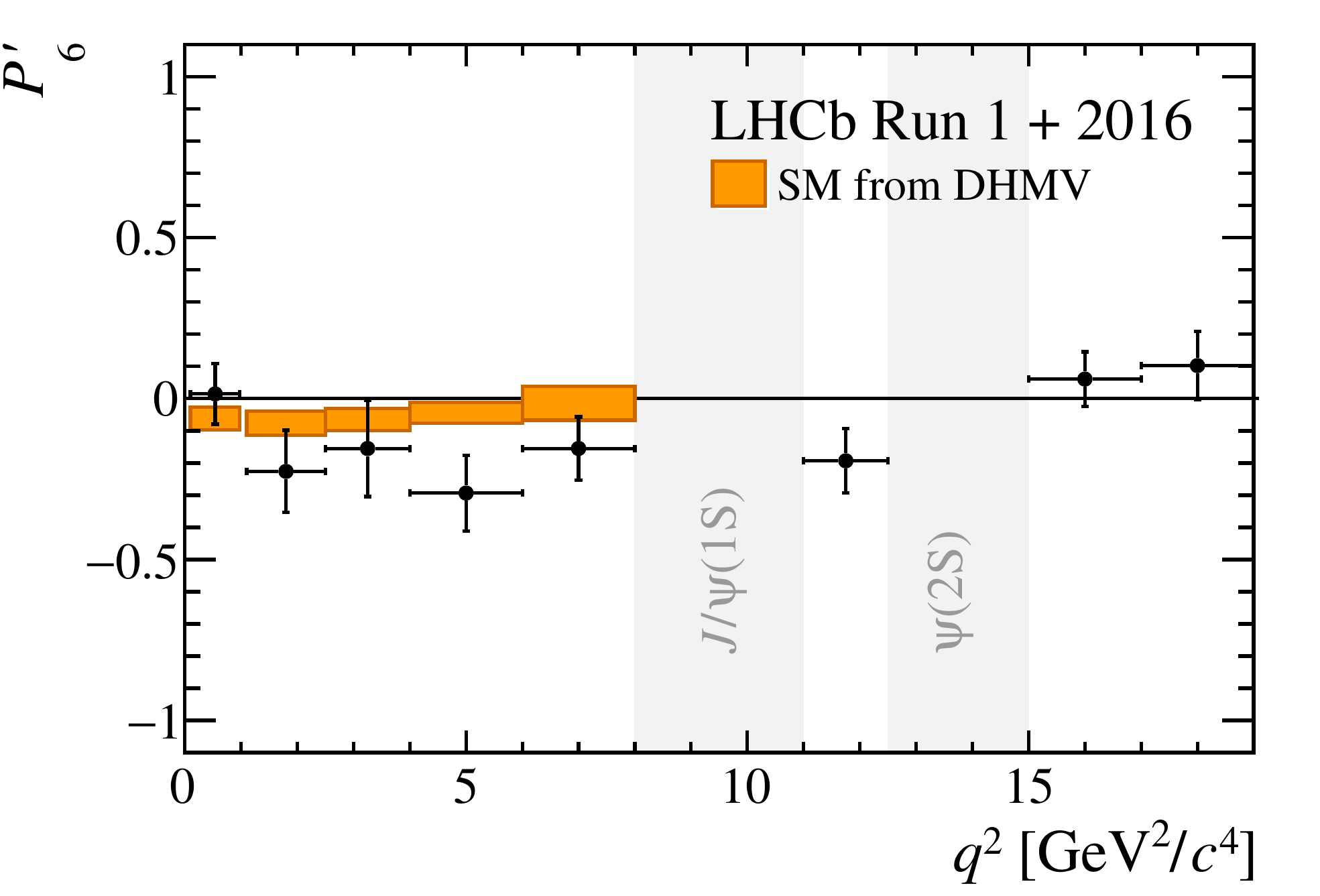} 
    \includegraphics[width=0.48\linewidth]{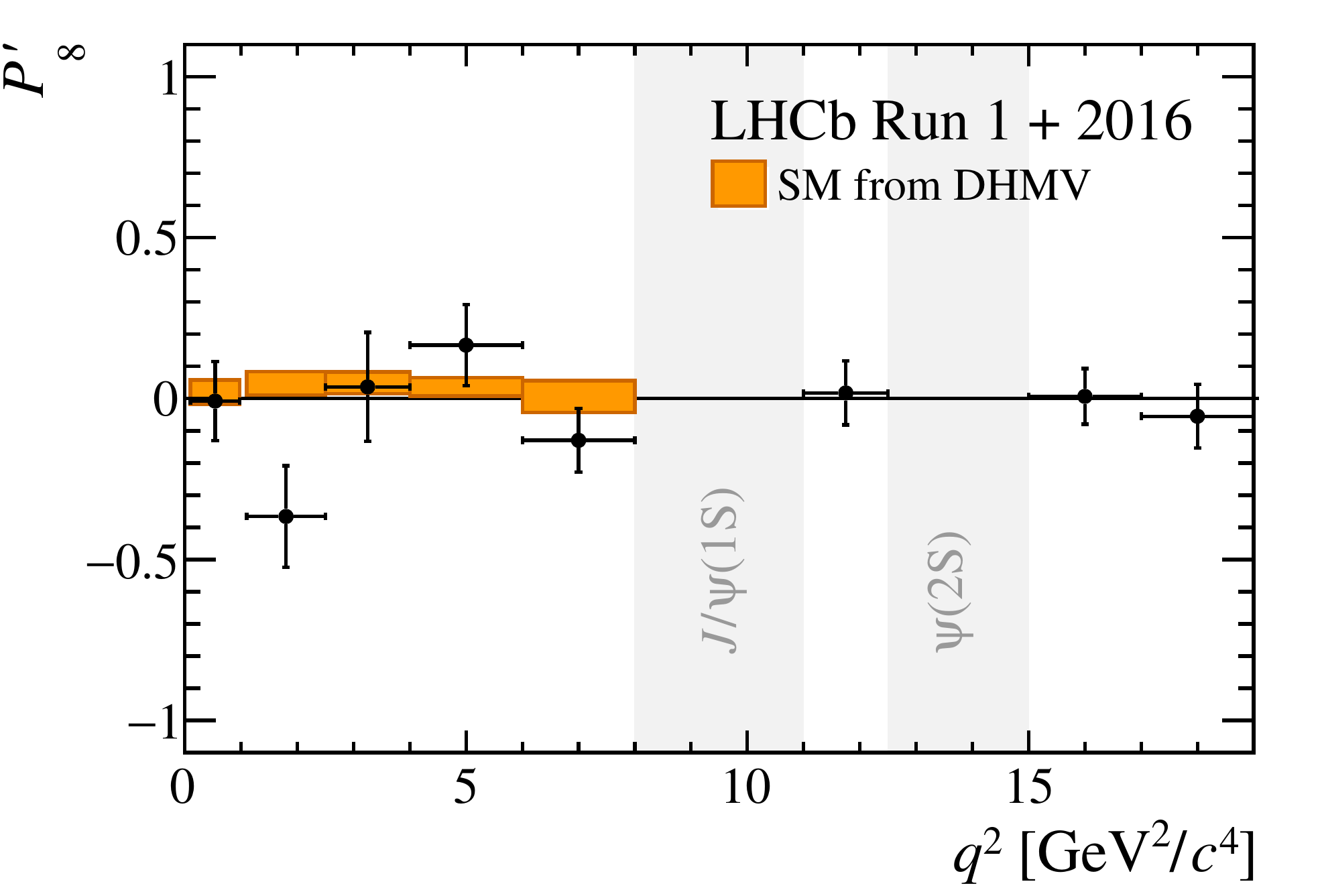} \\
    \caption{
    Results for the optimised angular observables $P_{1}$--$P_{3}$, $P_{4}'$, $P_{6}'$ and $P_{8}'$ in bins of \qsq. 
    The data are compared to SM predictions based on Refs.~\cite{Descotes-Genon:2014uoa, Khodjamirian:2010vf}.
    }
    \label{fig:sub:Pi}
\end{figure}

 The values of $S_3$, $S_4$ and $S_7$--$S_9$ obtained from the simultaneous fit are shown in Fig.~\ref{fig:sub:Si}. 
The data are compared to theoretical predictions based on the prescription of  Ref.~\cite{Altmannshofer:2014rta}. 
The predictions combine light-cone sum rule calculations~\cite{Straub:2015ica} with lattice determinations~\cite{Horgan:2013hoa,Horgan:2015vla} of the $\Bz\to\Kstarz$ form factors. 
Figure~\ref{fig:sub:Pi} shows the values of the optimised observables, $P_{i}^{(\prime)}$, obtained from the fit. 
The data are compared to predictions based on the prescription in Ref.~\cite{Descotes-Genon:2014uoa}. These predictions use form factors from Ref.~\cite{Khodjamirian:2010vf}. 
The values of the observables in the standard and optimised basis are given in Tables~\ref{tab:sub:siresults} and \ref{tab:sub:piresults}, respectively. 
The statistical correlation between the observables in each \qsq bin is provided in Tables~\ref{appendix:likelihood:correlation:average:1}--\ref{appendix:likelihood:correlation:average:10} and Tables~\ref{appendix:likelihood:correlation:optimised:1}--\ref{appendix:likelihood:correlation:optimised:10}.

\clearpage

%\vspace{-2cm}
\begin{table}[h]
\caption{Results for the \CP-averaged observables $F_{\rm L}$, $A_{\rm FB}$ and $S_3$--$S_9$. The first uncertainties are statistical and the second systematic.\label{tab:sub:siresults}}
\begin{center}\footnotesize
\begin{tabular}{lr}
\multicolumn{2}{c}{$0.10 < q^{2} < 0.98 \gevgevcccc$}\\ \hline
$F_{\rm L}$ & $0.255\pm0.032\pm0.007$\\
$S_{3}$ & $0.034\pm0.044\pm0.003$\\
$S_{4}$ & $0.059\pm0.050\pm0.004$\\
$S_{5}$ & $0.227\pm0.041\pm0.008$\\
$A_{\rm FB}$ & $-0.004\pm0.040\pm0.004$\\
$S_{7}$ & $0.006\pm0.042\pm0.002$\\
$S_{8}$ & $-0.003\pm0.051\pm0.001$\\
$S_{9}$ & $-0.055\pm0.041\pm0.002$\\
\end{tabular}
\begin{tabular}{lr}
\multicolumn{2}{c}{$1.1 < q^{2} < 2.5 \gevgevcccc$}\\ \hline
$F_{\rm L}$ & $0.655\pm0.046\pm0.017$\\
$S_{3}$ & $-0.107\pm0.052\pm0.003$\\
$S_{4}$ & $-0.038\pm0.070\pm0.011$\\
$S_{5}$ & $0.174\pm0.060\pm0.007$\\
$A_{\rm FB}$ & $-0.229\pm0.046\pm0.009$\\
$S_{7}$ & $-0.107\pm0.063\pm0.004$\\
$S_{8}$ & $-0.174\pm0.075\pm0.002$\\
$S_{9}$ & $-0.112\pm0.054\pm0.005$\\
\end{tabular}
\begin{tabular}{lr}
\multicolumn{2}{c}{$2.5 < q^{2} < 4.0 \gevgevcccc$}\\ \hline
$F_{\rm L}$ & $0.756\pm0.047\pm0.023$\\
$S_{3}$ & $0.020\pm0.053\pm0.002$\\
$S_{4}$ & $-0.187\pm0.074\pm0.008$\\
$S_{5}$ & $-0.064\pm0.068\pm0.010$\\
$A_{\rm FB}$ & $-0.070\pm0.043\pm0.006$\\
$S_{7}$ & $-0.066\pm0.065\pm0.004$\\
$S_{8}$ & $0.016\pm0.074\pm0.002$\\
$S_{9}$ & $-0.012\pm0.055\pm0.003$\\
\end{tabular}\\[0.6em]
\begin{tabular}{lr}
\multicolumn{2}{c}{$4.0 < q^{2} < 6.0 \gevgevcccc$}\\ \hline
$F_{\rm L}$ & $0.684\pm0.035\pm0.015$\\
$S_{3}$ & $0.014\pm0.038\pm0.003$\\
$S_{4}$ & $-0.145\pm0.057\pm0.004$\\
$S_{5}$ & $-0.204\pm0.051\pm0.013$\\
$A_{\rm FB}$ & $0.050\pm0.033\pm0.002$\\
$S_{7}$ & $-0.136\pm0.053\pm0.002$\\
$S_{8}$ & $0.077\pm0.062\pm0.001$\\
$S_{9}$ & $0.029\pm0.045\pm0.002$\\
\end{tabular}
\begin{tabular}{lr}
\multicolumn{2}{c}{$6.0 < q^{2} < 8.0 \gevgevcccc$}\\ \hline
$F_{\rm L}$ & $0.645\pm0.030\pm0.011$\\
$S_{3}$ & $-0.013\pm0.038\pm0.004$\\
$S_{4}$ & $-0.275\pm0.045\pm0.006$\\
$S_{5}$ & $-0.279\pm0.043\pm0.013$\\
$A_{\rm FB}$ & $0.110\pm0.027\pm0.005$\\
$S_{7}$ & $-0.074\pm0.046\pm0.003$\\
$S_{8}$ & $-0.062\pm0.047\pm0.001$\\
$S_{9}$ & $0.024\pm0.035\pm0.002$\\
\end{tabular}
\begin{tabular}{lr}
\multicolumn{2}{c}{$11.0 < q^{2} < 12.5 \gevgevcccc$}\\ \hline
$F_{\rm L}$ & $0.461\pm0.031\pm0.010$\\
$S_{3}$ & $-0.124\pm0.037\pm0.003$\\
$S_{4}$ & $-0.245\pm0.047\pm0.007$\\
$S_{5}$ & $-0.310\pm0.043\pm0.011$\\
$A_{\rm FB}$ & $0.333\pm0.030\pm0.008$\\
$S_{7}$ & $-0.096\pm0.050\pm0.003$\\
$S_{8}$ & $0.009\pm0.049\pm0.001$\\
$S_{9}$ & $0.042\pm0.040\pm0.003$\\
\end{tabular}\\[0.6em]
\begin{tabular}{lr}
\multicolumn{2}{c}{$15.0 < q^{2} < 17.0 \gevgevcccc$}\\ \hline
$F_{\rm L}$ & $0.352\pm0.026\pm0.009$\\
$S_{3}$ & $-0.166\pm0.034\pm0.007$\\
$S_{4}$ & $-0.299\pm0.033\pm0.008$\\
$S_{5}$ & $-0.341\pm0.034\pm0.009$\\
$A_{\rm FB}$ & $0.385\pm0.024\pm0.007$\\
$S_{7}$ & $0.029\pm0.039\pm0.001$\\
$S_{8}$ & $0.003\pm0.042\pm0.002$\\
$S_{9}$ & $0.000\pm0.037\pm0.002$\\
\end{tabular}
\begin{tabular}{lr}
\multicolumn{2}{c}{$17.0 < q^{2} < 19.0 \gevgevcccc$}\\ \hline
$F_{\rm L}$ & $0.344\pm0.032\pm0.025$\\
$S_{3}$ & $-0.250\pm0.050\pm0.025$\\
$S_{4}$ & $-0.307\pm0.041\pm0.008$\\
$S_{5}$ & $-0.280\pm0.040\pm0.014$\\
$A_{\rm FB}$ & $0.323\pm0.032\pm0.019$\\
$S_{7}$ & $0.049\pm0.049\pm0.007$\\
$S_{8}$ & $-0.026\pm0.046\pm0.002$\\
$S_{9}$ & $-0.056\pm0.045\pm0.002$\\
\end{tabular}
\begin{tabular}{lr}
\multicolumn{2}{c}{$1.1 < q^{2} < 6.0 \gevgevcccc$}\\ \hline
$F_{\rm L}$ & $0.700\pm0.025\pm0.013$\\
$S_{3}$ & $-0.012\pm0.025\pm0.003$\\
$S_{4}$ & $-0.136\pm0.039\pm0.003$\\
$S_{5}$ & $-0.052\pm0.034\pm0.007$\\
$A_{\rm FB}$ & $-0.073\pm0.021\pm0.002$\\
$S_{7}$ & $-0.090\pm0.034\pm0.002$\\
$S_{8}$ & $-0.009\pm0.037\pm0.002$\\
$S_{9}$ & $-0.025\pm0.026\pm0.002$\\
\end{tabular}\\[0.6em]
\begin{tabular}{lr}
\multicolumn{2}{c}{$15.0 < q^{2} < 19.0 \gevgevcccc$}\\ \hline
$F_{\rm L}$ & $0.345\pm0.020\pm0.007$\\
$S_{3}$ & $-0.189\pm0.030\pm0.009$\\
$S_{4}$ & $-0.303\pm0.024\pm0.008$\\
$S_{5}$ & $-0.317\pm0.024\pm0.011$\\
$A_{\rm FB}$ & $0.353\pm0.020\pm0.010$\\
$S_{7}$ & $0.035\pm0.030\pm0.003$\\
$S_{8}$ & $0.005\pm0.031\pm0.001$\\
$S_{9}$ & $-0.031\pm0.029\pm0.001$\\
\end{tabular}
\end{center}
\end{table}

\begin{table}
\caption{Results for the optimised observables $P_i^{(\prime)}$. The first uncertainties are statistical and the second systematic.\label{tab:sub:piresults}}
\begin{center}\footnotesize
\begin{tabular}{lr}
\multicolumn{2}{c}{$0.10 < q^{2} < 0.98 \gevgevcccc$}\\ \hline
$P_{1}$ & $0.090\pm0.119\pm0.009$\\
$P_{2}$ & $-0.003\pm0.038\pm0.003$\\
$P_{3}$ & $0.073\pm0.057\pm0.003$\\
$P_{4}'$ & $0.135\pm0.118\pm0.010$\\
$P_{5}'$ & $0.521\pm0.095\pm0.024$\\
$P_{6}'$ & $0.015\pm0.094\pm0.007$\\
$P_{8}'$ & $-0.007\pm0.122\pm0.002$\\
\end{tabular}
\begin{tabular}{lr}
\multicolumn{2}{c}{$1.1 < q^{2} < 2.5 \gevgevcccc$}\\ \hline
$P_{1}$ & $-0.617\pm0.296\pm0.023$\\
$P_{2}$ & $-0.443\pm0.100\pm0.027$\\
$P_{3}$ & $0.324\pm0.147\pm0.014$\\
$P_{4}'$ & $-0.080\pm0.142\pm0.019$\\
$P_{5}'$ & $0.365\pm0.122\pm0.013$\\
$P_{6}'$ & $-0.226\pm0.128\pm0.005$\\
$P_{8}'$ & $-0.366\pm0.158\pm0.005$\\
\end{tabular}
\begin{tabular}{lr}
\multicolumn{2}{c}{$2.5 < q^{2} < 4.0 \gevgevcccc$}\\ \hline
$P_{1}$ & $0.168\pm0.371\pm0.043$\\
$P_{2}$ & $-0.191\pm0.116\pm0.043$\\
$P_{3}$ & $0.049\pm0.195\pm0.014$\\
$P_{4}'$ & $-0.435\pm0.169\pm0.035$\\
$P_{5}'$ & $-0.150\pm0.144\pm0.032$\\
$P_{6}'$ & $-0.155\pm0.148\pm0.024$\\
$P_{8}'$ & $0.037\pm0.169\pm0.007$\\
\end{tabular} \\[0.6em]
\begin{tabular}{lr}
\multicolumn{2}{c}{$4.0 < q^{2} < 6.0 \gevgevcccc$}\\ \hline
$P_{1}$ & $0.088\pm0.235\pm0.029$\\
$P_{2}$ & $0.105\pm0.068\pm0.009$\\
$P_{3}$ & $-0.090\pm0.139\pm0.006$\\
$P_{4}'$ & $-0.312\pm0.115\pm0.013$\\
$P_{5}'$ & $-0.439\pm0.111\pm0.036$\\
$P_{6}'$ & $-0.293\pm0.117\pm0.004$\\
$P_{8}'$ & $0.166\pm0.127\pm0.004$\\
\end{tabular}
\begin{tabular}{lr}
\multicolumn{2}{c}{$6.0 < q^{2} < 8.0 \gevgevcccc$}\\ \hline
$P_{1}$ & $-0.071\pm0.211\pm0.020$\\
$P_{2}$ & $0.207\pm0.048\pm0.013$\\
$P_{3}$ & $-0.068\pm0.104\pm0.007$\\
$P_{4}'$ & $-0.574\pm0.091\pm0.018$\\
$P_{5}'$ & $-0.583\pm0.090\pm0.030$\\
$P_{6}'$ & $-0.155\pm0.098\pm0.009$\\
$P_{8}'$ & $-0.129\pm0.098\pm0.005$\\
\end{tabular}
\begin{tabular}{lr}
\multicolumn{2}{c}{$11.0 < q^{2} < 12.5 \gevgevcccc$}\\ \hline
$P_{1}$ & $-0.460\pm0.132\pm0.015$\\
$P_{2}$ & $0.411\pm0.033\pm0.008$\\
$P_{3}$ & $-0.078\pm0.077\pm0.007$\\
$P_{4}'$ & $-0.491\pm0.095\pm0.013$\\
$P_{5}'$ & $-0.622\pm0.088\pm0.017$\\
$P_{6}'$ & $-0.193\pm0.100\pm0.003$\\
$P_{8}'$ & $0.018\pm0.099\pm0.009$\\
\end{tabular} \\[0.6em]
\begin{tabular}{lr}
\multicolumn{2}{c}{$15.0 < q^{2} < 17.0 \gevgevcccc$}\\ \hline
$P_{1}$ & $-0.511\pm0.096\pm0.020$\\
$P_{2}$ & $0.396\pm0.022\pm0.004$\\
$P_{3}$ & $-0.000\pm0.056\pm0.003$\\
$P_{4}'$ & $-0.626\pm0.069\pm0.018$\\
$P_{5}'$ & $-0.714\pm0.074\pm0.021$\\
$P_{6}'$ & $0.061\pm0.085\pm0.003$\\
$P_{8}'$ & $0.007\pm0.086\pm0.002$\\
\end{tabular}
\begin{tabular}{lr}
\multicolumn{2}{c}{$17.0 < q^{2} < 19.0 \gevgevcccc$}\\ \hline
$P_{1}$ & $-0.763\pm0.152\pm0.094$\\
$P_{2}$ & $0.328\pm0.032\pm0.017$\\
$P_{3}$ & $0.085\pm0.068\pm0.004$\\
$P_{4}'$ & $-0.647\pm0.086\pm0.057$\\
$P_{5}'$ & $-0.590\pm0.084\pm0.059$\\
$P_{6}'$ & $0.103\pm0.105\pm0.016$\\
$P_{8}'$ & $-0.055\pm0.099\pm0.006$\\
\end{tabular}
\begin{tabular}{lr}
\multicolumn{2}{c}{$1.1 < q^{2} < 6.0 \gevgevcccc$}\\ \hline
$P_{1}$ & $-0.079\pm0.159\pm0.021$\\
$P_{2}$ & $-0.162\pm0.050\pm0.012$\\
$P_{3}$ & $0.085\pm0.090\pm0.005$\\
$P_{4}'$ & $-0.298\pm0.087\pm0.016$\\
$P_{5}'$ & $-0.114\pm0.068\pm0.026$\\
$P_{6}'$ & $-0.197\pm0.075\pm0.009$\\
$P_{8}'$ & $-0.020\pm0.089\pm0.009$\\
\end{tabular} \\[0.6em]
\begin{tabular}{lr}
\multicolumn{2}{c}{$15.0 < q^{2} < 19.0 \gevgevcccc$}\\ \hline
$P_{1}$ & $-0.577\pm0.090\pm0.031$\\
$P_{2}$ & $0.359\pm0.018\pm0.009$\\
$P_{3}$ & $0.048\pm0.045\pm0.002$\\
$P_{4}'$ & $-0.638\pm0.055\pm0.020$\\
$P_{5}'$ & $-0.667\pm0.053\pm0.029$\\
$P_{6}'$ & $0.073\pm0.067\pm0.006$\\
$P_{8}'$ & $0.011\pm0.069\pm0.003$\\
\end{tabular}
\end{center}
\end{table}

\clearpage
\section{Systematic uncertainties}
\label{sec:supp:sys}
A summary of the sources of systematic uncertainty on the angular observables is shown in Table~\ref{tab:systematics}. Details of how the systematic uncertainties are estimated are given in the letter. The dominant systematic uncertainties arise from the peaking backgrounds that are neglected in the analysis (\emph{peaking backgrounds} in Table~\ref{tab:systematics}) and, for the narrow \qsq bins, from
the uncertainty associated with evaluating the acceptance at a fixed point in \qsq (\emph{acceptance variation with $q^2$} in Table~\ref{tab:systematics}). The \emph{bias correction} in Table~\ref{tab:systematics} refers to the biases observed when generating pseudoexperiments using the result of the best fit to data, as discussed in the letter. The systematic uncertainty associated with the \emph{background model} is calculated by increasing the polynomial order to four. 

\begin{table}[ht]
\caption{Summary of the different sources of systematic uncertainty on the angular observables.
%Upper limits are quoted for the different groups of observables. 
\label{tab:systematics}}
\begin{center}
\scalebox{1.00}{
\setlength\extrarowheight{2pt}
\begin{tabular}{r|ccccc}
Source                                 & $F_{\rm L}$     & $A_{\rm FB}$, $S_3$--$S_9$  & $P^{}_1$--$P'_{8}$     \\
\hline
Acceptance stat. uncertainty           & $<0.01$         & $<0.01$       &   $<0.01$                                 \\
Acceptance polynomial order            &   $<0.01$       & $<0.01$       &   $<0.02$               \\
Data-simulation differences            & $<0.01$  & $<0.01$              &  $<0.01$                            \\
Acceptance variation with $q^2$        & $<0.03$         & $<0.03$       &   $<0.09$                         \\
$m(\Kp\pim)$ model                     & $<0.01$         & $<0.01$       &    $<0.02$                    \\
Background model                       & $<0.01$         & $<0.01$       &  $<0.03 $                \\
Peaking backgrounds                    & $<0.02$         & $<0.02 $       &   $<0.03$                    \\
$m(\Kp\pim\mumu)$ model                & $<0.01$         & $<0.01$       &    $<0.02 $                    \\
\Kp\mumu veto                & $<0.01$         & $<0.01$       &    $<0.01$                    \\
Trigger                & $<0.01$         & $<0.01$       &    $<0.01$                    \\
Bias correction               & $<0.02$         & $<0.02 $       &    $<0.04$                    \\
\end{tabular}

}

\end{center}
\end{table}

\clearpage
\section[Correlation matrices for the CP-averaged observables]{Correlation matrices for the \boldmath{\CP}-averaged observables} 
\label{sec:appendix:likelihood:correlation}

Correlation matrices between the \CP-averaged observables in the different \qsq bins are provided in Tables~\ref{appendix:likelihood:correlation:average:1}--\ref{appendix:likelihood:correlation:average:10}.  The different $q^2$ bins are statistically independent.

\begin{table}[!htb]
\caption{
Correlation matrix for the \CP-averaged observables from the maximum-likelihood fit in the bin $0.10<q^2<0.98\gevgevcccc$. 
\label{appendix:likelihood:correlation:average:1}
}
\centering 
\begin{tabular}{c|rrrrrrrr}
      & $F_{\rm L}$ & $S_{3}$ & $S_{4}$ & $S_{5}$ & $A_{\rm FB}$ & $S_{7}$ & $S_{8}$ & $S_{9}$\\ \hline
                   $F_{\rm L}$ &  1.00 & $-0.00$ & $-0.03$ &  0.09 &  0.03 & $-0.01$ &  0.06 &  0.03 \\ 
                   $S_{3}$ &       &  1.00 &  0.02 &  0.14 &  0.02 & $-0.06$ &  0.01 & $-0.01$ \\ 
                   $S_{4}$ &       &       &  1.00 &  0.06 &  0.15 & $-0.03$ &  0.06 &  0.00 \\ 
                   $S_{5}$ &       &       &       &  1.00 &  0.04 & $-0.03$ & $-0.01$ &  0.00 \\ 
                   $A_{\rm FB}$ &       &       &       &       &  1.00 & $-0.02$ & $-0.01$ & $-0.02$ \\ 
                   $S_{7}$ &       &       &       &       &       &  1.00 & $-0.04$ &  0.10 \\ 
                   $S_{8}$ &       &       &       &       &       &       &  1.00 &  0.02 \\ 
                   $S_{9}$ &       &       &       &       &       &       &       &  1.00 \\ 
\end{tabular}
\end{table}

\begin{table}[!htb]
\caption{
Correlation matrix for the \CP-averaged observables from the maximum-likelihood fit in the bin $1.1<q^2<2.5\gevgevcccc$. 
\label{appendix:likelihood:correlation:average:2}
}
\centering 
\begin{tabular}{c|rrrrrrrr}
      & $F_{\rm L}$ & $S_{3}$ & $S_{4}$ & $S_{5}$ & $A_{\rm FB}$ & $S_{7}$ & $S_{8}$ & $S_{9}$\\ \hline
$                   F_{\rm L}$ & 1.00  & 0.05  & 0.04  & 0.16  & 0.11  & $-0.08$ & $-0.06$ & 0.05  \\ 
$                   S_{3}$ &       & 1.00  & 0.00  & 0.04  & 0.05  & 0.08  & 0.08  & 0.18  \\ 
$                   S_{4}$ &       &       & 1.00  & $-0.20$ & $-0.01$ & 0.02  & $-0.09$ & $-0.07$ \\ 
$                   S_{5}$ &       &       &       & 1.00  & $-0.09$ & $-0.11$ & $-0.02$ & $-0.12$ \\ 
$                   A_{\rm FB}$ &       &       &       &       & 1.00  & $-0.03$ & 0.08  & $-0.04$ \\ 
$                   S_{7}$ &       &       &       &       &       & 1.00  & $-0.16$ & 0.14  \\ 
$                   S_{8}$ &       &       &       &       &       &       & 1.00  & $-0.04$ \\ 
$                   S_{9}$ &       &       &       &       &       &       &       & 1.00  \\ 
\end{tabular}
\end{table}

\begin{table}[!htb]
\caption{
Correlation matrix for the \CP-averaged observables from the maximum-likelihood fit in the bin $2.5<q^2<4.0\gevgevcccc$.
\label{appendix:likelihood:correlation:average:3}
}
\centering 
\begin{tabular}{c|rrrrrrrr}
      & $F_{\rm L}$ & $S_{3}$ & $S_{4}$ & $S_{5}$ & $A_{\rm FB}$ & $S_{7}$ & $S_{8}$ & $S_{9}$\\ \hline
$                   F_{\rm L}$ & 1.00  & $-0.02$ & $-0.03$ & $-0.02$ & $-0.03$ & $-0.01$ & $-0.08$ & 0.06  \\ 
$                   S_{3}$ &       & 1.00  & $-0.05$ & $-0.03$ & 0.05  & 0.02  & $-0.07$ & 0.02  \\ 
$                   S_{4}$ &       &       & 1.00  & $-0.13$ & $-0.10$ & 0.01  & 0.03  & $-0.03$ \\ 
$                   S_{5}$ &       &       &       & 1.00  & $-0.08$ & 0.01  & 0.02  & 0.03  \\ 
$                   A_{\rm FB}$ &       &       &       &       & 1.00  & 0.06  & $-0.05$ & $-0.08$ \\ 
$                   S_{7}$ &       &       &       &       &       & 1.00  & 0.01  & 0.03  \\ 
$                   S_{8}$ &       &       &       &       &       &       & 1.00  & $-0.08$ \\ 
$                   S_{9}$ &       &       &       &       &       &       &       & 1.00  \\ 
\end{tabular}
\end{table}

\begin{table}[!htb]
\caption{
Correlation matrix for the \CP-averaged observables from the maximum-likelihood fit in the bin $4.0 <q^2< 6.0\gevgevcccc$.
\label{appendix:likelihood:correlation:average:4}
}
\centering 
\begin{tabular}{c|rrrrrrrr}
      & $F_{\rm L}$ & $S_{3}$ & $S_{4}$ & $S_{5}$ & $A_{\rm FB}$ & $S_{7}$ & $S_{8}$ & $S_{9}$\\ \hline
$                   F_{\rm L}$ & 1.00  & $-0.01$ & 0.05  & $-0.02$ & $-0.14$ & $-0.10$ & 0.09  & 0.04  \\ 
$                   S_{3}$ &       & 1.00  & $-0.06$ & $-0.10$ & 0.06  & $-0.02$ & 0.02  & $-0.08$ \\ 
$                   S_{4}$ &       &       & 1.00  & 0.01  & $-0.14$ & 0.03  & 0.02  & 0.01  \\ 
$                   S_{5}$ &       &       &       & 1.00  & $-0.08$ & 0.07  & 0.02  & $-0.05$ \\ 
$                   A_{\rm FB}$ &       &       &       &       & 1.00  & $-0.01$ & $-0.03$ & 0.01  \\ 
$                   S_{7}$ &       &       &       &       &       & 1.00  & 0.03  & $-0.18$ \\ 
$                   S_{8}$ &       &       &       &       &       &       & 1.00  & $-0.00$ \\ 
$                   S_{9}$ &       &       &       &       &       &       &       & 1.00  \\ 
\end{tabular}
\end{table}

\begin{table}[!htb]
\caption{
Correlation matrix for the \CP-averaged observables from the maximum-likelihood fit in the bin $6.0<q^2<8.0\gevgevcccc$.
\label{appendix:likelihood:correlation:average:5}
}
\centering 
\begin{tabular}{c|rrrrrrrr}
      & $F_{\rm L}$ & $S_{3}$ & $S_{4}$ & $S_{5}$ & $A_{\rm FB}$ & $S_{7}$ & $S_{8}$ & $S_{9}$\\ \hline
$                   F_{\rm L}$ & 1.00  & 0.00  & $-0.01$ & $-0.06$ & $-0.20$ & $-0.05$ & 0.00  & $-0.06$ \\ 
$                   S_{3}$ &       & 1.00  & $-0.12$ & $-0.24$ & 0.01  & 0.05  & 0.04  & $-0.10$ \\ 
$                   S_{4}$ &       &       & 1.00  & 0.13  & $-0.10$ & 0.02  & $-0.04$ & $-0.04$ \\ 
$                   S_{5}$ &       &       &       & 1.00  & $-0.16$ & $-0.01$ & 0.02  & $-0.06$ \\ 
$                   A_{\rm FB}$ &       &       &       &       & 1.00  & $-0.03$ & 0.02  & 0.02  \\ 
$                   S_{7}$ &       &       &       &       &       & 1.00  & 0.08  & $-0.09$ \\ 
$                   S_{8}$ &       &       &       &       &       &       & 1.00  & $-0.08$ \\ 
$                   S_{9}$ &       &       &       &       &       &       &       & 1.00  \\ 
\end{tabular}
\end{table}

\begin{table}[!htb]
\caption{
Correlation matrix for the \CP-averaged observables from the maximum-likelihood fit in the bin $11.0 <q^2< 12.5 \gevgevcccc$.
\label{appendix:likelihood:correlation:average:6}
}
\centering 
\begin{tabular}{c|rrrrrrrr}
      & $F_{\rm L}$ & $S_{3}$ & $S_{4}$ & $S_{5}$ & $A_{\rm FB}$ & $S_{7}$ & $S_{8}$ & $S_{9}$\\ \hline
$                   F_{\rm L}$ & 1.00  & 0.14  & 0.02  & $-0.09$ & $-0.56$ & 0.02  & 0.01  & 0.01  \\ 
$                   S_{3}$ &       & 1.00  & 0.08  & $-0.08$ & $-0.15$ & 0.02  & 0.06  & $-0.10$ \\ 
$                   S_{4}$ &       &       & 1.00  & 0.08  & $-0.12$ & 0.03  & $-0.02$ & $-0.02$ \\ 
$                   S_{5}$ &       &       &       & 1.00  & $-0.13$ & 0.03  & $-0.00$ & $-0.17$ \\ 
$                   A_{\rm FB}$ &       &       &       &       & 1.00  & $-0.05$ & $-0.10$ & 0.12  \\ 
$                   S_{7}$ &       &       &       &       &       & 1.00  & 0.27  & $-0.10$ \\ 
$                   S_{8}$ &       &       &       &       &       &       & 1.00  & $-0.01$ \\ 
$                   S_{9}$ &       &       &       &       &       &       &       & 1.00  \\ 
\end{tabular}
\end{table}

\begin{table}[!htb]
\caption{
Correlation matrix for the \CP-averaged observables from the maximum-likelihood fit in the bin $15.0 <q^2< 17.0 \gevgevcccc$.
\label{appendix:likelihood:correlation:average:7}
}
\centering 
\begin{tabular}{c|rrrrrrrr}
      & $F_{\rm L}$ & $S_{3}$ & $S_{4}$ & $S_{5}$ & $A_{\rm FB}$ & $S_{7}$ & $S_{8}$ & $S_{9}$\\ \hline
$                   F_{\rm L}$ & 1.00  & 0.27  & 0.02  & 0.07  & $-0.53$ & 0.00  & $-0.04$ & 0.06  \\ 
$                   S_{3}$ &       & 1.00  & $-0.05$ & 0.01  & $-0.12$ & $-0.02$ & $-0.04$ & 0.10  \\ 
$                   S_{4}$ &       &       & 1.00  & 0.29  & $-0.15$ & 0.02  & 0.06  & 0.03  \\ 
$                   S_{5}$ &       &       &       & 1.00  & $-0.28$ & 0.06  & 0.03  & 0.04  \\ 
$                   A_{\rm FB}$ &       &       &       &       & 1.00  & 0.01  & $-0.00$ & 0.01  \\ 
$                   S_{7}$ &       &       &       &       &       & 1.00  & 0.31  & $-0.23$ \\ 
$                   S_{8}$ &       &       &       &       &       &       & 1.00  & $-0.13$ \\ 
$                   S_{9}$ &       &       &       &       &       &       &       & 1.00  \\ 
\end{tabular}
\end{table}

\begin{table}[!htb]
\caption{
Correlation matrix for the \CP-averaged observables from the maximum-likelihood fit in the bin $17.0 <q^2< 19.0\gevgevcccc$.
\label{appendix:likelihood:correlation:average:8}
}
\centering 
\begin{tabular}{c|rrrrrrrr}
      & $F_{\rm L}$ & $S_{3}$ & $S_{4}$ & $S_{5}$ & $A_{\rm FB}$ & $S_{7}$ & $S_{8}$ & $S_{9}$\\ \hline
$                   F_{\rm L}$ & 1.00  & 0.14  & 0.06  & 0.00  & $-0.35$ & 0.02  & $-0.02$ & 0.08  \\ 
$                   S_{3}$ &       & 1.00  & $-0.04$ & $-0.15$ & $-0.12$ & $-0.04$ & 0.03  & $-0.04$ \\ 
$                   S_{4}$ &       &       & 1.00  & 0.25  & $-0.14$ & $-0.10$ & 0.08  & 0.02  \\ 
$                   S_{5}$ &       &       &       & 1.00  & $-0.25$ & $-0.07$ & $-0.08$ & 0.05  \\ 
$                   A_{\rm FB}$ &       &       &       &       & 1.00  & $-0.00$ & $-0.03$ & $-0.09$ \\ 
$                   S_{7}$ &       &       &       &       &       & 1.00  & 0.33  & $-0.09$ \\ 
$                   S_{8}$ &       &       &       &       &       &       & 1.00  & $-0.13$ \\ 
$                   S_{9}$ &       &       &       &       &       &       &       & 1.00  \\ 
\end{tabular}
\end{table}

\begin{table}[!htb]
\caption{
Correlation matrix for the \CP-averaged observables from the maximum-likelihood fit in the bin $1.1 <q^2< 6.0\gevgevcccc$.
\label{appendix:likelihood:correlation:average:9}
}
\centering 
\begin{tabular}{c|rrrrrrrr}
      & $F_{\rm L}$ & $S_{3}$ & $S_{4}$ & $S_{5}$ & $A_{\rm FB}$ & $S_{7}$ & $S_{8}$ & $S_{9}$\\ \hline
$                   F_{\rm L}$ & 1.00  & $-0.01$ & $-0.02$ & 0.00  & 0.01  & $-0.08$ & 0.02  & 0.03  \\ 
$                   S_{3}$ &       & 1.00  & $-0.04$ & $-0.01$ & 0.04  & 0.03  & 0.00  & $-0.02$ \\ 
$                   S_{4}$ &       &       & 1.00  & $-0.07$ & $-0.09$ & 0.01  & 0.01  & $-0.03$ \\ 
$                   S_{5}$ &       &       &       & 1.00  & $-0.07$ & 0.00  & 0.01  & $-0.04$ \\ 
$                   A_{\rm FB}$ &       &       &       &       & 1.00  & $-0.01$ & $-0.03$ & $-0.03$ \\ 
$                   S_{7}$ &       &       &       &       &       & 1.00  & $-0.02$ & $-0.04$ \\ 
$                   S_{8}$ &       &       &       &       &       &       & 1.00  & $-0.08$ \\ 
$                   S_{9}$ &       &       &       &       &       &       &       & 1.00  \\ 
\end{tabular}
\end{table}

\begin{table}[!htb]
\caption{
Correlation matrix for the \CP-averaged observables from the maximum-likelihood fit in the bin $15.0 <q^2< 19.0\gevgevcccc$.
\label{appendix:likelihood:correlation:average:10}
}
\centering 
\begin{tabular}{c|rrrrrrrr}
      & $F_{\rm L}$ & $S_{3}$ & $S_{4}$ & $S_{5}$ & $A_{\rm FB}$ & $S_{7}$ & $S_{8}$ & $S_{9}$\\ \hline
$                   F_{\rm L}$ & 1.00  & 0.18  & $-0.06$ & $-0.07$ & $-0.37$ & 0.00  & $-0.03$ & 0.07  \\ 
$                   S_{3}$ &       & 1.00  & $-0.04$ & $-0.03$ & $-0.07$ & $-0.00$ & $-0.04$ & 0.02  \\ 
$                   S_{4}$ &       &       & 1.00  & 0.21  & $-0.13$ & $-0.03$ & 0.04  & 0.06  \\ 
$                   S_{5}$ &       &       &       & 1.00  & $-0.23$ & 0.02  & $-0.01$ & 0.04  \\ 
$                   A_{\rm FB}$ &       &       &       &       & 1.00  & 0.03  & $-0.01$ & 0.00  \\ 
$                   S_{7}$ &       &       &       &       &       & 1.00  & 0.28  & $-0.18$ \\ 
$                   S_{8}$ &       &       &       &       &       &       & 1.00  & $-0.14$ \\ 
$                   S_{9}$ &       &       &       &       &       &       &       & 1.00  \\ 
\end{tabular}
\end{table}

\clearpage

\section{Correlation matrices for the optimised angular observables} 
\label{sec:appendix:correlation:optimised}

Correlation matrices between the optimised $P^{(\prime)}_{i}$ basis of observables in the different \qsq bins are provided in Tables~\ref{appendix:likelihood:correlation:optimised:1}--\ref{appendix:likelihood:correlation:optimised:10}.

\begin{table}[!htb]
\caption{
Correlation matrix for the optimised angular observables from the maximum-likelihood fit in the bin $0.10<q^2<0.98 \gevgevcccc$. 
\label{appendix:likelihood:correlation:optimised:1}
}
\centering
\begin{tabular}{c|rrrrrrrr}
      & $F_{\rm L}$ & $P_{1}$ & $P_{2}$ & $P_{3}$ & $P_{4}'$ & $P_{5}'$ & $P_{6}'$ & $P_{8}'$\\ \hline
                   $F_{\rm L}$ &  1.00 &  0.03 &  0.02 &  0.03 & $-0.08$ & $-0.13$ & $-0.02$ &  0.06 \\ 
                   $P_{1}$ &       &  1.00 &  0.02 &  0.01 &  0.02 &  0.14 & $-0.06$ &  0.01 \\ 
                   $P_{2}$ &       &       &  1.00 &  0.02 &  0.14 &  0.03 & $-0.02$ & $-0.01$ \\ 
                   $P_{3}$ &       &       &       &  1.00 & $-0.01$ & $-0.00$ & $-0.10$ & $-0.02$ \\ 
                   $P_{4}'$ &       &       &       &       &  1.00 &  0.07 & $-0.03$ &  0.06 \\ 
                   $P_{5}'$ &       &       &       &       &       &  1.00 & $-0.03$ & $-0.02$ \\ 
                   $P_{6}'$ &       &       &       &       &       &       &  1.00 & $-0.04$ \\ 
                   $P_{8}'$ &       &       &       &       &       &       &       &  1.00 \\ 
\end{tabular}
\end{table}

\begin{table}[!htb]
\caption{
Correlation matrix for the optimised angular observables from the maximum-likelihood fit in the bin $1.1<q^2<2.5 \gevgevcccc$. 
\label{appendix:likelihood:correlation:optimised:2}
}
\centering
\begin{tabular}{c|rrrrrrrr}
      & $F_{\rm L}$ & $P_{1}$ & $P_{2}$ & $P_{3}$ & $P_{4}'$ & $P_{5}'$ & $P_{6}'$ & $P_{8}'$\\ \hline
$                   F_{\rm L}$ & 1.00  & $-0.23$ & $-0.51$ & 0.26  & 0.03  & 0.24  & $-0.13$ & $-0.13$ \\ 
$                   P_{1}$ &       & 1.00  & 0.15  & $-0.23$ & $-0.00$ & $-0.02$ & 0.11  & 0.11  \\ 
$                   P_{2}$ &       &       & 1.00  & $-0.09$ & $-0.03$ & $-0.22$ & 0.05  & 0.14  \\ 
$                   P_{3}$ &       &       &       & 1.00  & 0.07  & 0.19  & $-0.17$ & $-0.00$ \\ 
$                   P_{4}'$ &       &       &       &       & 1.00  & $-0.20$ & 0.02  & $-0.09$ \\ 
$                   P_{5}'$ &       &       &       &       &       & 1.00  & $-0.12$ & $-0.04$ \\ 
$                   P_{6}'$ &       &       &       &       &       &       & 1.00  & $-0.14$ \\ 
$                   P_{8}'$ &       &       &       &       &       &       &       & 1.00  \\ 
\end{tabular}
\end{table}

\begin{table}[!htb]
\caption{
Correlation matrix for the optimised angular observables from the maximum-likelihood fit in the bin $2.5<q^2<4.0 \gevgevcccc$. 
\label{appendix:likelihood:correlation:optimised:3}
}
\centering
\begin{tabular}{c|rrrrrrrr}
      & $F_{\rm L}$ & $P_{1}$ & $P_{2}$ & $P_{3}$ & $P_{4}'$ & $P_{5}'$ & $P_{6}'$ & $P_{8}'$\\ \hline
$                   F_{\rm L}$ & 1.00  & 0.08  & $-0.34$ & 0.01  & $-0.21$ & $-0.09$ & $-0.08$ & $-0.06$ \\ 
$                   P_{1}$ &       & 1.00  & 0.02  & $-0.02$ & $-0.07$ & $-0.03$ & 0.00  & $-0.08$ \\ 
$                   P_{2}$ &       &       & 1.00  & 0.07  & $-0.02$ & $-0.05$ & 0.08  & $-0.03$ \\ 
$                   P_{3}$ &       &       &       & 1.00  & 0.02  & $-0.04$ & $-0.04$ & 0.07  \\ 
$                   P_{4}'$ &       &       &       &       & 1.00  & $-0.10$ & 0.02  & 0.04  \\ 
$                   P_{5}'$ &       &       &       &       &       & 1.00  & 0.01  & 0.02  \\ 
$                   P_{6}'$ &       &       &       &       &       &       & 1.00  & 0.01  \\ 
$                   P_{8}'$ &       &       &       &       &       &       &       & 1.00  \\ 
\end{tabular}
\end{table}

\begin{table}[!htb]
\caption{
Correlation matrix for the optimised angular observables from the maximum-likelihood fit in the bin $4.0 <q^2< 6.0 \gevgevcccc$. 
\label{appendix:likelihood:correlation:optimised:4}
}
\centering
\begin{tabular}{c|rrrrrrrr}
      & $F_{\rm L}$ & $P_{1}$ & $P_{2}$ & $P_{3}$ & $P_{4}'$ & $P_{5}'$ & $P_{6}'$ & $P_{8}'$\\ \hline
$                   F_{\rm L}$ & 1.00  & 0.04  & 0.05  & $-0.10$ & $-0.04$ & $-0.14$ & $-0.17$ & 0.14  \\ 
$                   P_{1}$ &       & 1.00  & 0.06  & 0.07  & $-0.06$ & $-0.10$ & $-0.03$ & 0.02  \\ 
$                   P_{2}$ &       &       & 1.00  & $-0.02$ & $-0.14$ & $-0.09$ & $-0.03$ & $-0.01$ \\ 
$                   P_{3}$ &       &       &       & 1.00  & $-0.01$ & 0.07  & 0.19  & $-0.01$ \\ 
$                   P_{4}'$ &       &       &       &       & 1.00  & 0.02  & 0.04  & 0.01  \\ 
$                   P_{5}'$ &       &       &       &       &       & 1.00  & 0.09  & 0.00  \\ 
$                   P_{6}'$ &       &       &       &       &       &       & 1.00  & 0.02  \\ 
$                   P_{8}'$ &       &       &       &       &       &       &       & 1.00  \\ 
\end{tabular}
\end{table}

\begin{table}[!htb]
\caption{
Correlation matrix for the optimised angular observables from the maximum-likelihood fit in the bin $6.0 <q^2< 8.0 \gevgevcccc$. 
\label{appendix:likelihood:correlation:optimised:5}
}
\centering
\begin{tabular}{c|rrrrrrrr}
      & $F_{\rm L}$ & $P_{1}$ & $P_{2}$ & $P_{3}$ & $P_{4}'$ & $P_{5}'$ & $P_{6}'$ & $P_{8}'$\\ \hline
$                   F_{\rm L}$ & 1.00  & $-0.02$ & 0.17  & 0.01  & $-0.14$ & $-0.18$ & $-0.08$ & $-0.02$ \\ 
$                   P_{1}$ &       & 1.00  & 0.01  & 0.10  & $-0.12$ & $-0.23$ & 0.04  & 0.04  \\ 
$                   P_{2}$ &       &       & 1.00  & $-0.00$ & $-0.13$ & $-0.21$ & $-0.06$ & 0.02  \\ 
$                   P_{3}$ &       &       &       & 1.00  & 0.03  & 0.06  & 0.09  & 0.08  \\ 
$                   P_{4}'$ &       &       &       &       & 1.00  & 0.15  & 0.03  & $-0.03$ \\ 
$                   P_{5}'$ &       &       &       &       &       & 1.00  & 0.00  & 0.02  \\ 
$                   P_{6}'$ &       &       &       &       &       &       & 1.00  & 0.08  \\ 
$                   P_{8}'$ &       &       &       &       &       &       &       & 1.00  \\ 
\end{tabular}
\end{table}

\begin{table}[!htb]
\caption{
Correlation matrix for the optimised angular observables from the maximum-likelihood fit in the bin $11.0 <q^2< 12.5 \gevgevcccc$. 
\label{appendix:likelihood:correlation:optimised:6}
}
\centering
\begin{tabular}{c|rrrrrrrr}
      & $F_{\rm L}$ & $P_{1}$ & $P_{2}$ & $P_{3}$ & $P_{4}'$ & $P_{5}'$ & $P_{6}'$ & $P_{8}'$\\ \hline
$                   F_{\rm L}$ & 1.00  & $-0.07$ & 0.13  & $-0.07$ & 0.04  & $-0.07$ & 0.03  & 0.00  \\ 
$                   P_{1}$ &       & 1.00  & $-0.09$ & 0.10  & 0.07  & $-0.06$ & 0.01  & 0.05  \\ 
$                   P_{2}$ &       &       & 1.00  & $-0.16$ & $-0.12$ & $-0.23$ & $-0.05$ & $-0.11$ \\ 
$                   P_{3}$ &       &       &       & 1.00  & 0.01  & 0.18  & 0.10  & 0.00  \\ 
$                   P_{4}'$ &       &       &       &       & 1.00  & 0.08  & 0.03  & $-0.02$ \\ 
$                   P_{5}'$ &       &       &       &       &       & 1.00  & 0.03  & 0.00  \\ 
$                   P_{6}'$ &       &       &       &       &       &       & 1.00  & 0.27  \\ 
$                   P_{8}'$ &       &       &       &       &       &       &       & 1.00  \\ 
\end{tabular}
\end{table}

\begin{table}[!htb]
\caption{
Correlation matrix for the optimised angular observables from the maximum-likelihood fit in the bin $15.0 <q^2< 17.0 \gevgevcccc$. 
\label{appendix:likelihood:correlation:optimised:7}
}
\centering
\begin{tabular}{c|rrrrrrrr}
      & $F_{\rm L}$ & $P_{1}$ & $P_{2}$ & $P_{3}$ & $P_{4}'$ & $P_{5}'$ & $P_{6}'$ & $P_{8}'$\\ \hline
$                   F_{\rm L}$ & 1.00  & 0.06  & 0.14  & $-0.06$ & 0.18  & 0.23  & $-0.01$ & $-0.04$ \\ 
$                   P_{1}$ &       & 1.00  & 0.03  & $-0.09$ & -0.04 & 0.00  & $-0.03$ & $-0.04$ \\ 
$                   P_{2}$ &       &       & 1.00  & $-0.06$ & $-0.13$ & $-0.25$ & 0.01  & $-0.03$ \\ 
$                   P_{3}$ &       &       &       & 1.00  & $-0.04$ & $-0.05$ & 0.23  & 0.13  \\ 
$                   P_{4}'$ &       &       &       &       & 1.00  & 0.32  & 0.02  & 0.06  \\ 
$                   P_{5}'$ &       &       &       &       &       & 1.00  & 0.06  & 0.03  \\ 
$                   P_{6}'$ &       &       &       &       &       &       & 1.00  & 0.31  \\ 
$                   P_{8}'$ &       &       &       &       &       &       &       & 1.00  \\ 
\end{tabular}
\end{table}

\begin{table}[!htb]
\caption{
Correlation matrix for the optimised angular observables from the maximum-likelihood fit in the bin $17.0 <q^2< 19.0 \gevgevcccc$. 
\label{appendix:likelihood:correlation:optimised:8}
}
\centering
\begin{tabular}{c|rrrrrrrr}
      & $F_{\rm L}$ & $P_{1}$ & $P_{2}$ & $P_{3}$ & $P_{4}'$ & $P_{5}'$ & $P_{6}'$ & $P_{8}'$\\ \hline
$                   F_{\rm L}$ & 1.00  & $-0.10$ & 0.16  & $-0.01$ & 0.22  & 0.14  & $-0.01$ & $-0.01$ \\ 
$                   P_{1}$ &       & 1.00  & $-0.10$ & 0.05  & $-0.07$ & $-0.16$ & $-0.05$ & 0.03  \\ 
$                   P_{2}$ &       &       & 1.00  & 0.06  & $-0.09$ & $-0.23$ & 0.00  & $-0.05$ \\ 
$                   P_{3}$ &       &       &       & 1.00  & $-0.01$ & $-0.06$ & 0.09  & 0.14  \\ 
$                   P_{4}'$ &       &       &       &       & 1.00  & 0.27  & $-0.09$ & 0.08  \\ 
$                   P_{5}'$ &       &       &       &       &       & 1.00  & $-0.07$ & $-0.09$ \\ 
$                   P_{6}'$ &       &       &       &       &       &       & 1.00  & 0.34  \\ 
$                   P_{8}'$ &       &       &       &       &       &       &       & 1.00  \\ 
\end{tabular}
\end{table}

\begin{table}[!htb]
\caption{
Correlation matrix for the optimised angular observables from the maximum-likelihood fit in the bin $1.1 <q^2< 6.0 \gevgevcccc$. 
\label{appendix:likelihood:correlation:optimised:9}
}
\centering
\begin{tabular}{c|rrrrrrrr}
      & $F_{\rm L}$ & $P_{1}$ & $P_{2}$ & $P_{3}$ & $P_{4}'$ & $P_{5}'$ & $P_{6}'$ & $P_{8}'$\\ \hline
$                   F_{\rm L}$ & 1.00  & $-0.05$ & $-0.33$ & 0.09  & $-0.11$ & $-0.03$ & $-0.14$ & 0.02  \\ 
$                   P_{1}$ &       & 1.00  & 0.05  & 0.02  & $-0.04$ & $-0.00$ & 0.03  & 0.01  \\ 
$                   P_{2}$ &       &       & 1.00  & $-0.00$ & $-0.04$ & $-0.06$ & 0.03  & $-0.04$ \\ 
$                   P_{3}$ &       &       &       & 1.00  & 0.02  & 0.03  & 0.03  & 0.08  \\ 
$                   P_{4}'$ &       &       &       &       & 1.00  & $-0.06$ & 0.03  & 0.01  \\ 
$                   P_{5}'$ &       &       &       &       &       & 1.00  & 0.01  & 0.00  \\ 
$                   P_{6}'$ &       &       &       &       &       &       & 1.00  & $-0.02$ \\ 
$                   P_{8}'$ &       &       &       &       &       &       &       & 1.00  \\ 
\end{tabular} 
\end{table}

\begin{table}[!htb]
\caption{
Correlation matrix for the optimised angular observables from the maximum-likelihood fit in the bin $15.0 <q^2< 19.0 \gevgevcccc$. 
\label{appendix:likelihood:correlation:optimised:10}
}
\centering
\begin{tabular}{c|rrrrrrrr}
      & $F_{\rm L}$ & $P_{1}$ & $P_{2}$ & $P_{3}$ & $P_{4}'$ & $P_{5}'$ & $P_{6}'$ & $P_{8}'$\\ \hline
$                   F_{\rm L}$ & 1.00  & $-0.08$ & 0.19  & $-0.02$ & 0.11  & 0.09  & $-0.01$ & $-0.04$ \\ 
$                   P_{1}$ &       & 1.00  & $-0.01$ & $-0.00$ & $-0.04$ & $-0.02$ & 0.00  & $-0.04$ \\ 
$                   P_{2}$ &       &       & 1.00  & $-0.04$ & $-0.14$ & $-0.25$ & 0.03  & $-0.03$ \\ 
$                   P_{3}$ &       &       &       & 1.00  & $-0.06$ & $-0.04$ & 0.18  & 0.14  \\ 
$                   P_{4}'$ &       &       &       &       & 1.00  & 0.21  & $-0.03$ & 0.04  \\ 
$                   P_{5}'$ &       &       &       &       &       & 1.00  & 0.02  & $-0.01$ \\ 
$                   P_{6}'$ &       &       &       &       &       &       & 1.00  & 0.28  \\ 
$                   P_{8}'$ &       &       &       &       &       &       &       & 1.00  \\ 
\end{tabular}
\end{table}

\clearpage

\section{Fit projections of the signal channel}
\label{sec:sub:signalchannelprojections}

\begin{figure}[ht]
   \centering
 \includegraphics[width=0.32\textwidth]{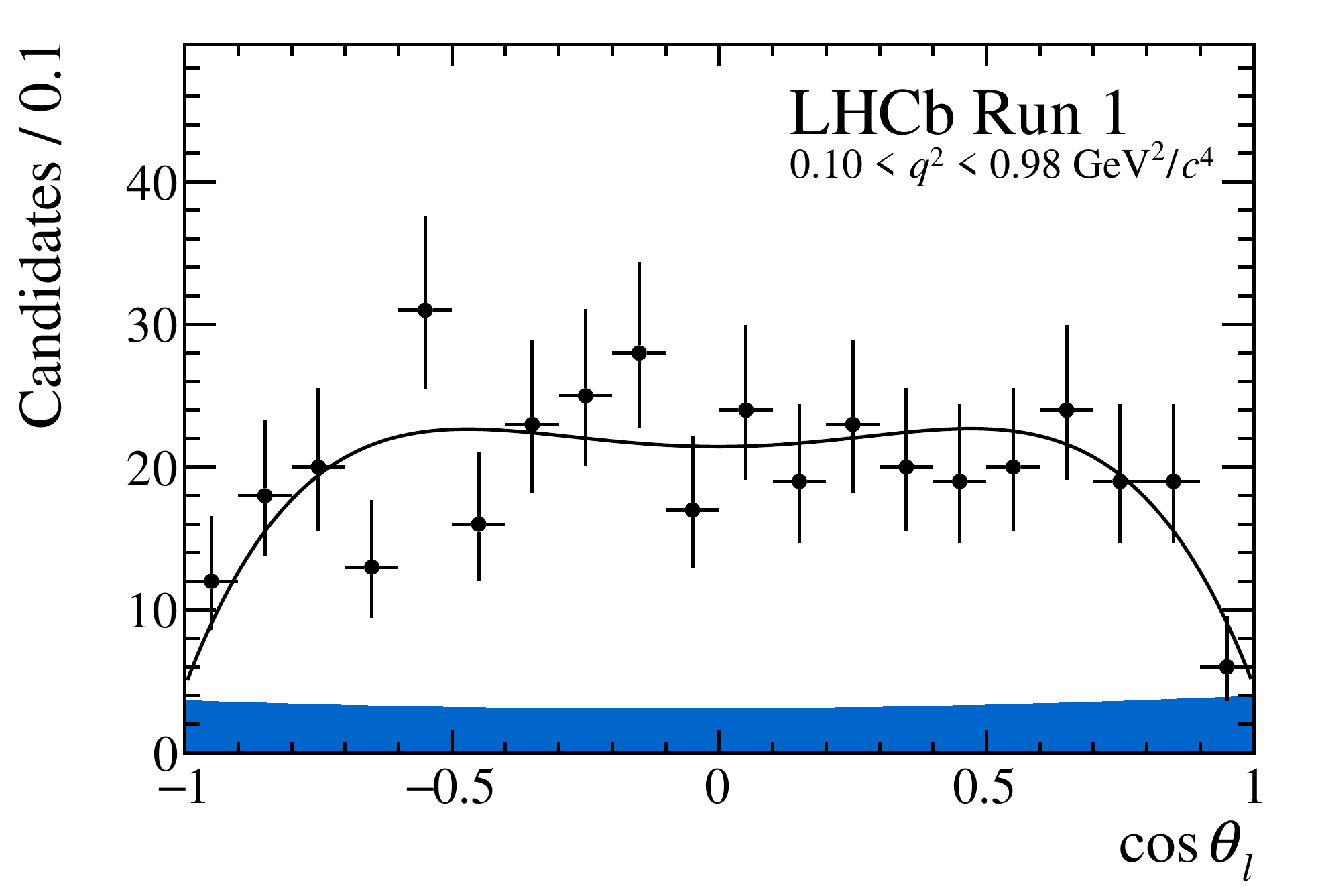}
 \includegraphics[width=0.32\textwidth]{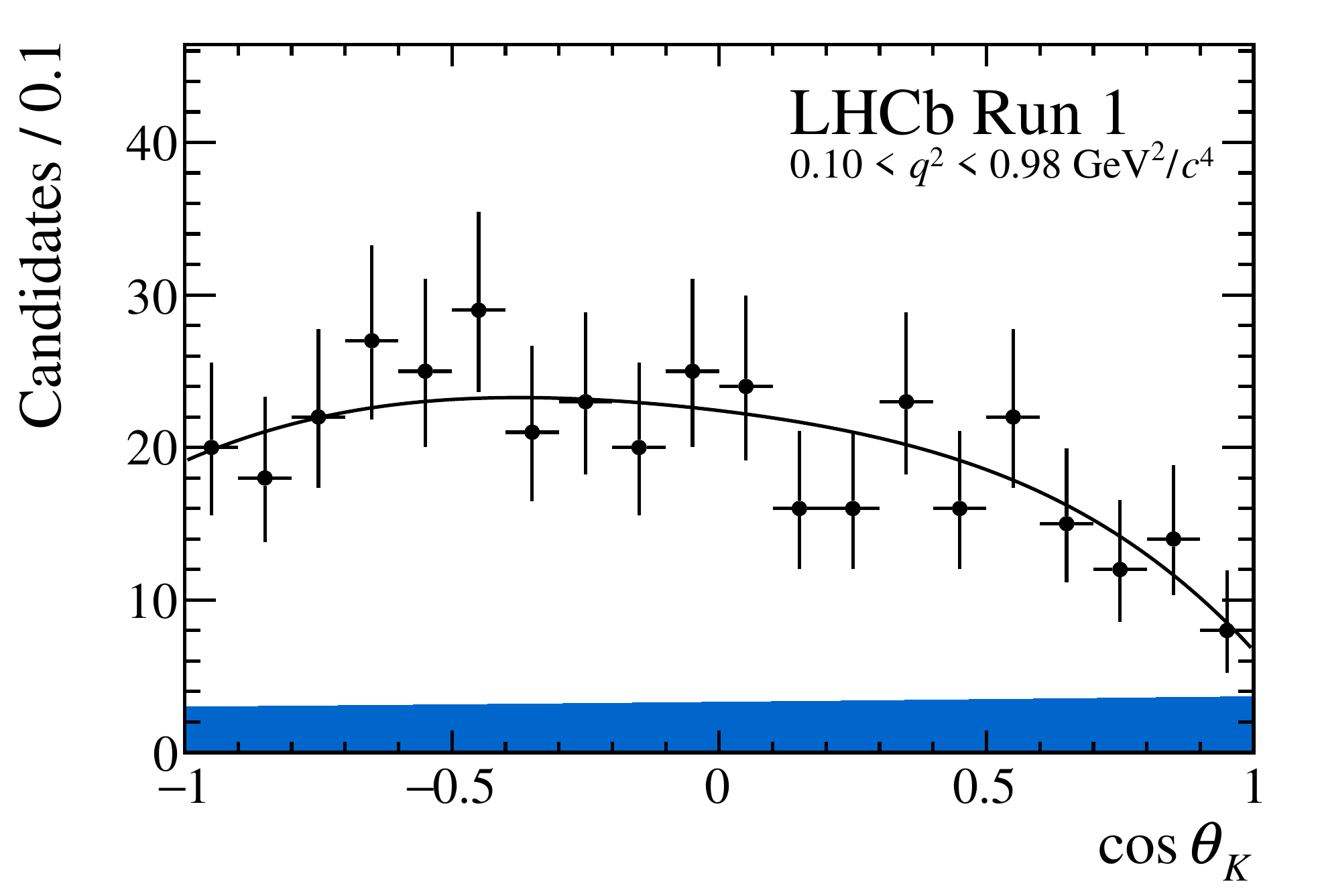}
 \includegraphics[width=0.32\textwidth]{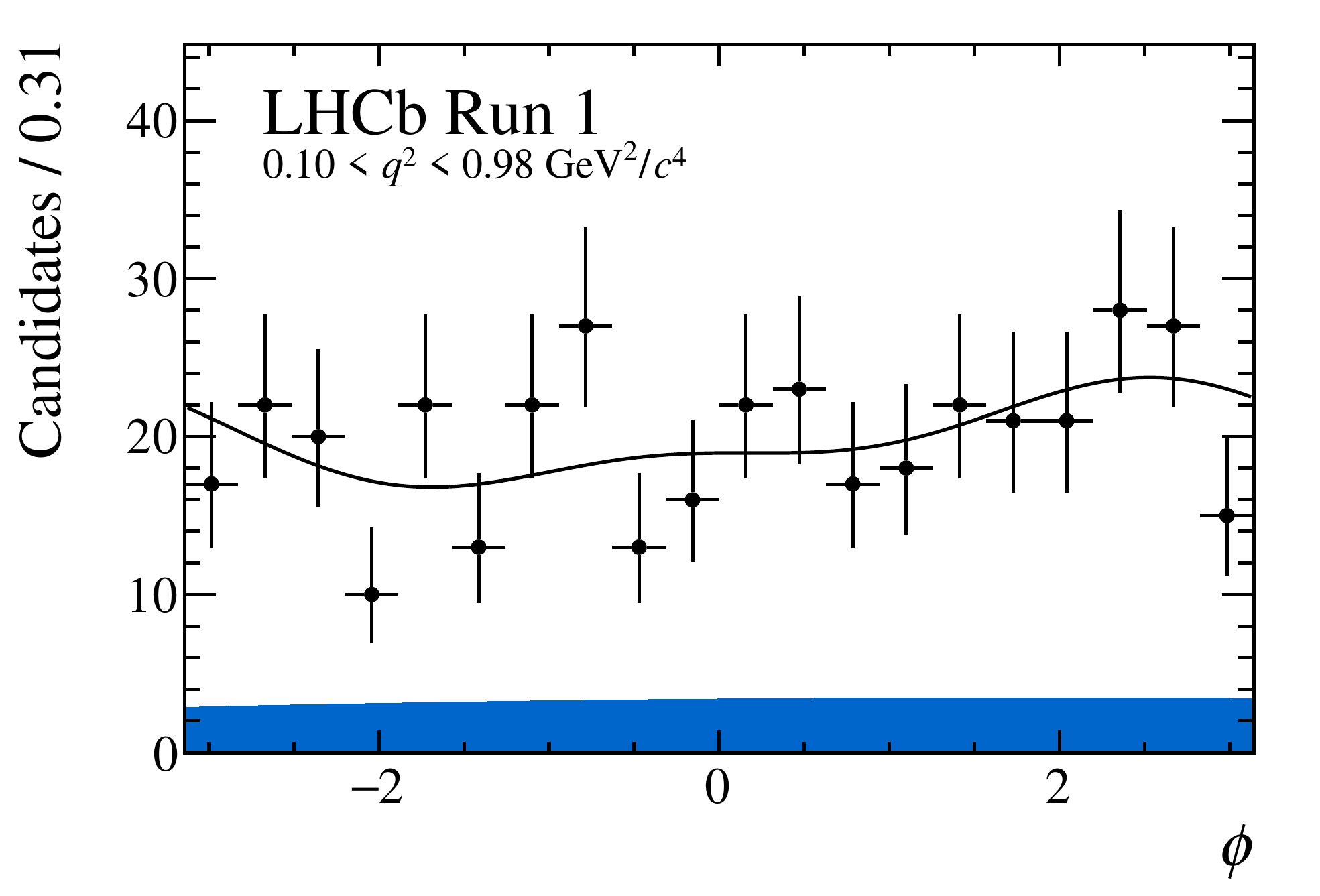}
 \includegraphics[width=0.32\textwidth]{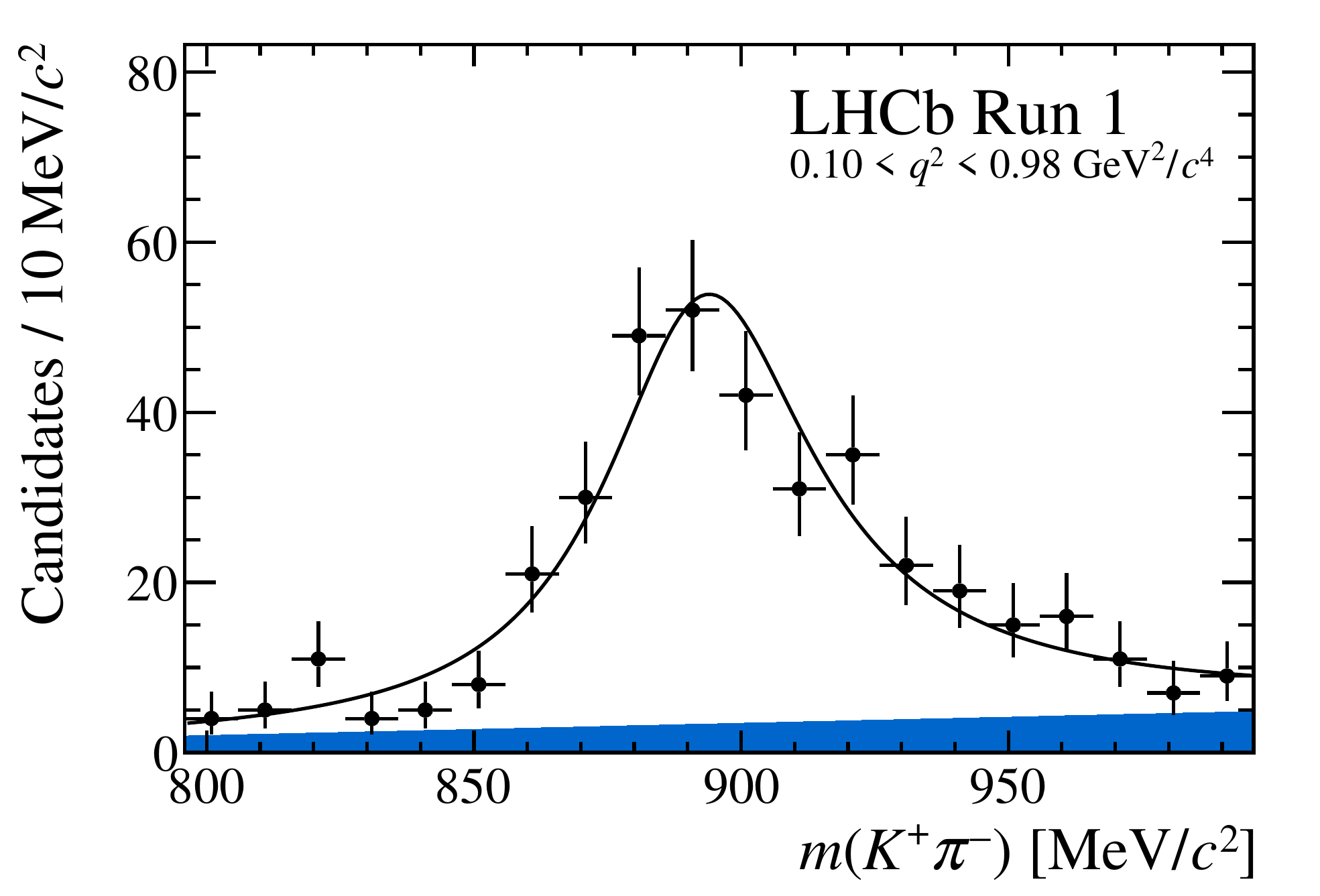}
 \includegraphics[width=0.32\textwidth]{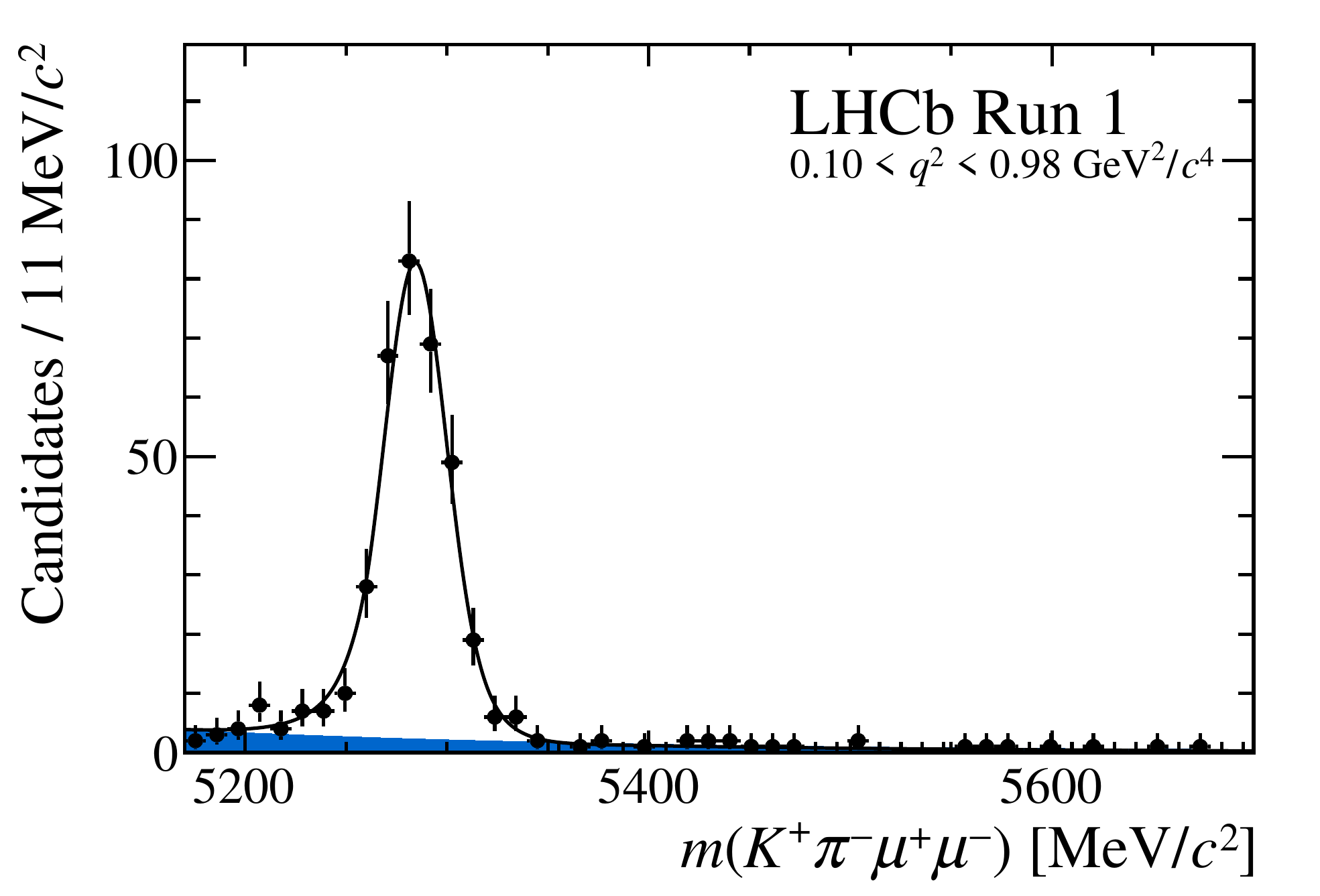}\\[0.5cm]
 \includegraphics[width=0.32\textwidth]{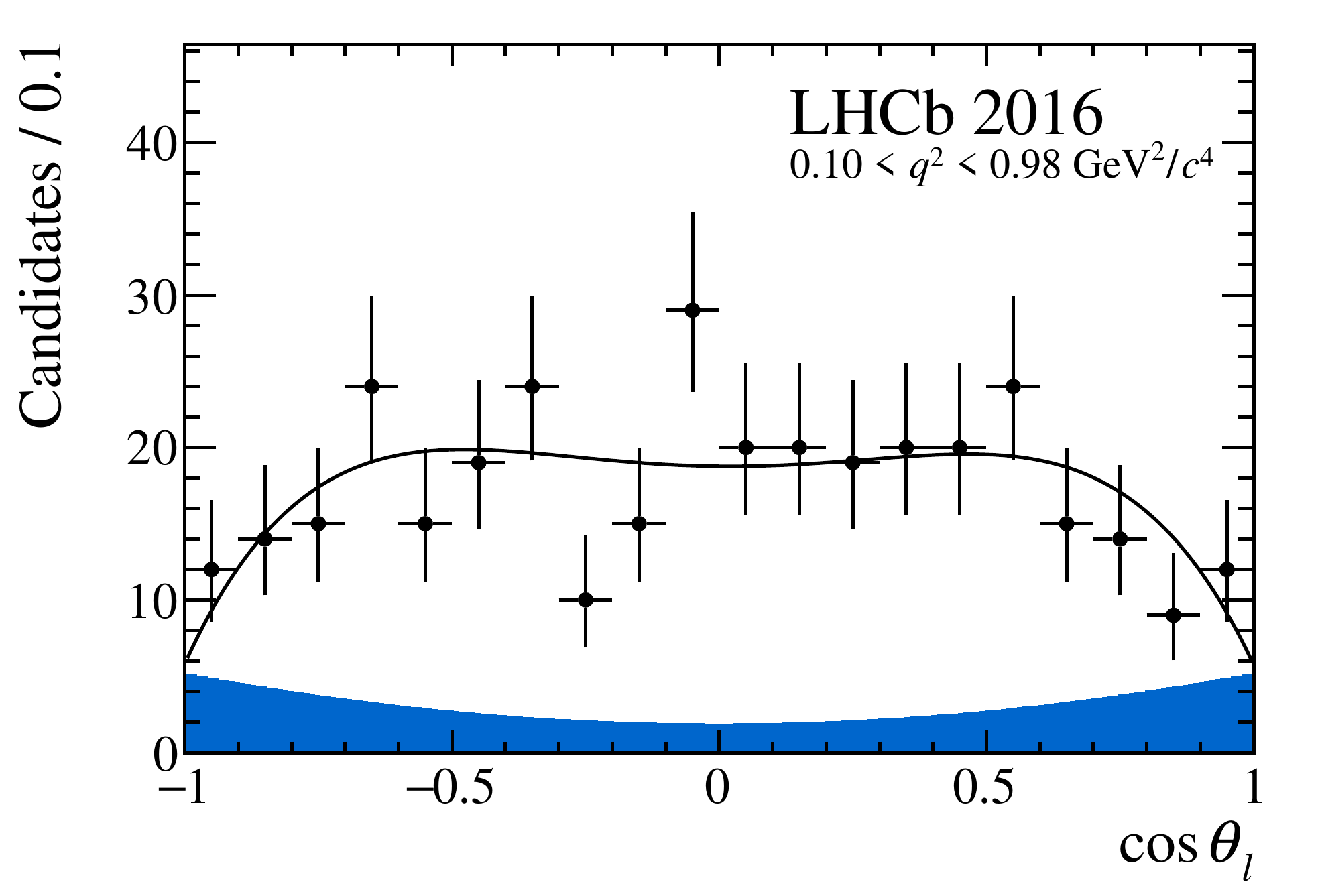}
 \includegraphics[width=0.32\textwidth]{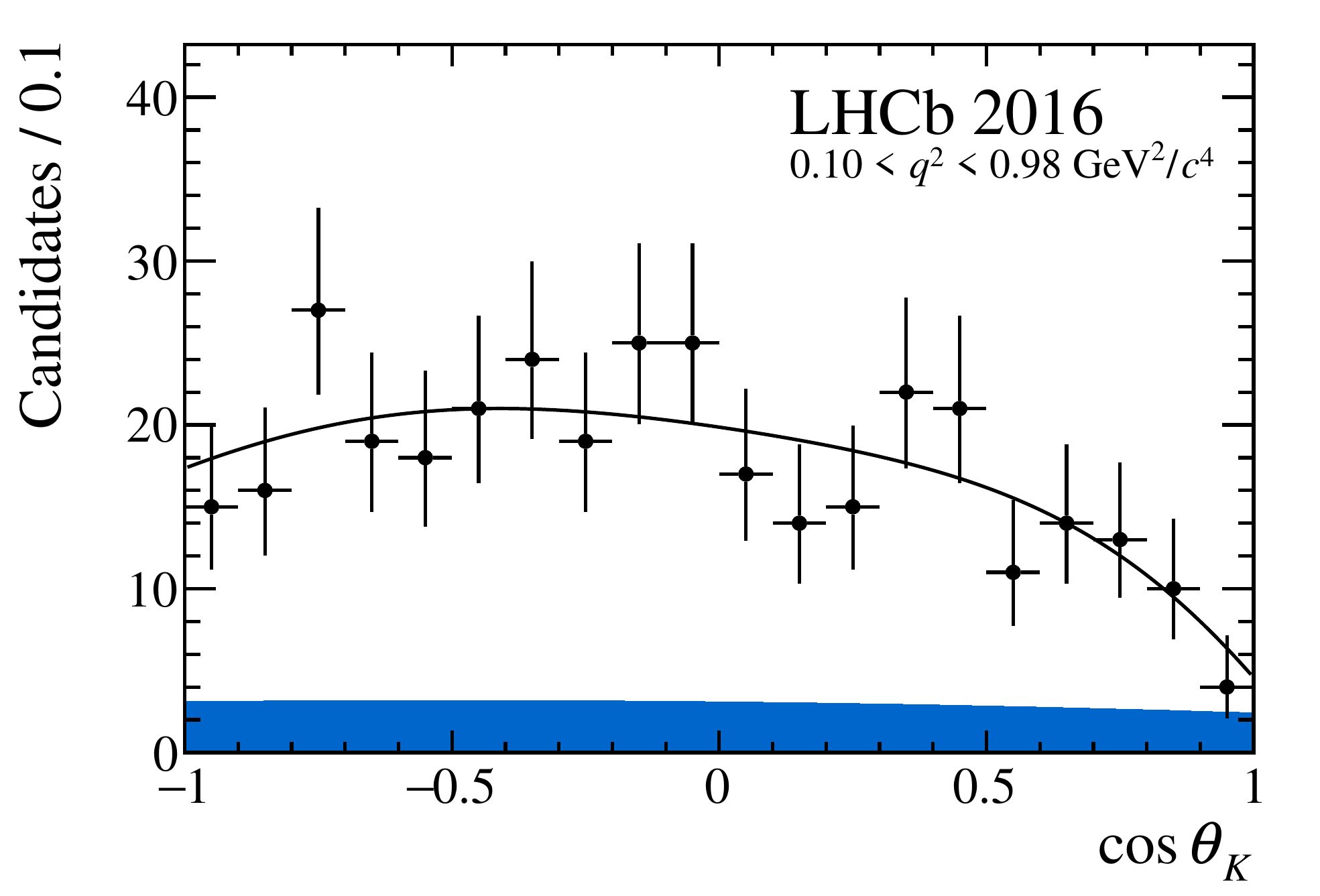}
 \includegraphics[width=0.32\textwidth]{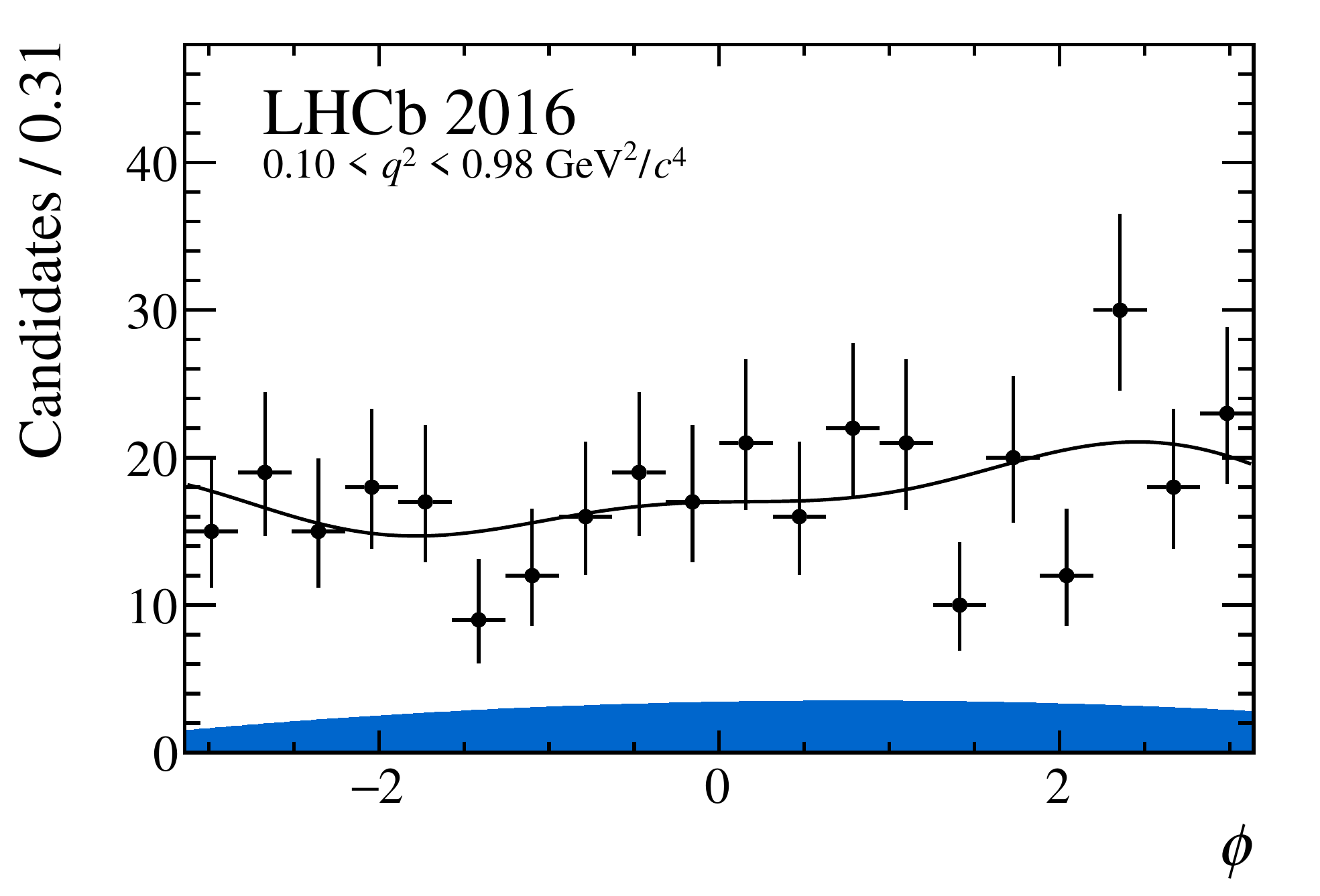}
 \includegraphics[width=0.32\textwidth]{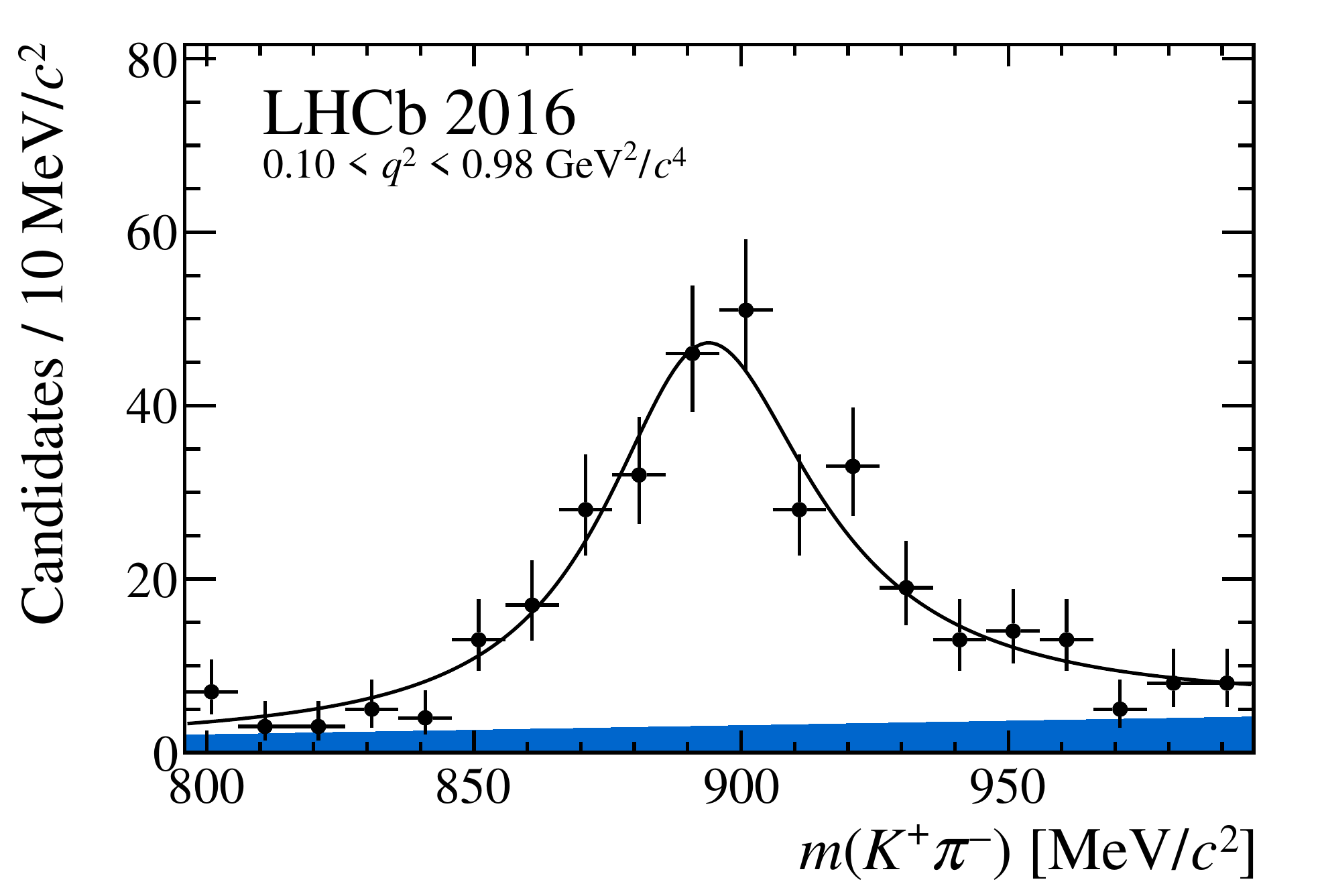}
 \includegraphics[width=0.32\textwidth]{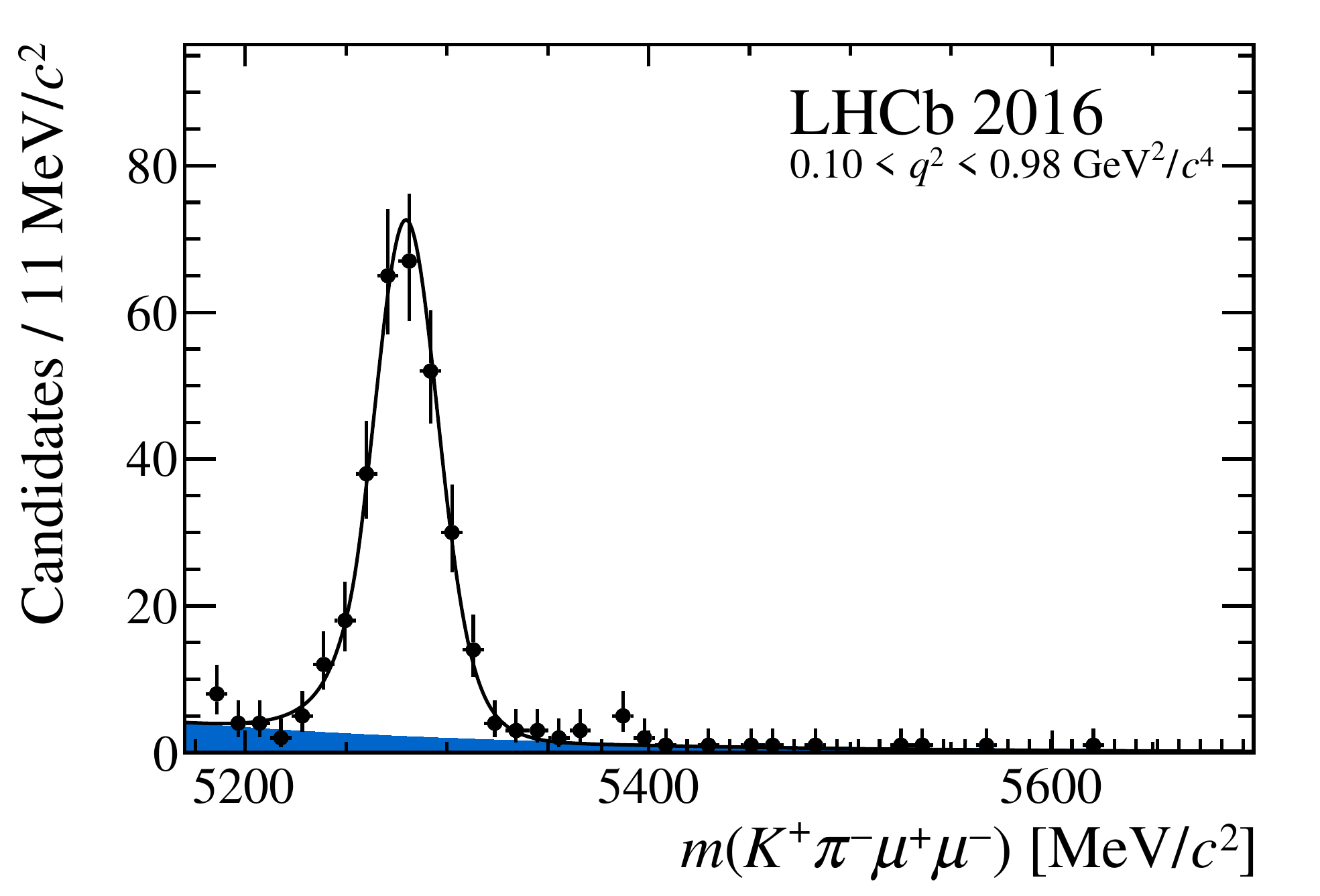}
 \caption{Projections of the fitted probability density function on the decay angles, \Mkpi and \Mkpimm for the bin $0.10<q^2<0.98\gevgevcccc$. The blue shaded region indicates background. \label{fig:projectionsa}}
 \end{figure}

\begin{figure}
   \centering
 \includegraphics[width=0.32\textwidth]{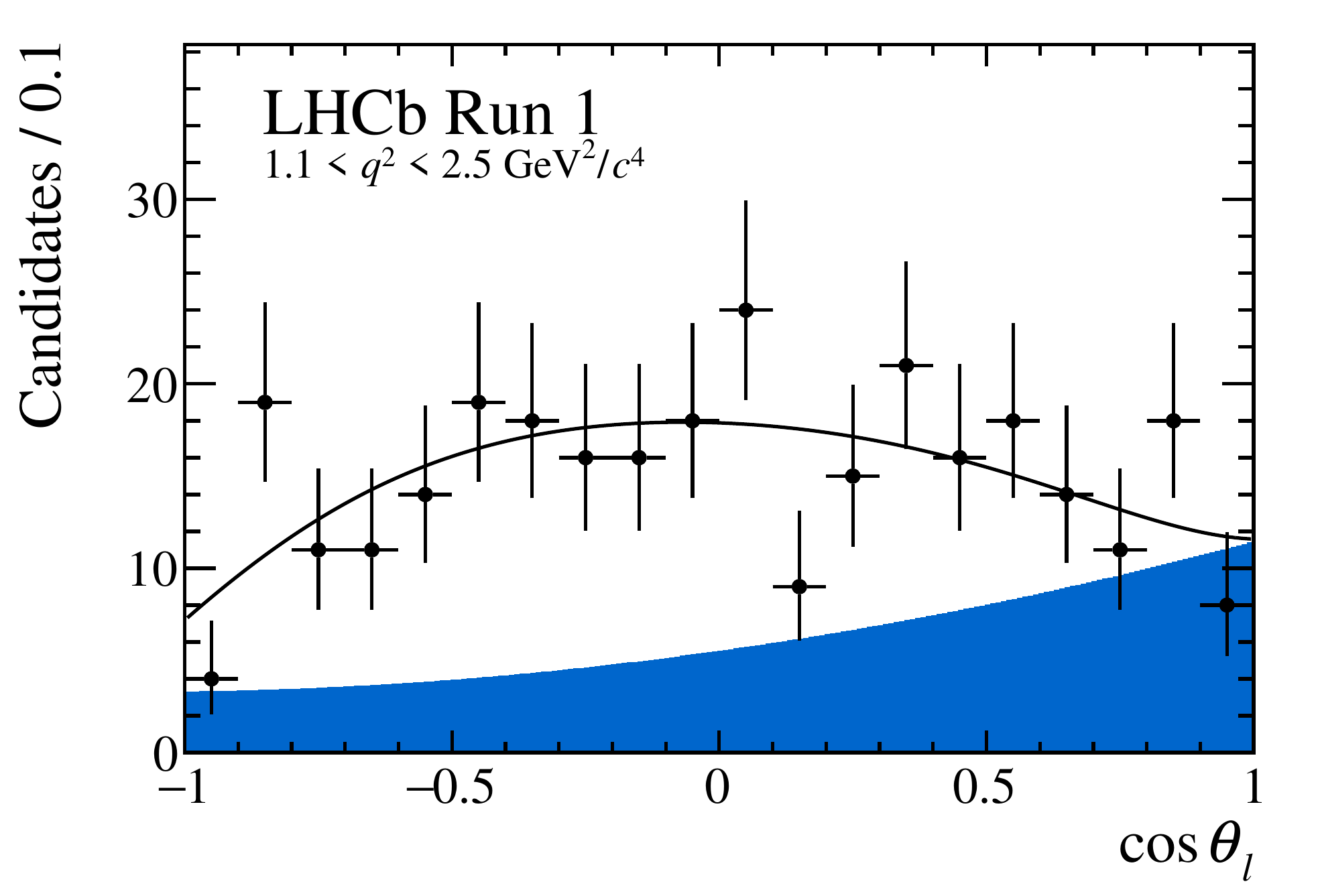}
 \includegraphics[width=0.32\textwidth]{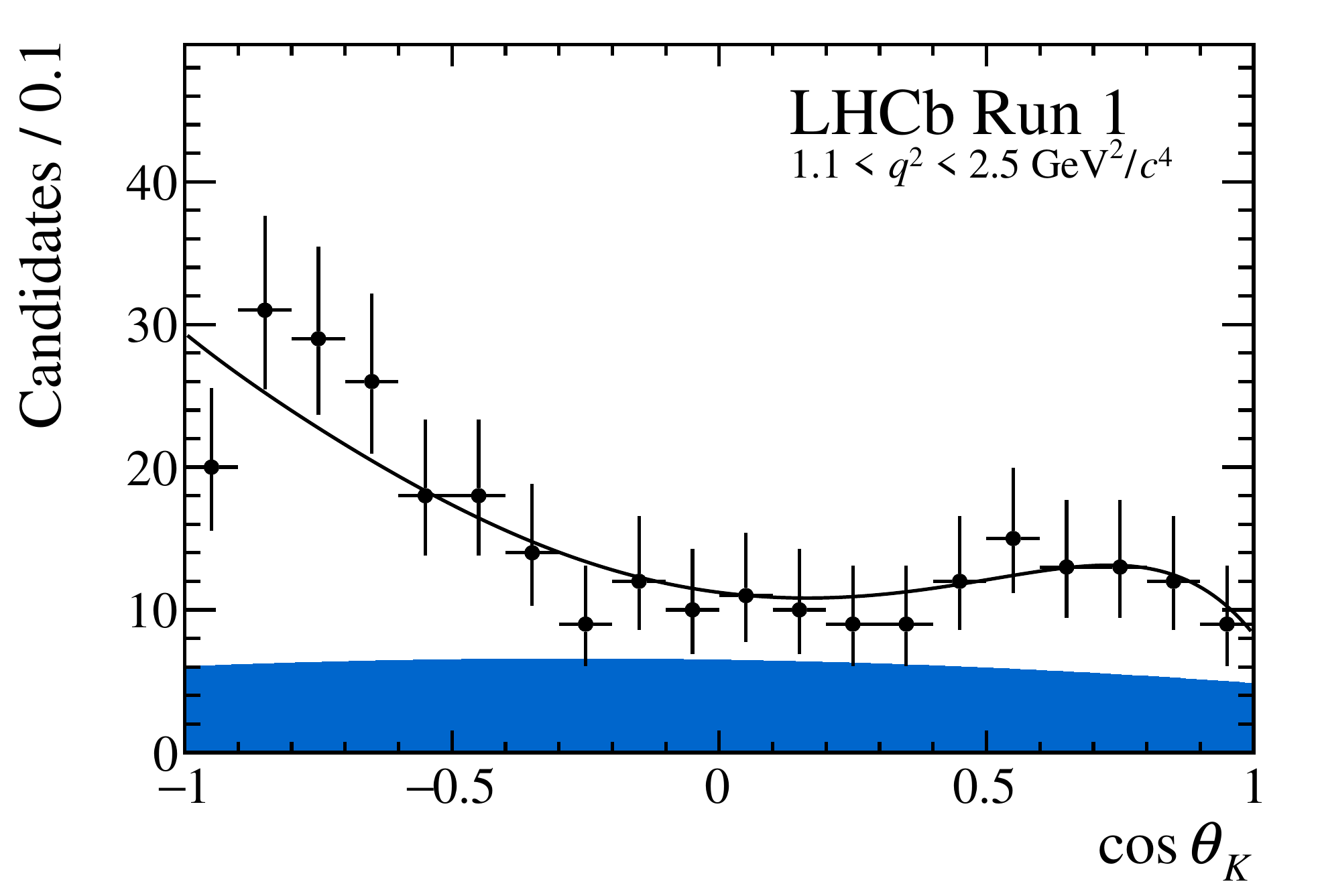}
 \includegraphics[width=0.32\textwidth]{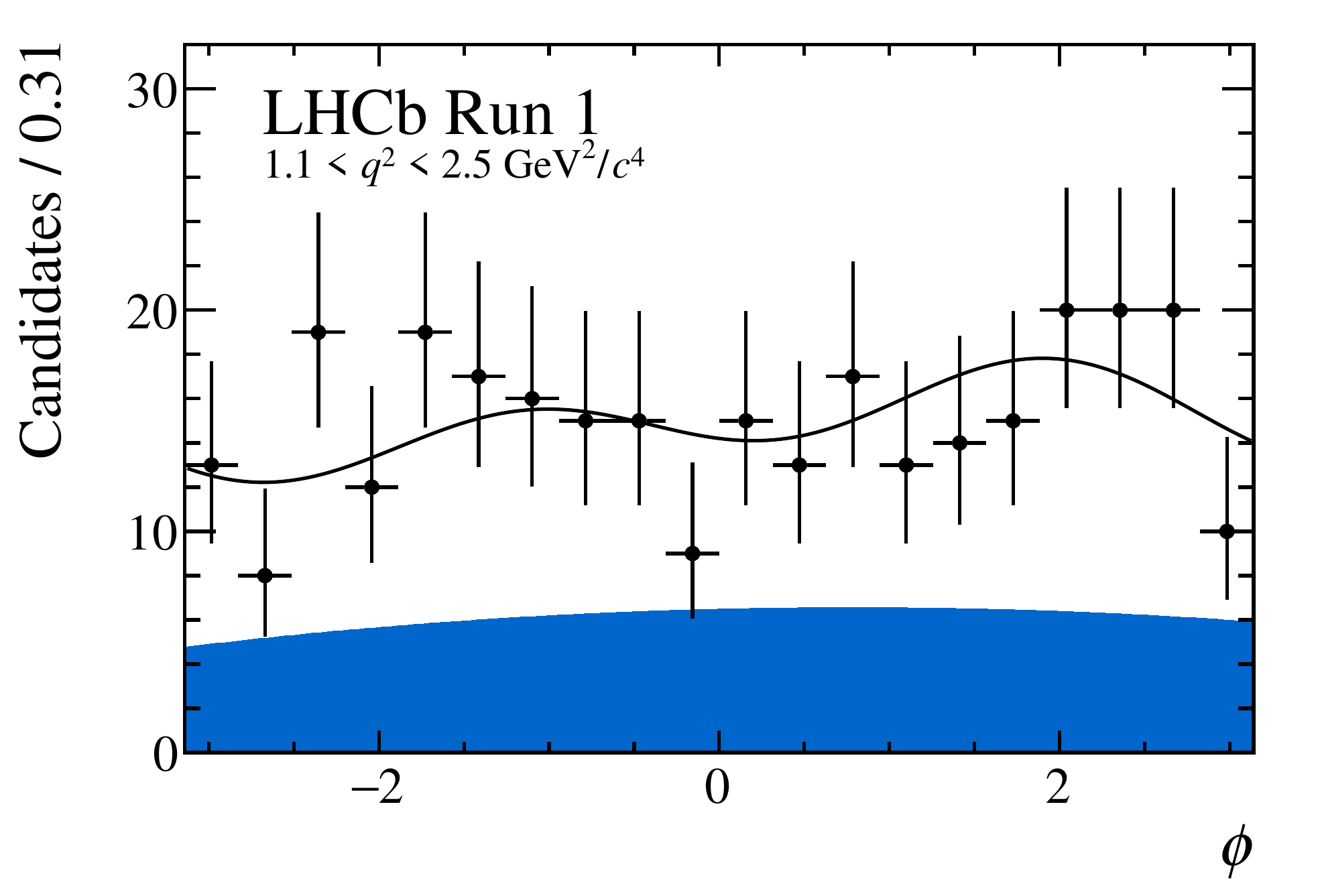}
 \includegraphics[width=0.32\textwidth]{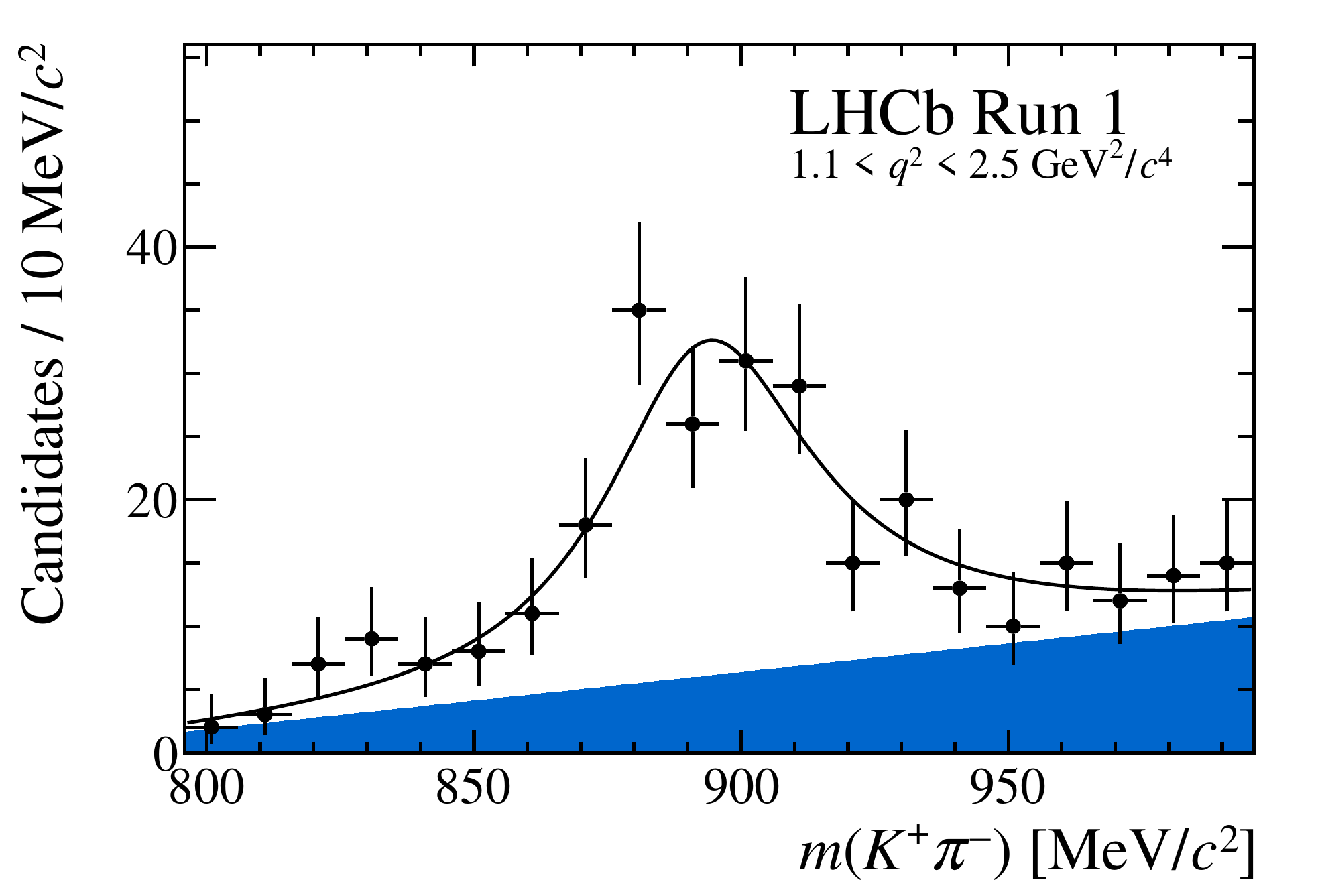}
 \includegraphics[width=0.32\textwidth]{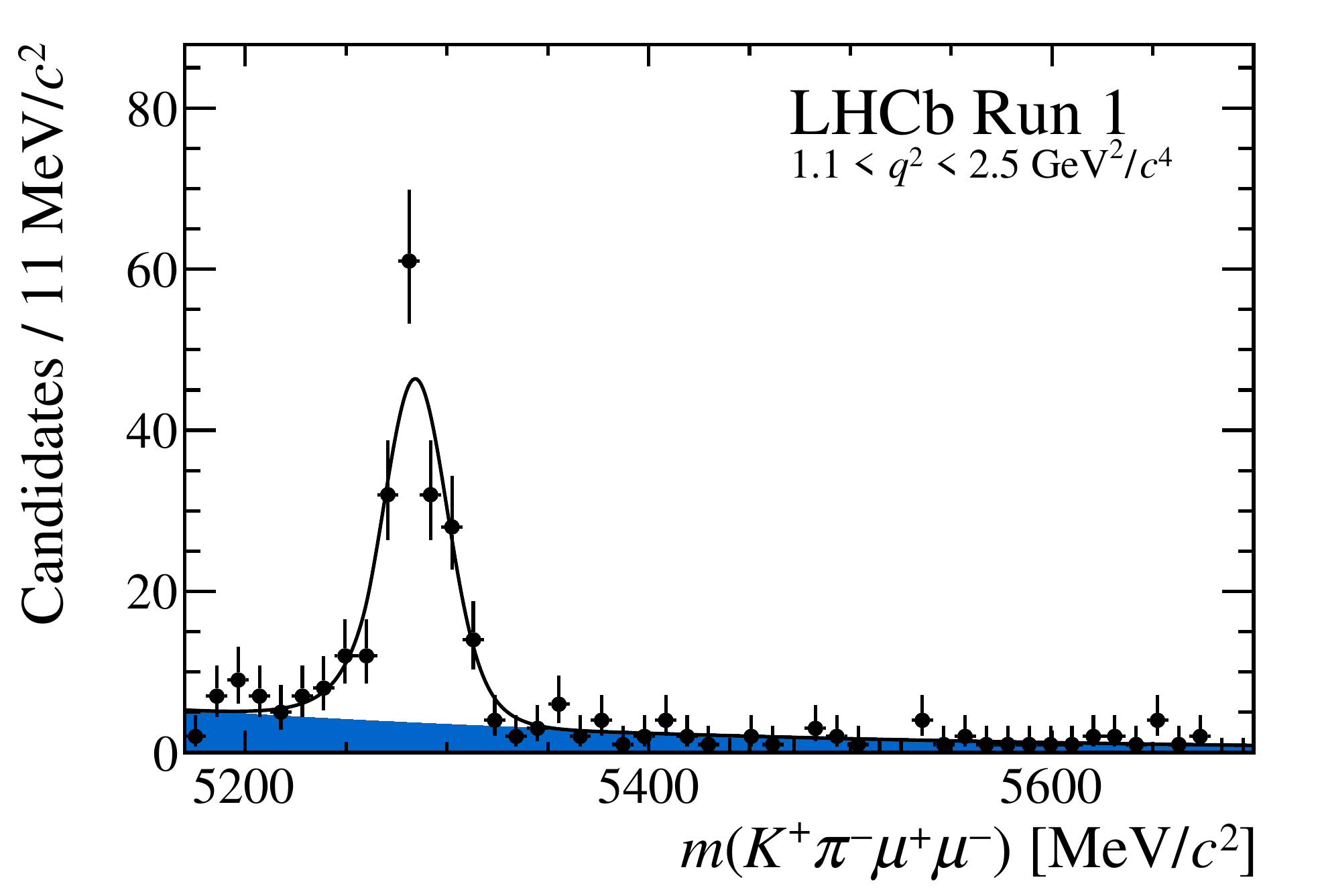}\\[0.5cm]
 \includegraphics[width=0.32\textwidth]{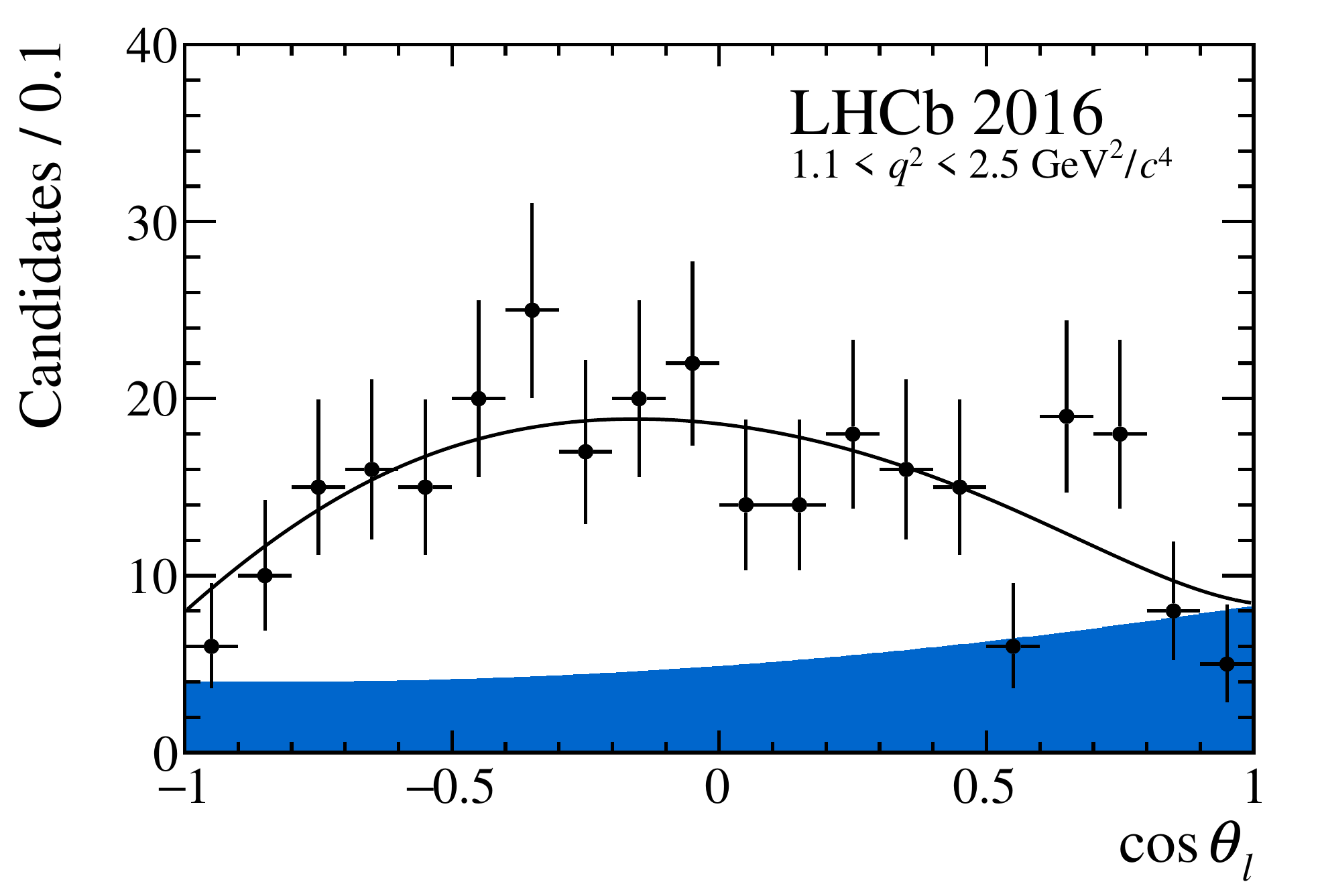}
 \includegraphics[width=0.32\textwidth]{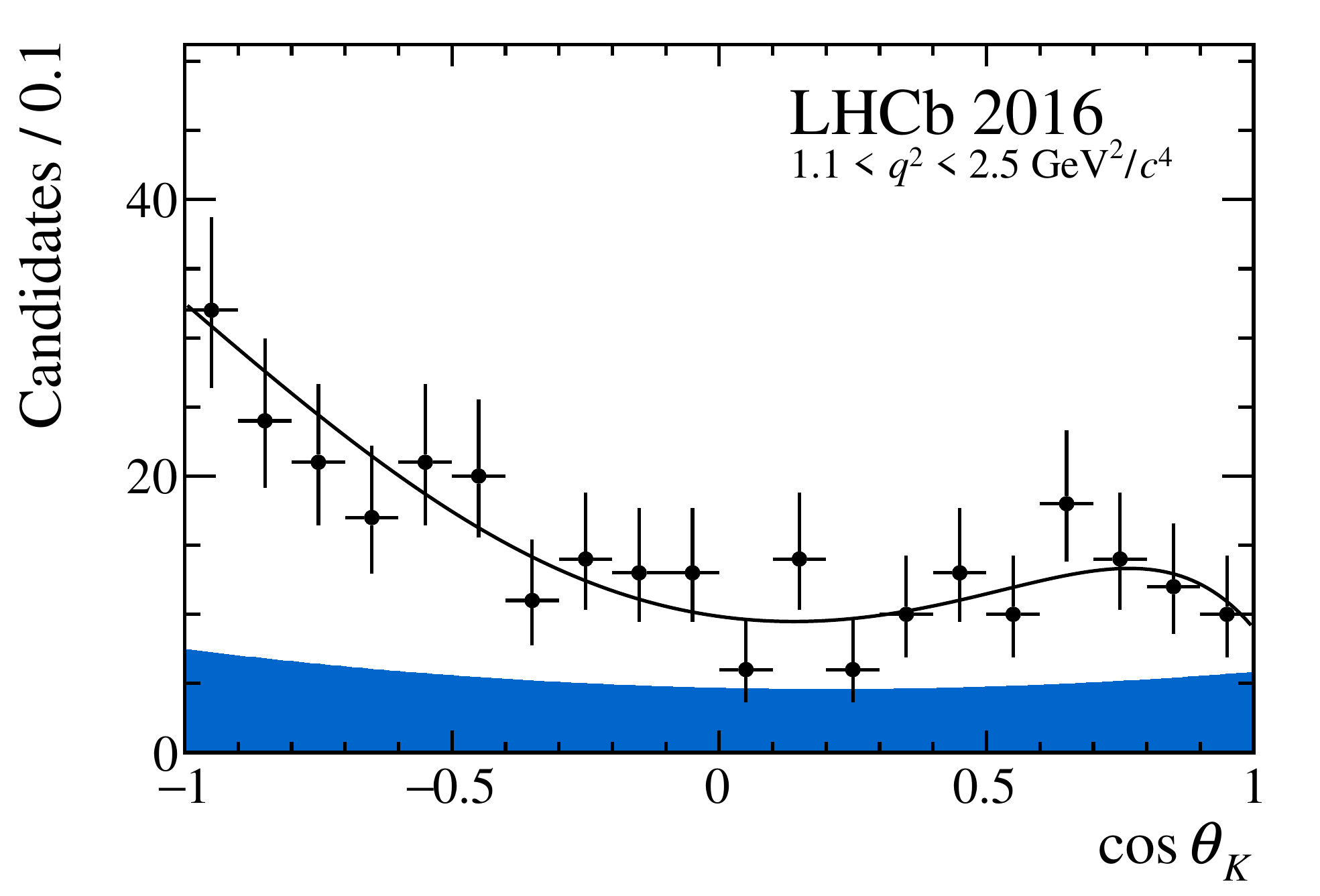}
 \includegraphics[width=0.32\textwidth]{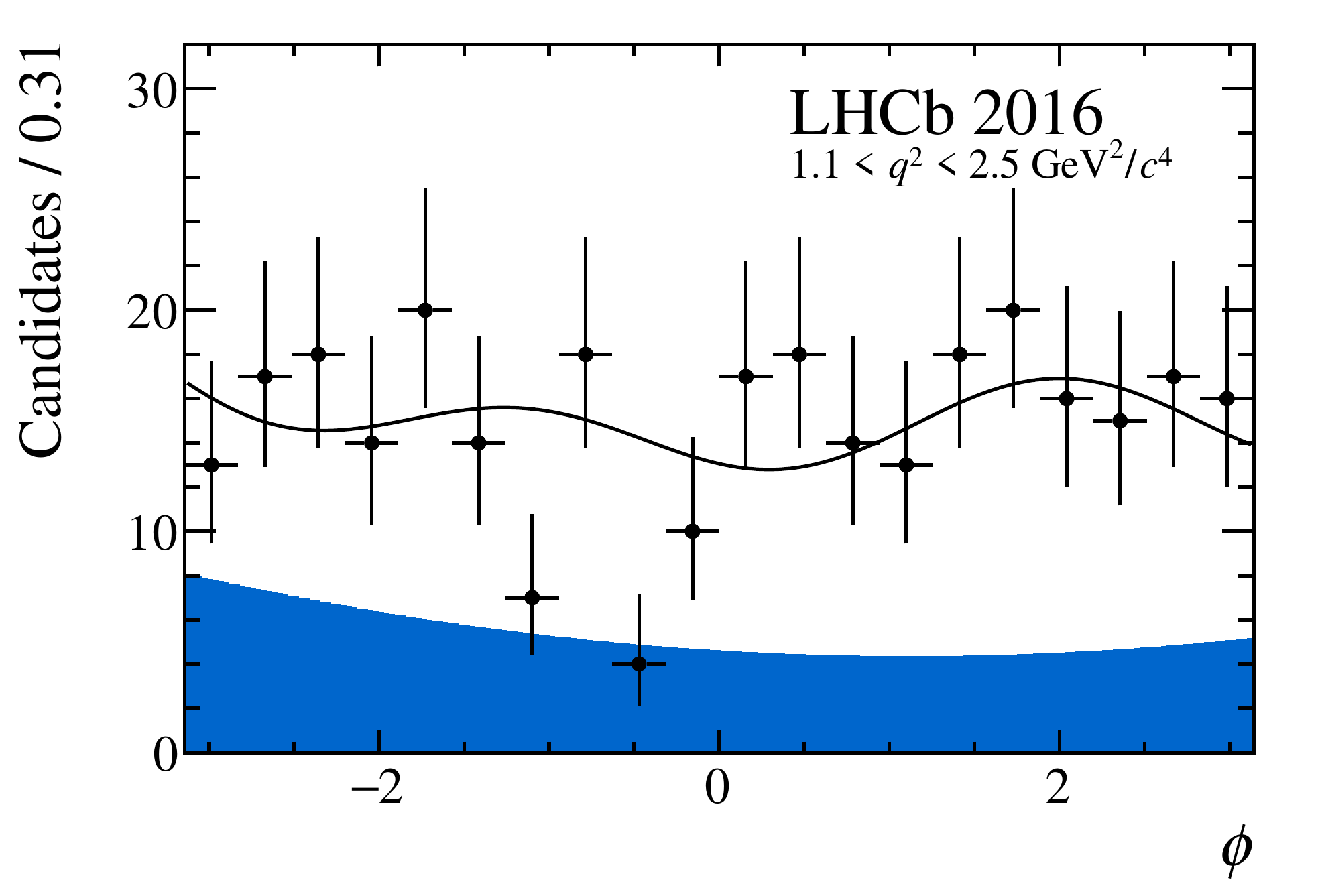}
 \includegraphics[width=0.32\textwidth]{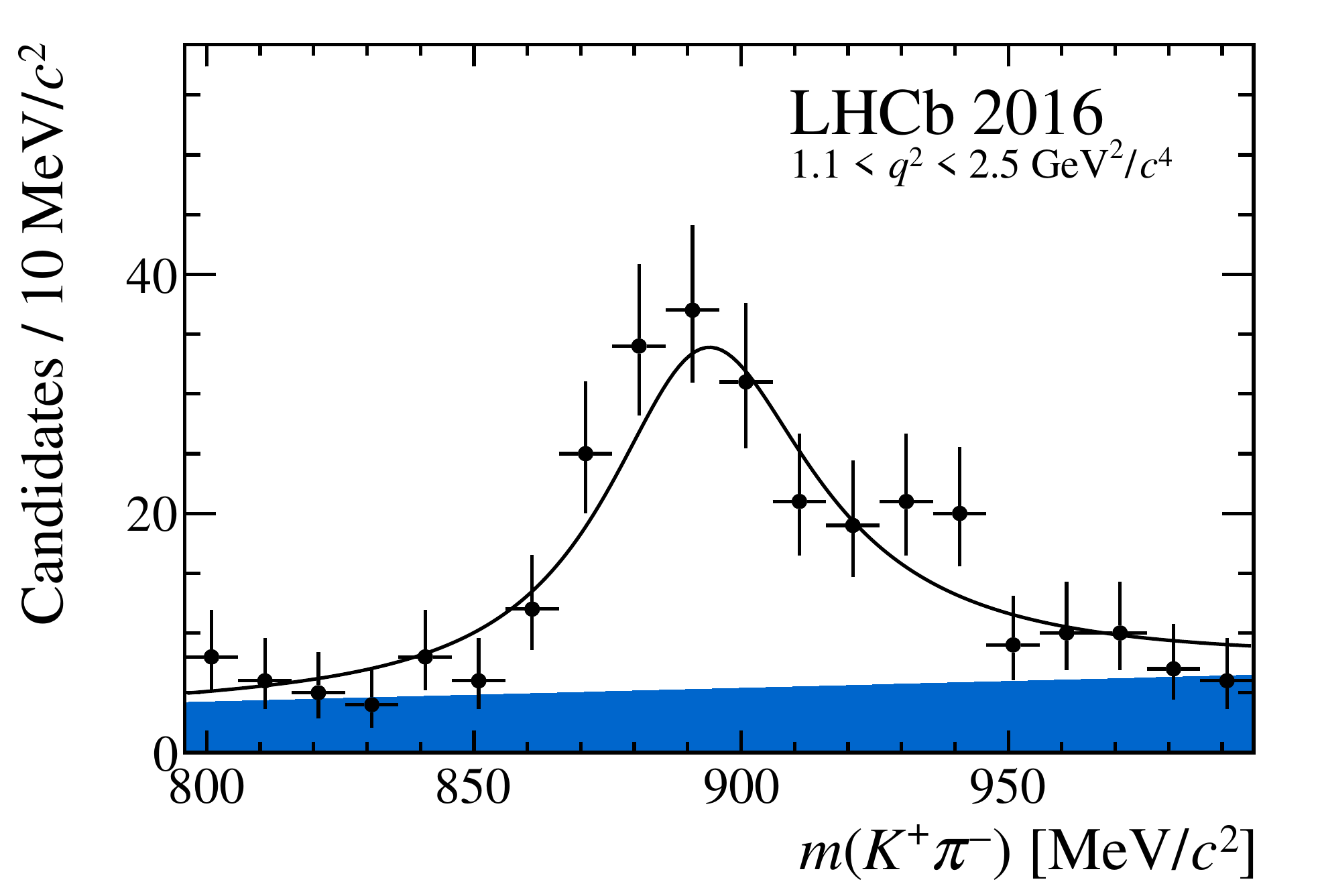}
 \includegraphics[width=0.32\textwidth]{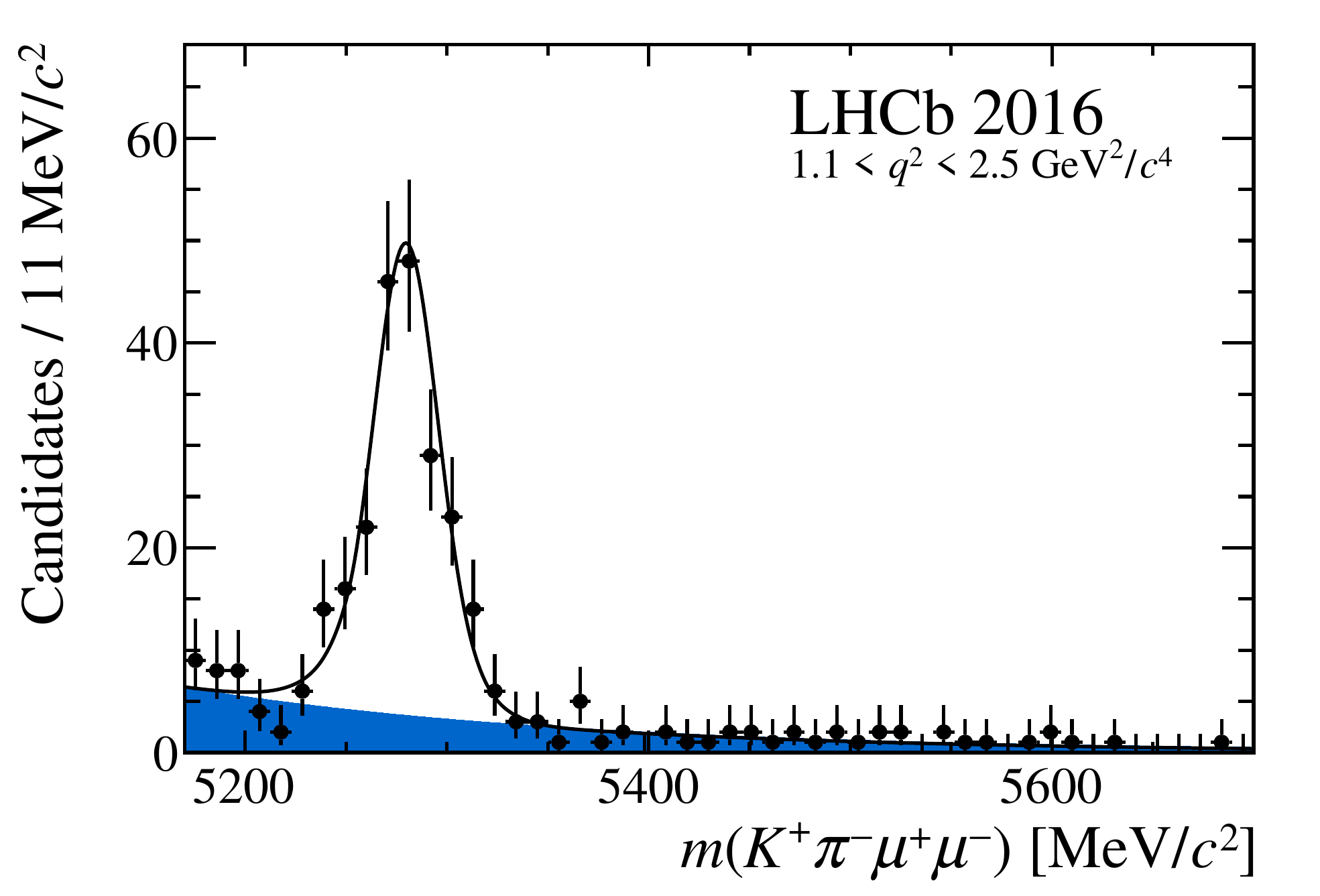}
 \caption{Projections of the fitted probability density function on the decay angles, \Mkpi and \Mkpimm for the bin $1.1<q^2<2.5\gevgevcccc$. The blue shaded region indicates background. \label{fig:projectionsb}}
 \end{figure}

\begin{figure}
   \centering
 \includegraphics[width=0.32\textwidth]{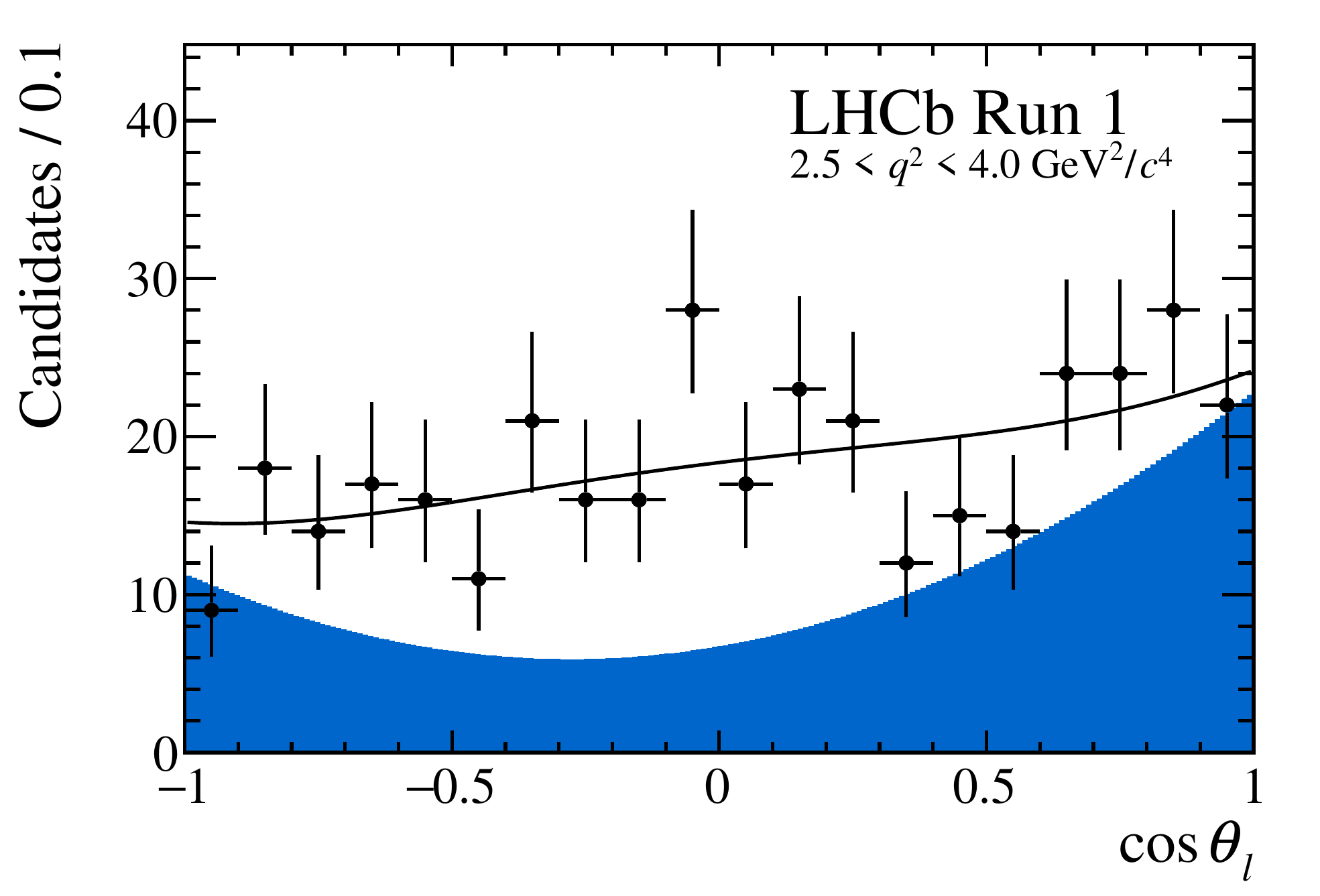}
 \includegraphics[width=0.32\textwidth]{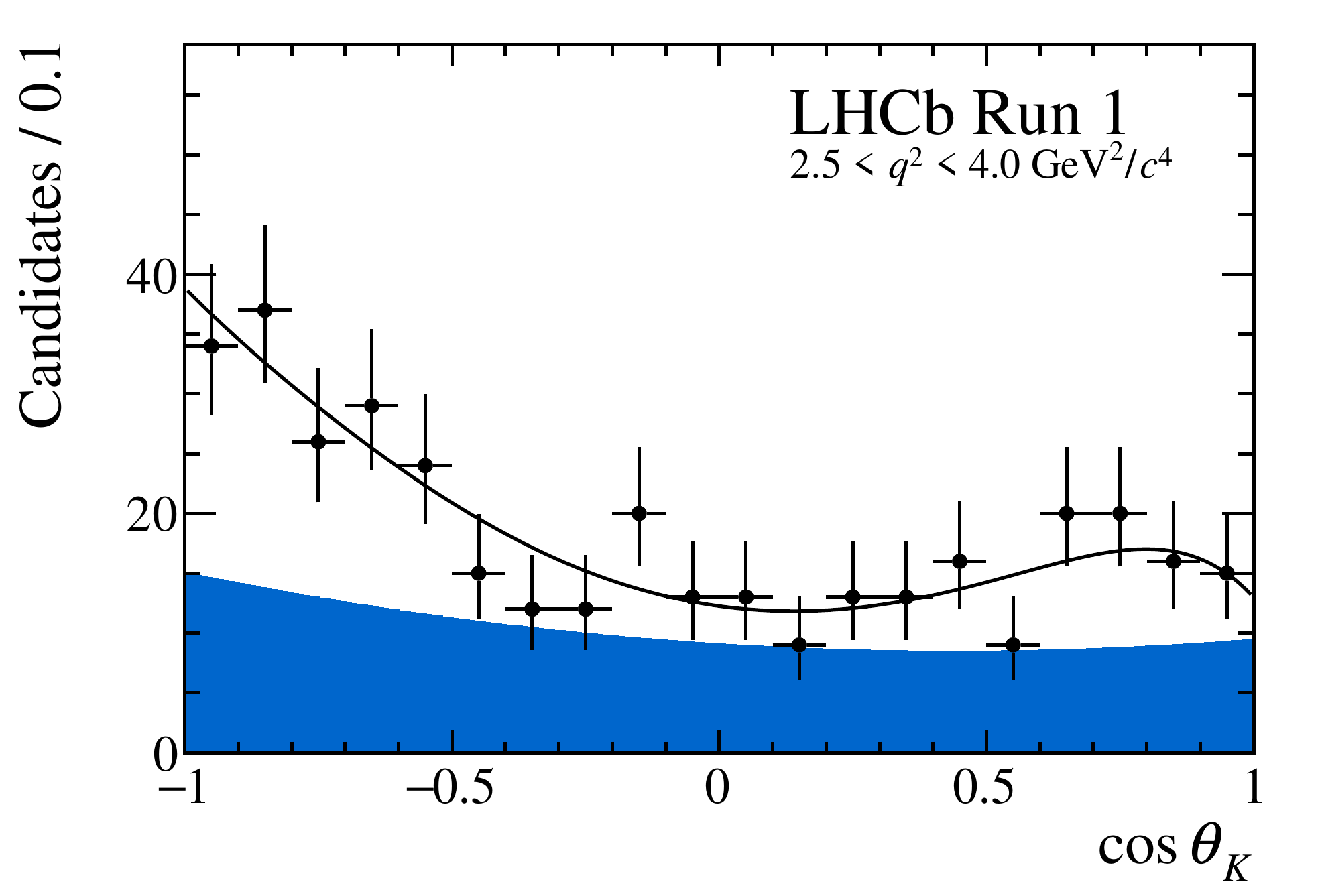}
 \includegraphics[width=0.32\textwidth]{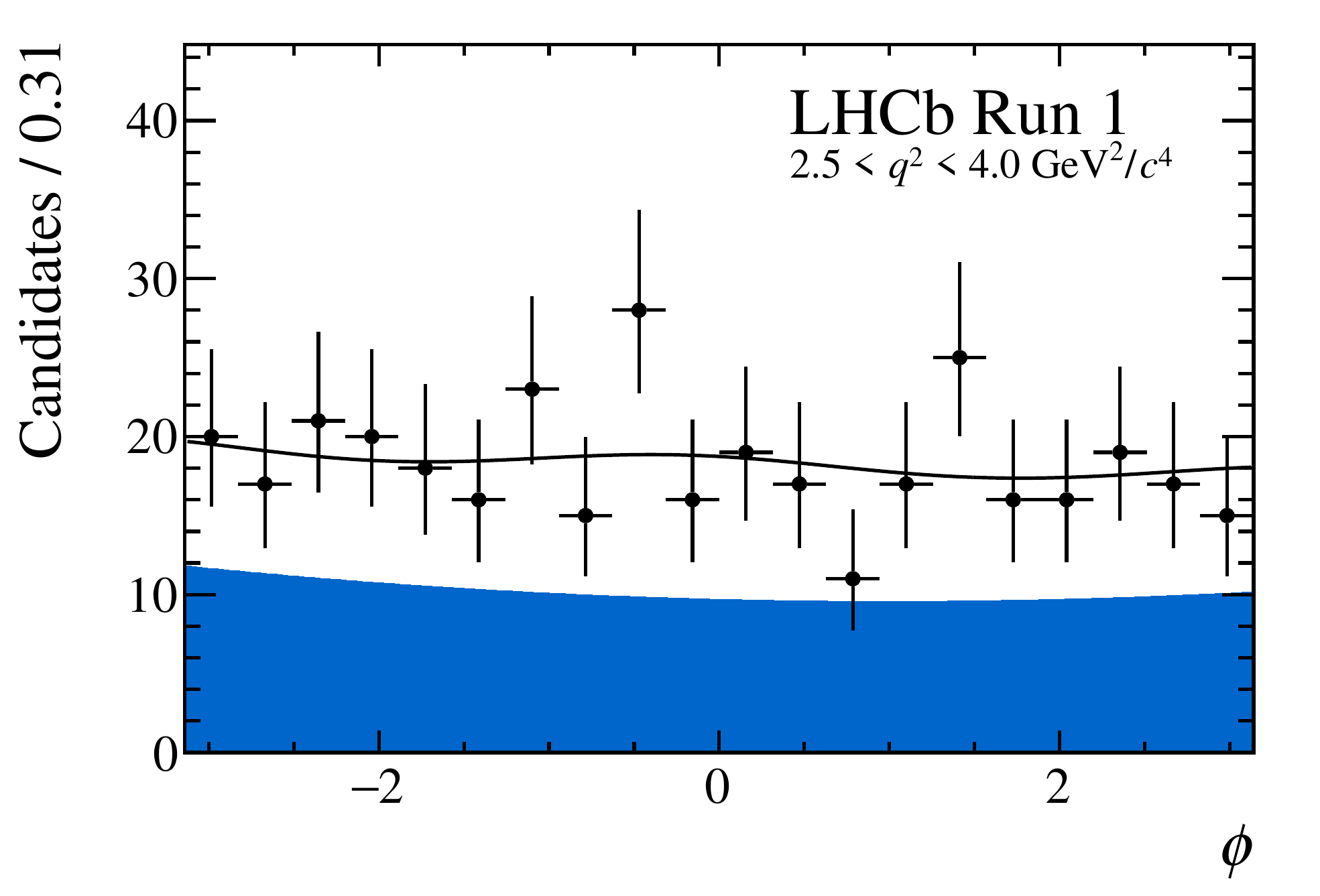}
 \includegraphics[width=0.32\textwidth]{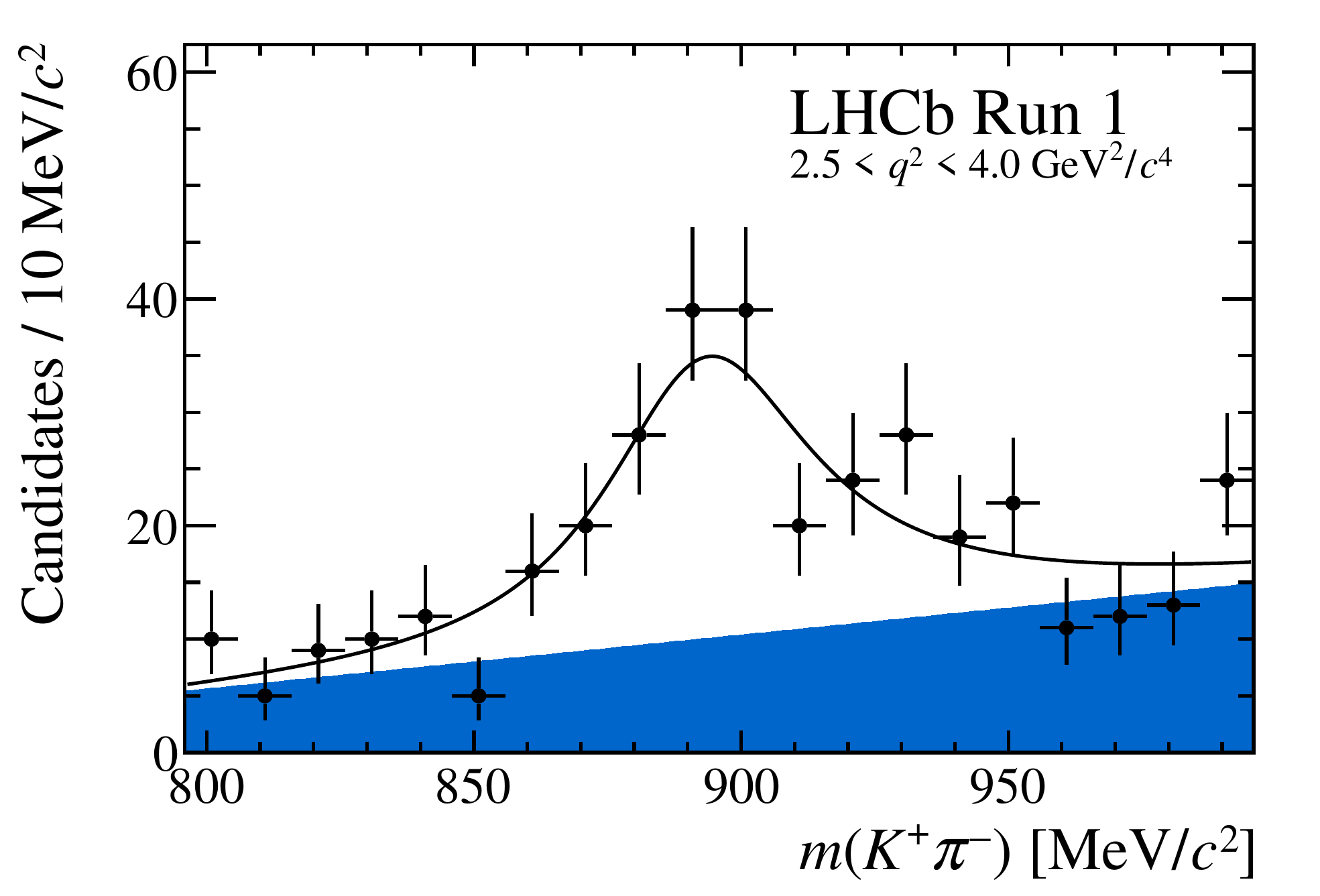}
 \includegraphics[width=0.32\textwidth]{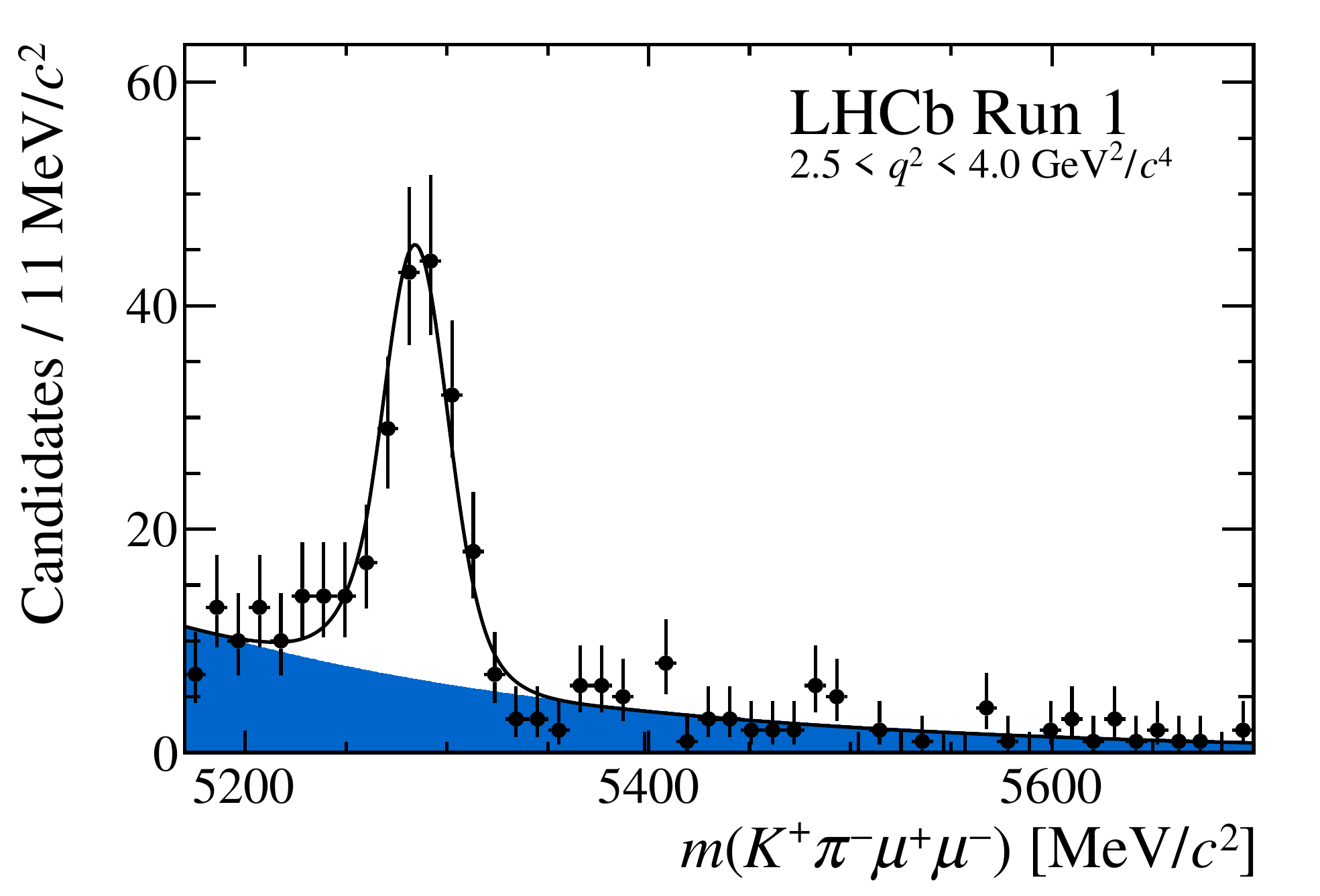}\\[0.5cm]
 \includegraphics[width=0.32\textwidth]{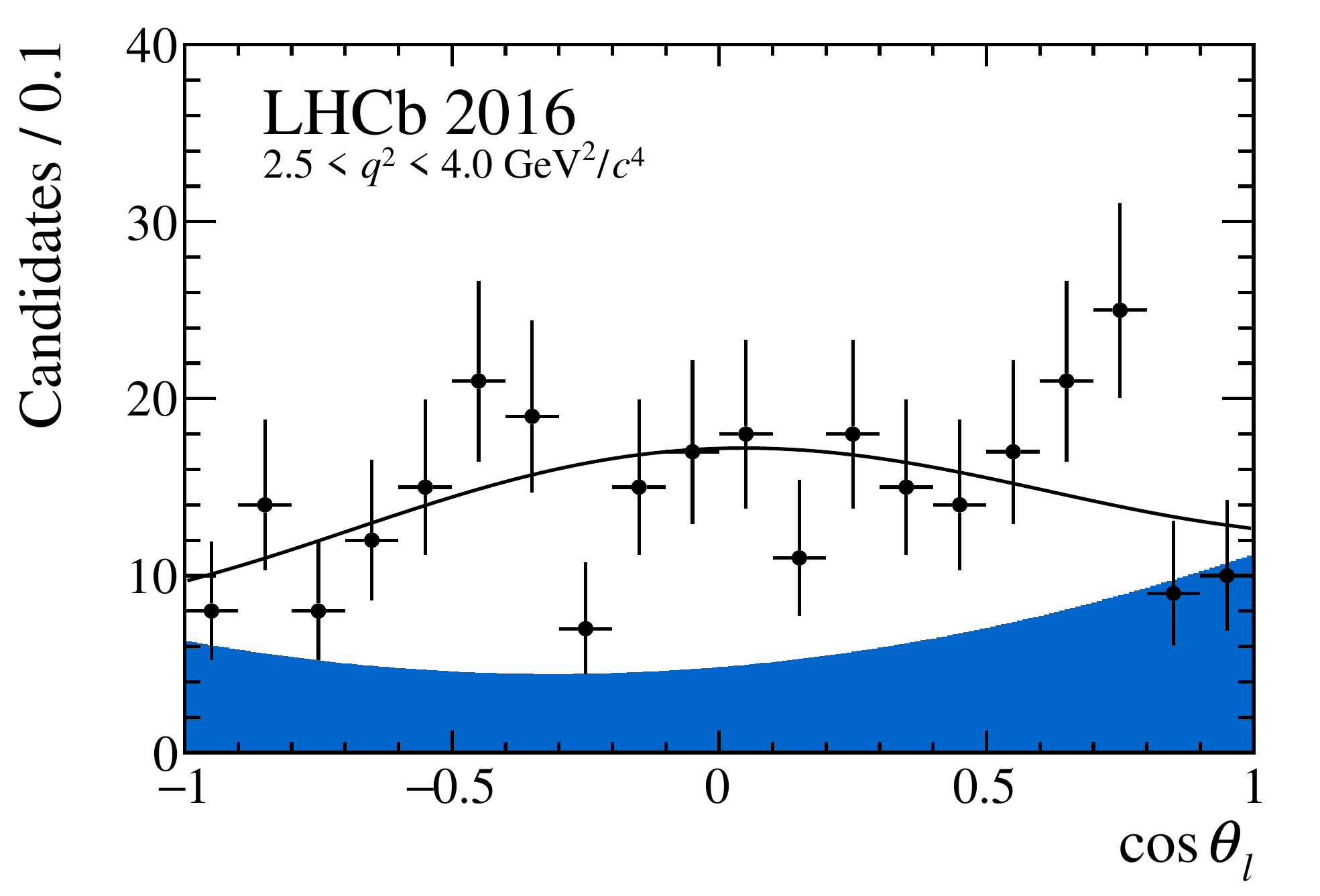}
 \includegraphics[width=0.32\textwidth]{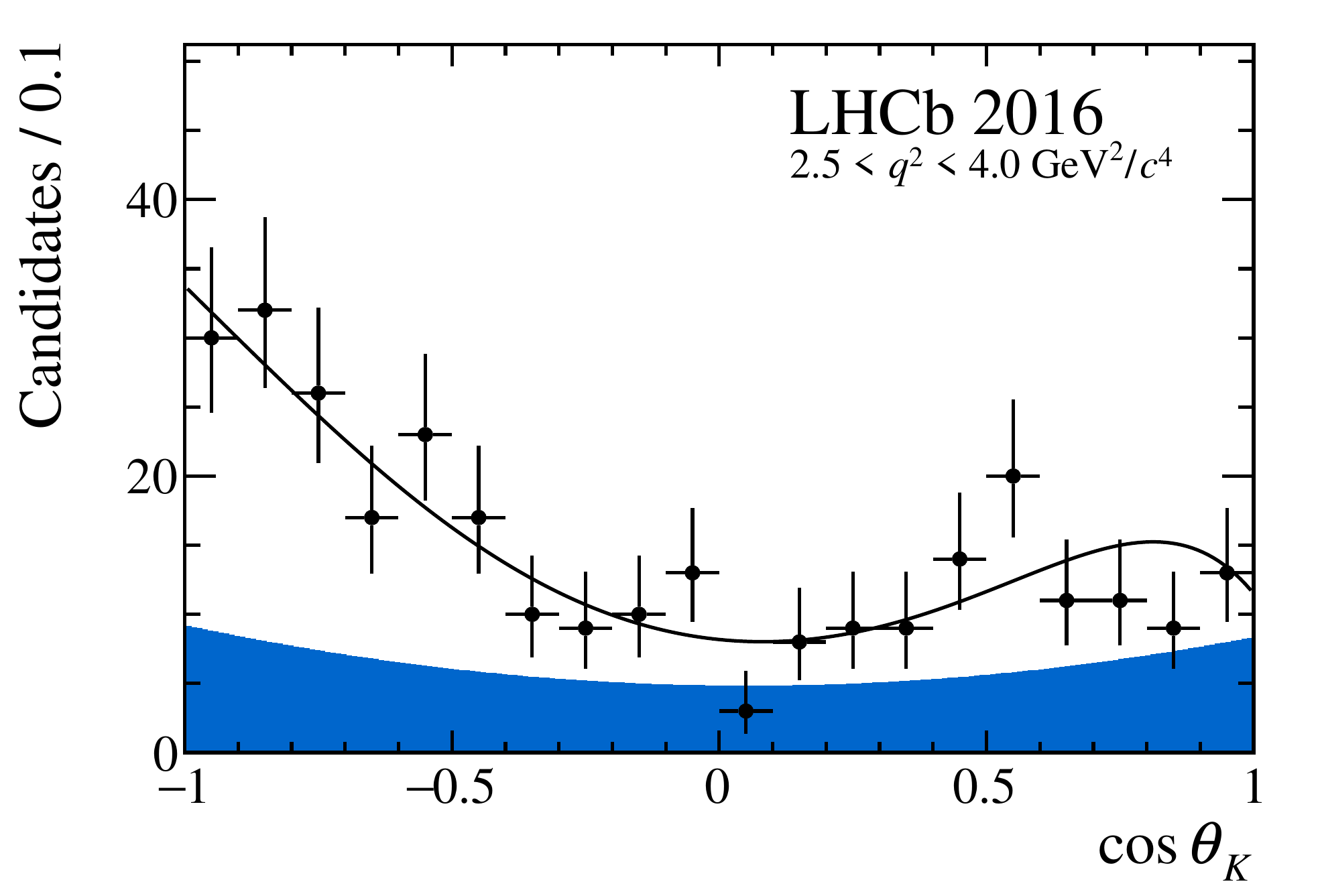}
 \includegraphics[width=0.32\textwidth]{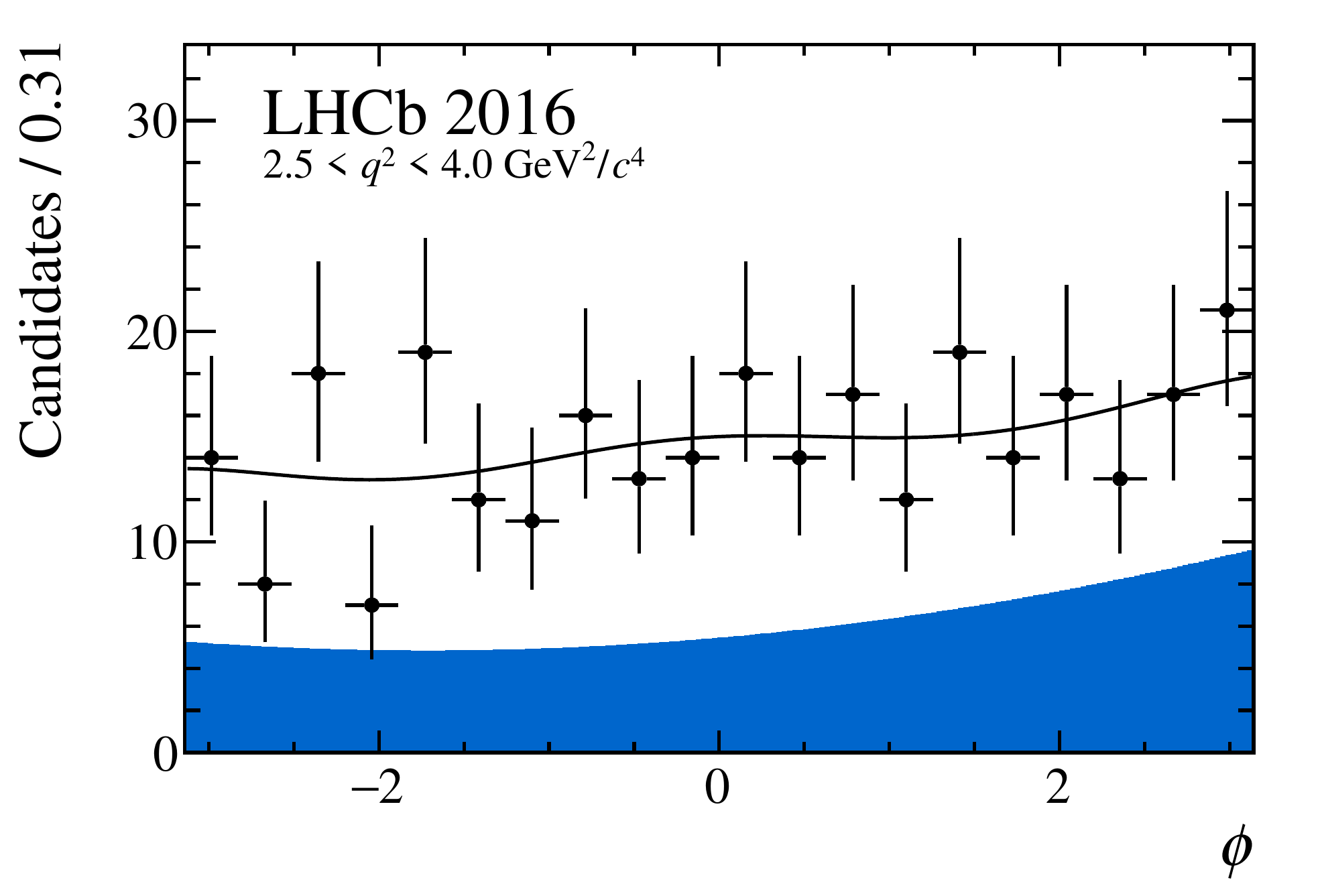}
 \includegraphics[width=0.32\textwidth]{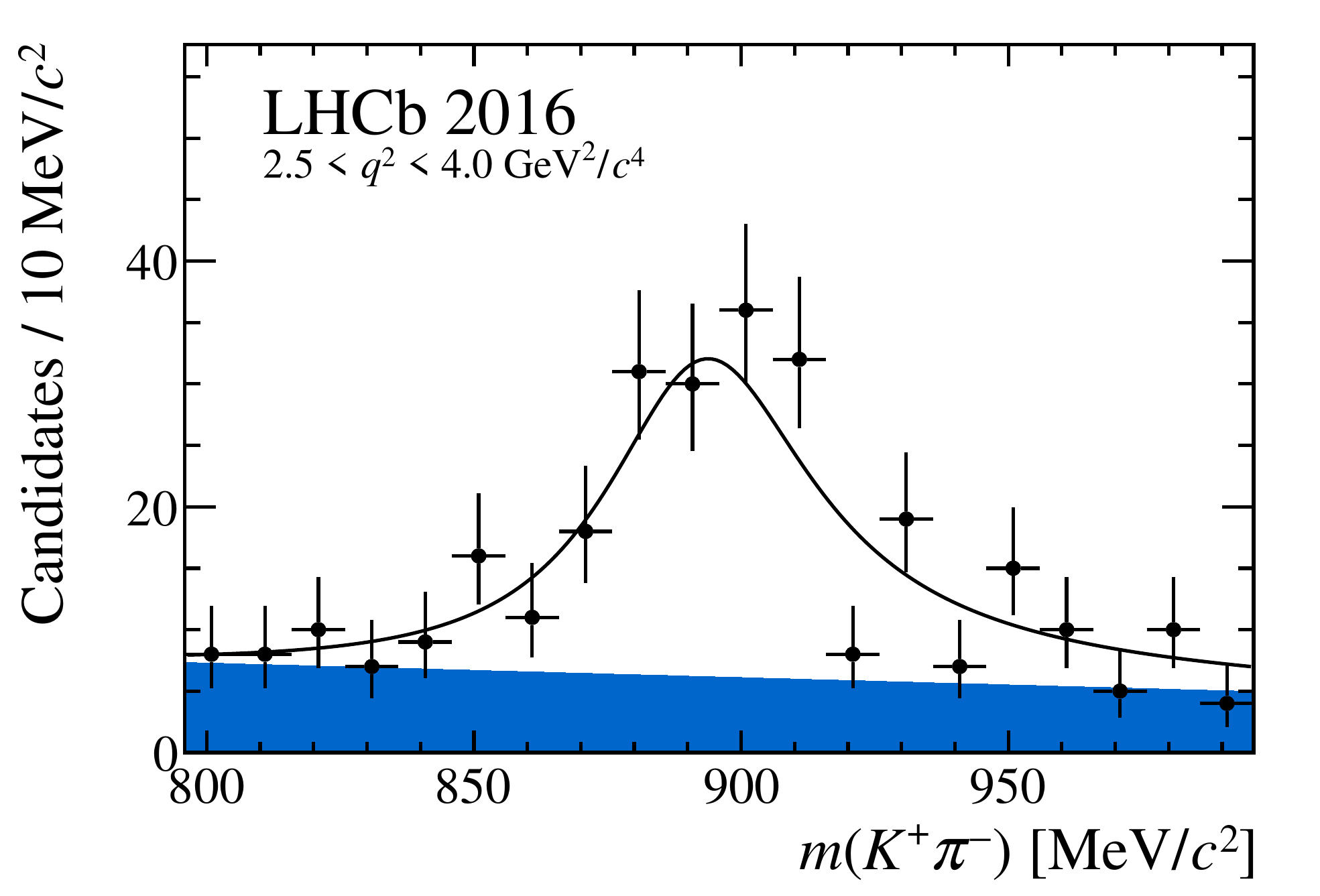}
 \includegraphics[width=0.32\textwidth]{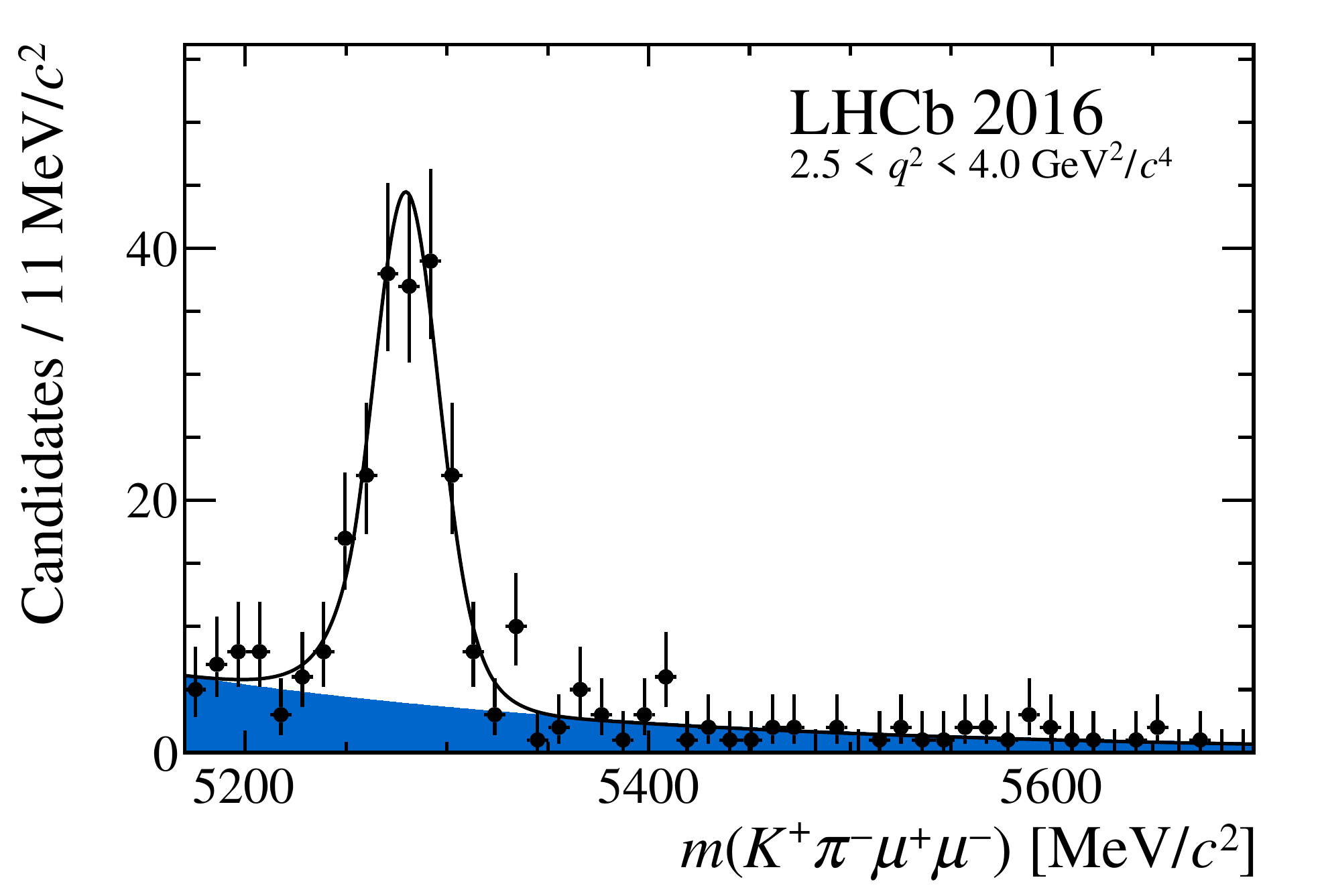}
 \caption{Projections of the fitted probability density function on the decay angles, \Mkpi and \Mkpimm for the bin $2.5<q^2<4.0\gevgevcccc$. The blue shaded region indicates background. \label{fig:projectionsc}}
 \end{figure}

\begin{figure}
   \centering
 \includegraphics[width=0.32\textwidth]{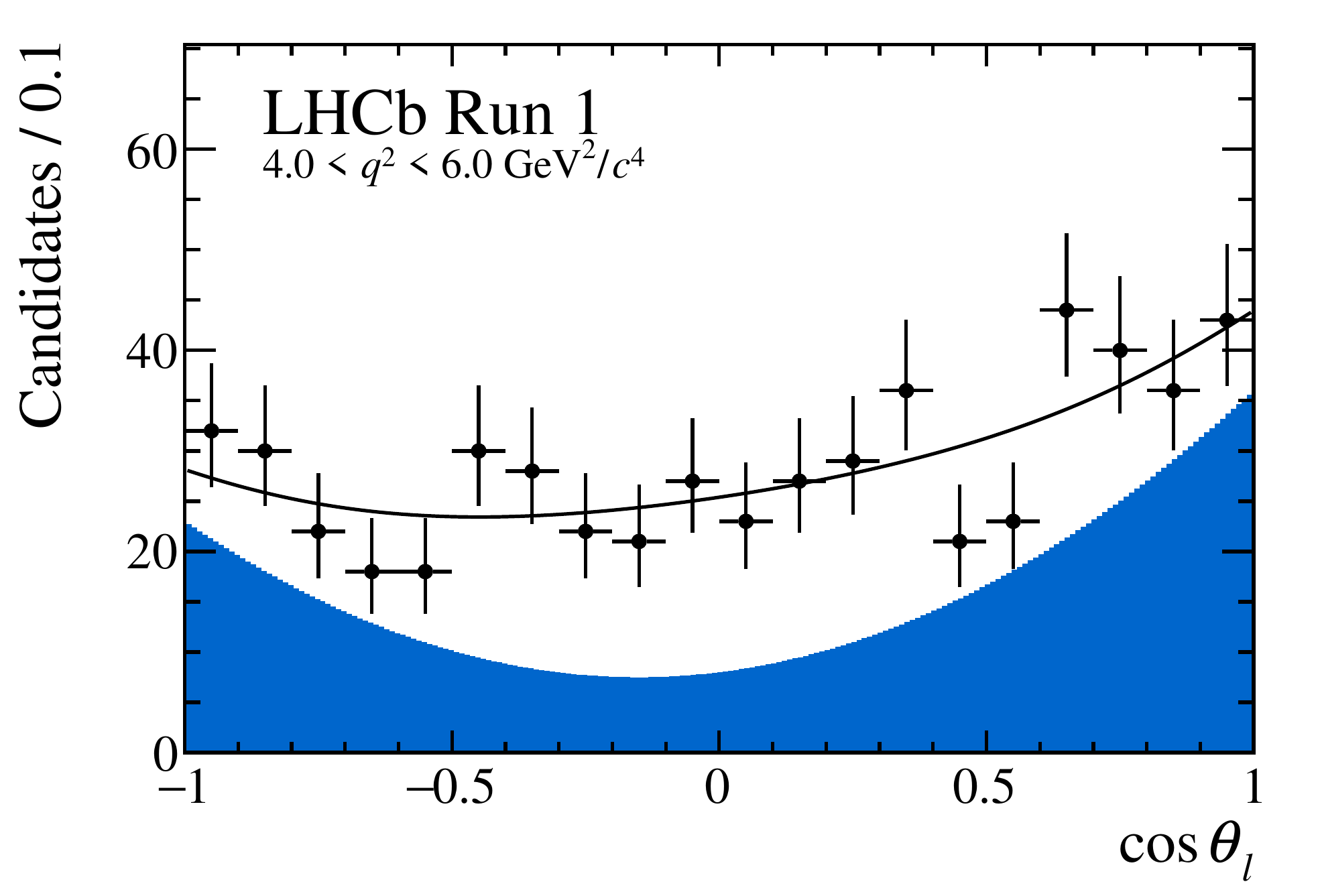}
 \includegraphics[width=0.32\textwidth]{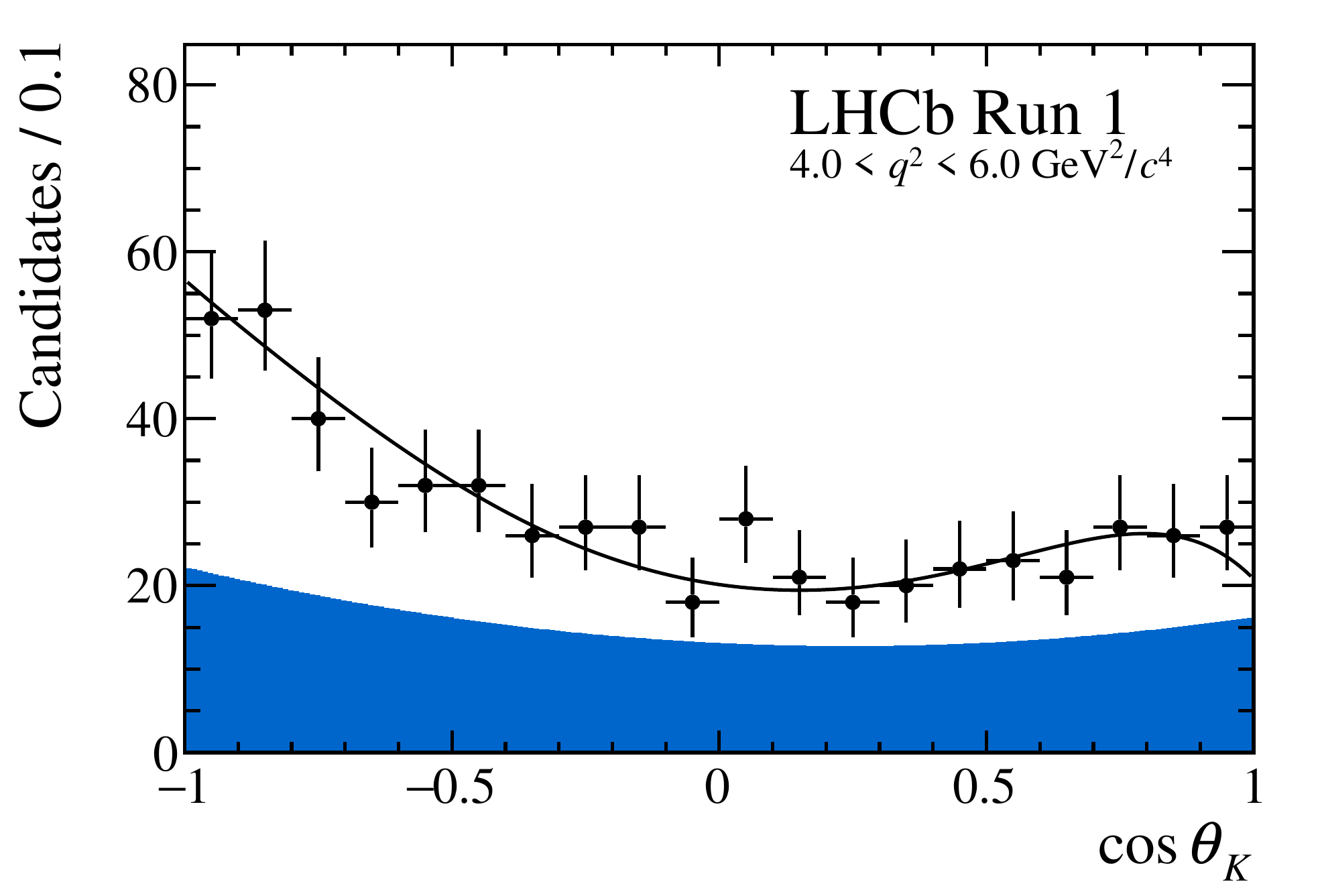}
 \includegraphics[width=0.32\textwidth]{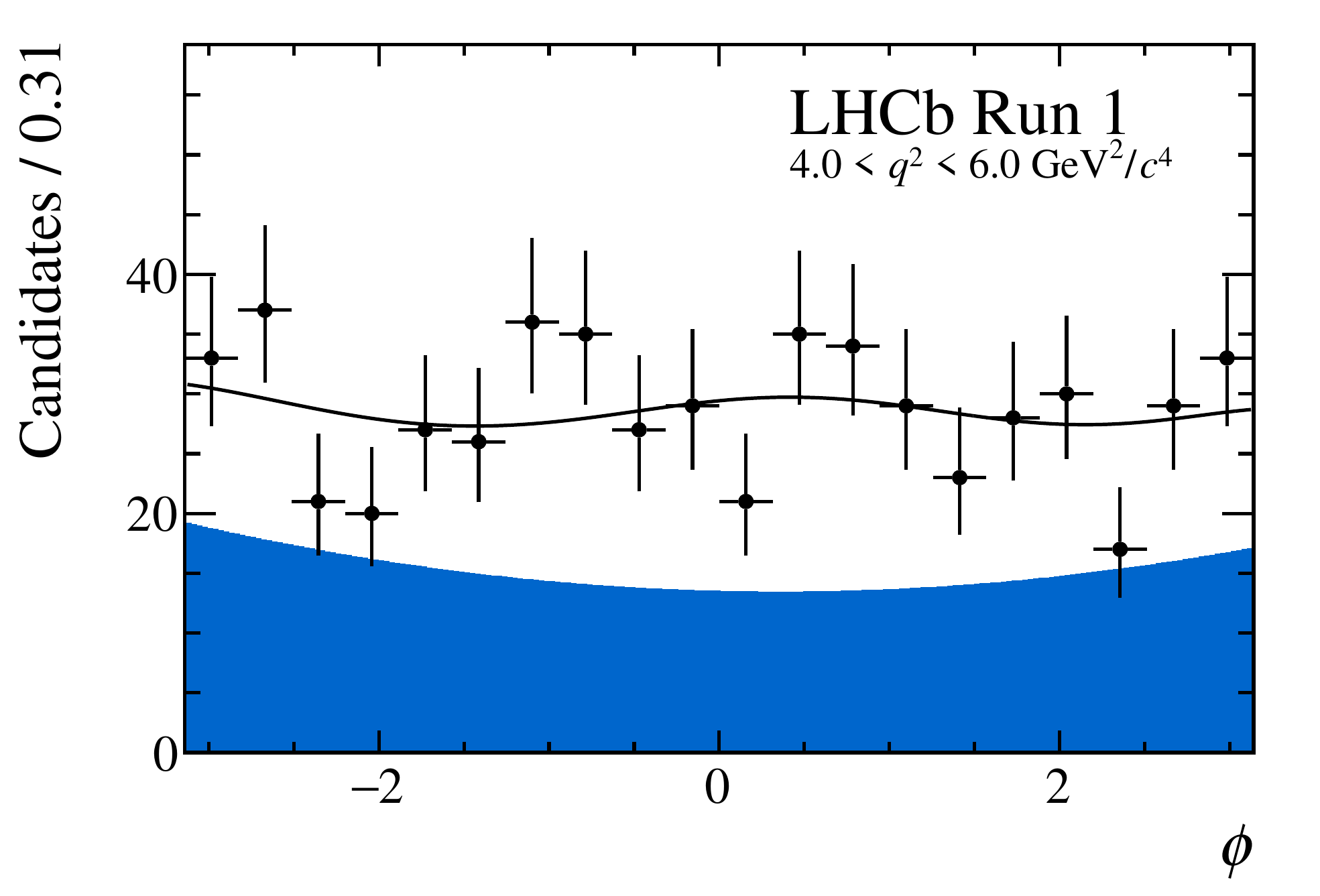}
 \includegraphics[width=0.32\textwidth]{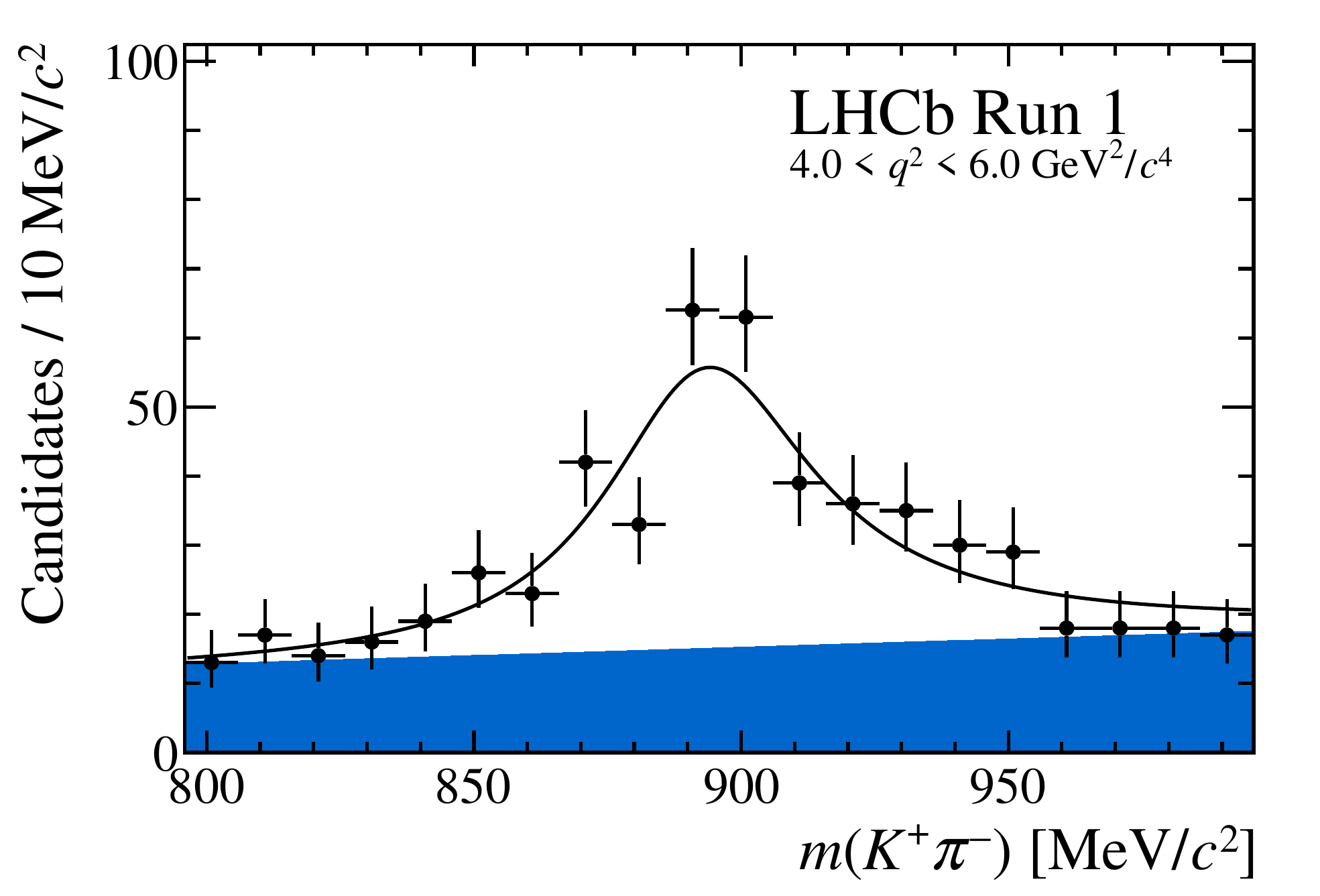}
 \includegraphics[width=0.32\textwidth]{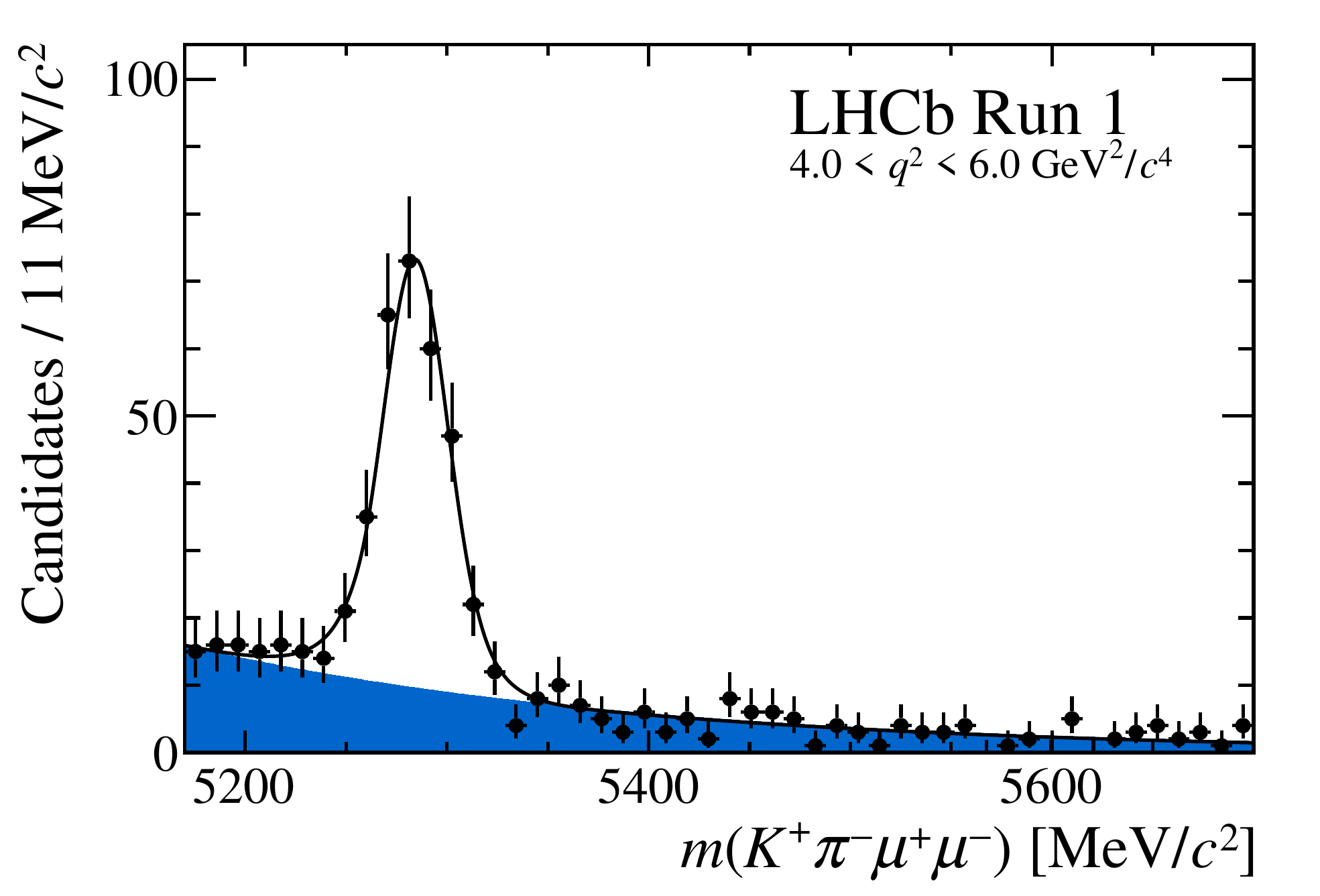}\\[0.5cm]
 \includegraphics[width=0.32\textwidth]{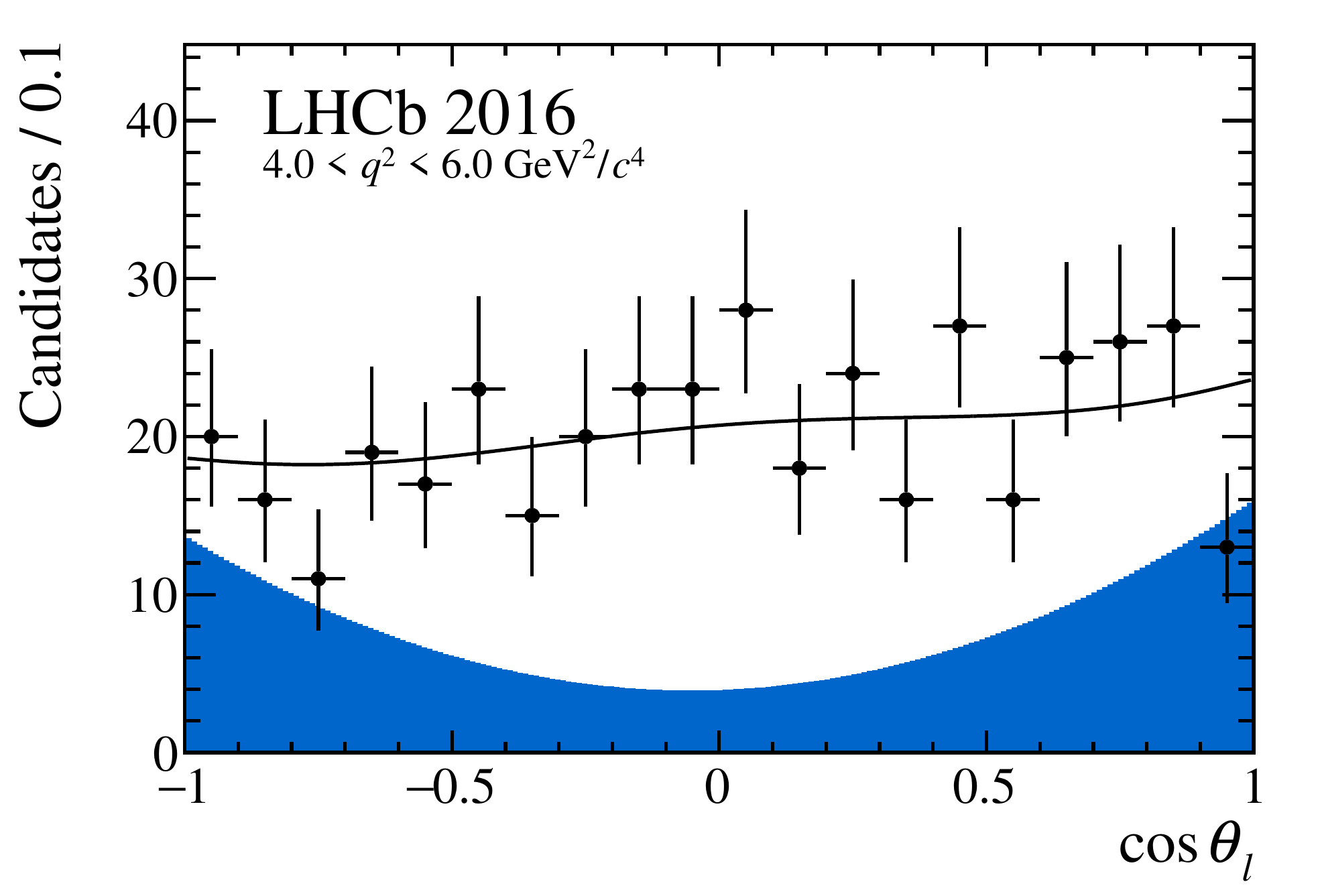}
 \includegraphics[width=0.32\textwidth]{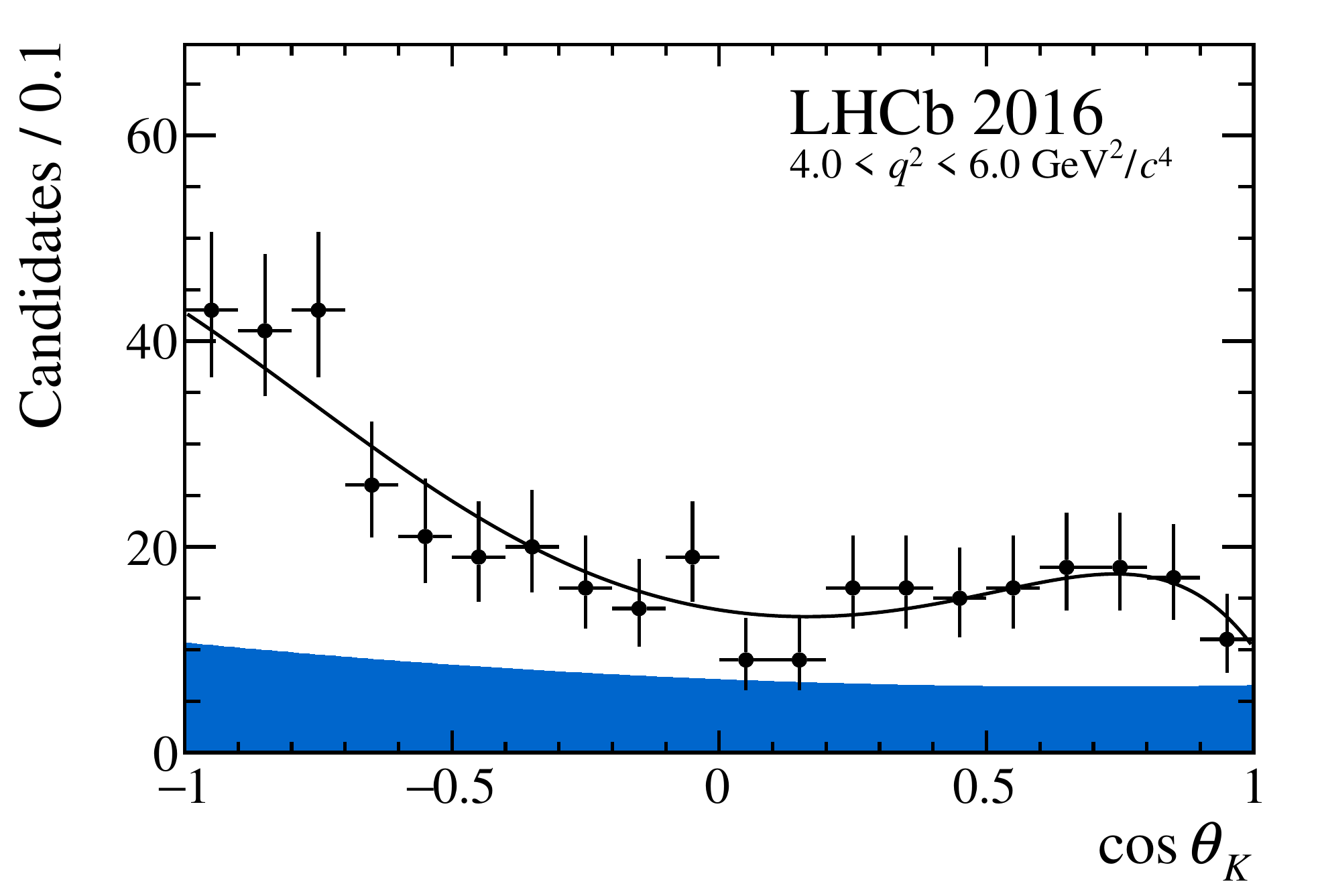}
 \includegraphics[width=0.32\textwidth]{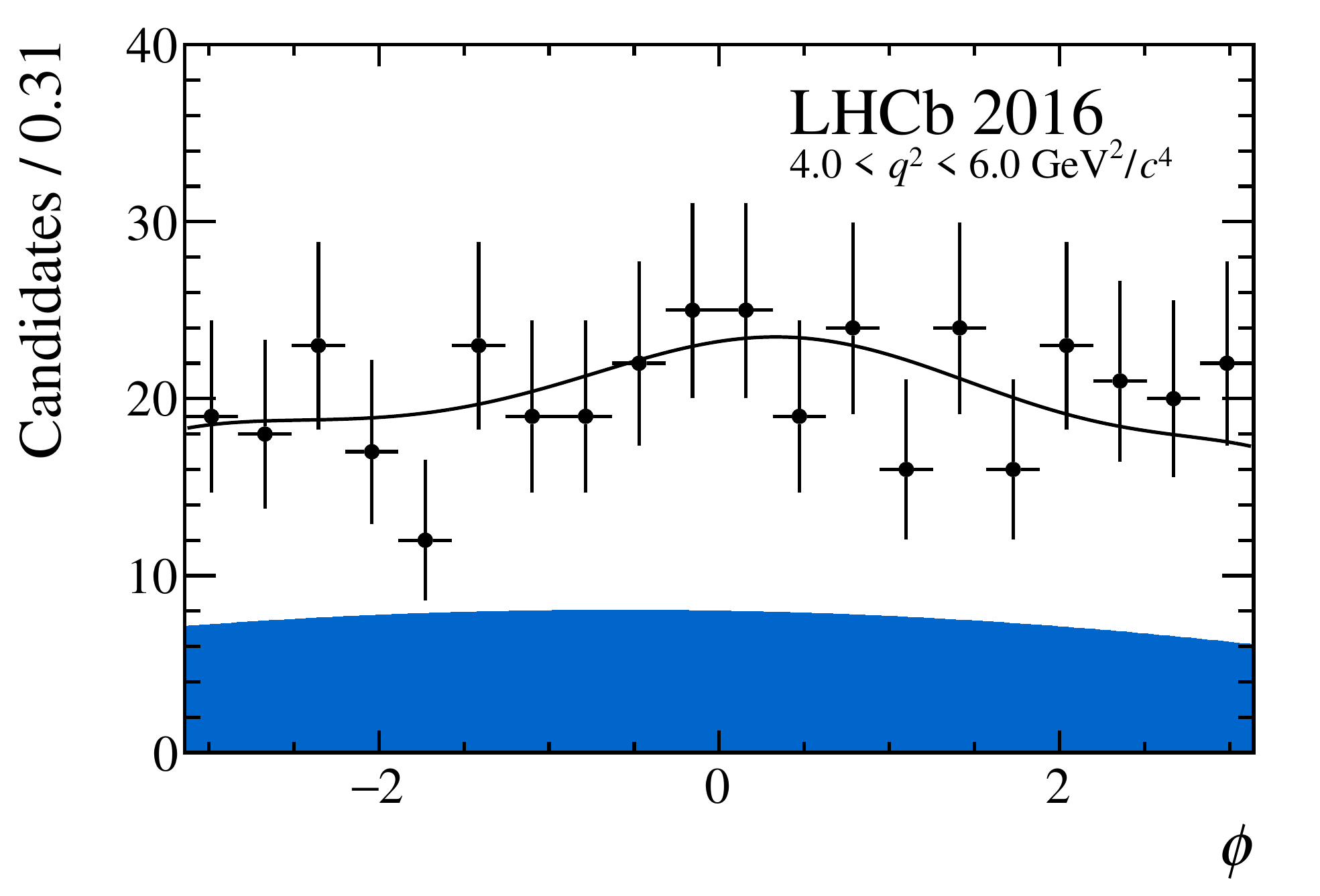}
 \includegraphics[width=0.32\textwidth]{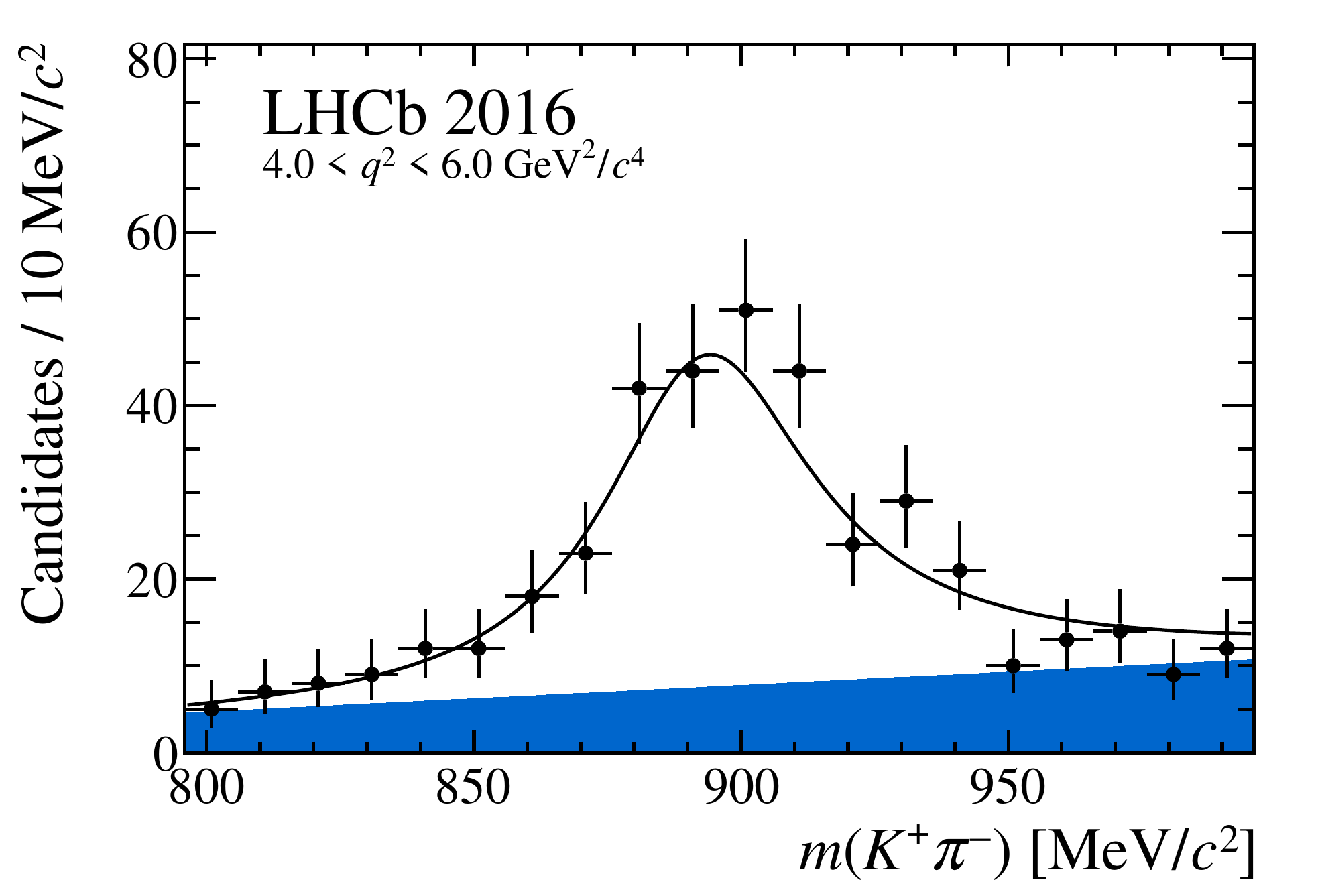}
 \includegraphics[width=0.32\textwidth]{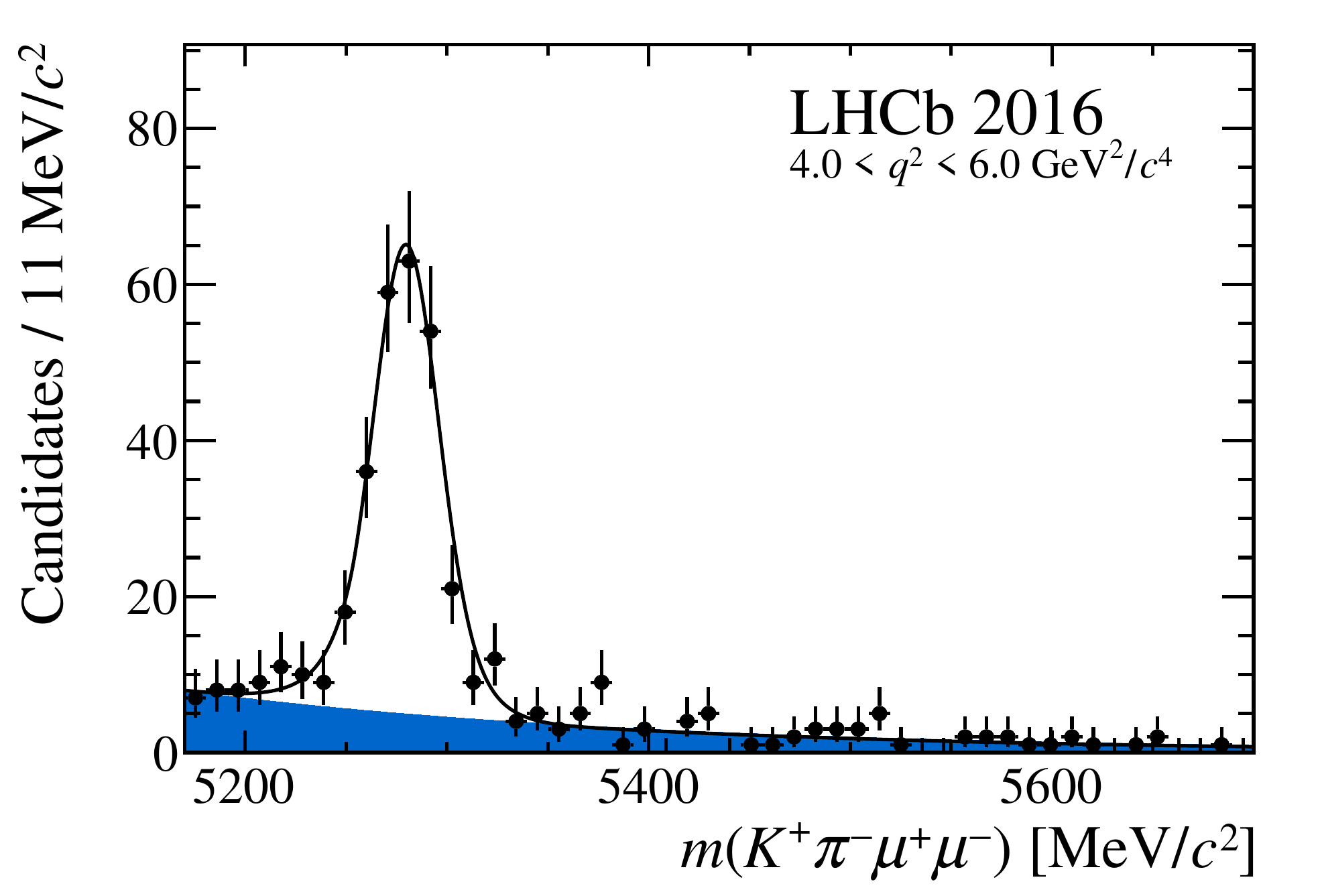}
 \caption{Projections of the fitted probability density function on the decay angles, \Mkpi and \Mkpimm for the bin $4.0<q^2<6.0 \gevgevcccc$. The blue shaded region indicates background. \label{fig:projectionsd}}
 \end{figure}

\begin{figure}
   \centering
 \includegraphics[width=0.32\textwidth]{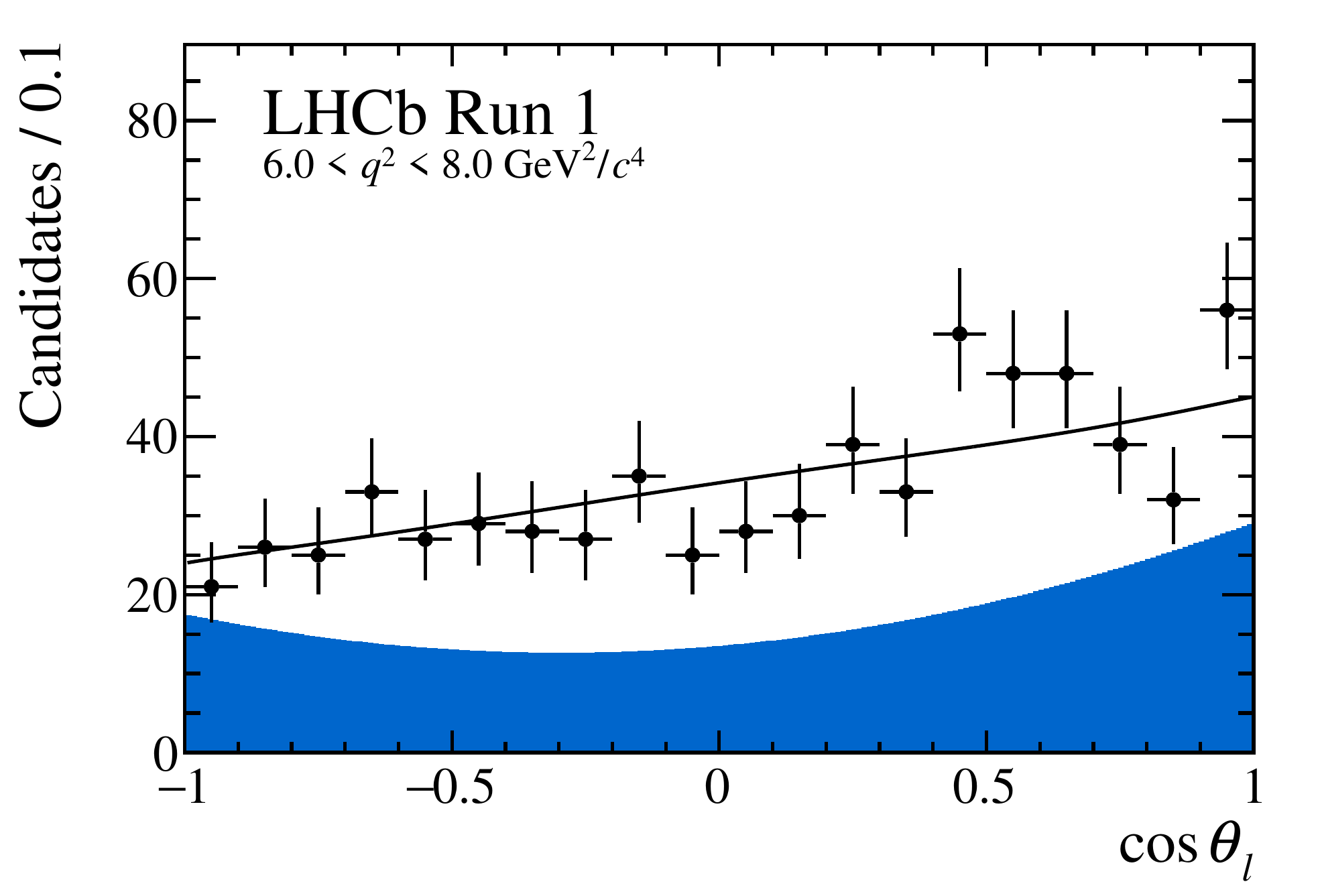}
 \includegraphics[width=0.32\textwidth]{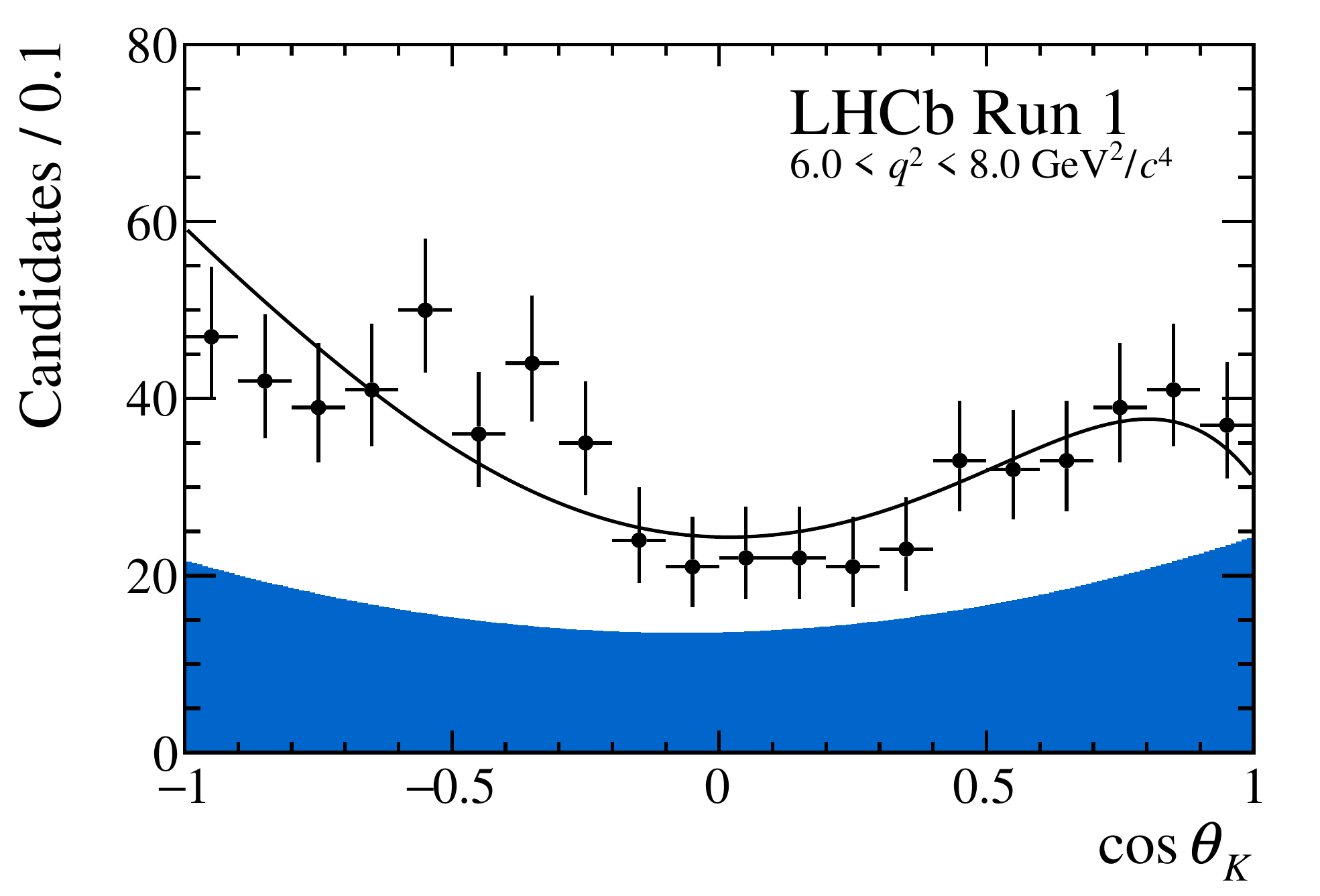}
 \includegraphics[width=0.32\textwidth]{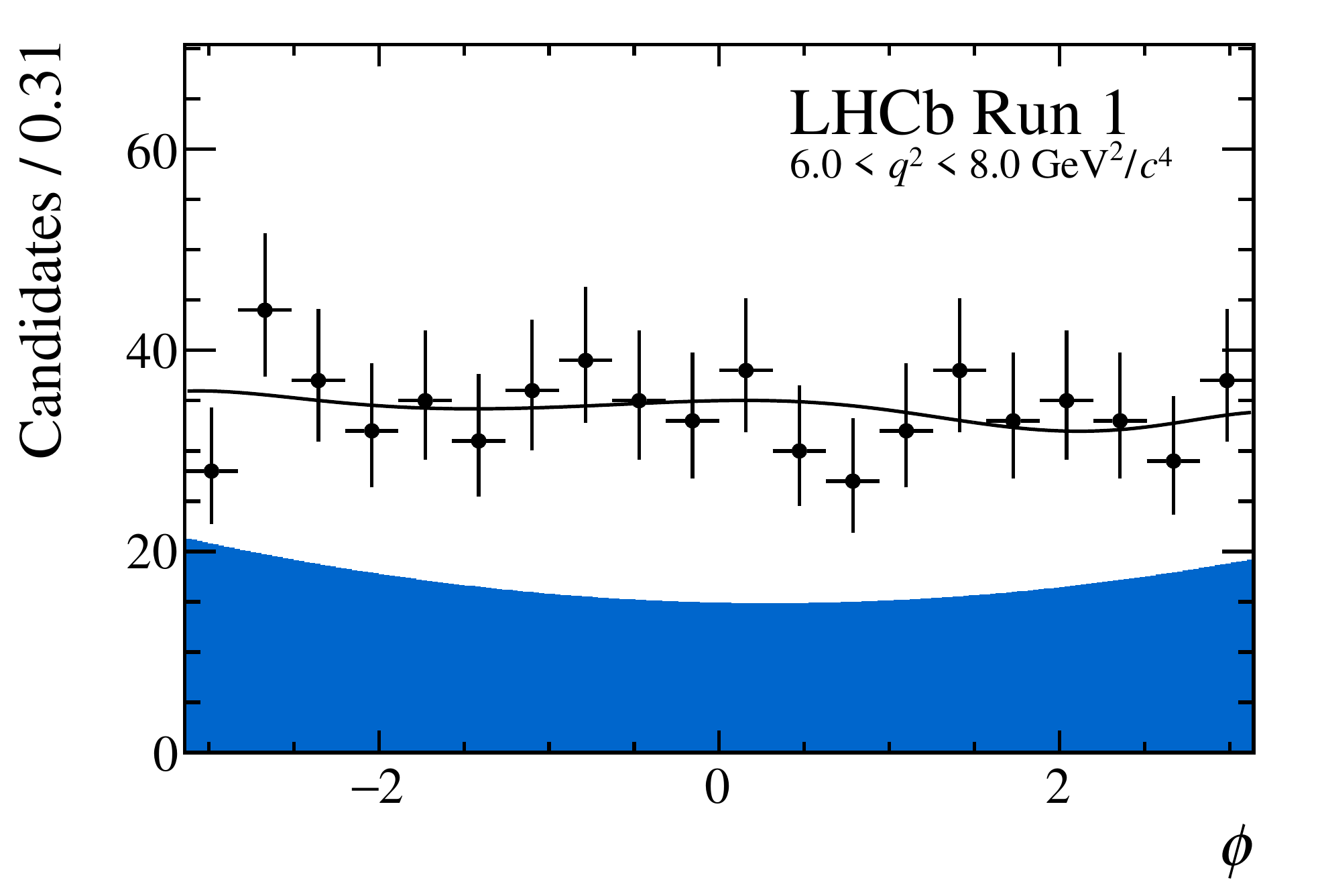}
 \includegraphics[width=0.32\textwidth]{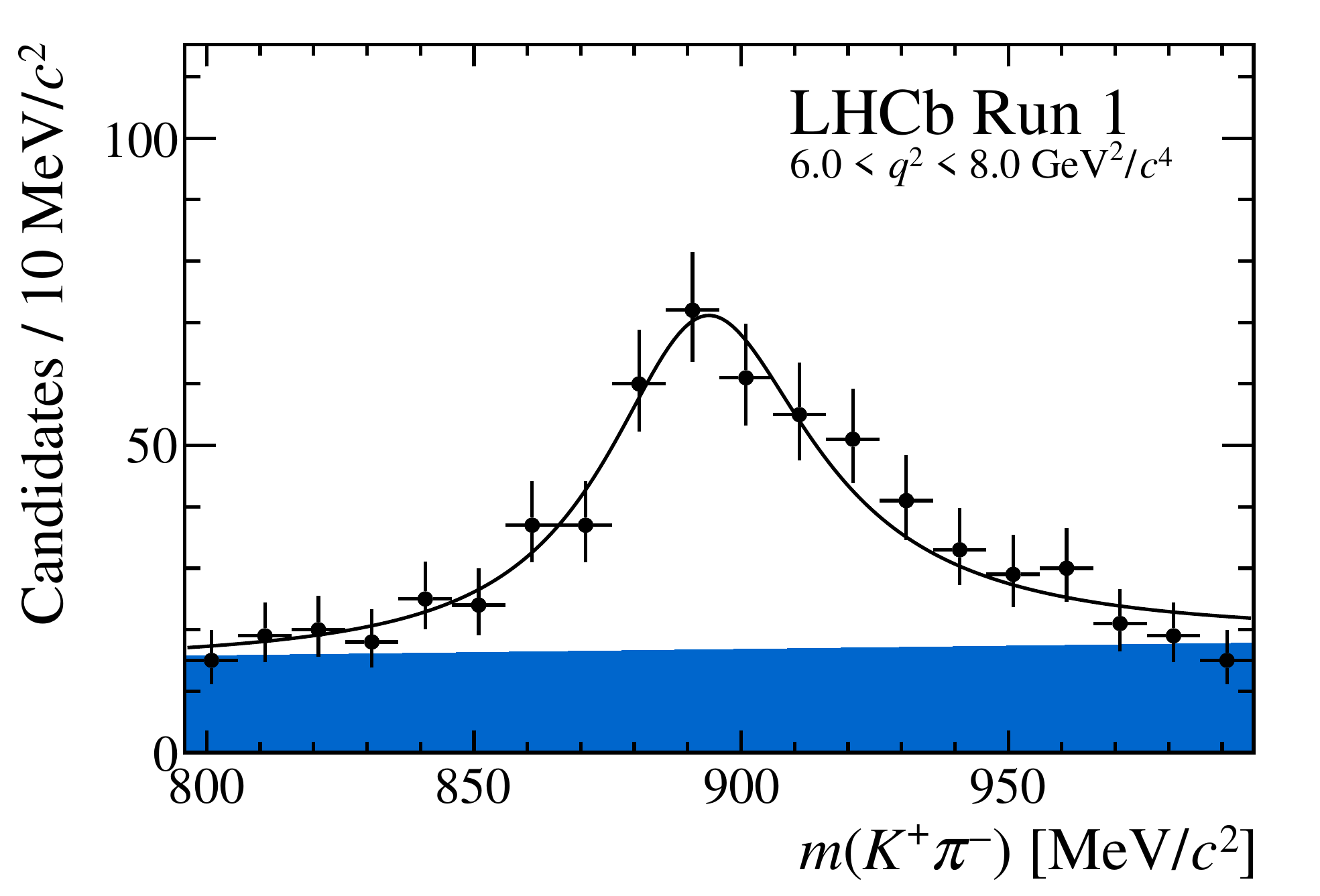}
 \includegraphics[width=0.32\textwidth]{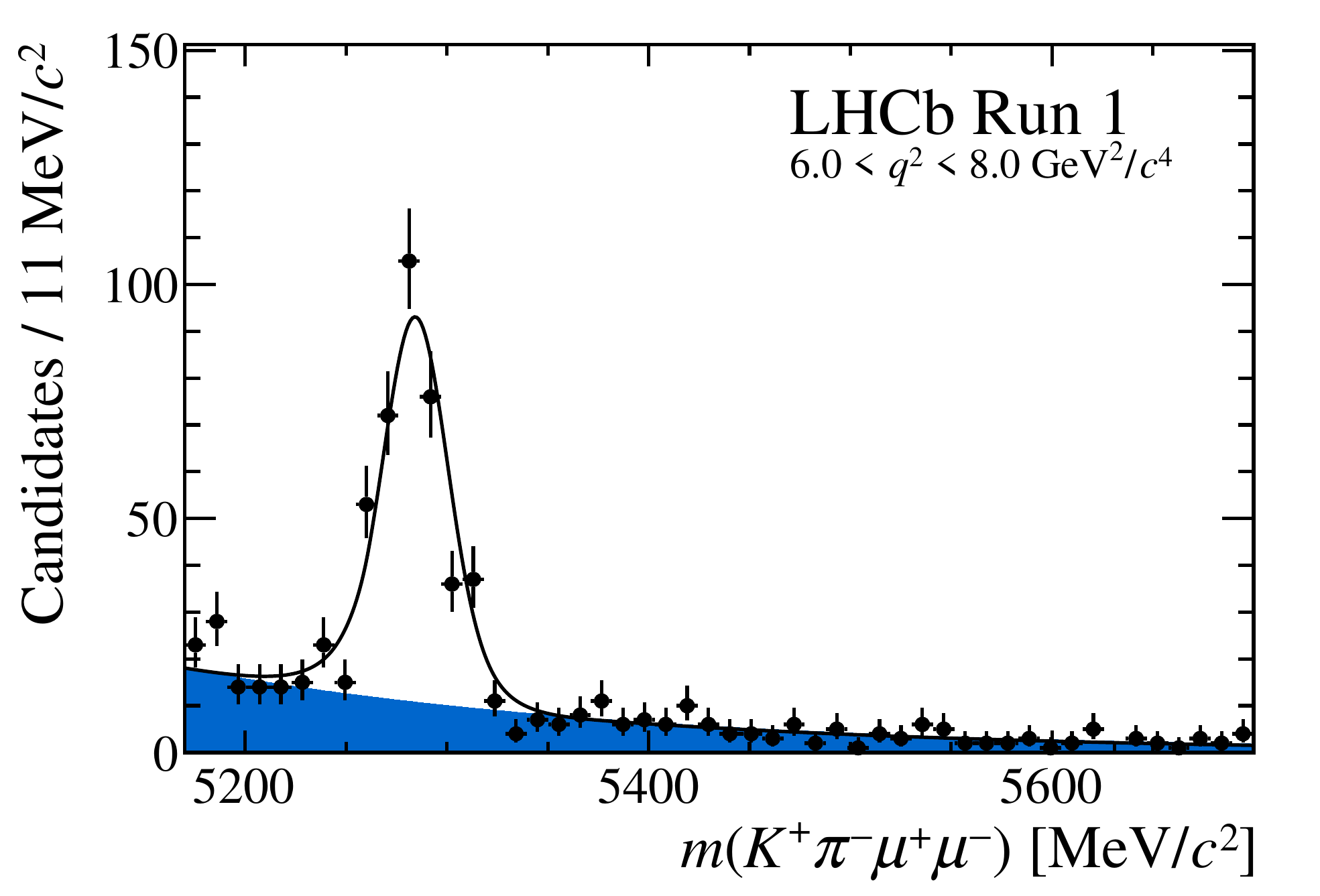}\\[0.5cm]
 \includegraphics[width=0.32\textwidth]{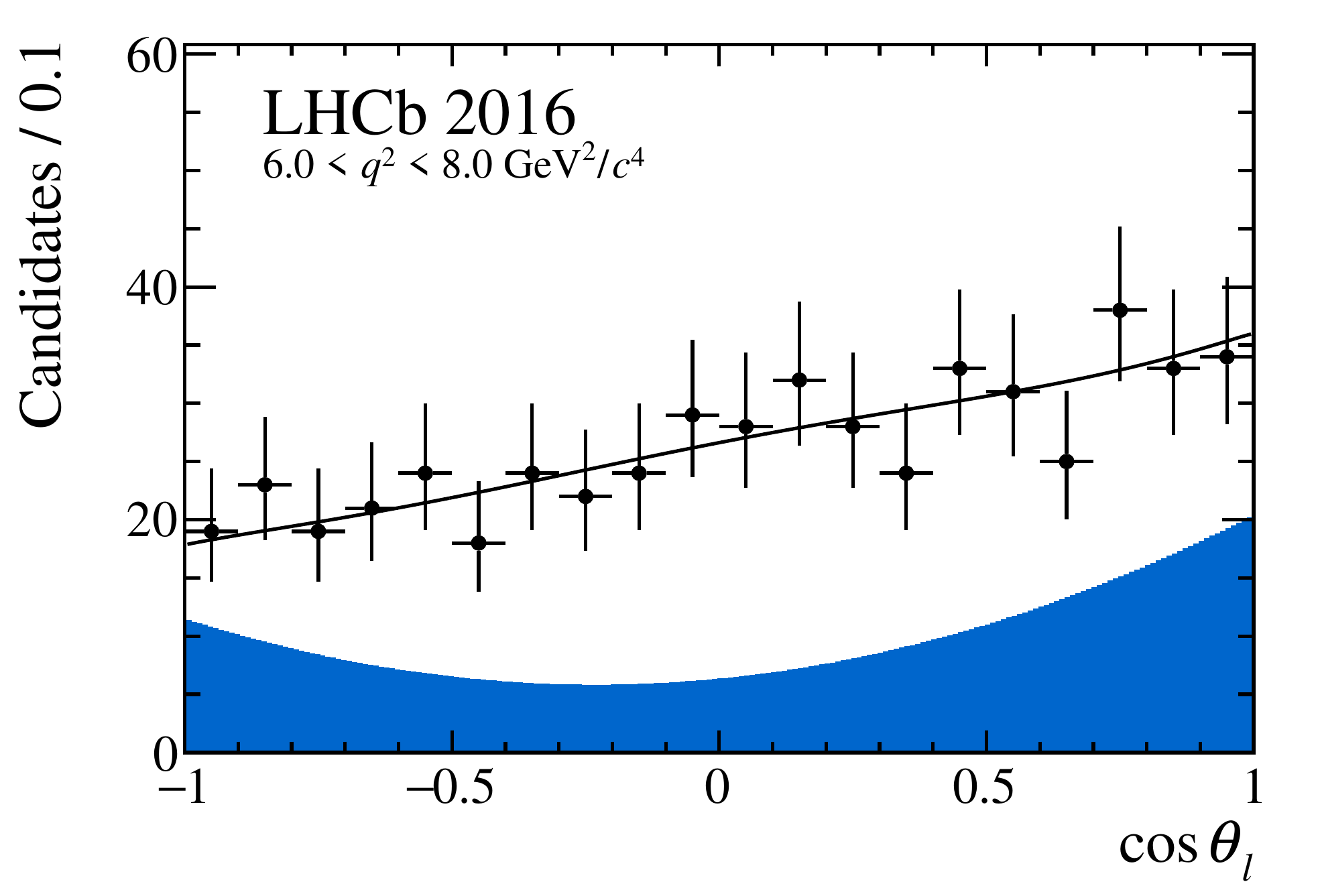}
 \includegraphics[width=0.32\textwidth]{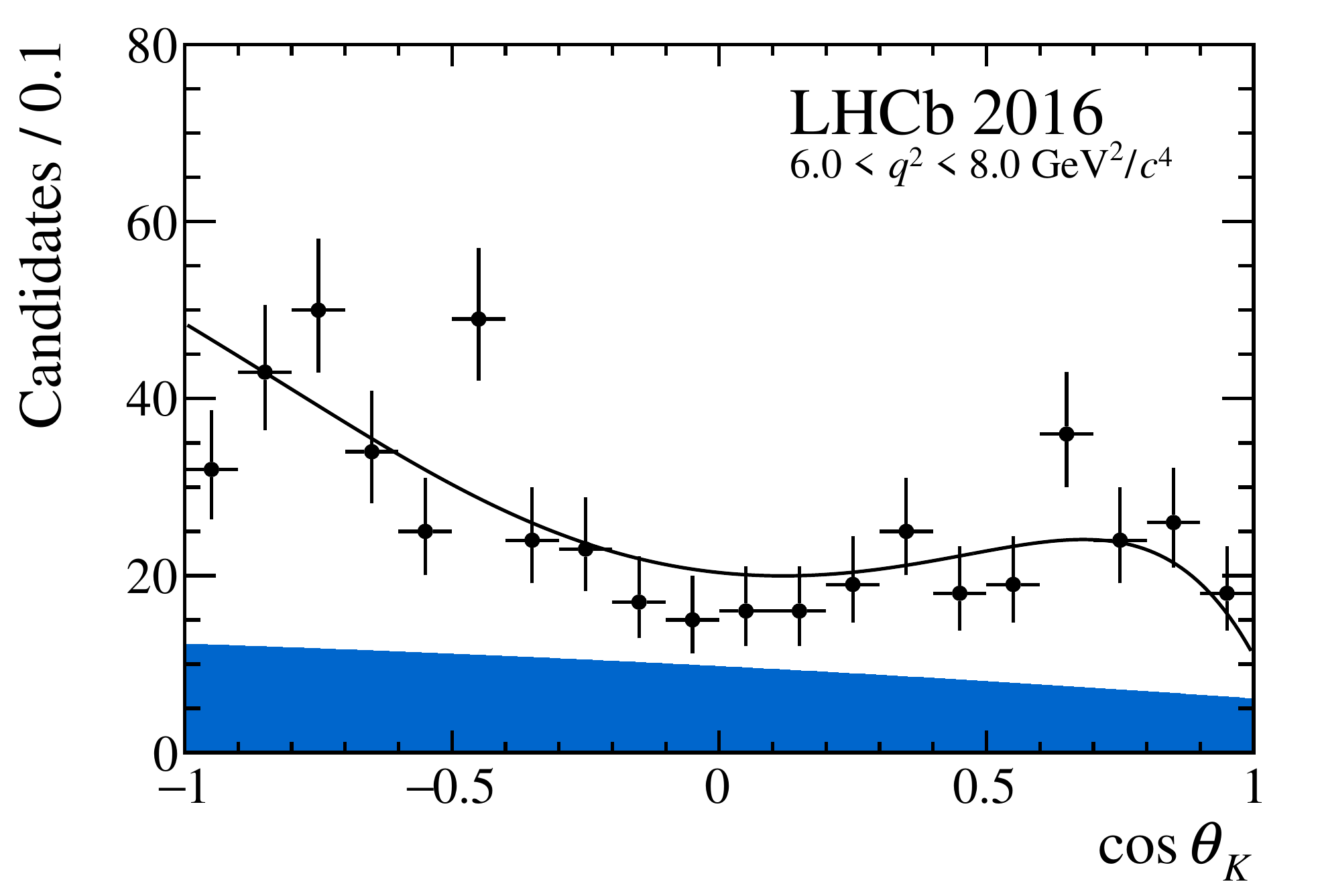}
 \includegraphics[width=0.32\textwidth]{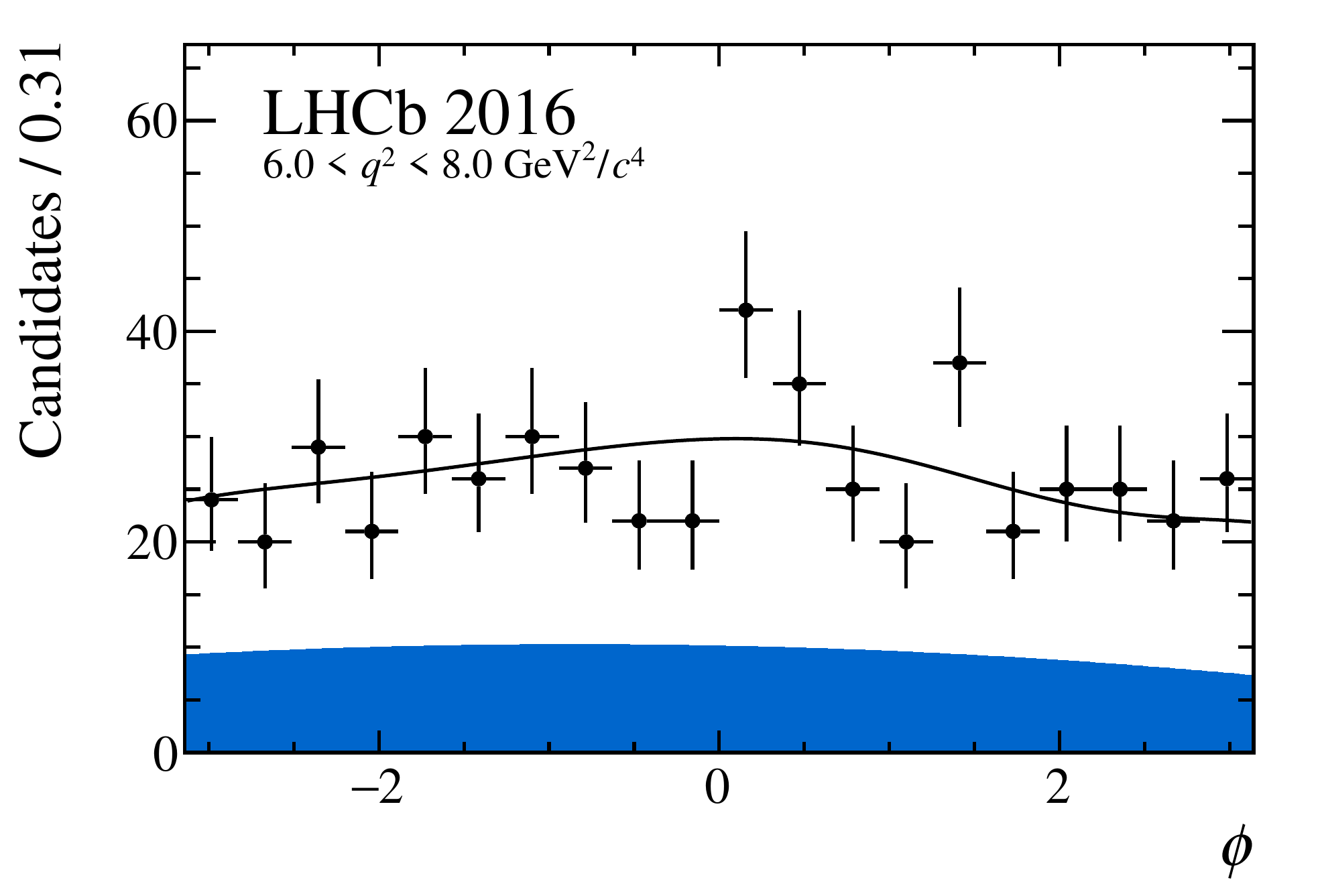}
 \includegraphics[width=0.32\textwidth]{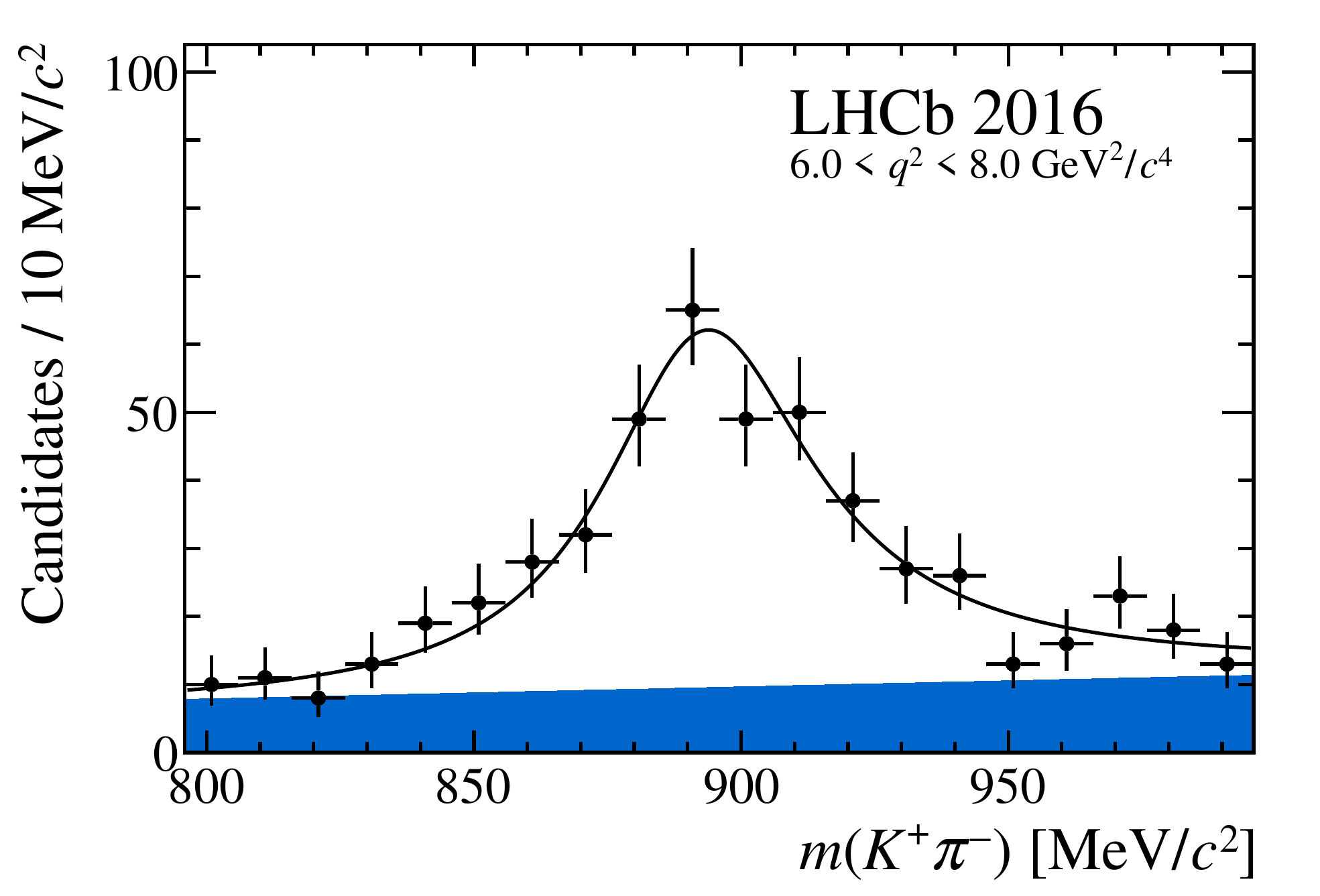}
 \includegraphics[width=0.32\textwidth]{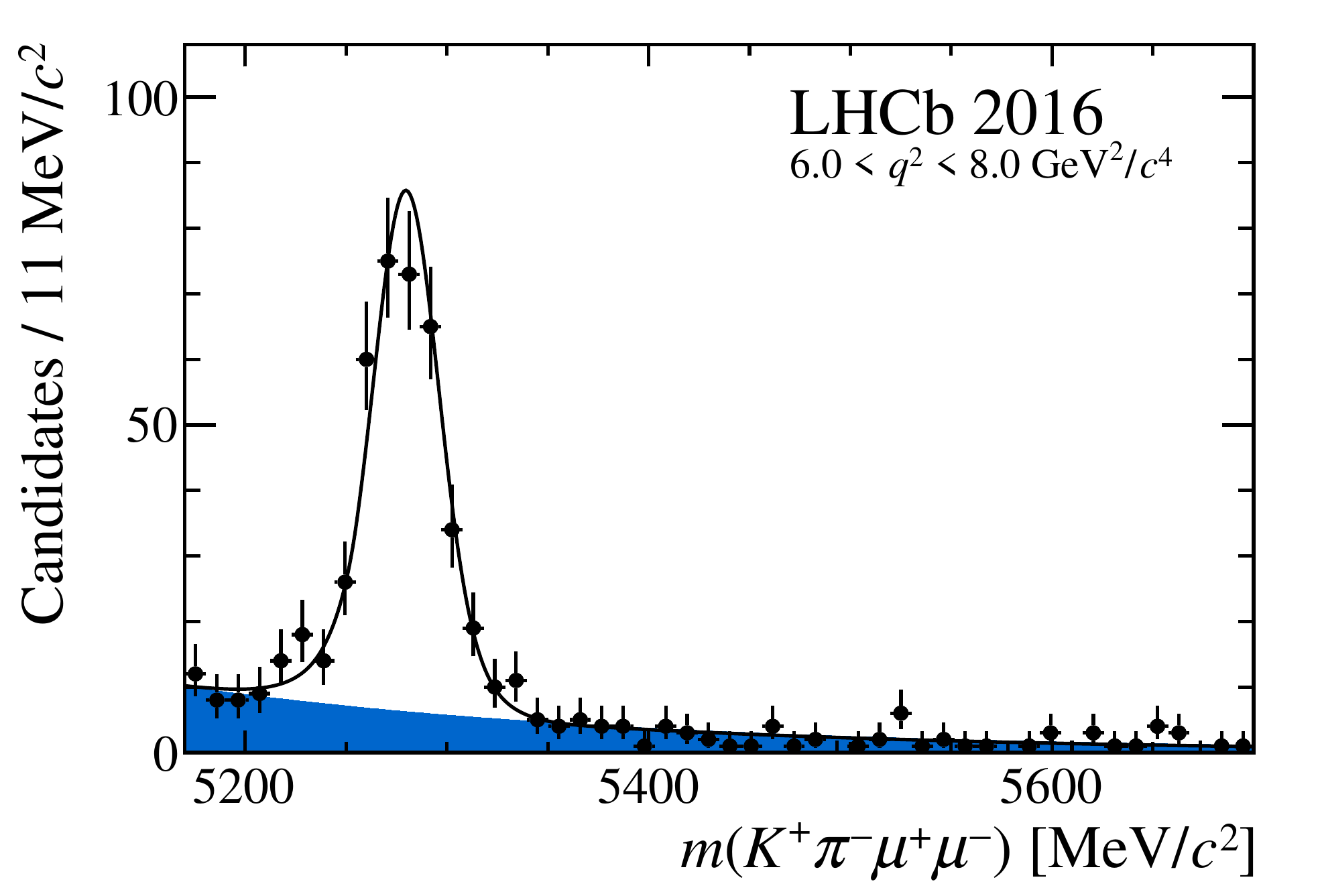}
 \caption{Projections of the fitted probability density function on the decay angles, \Mkpi and \Mkpimm for the bin $6.0<q^2<8.0\gevgevcccc$. The blue shaded region indicates background. \label{fig:projectionse}}
 \end{figure}

\begin{figure}
   \centering
 \includegraphics[width=0.32\textwidth]{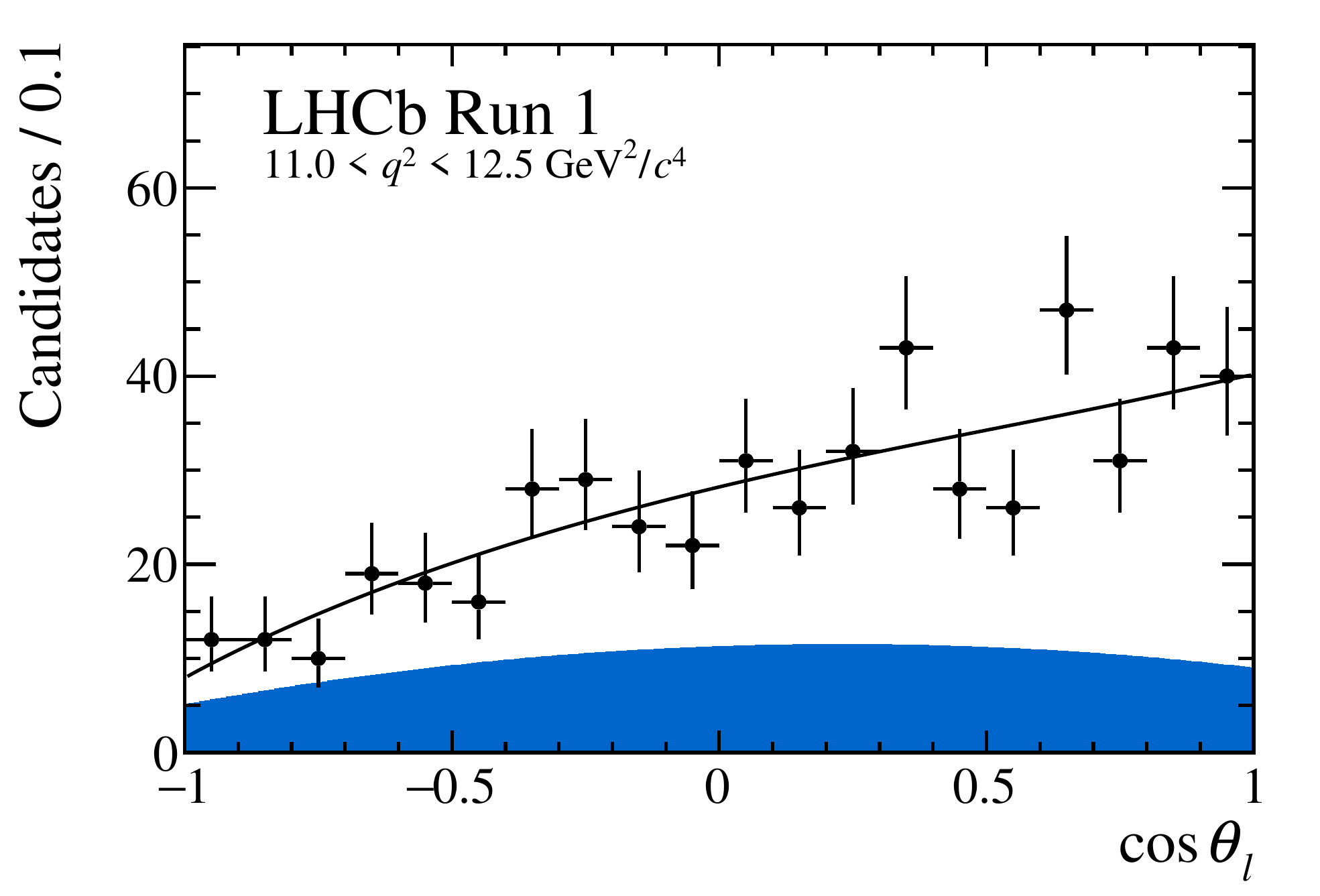}
 \includegraphics[width=0.32\textwidth]{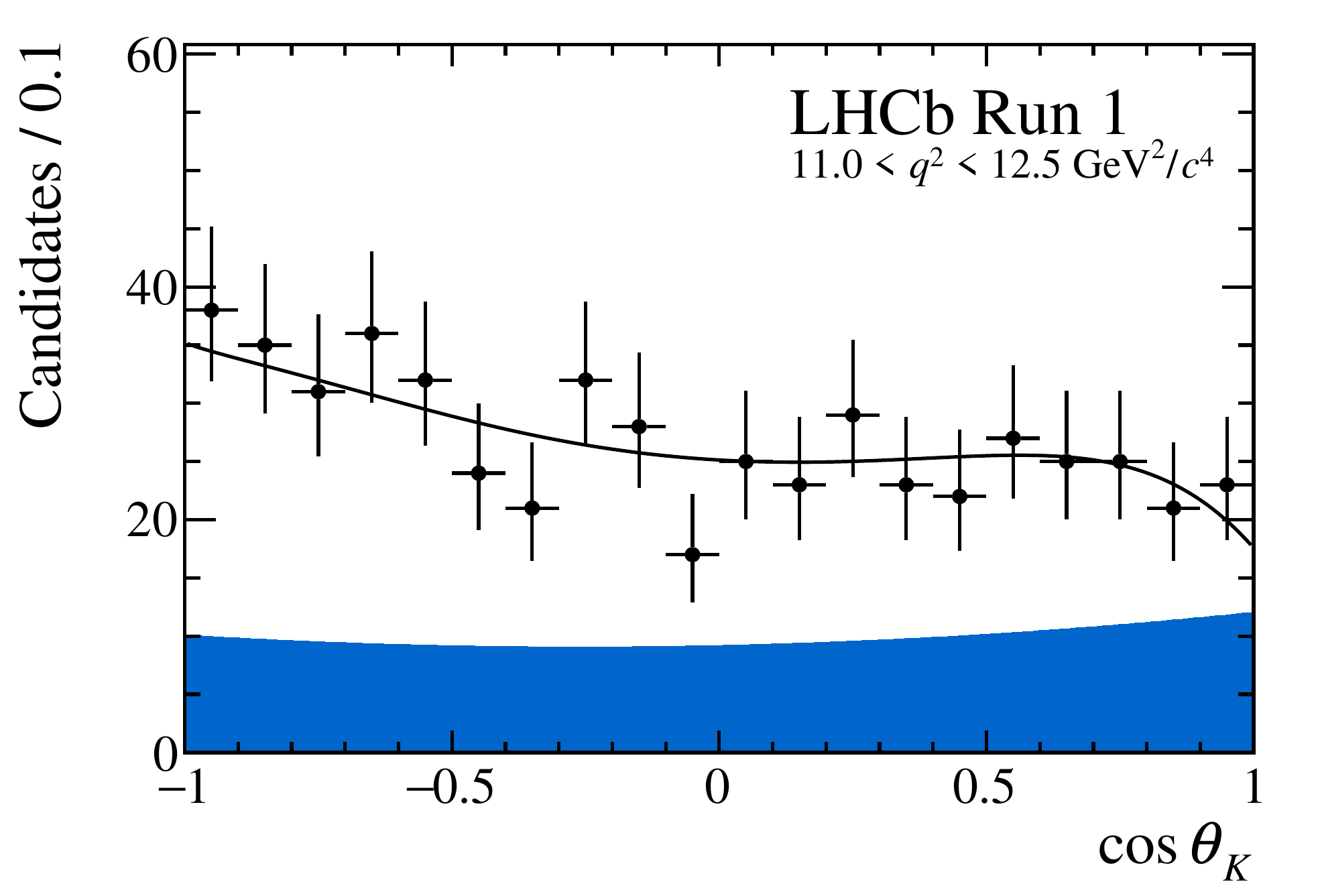}
 \includegraphics[width=0.32\textwidth]{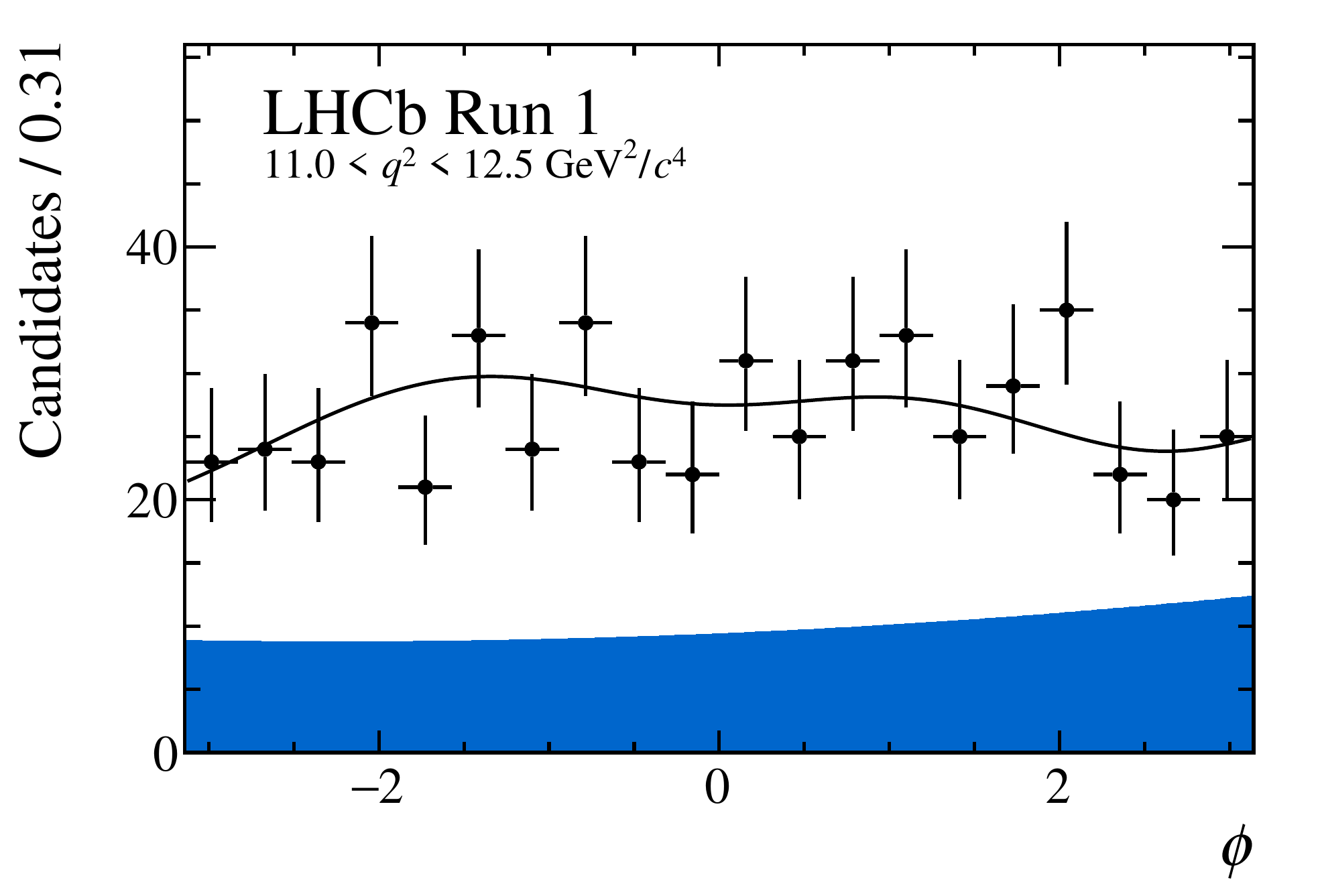}
 \includegraphics[width=0.32\textwidth]{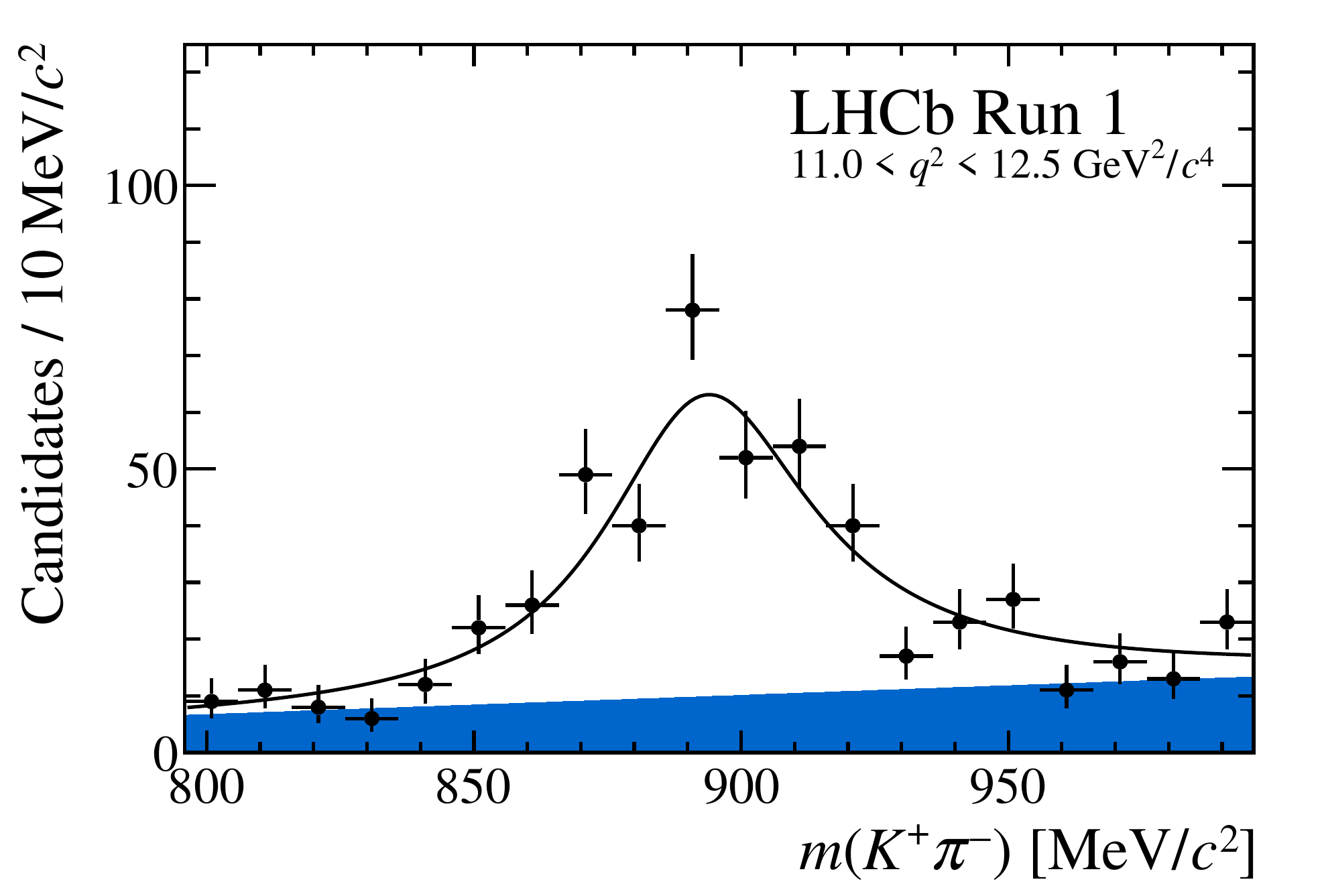}
 \includegraphics[width=0.32\textwidth]{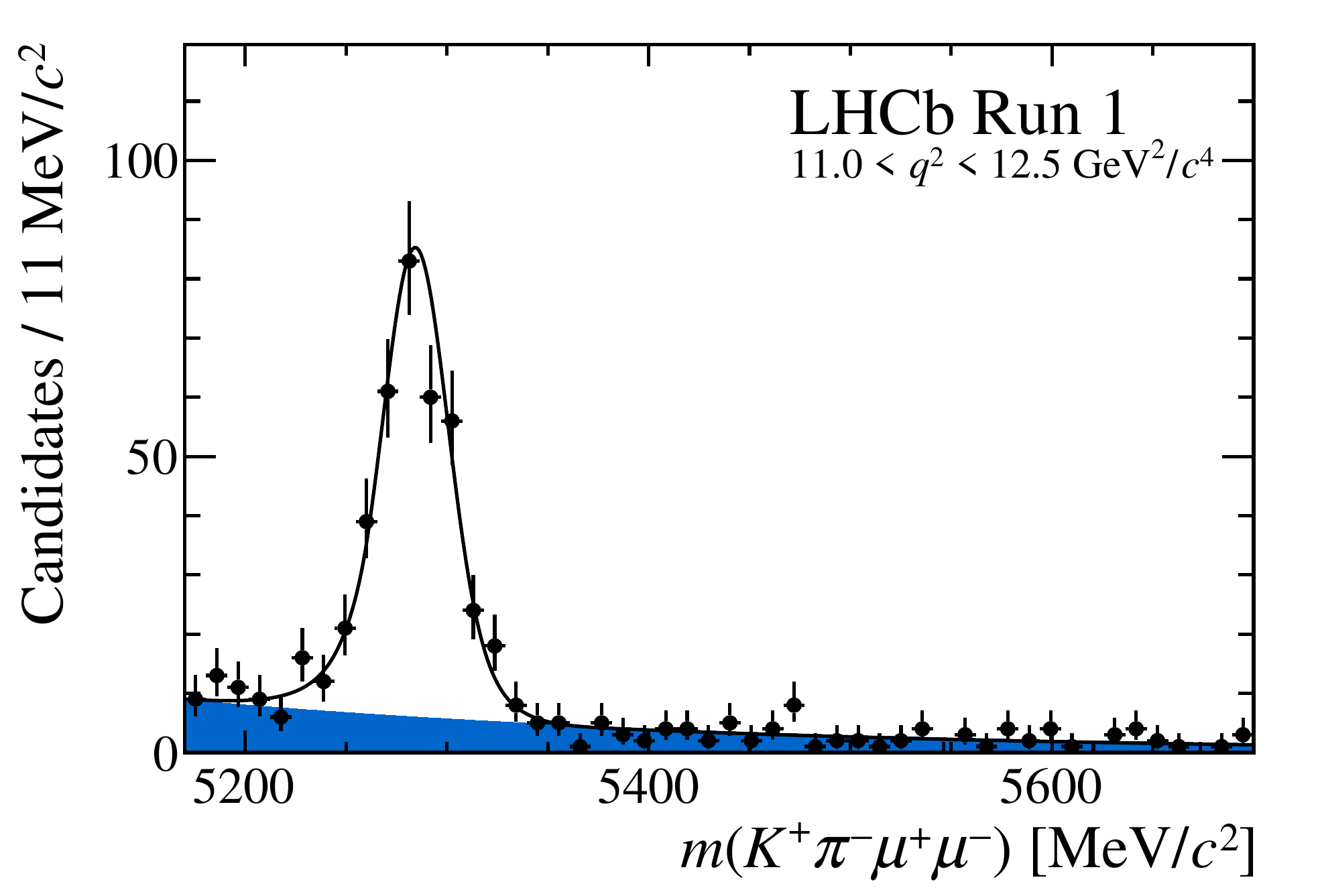}\\[0.5cm]
 \includegraphics[width=0.32\textwidth]{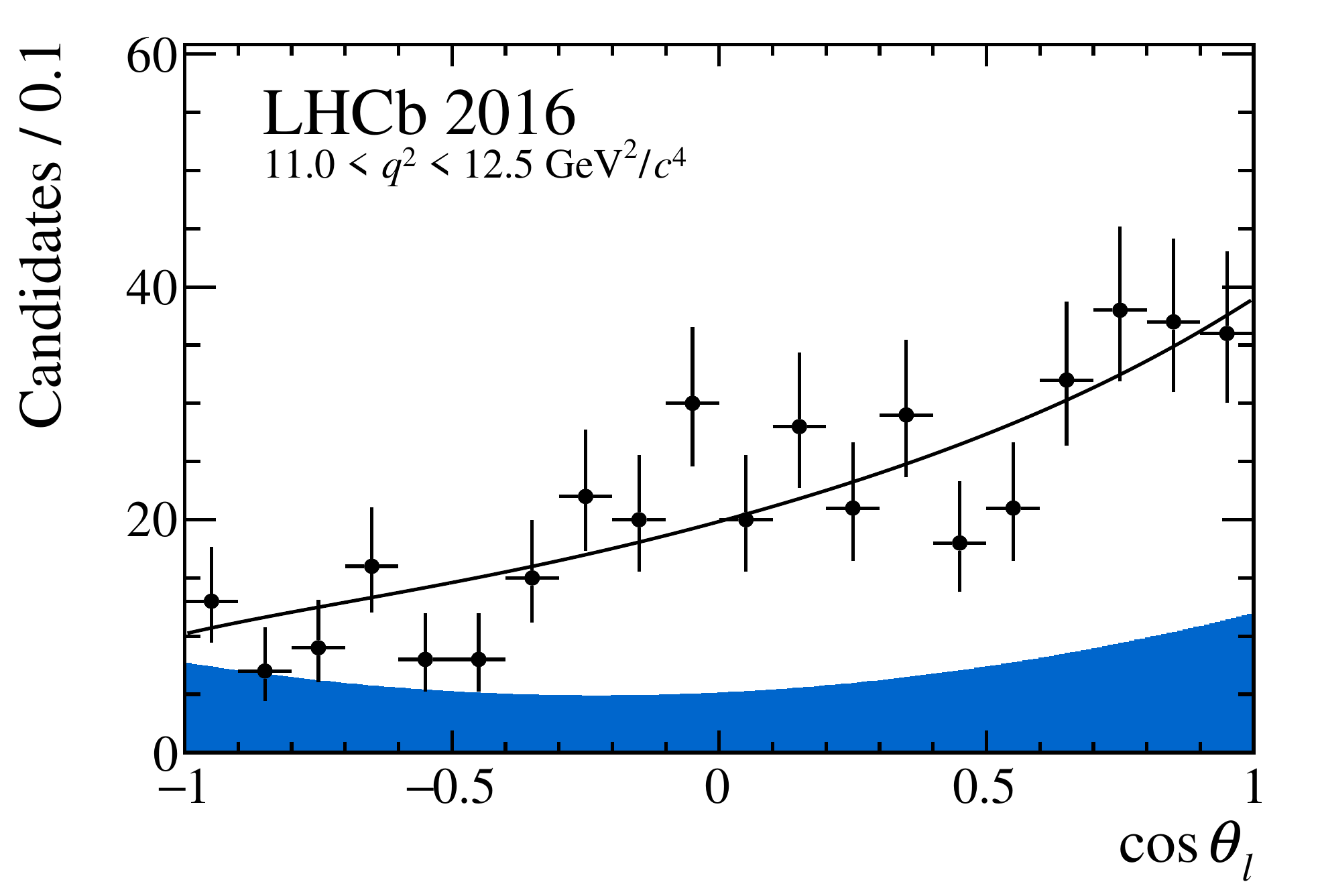}
 \includegraphics[width=0.32\textwidth]{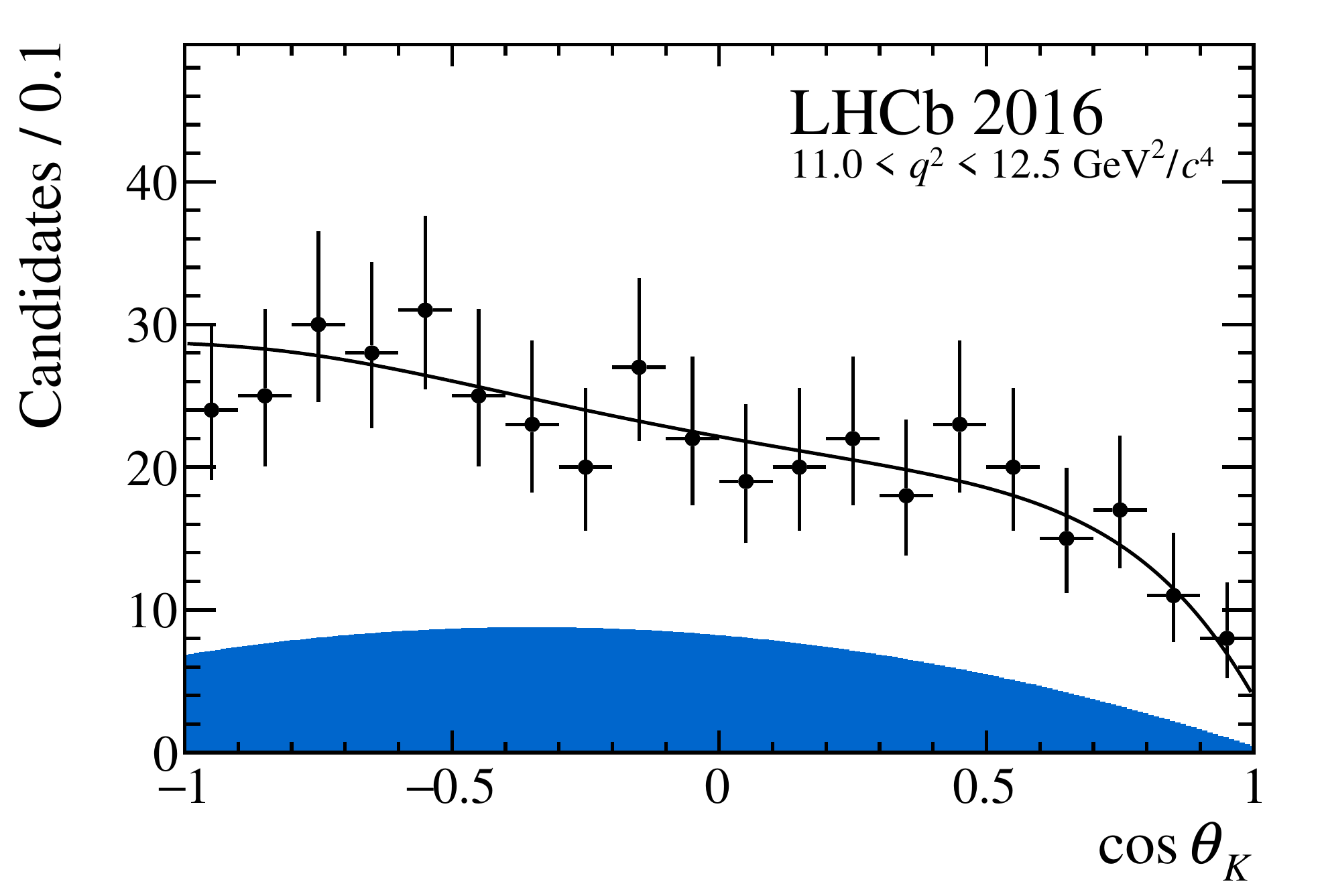}
 \includegraphics[width=0.32\textwidth]{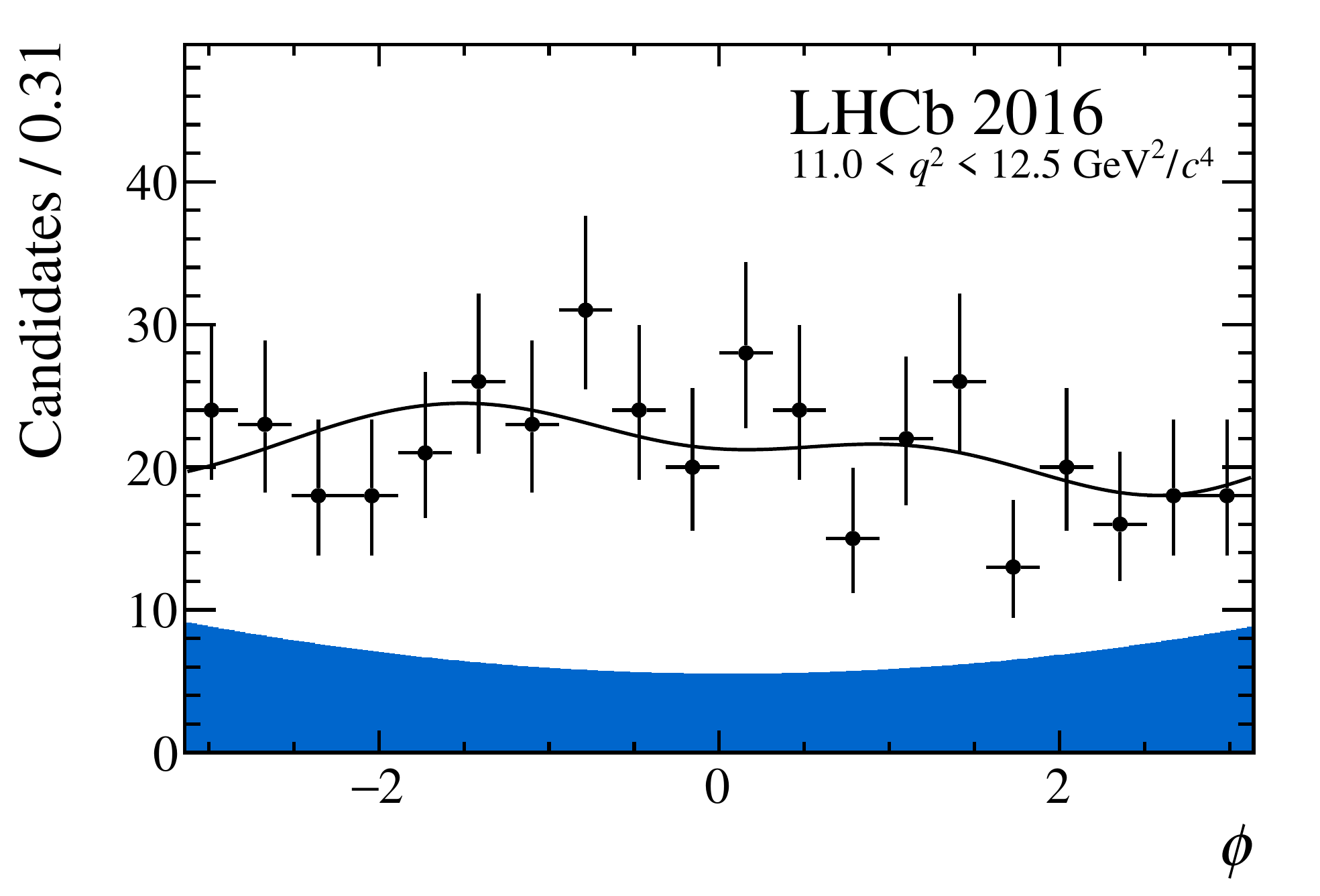}
 \includegraphics[width=0.32\textwidth]{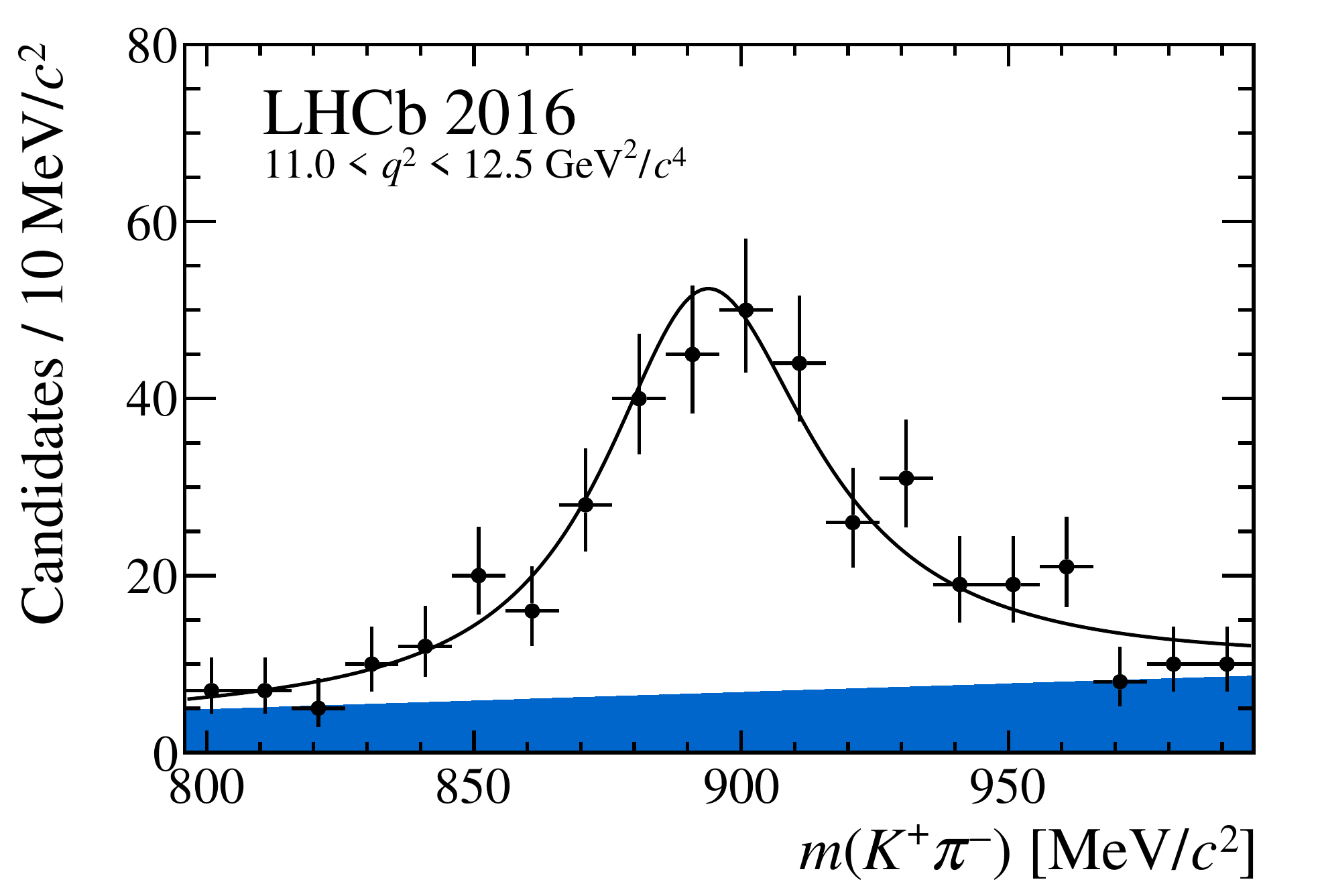}
 \includegraphics[width=0.32\textwidth]{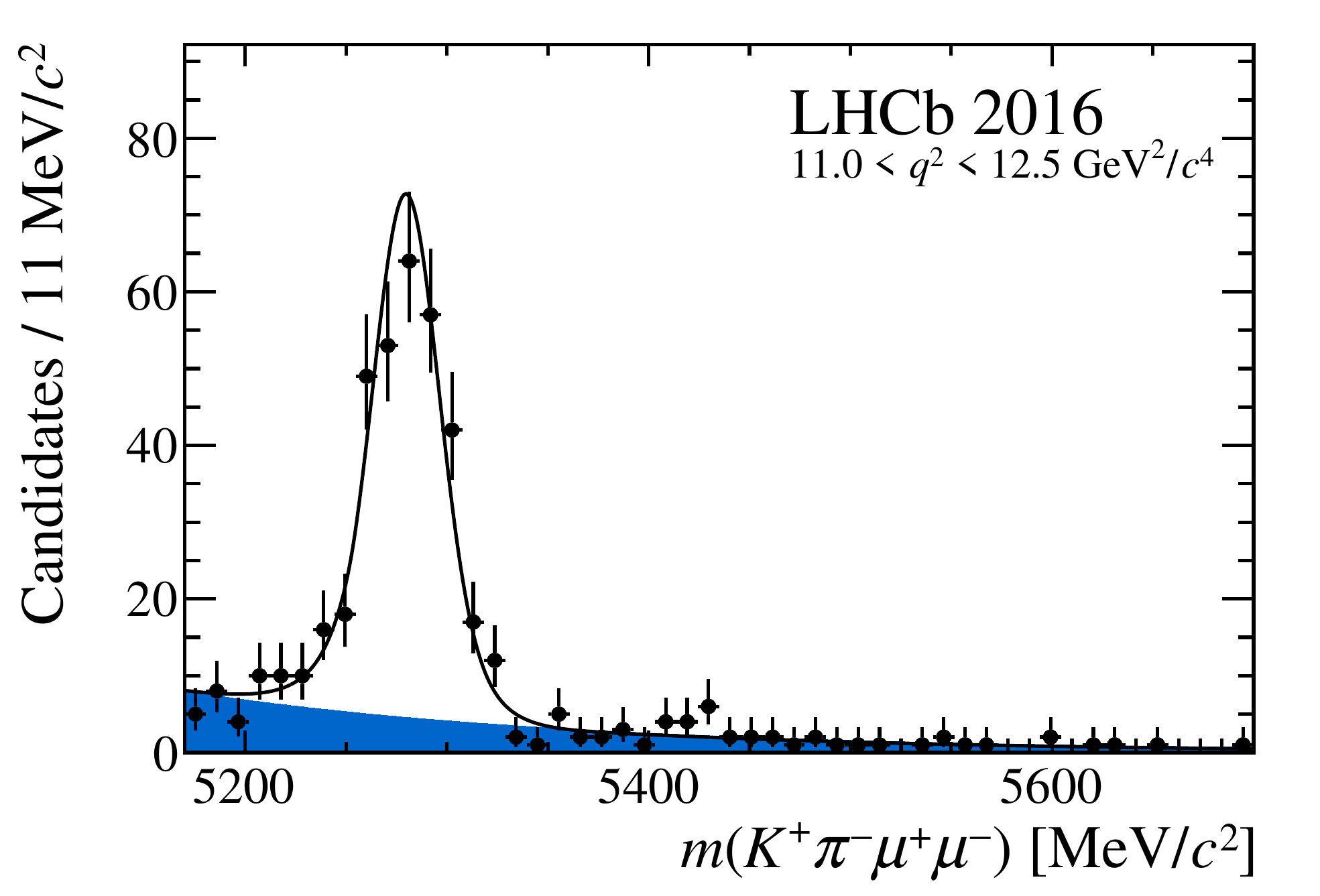}
 \caption{Projections of the fitted probability density function on the decay angles, \Mkpi and \Mkpimm for the bin $11.0<q^2<12.5\gevgevcccc$. The blue shaded region indicates background. \label{fig:projectionsf}}
 \end{figure}

\begin{figure}
   \centering
 \includegraphics[width=0.32\textwidth]{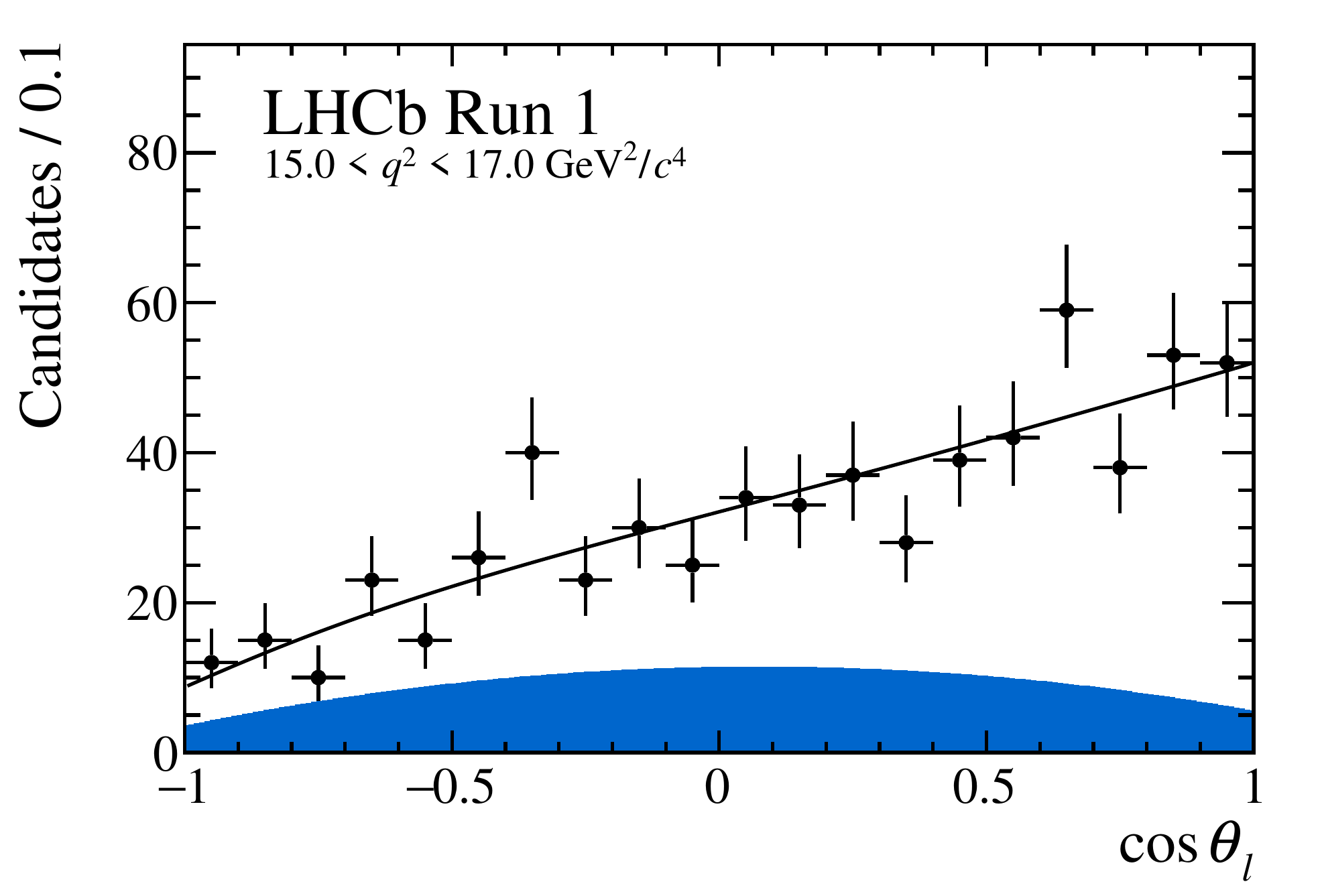}
 \includegraphics[width=0.32\textwidth]{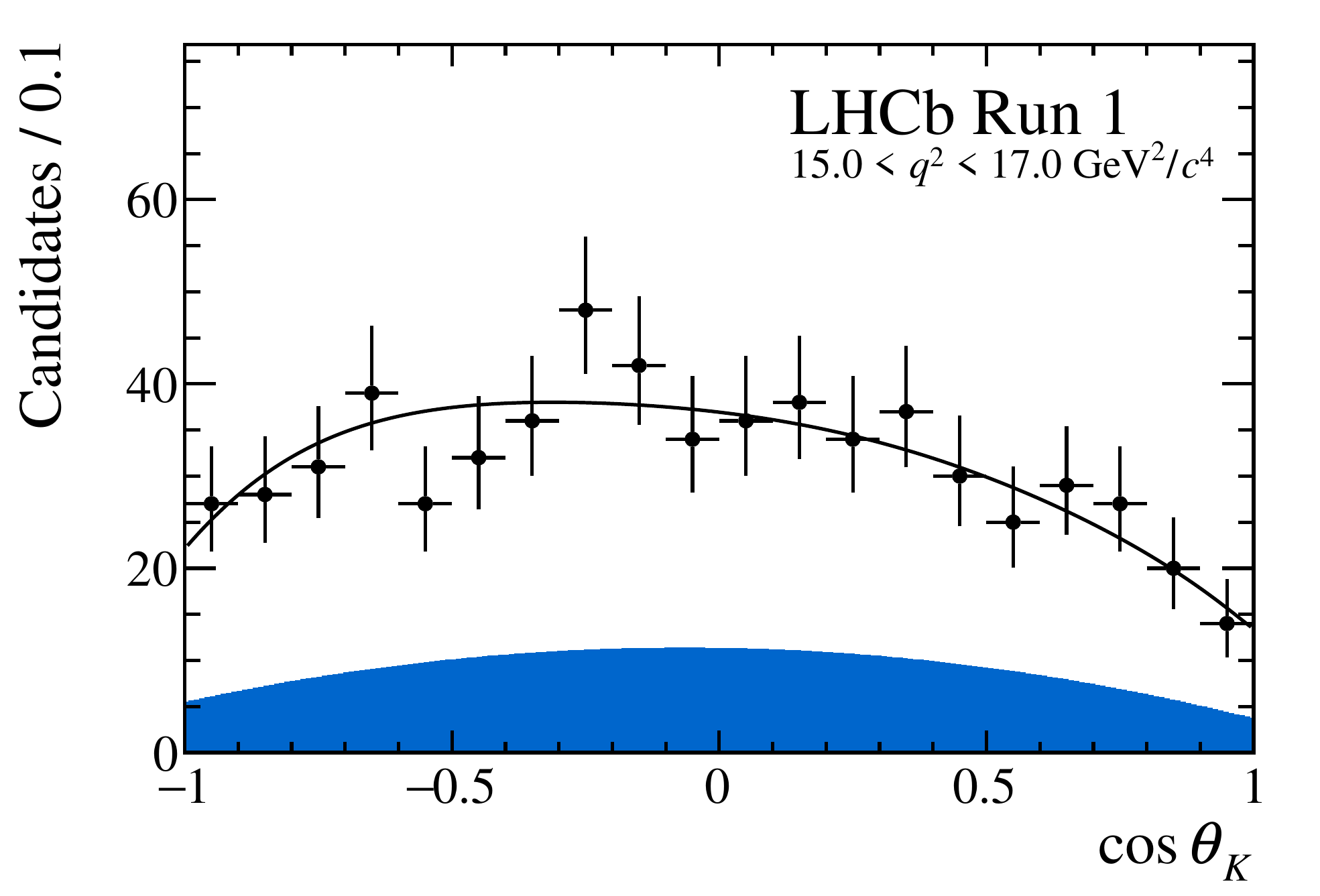}
 \includegraphics[width=0.32\textwidth]{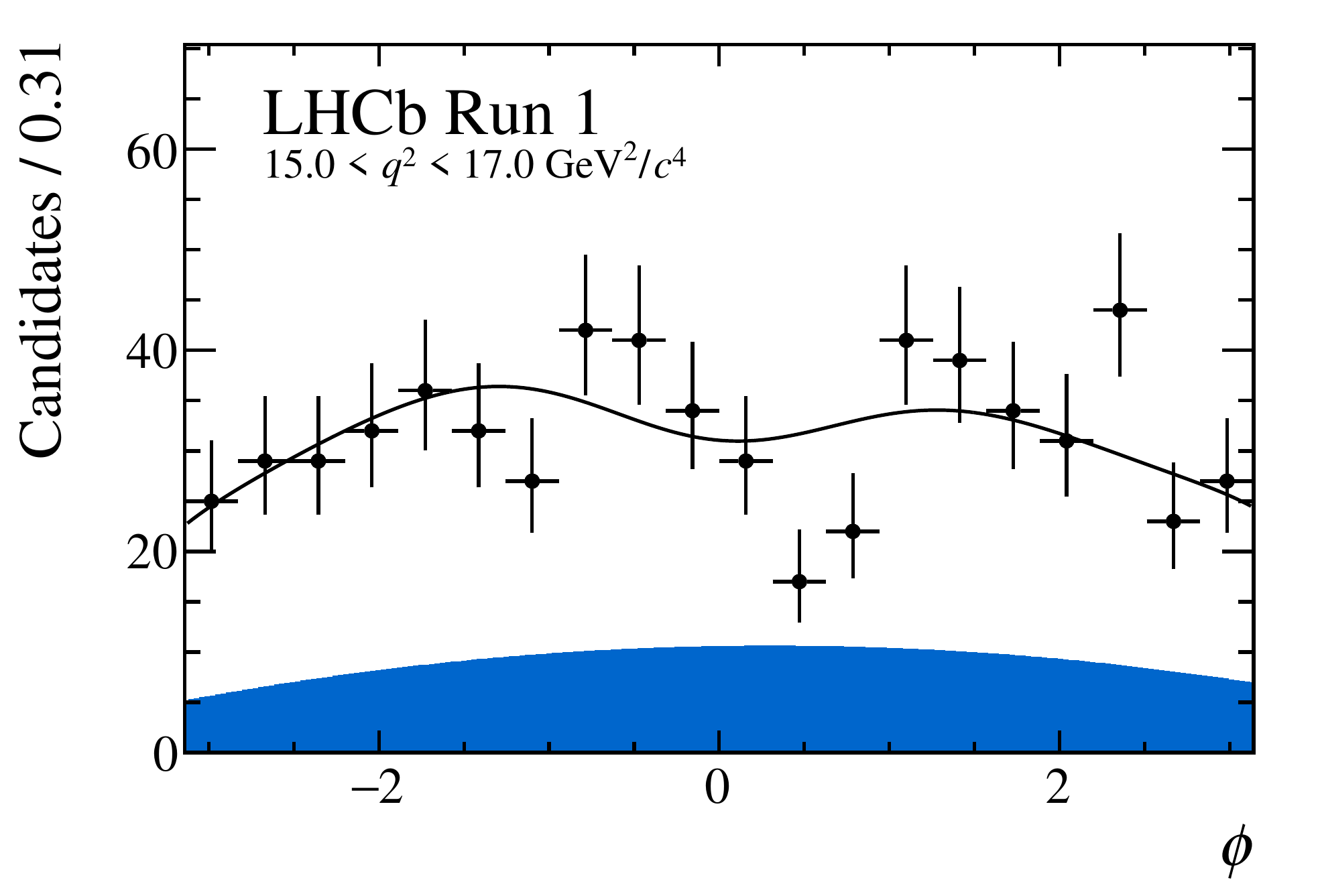}
 \includegraphics[width=0.32\textwidth]{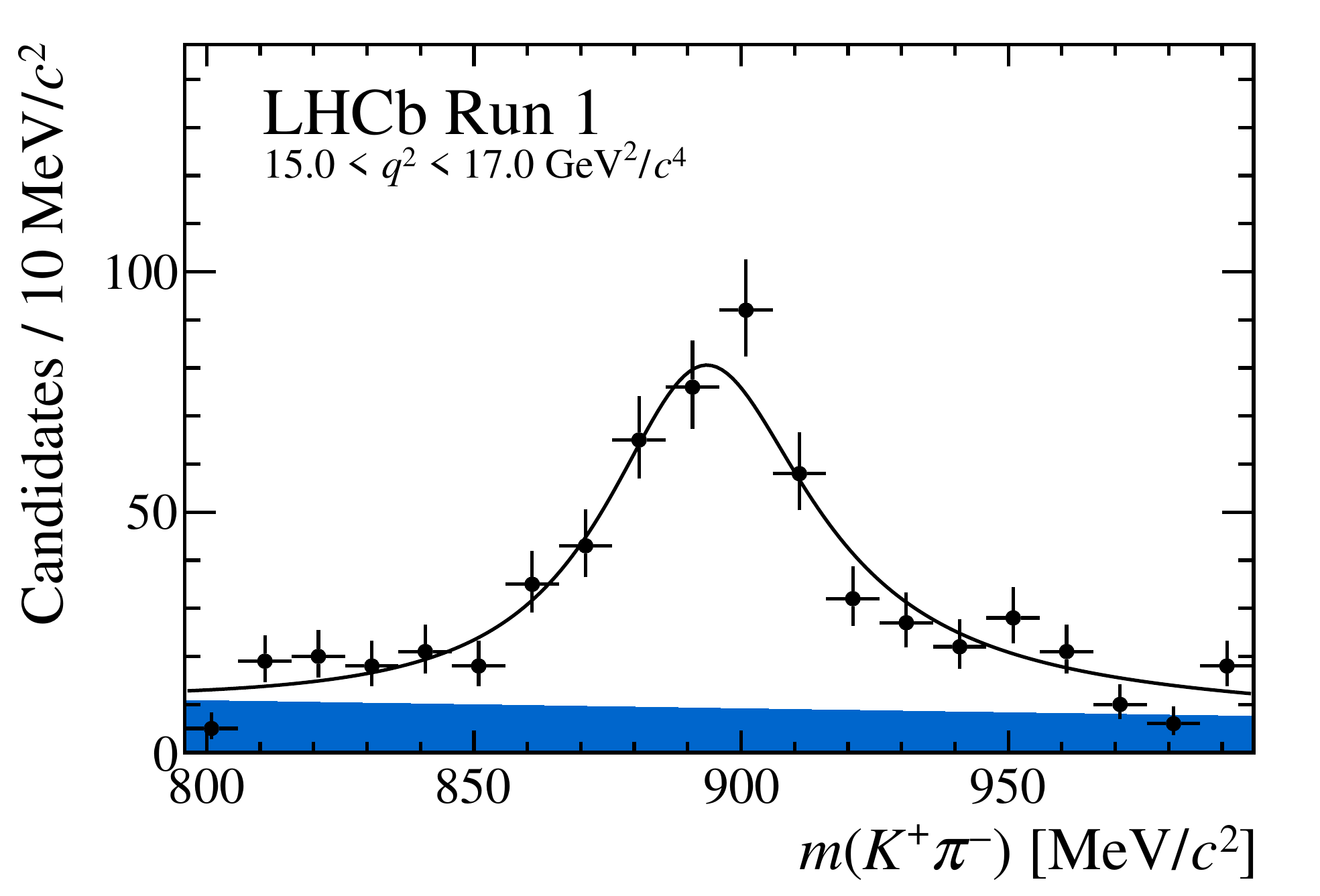}
 \includegraphics[width=0.32\textwidth]{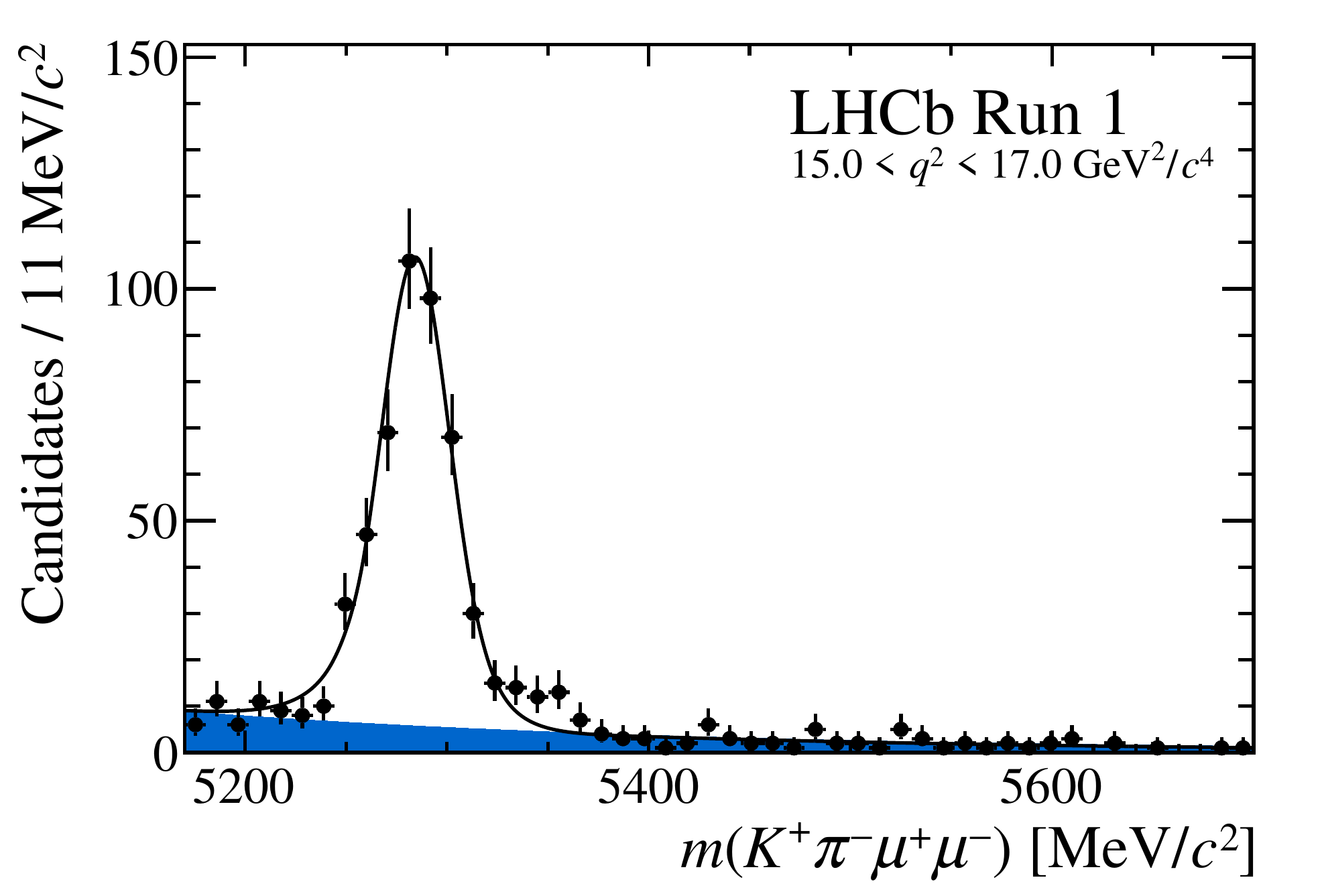}\\[0.5cm]
 \includegraphics[width=0.32\textwidth]{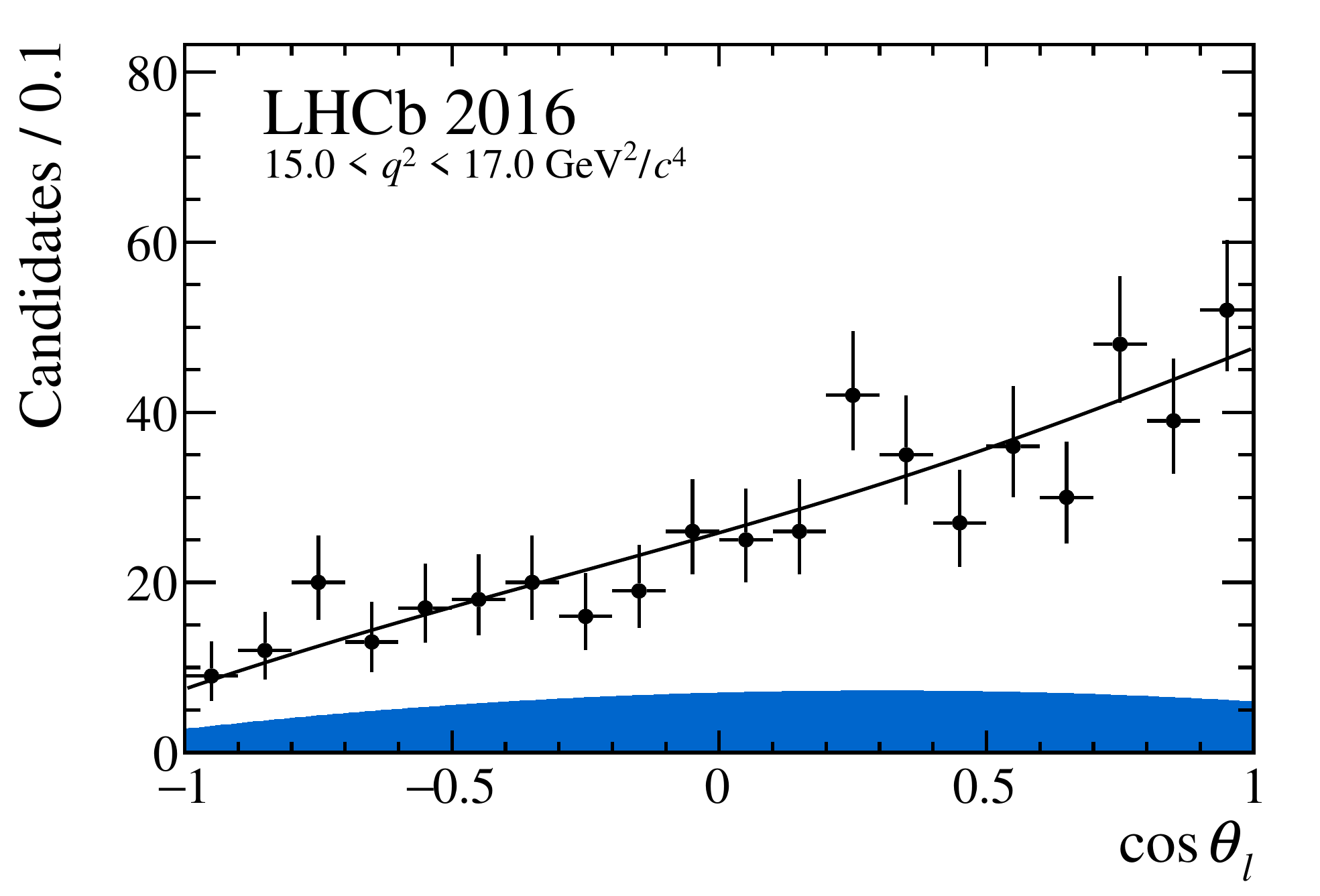}
 \includegraphics[width=0.32\textwidth]{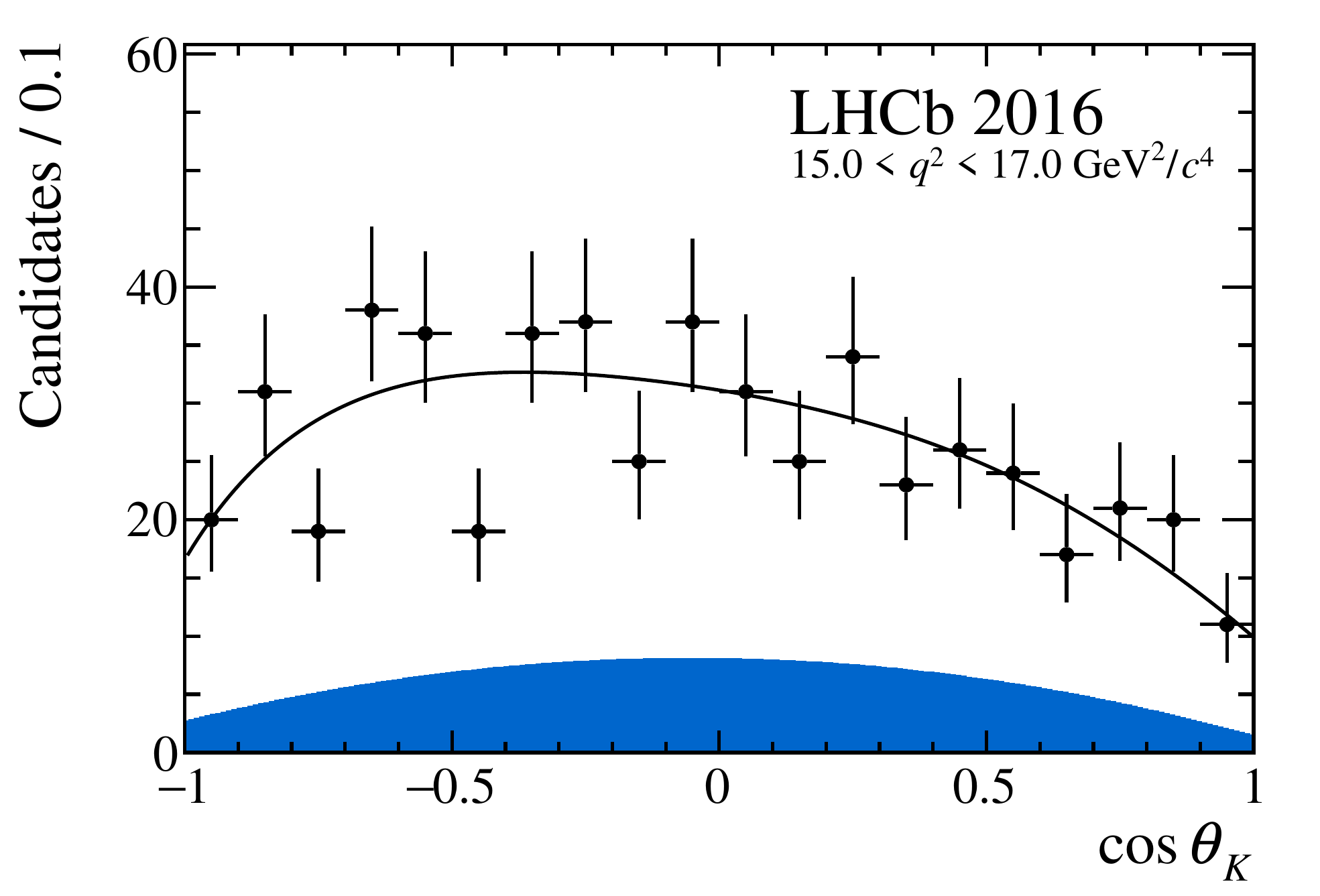}
 \includegraphics[width=0.32\textwidth]{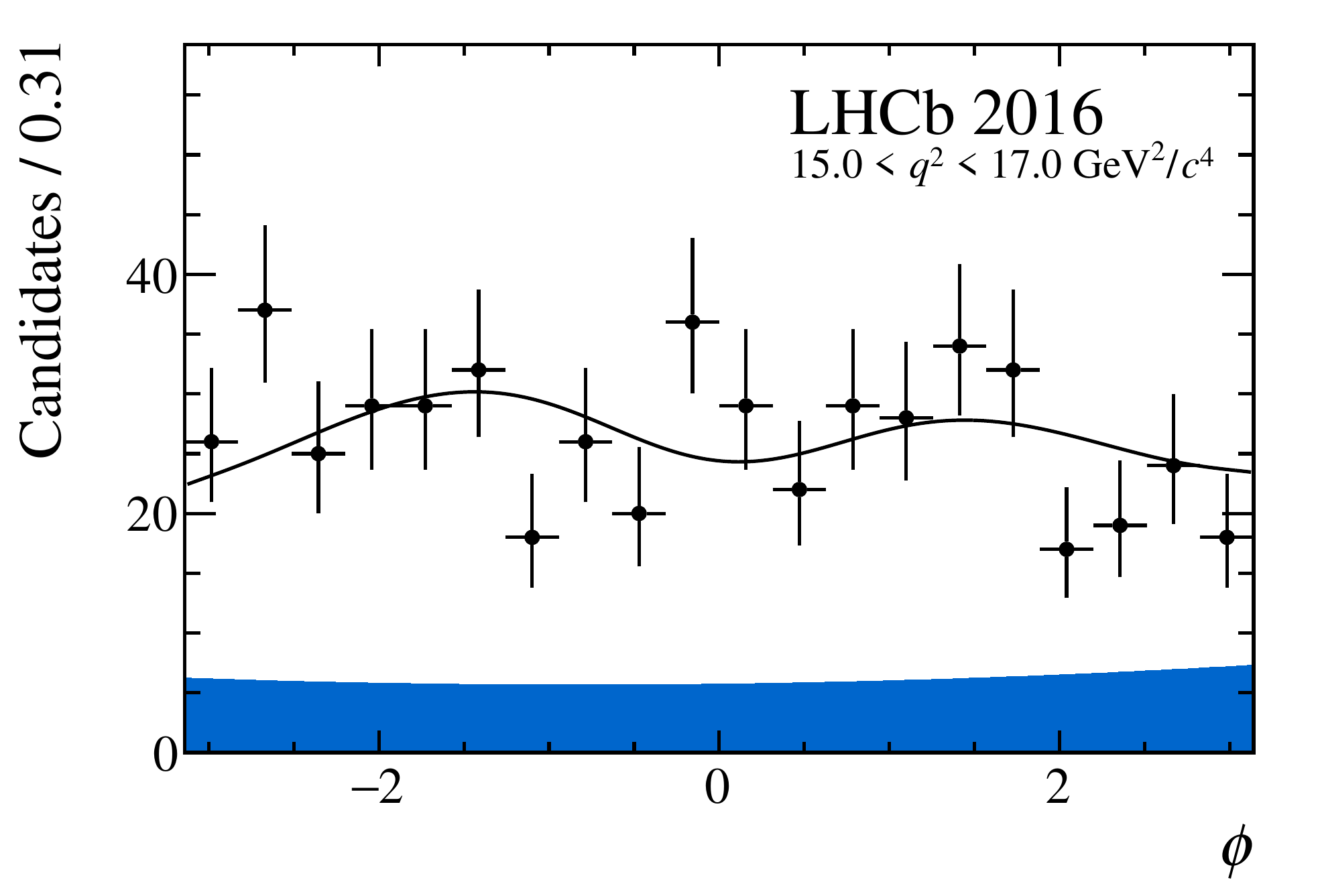}
 \includegraphics[width=0.32\textwidth]{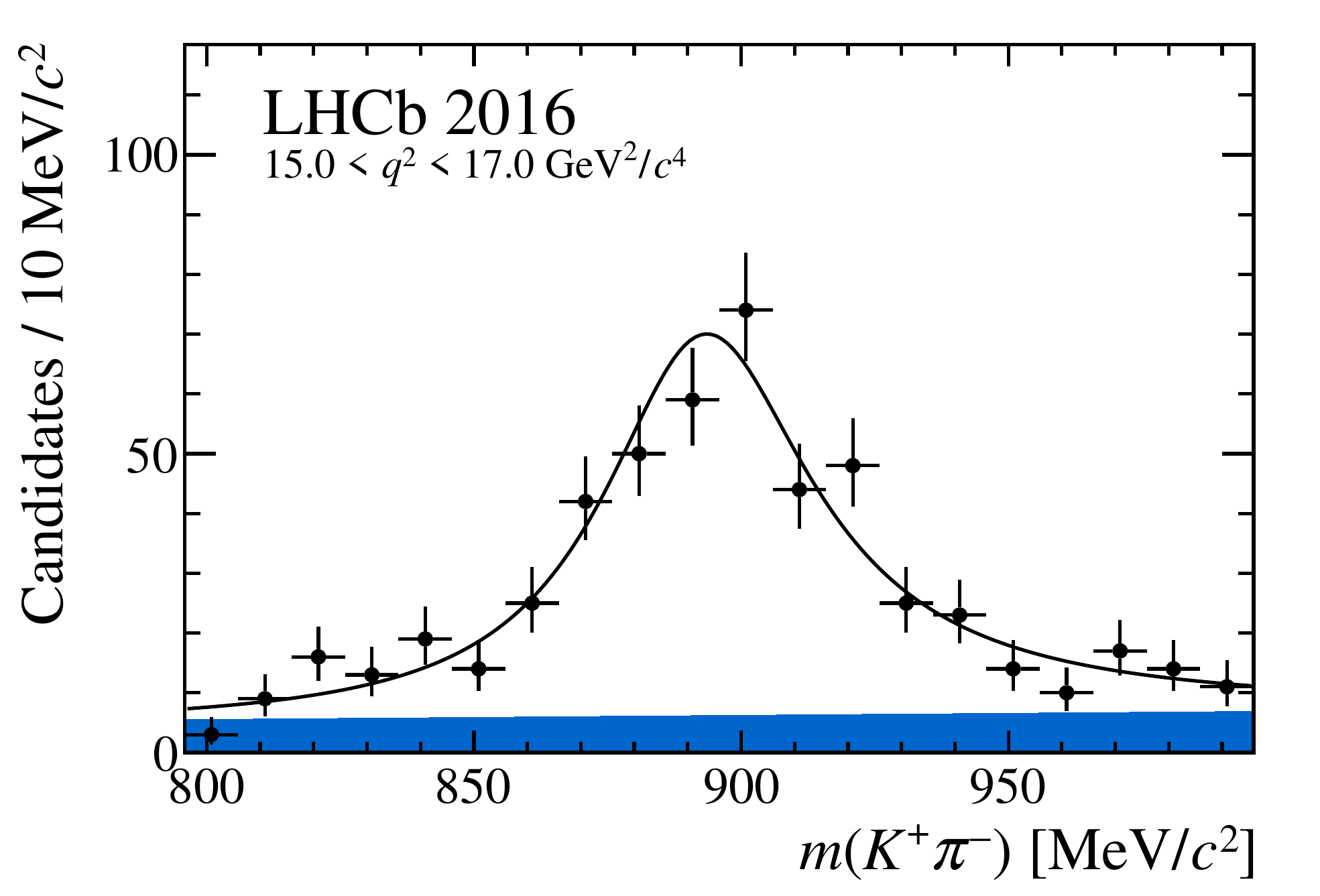}
 \includegraphics[width=0.32\textwidth]{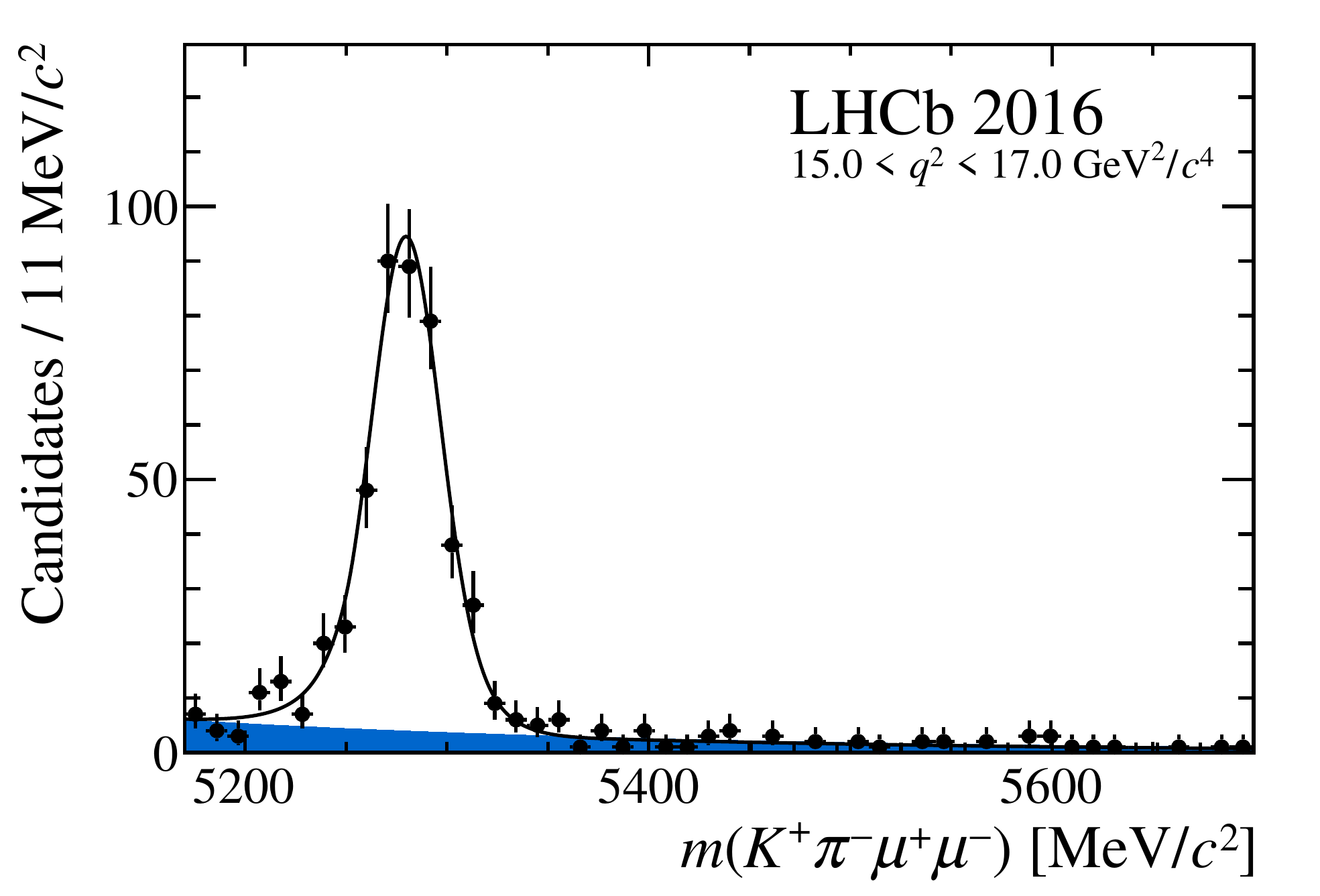}
 \caption{Projections of the fitted probability density function on the decay angles, \Mkpi and \Mkpimm for the bin $15.0<q^2<17.0\gevgevcccc$. The blue shaded region indicates background. \label{fig:projectionsg}}
 \end{figure}

\begin{figure}
   \centering
 \includegraphics[width=0.32\textwidth]{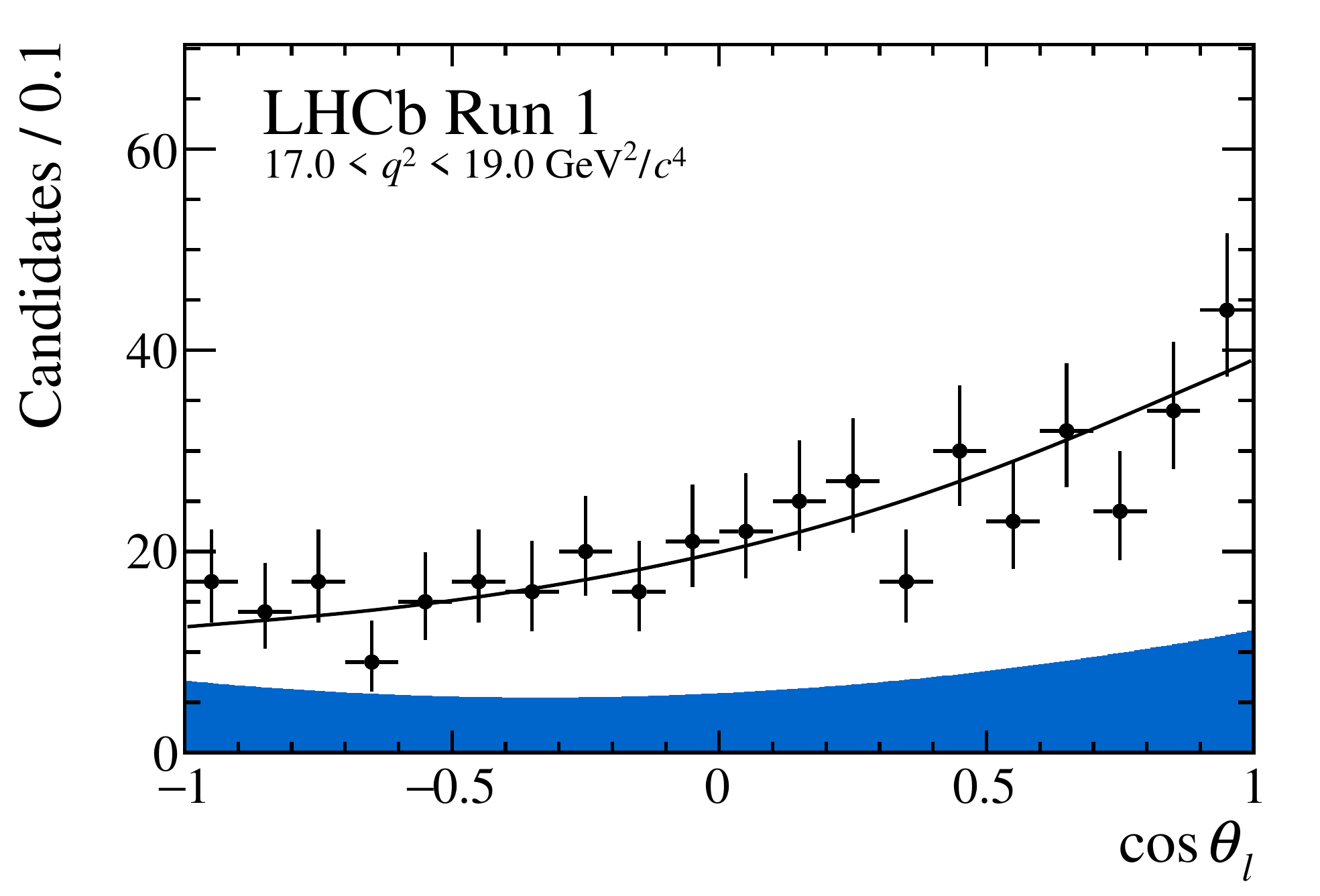}
 \includegraphics[width=0.32\textwidth]{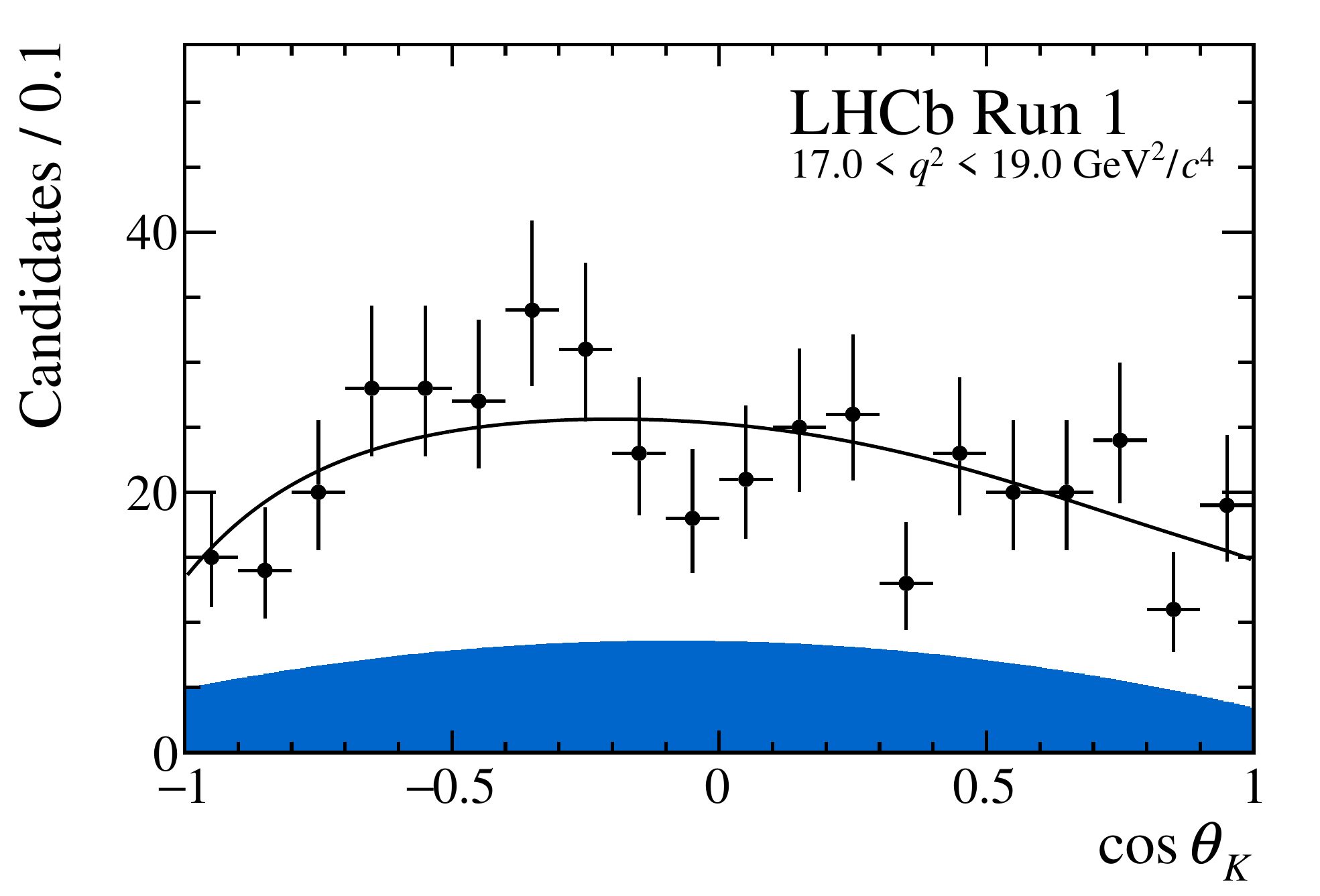}
 \includegraphics[width=0.32\textwidth]{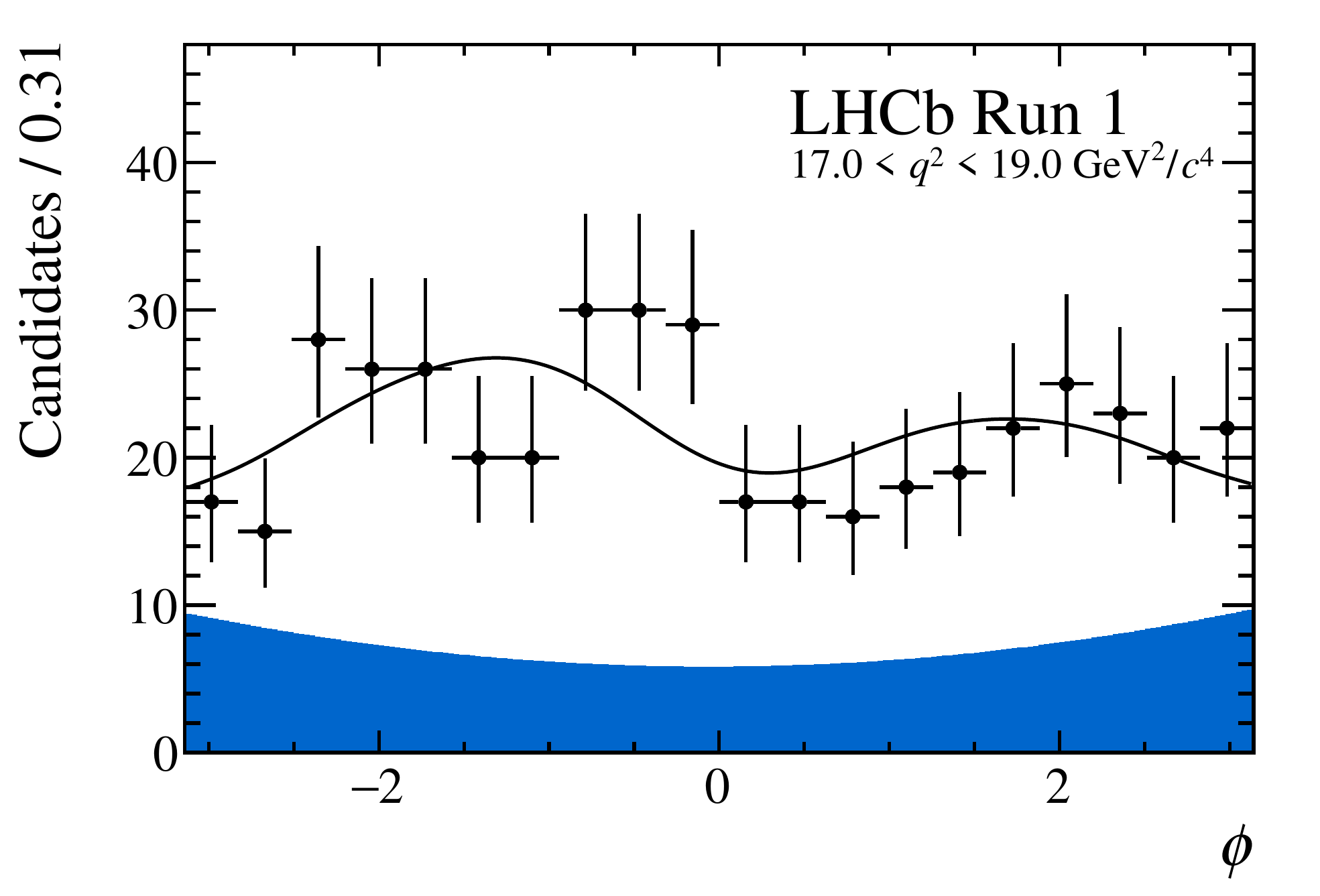}
 \includegraphics[width=0.32\textwidth]{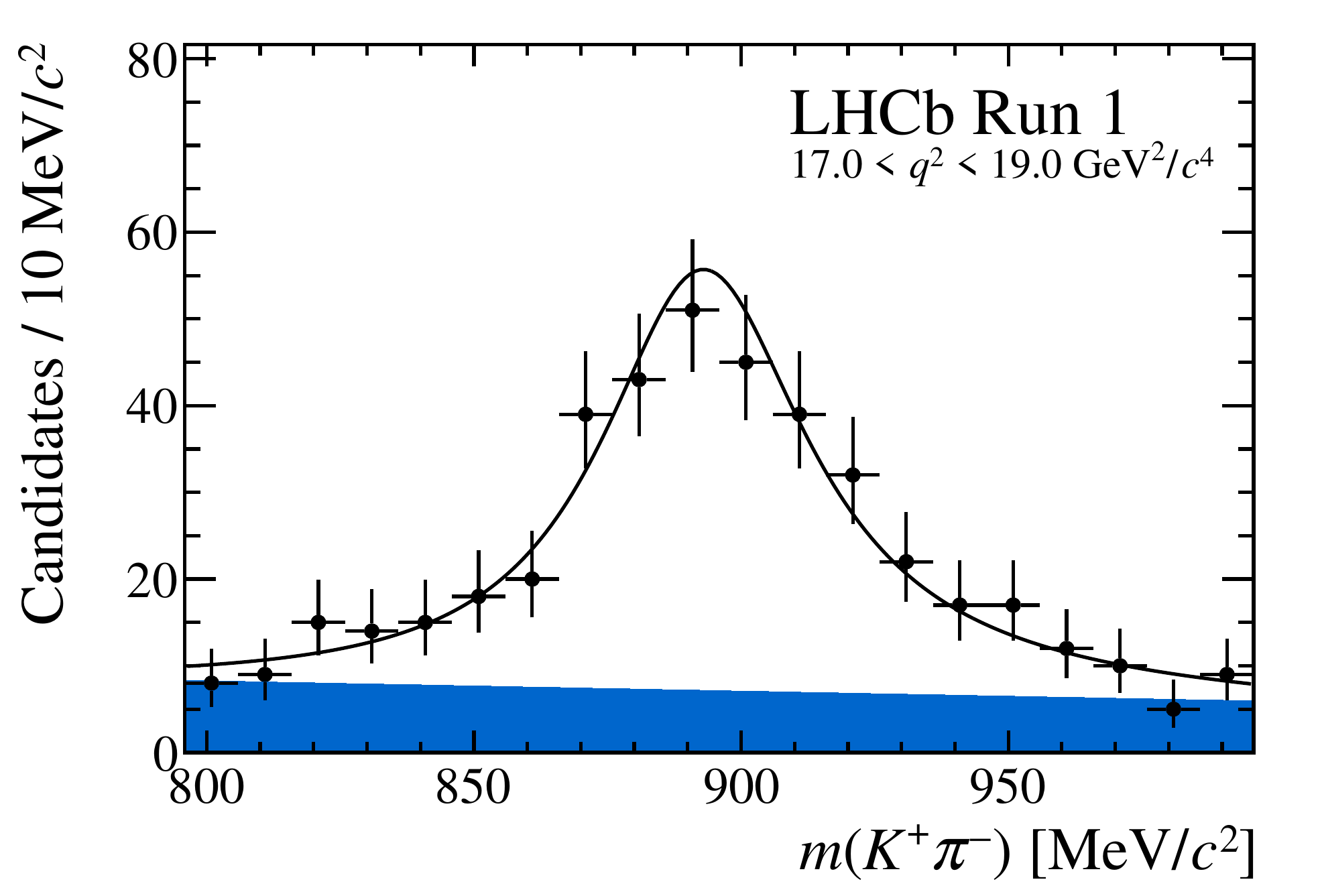}
 \includegraphics[width=0.32\textwidth]{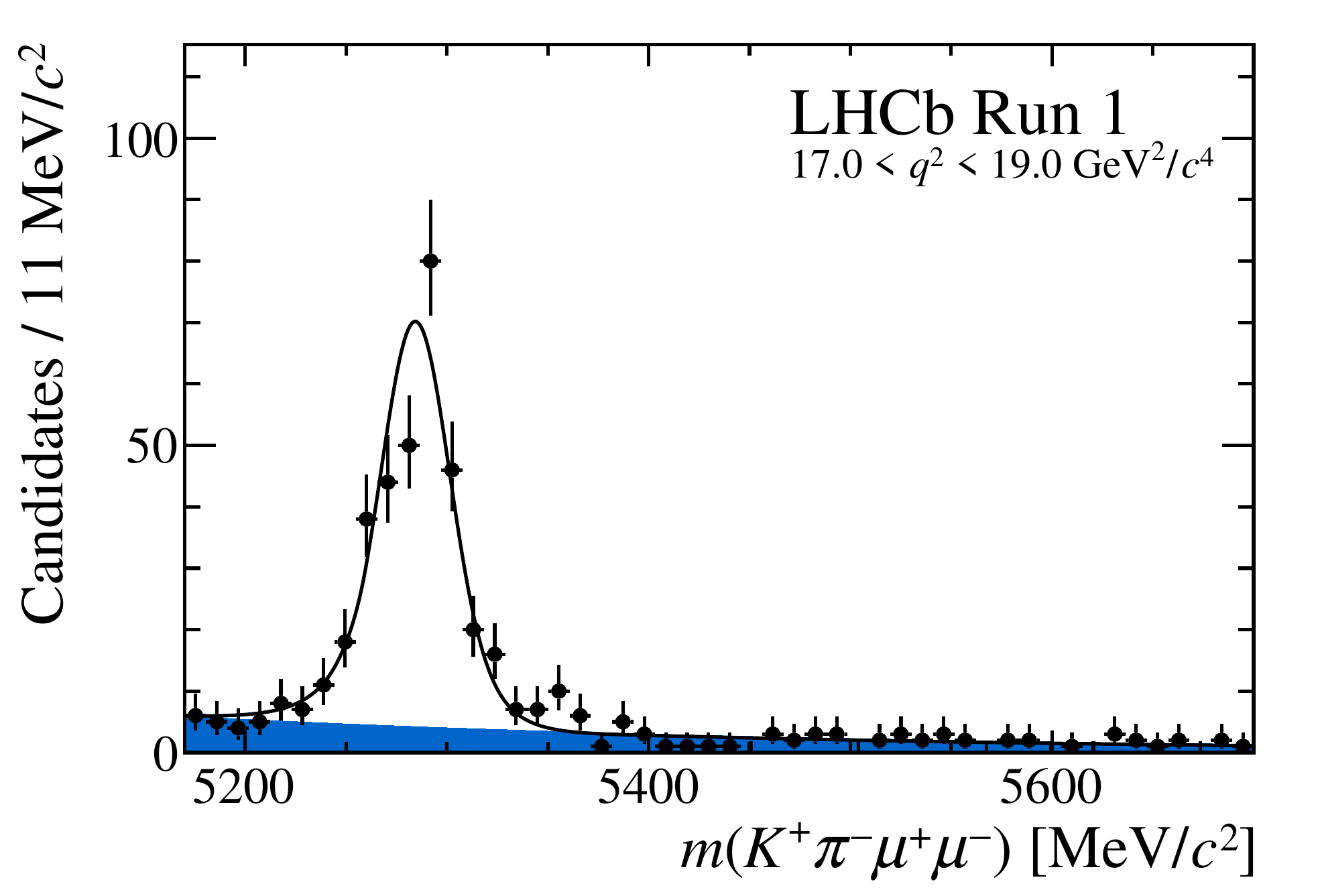}\\[0.5cm]
 \includegraphics[width=0.32\textwidth]{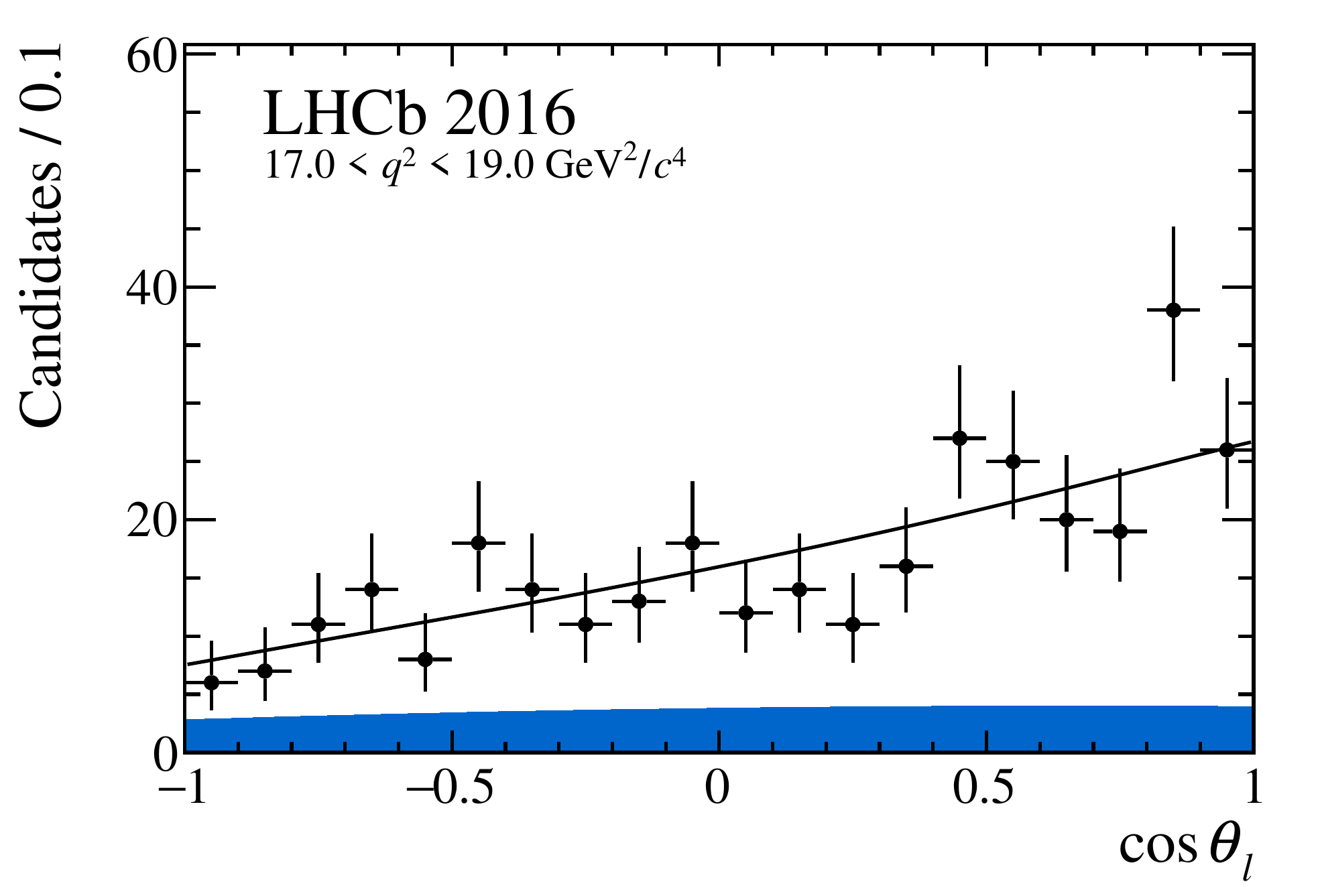}
 \includegraphics[width=0.32\textwidth]{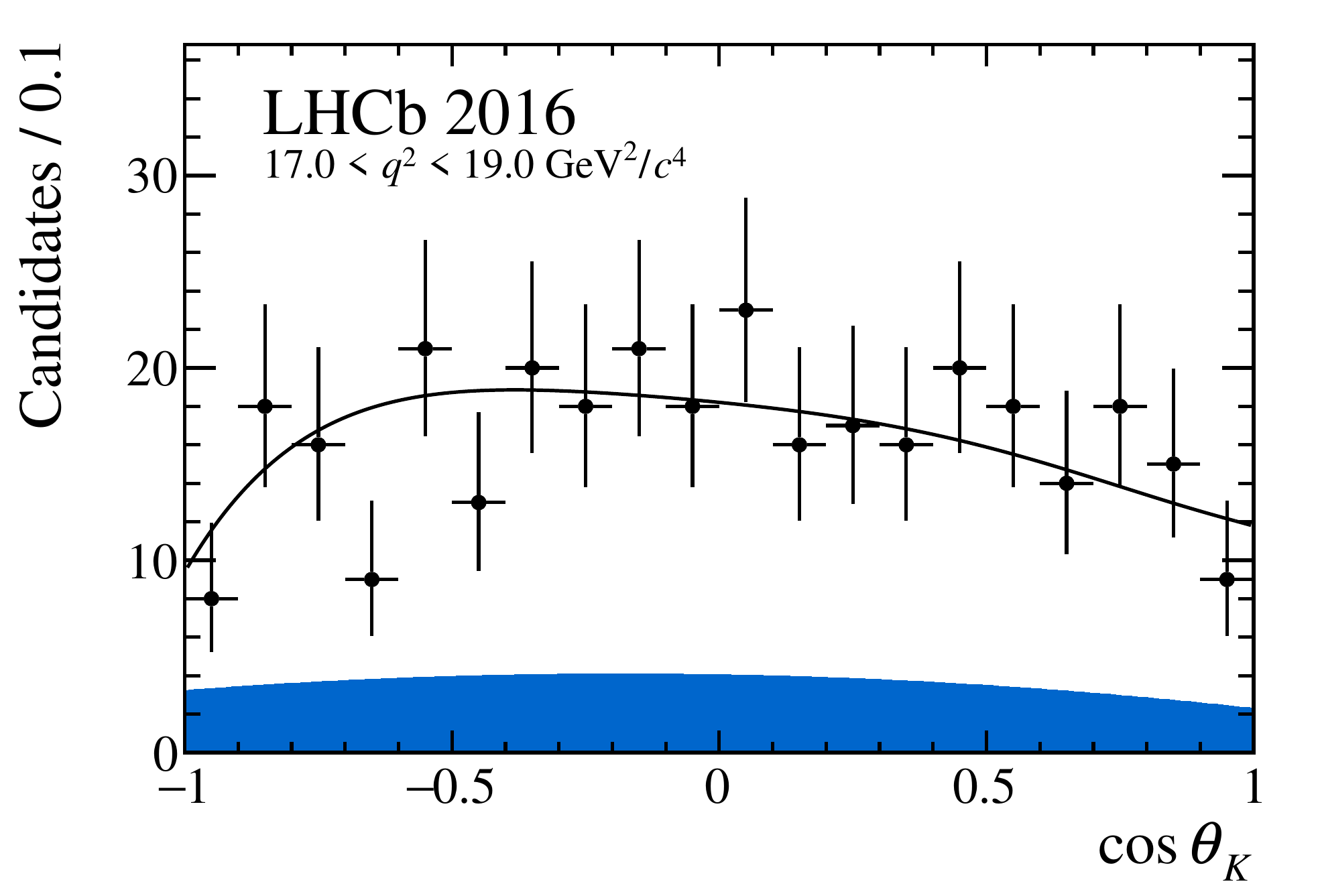}
 \includegraphics[width=0.32\textwidth]{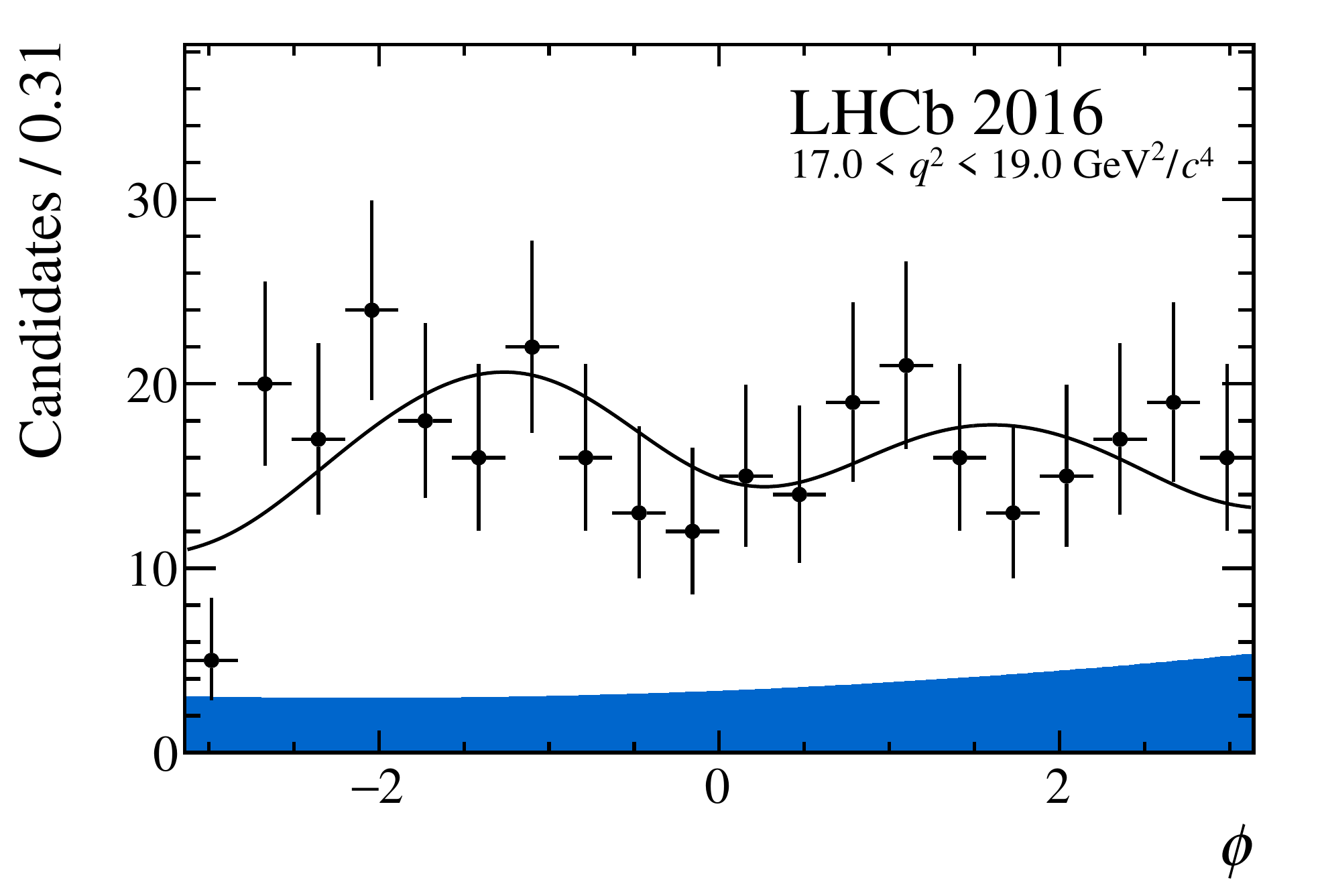}
 \includegraphics[width=0.32\textwidth]{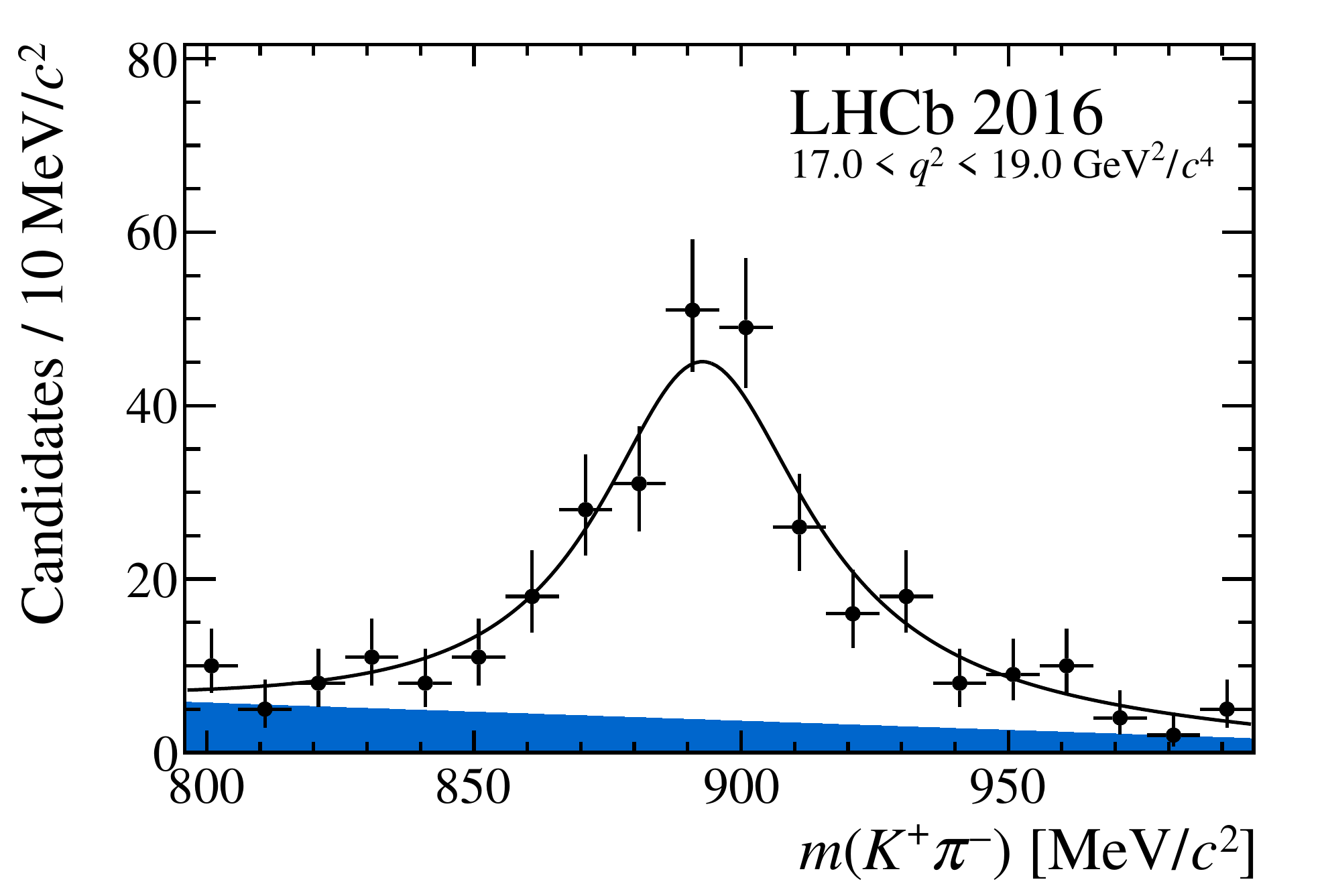}
 \includegraphics[width=0.32\textwidth]{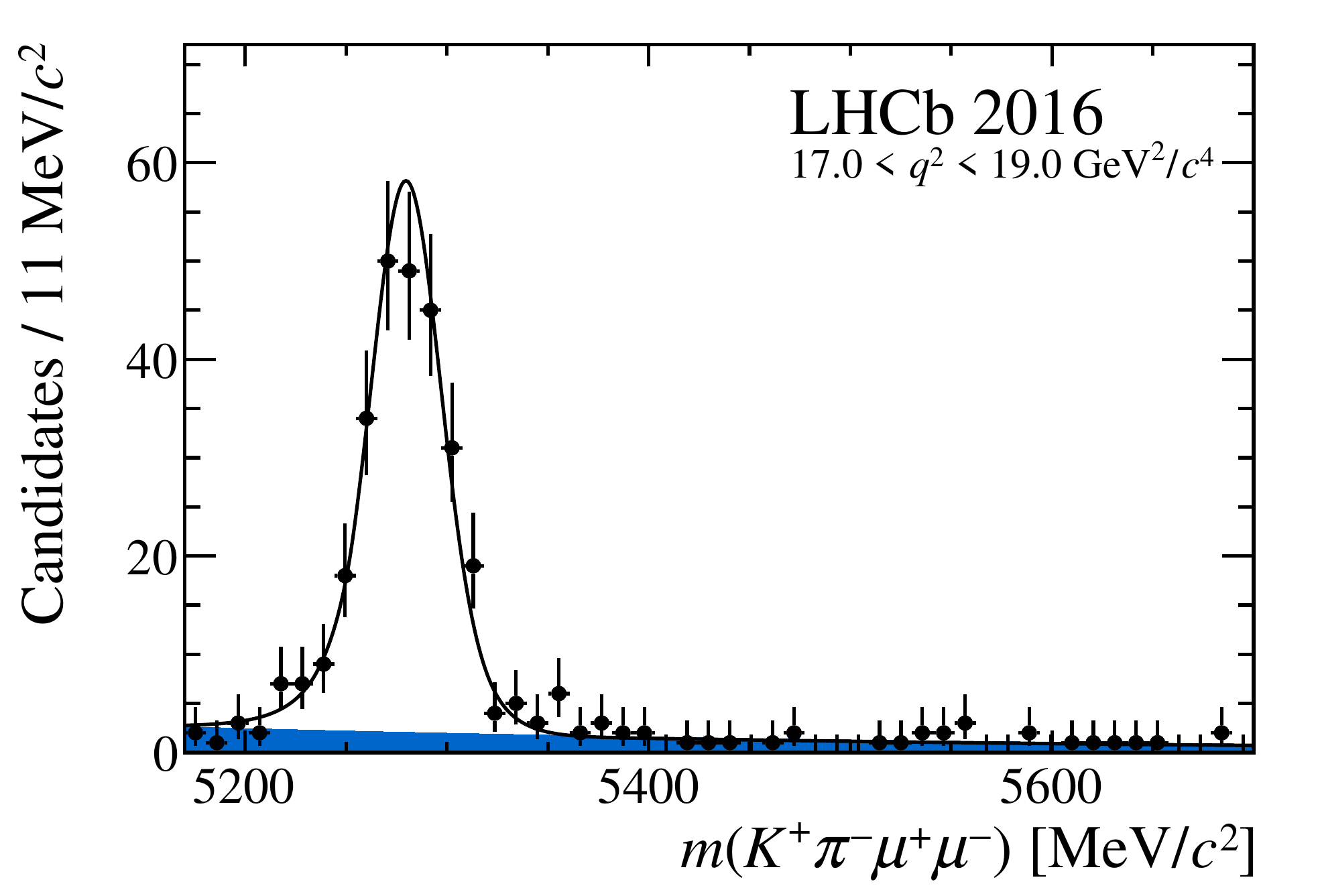}
 \caption{Projections of the fitted probability density function on the decay angles, \Mkpi and \Mkpimm for the bin $17.0<q^2<19.0\gevgevcccc$. The blue shaded region indicates background. \label{fig:projectionsh}}
 \end{figure}

\begin{figure}
   \centering
 \includegraphics[width=0.32\textwidth]{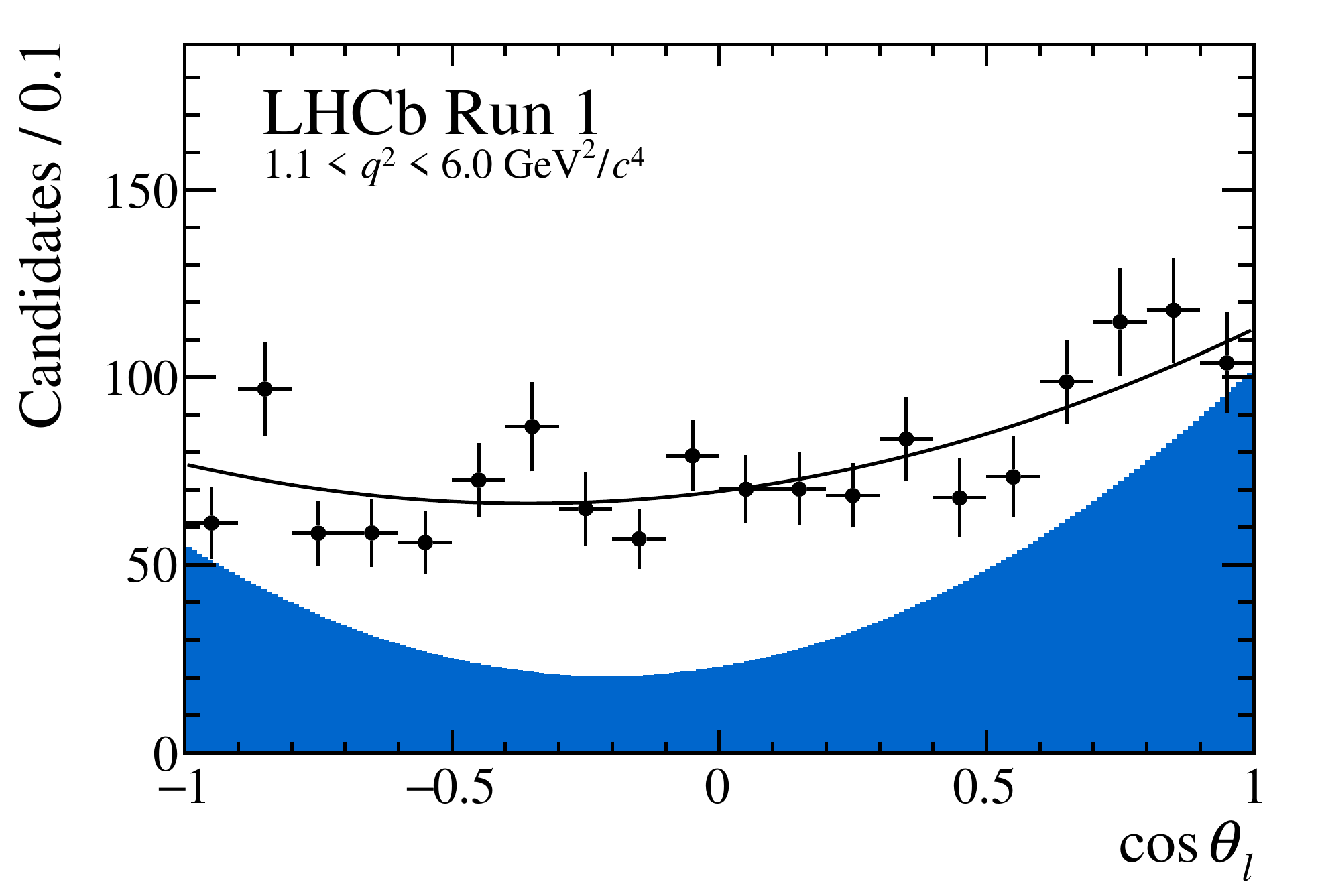}
 \includegraphics[width=0.32\textwidth]{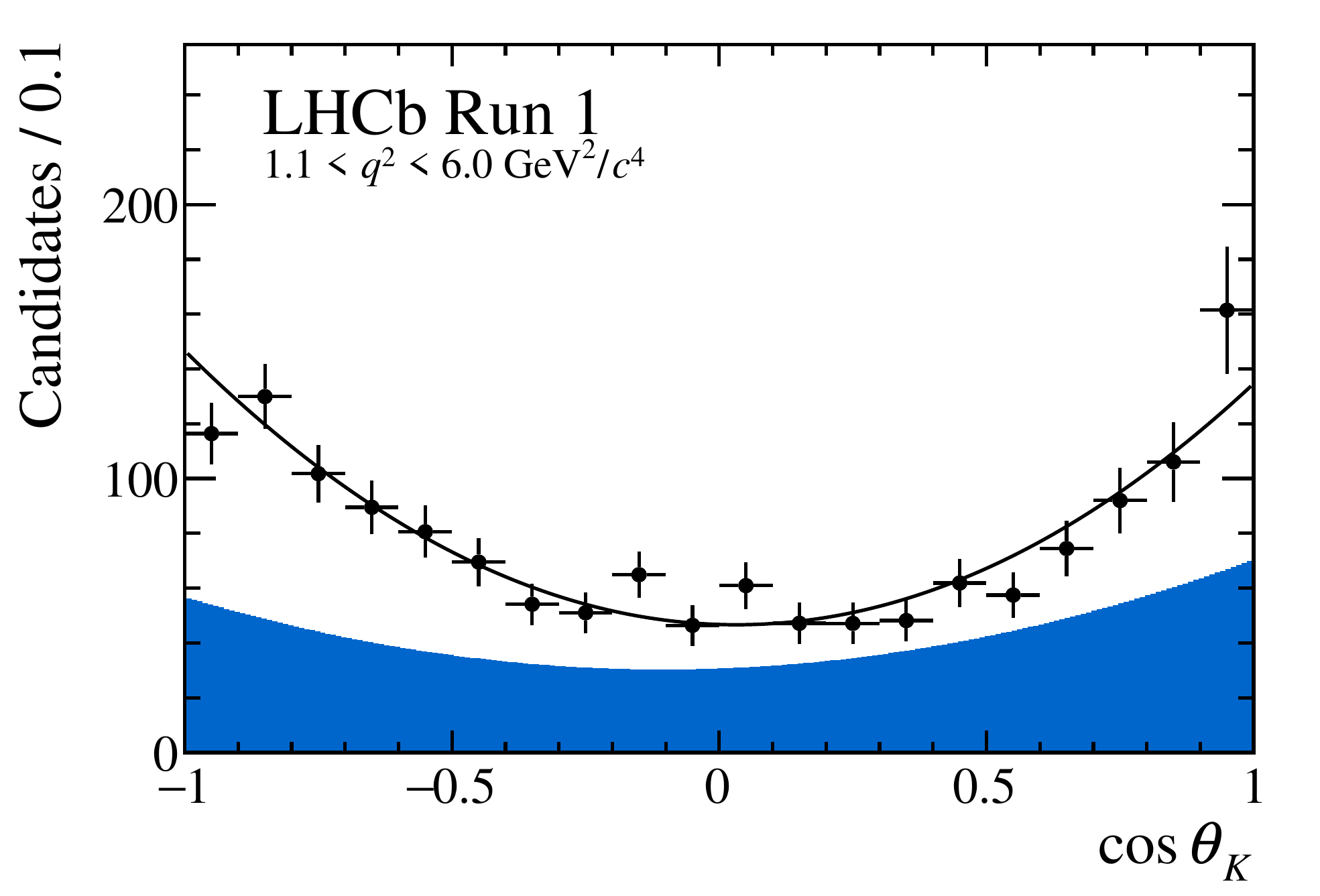}
 \includegraphics[width=0.32\textwidth]{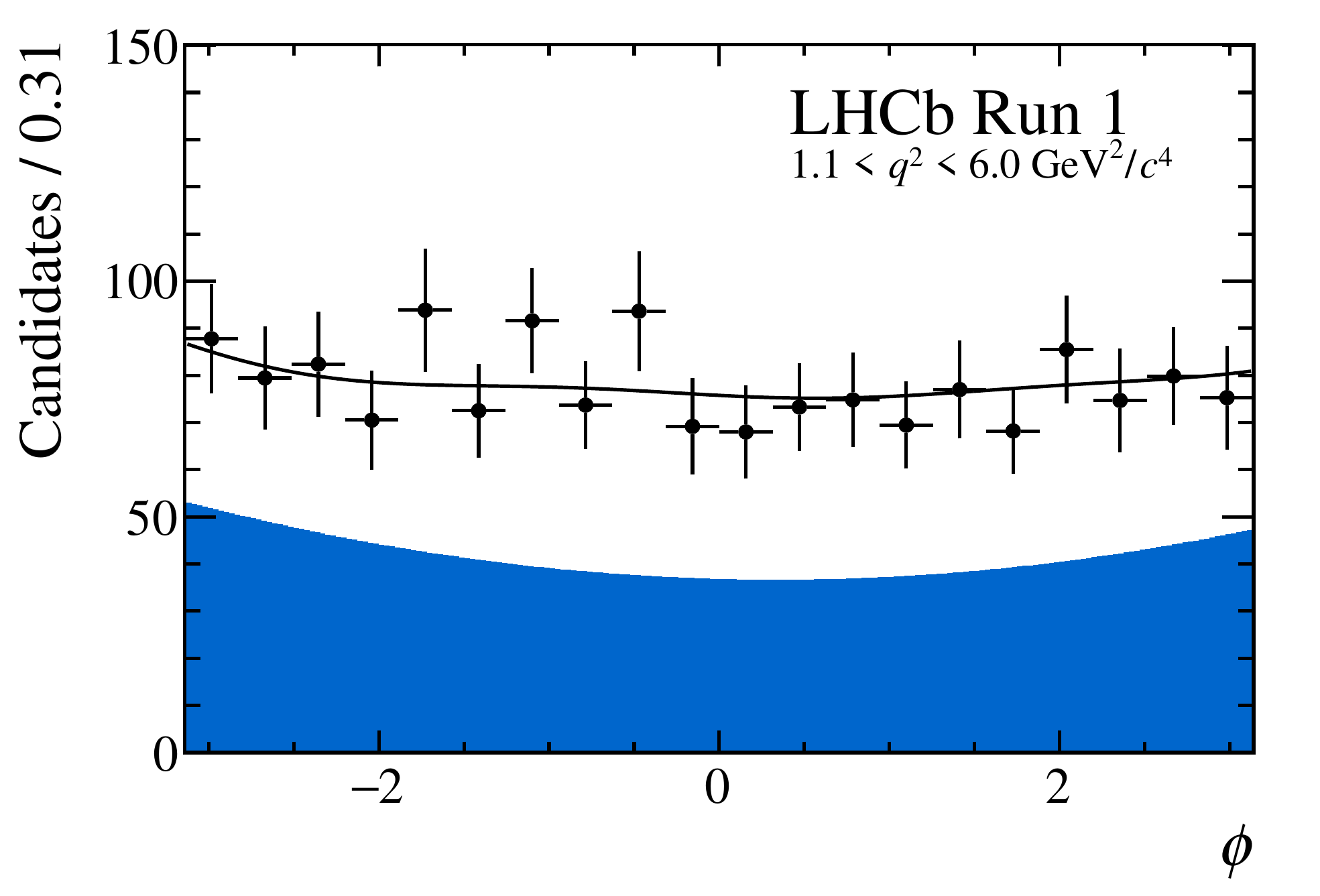}
 \includegraphics[width=0.32\textwidth]{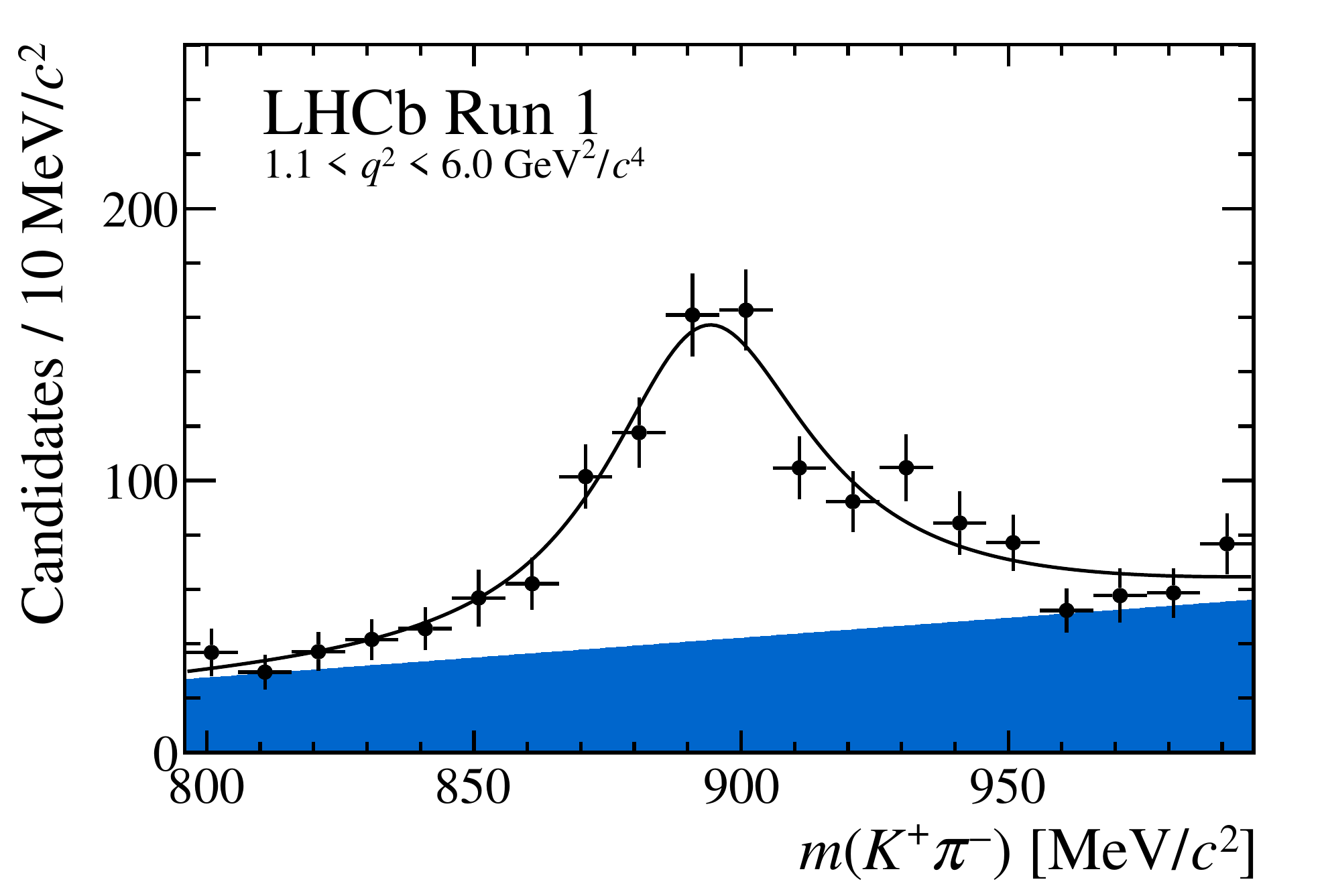}
 \includegraphics[width=0.32\textwidth]{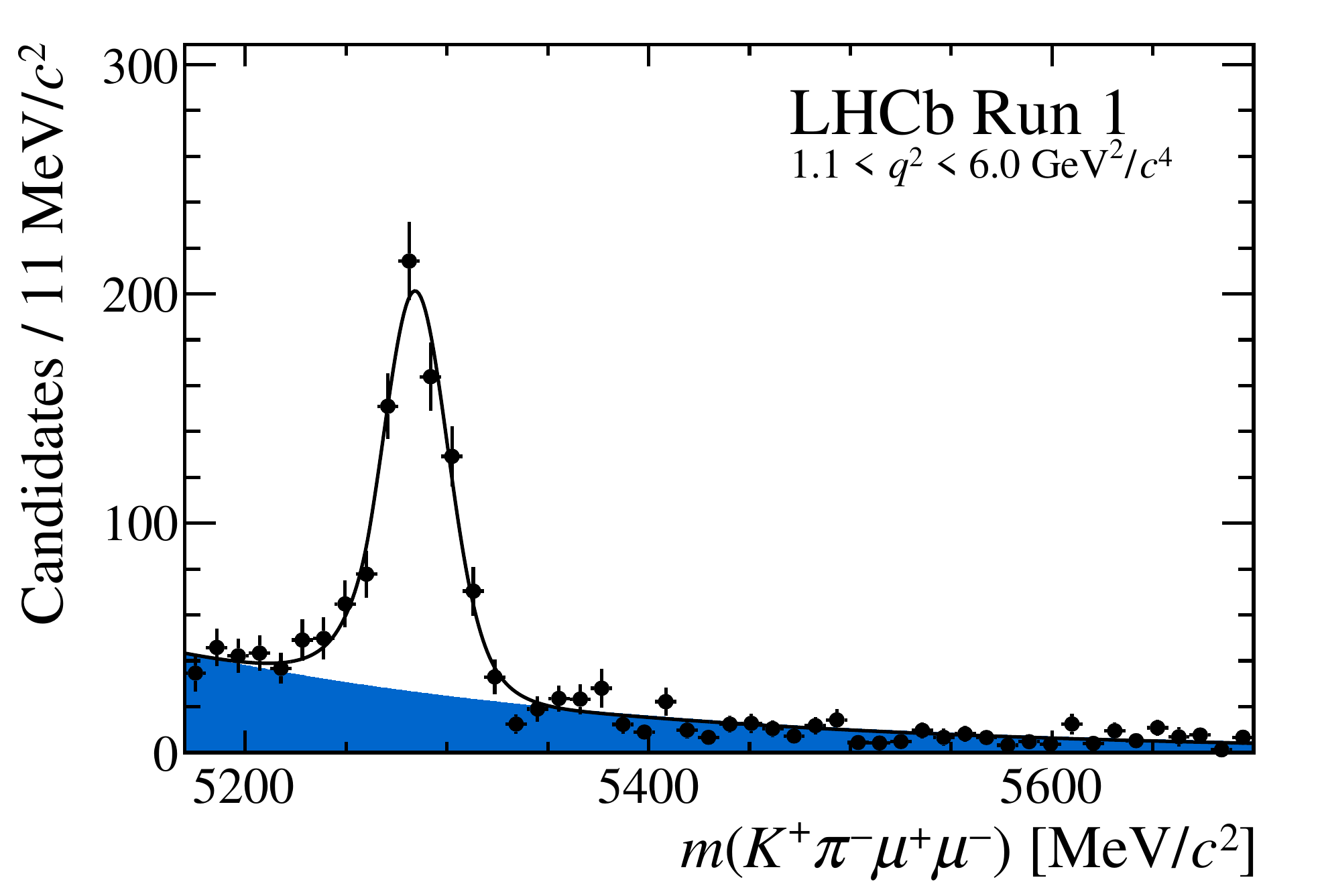}\\[0.5cm]
 \includegraphics[width=0.32\textwidth]{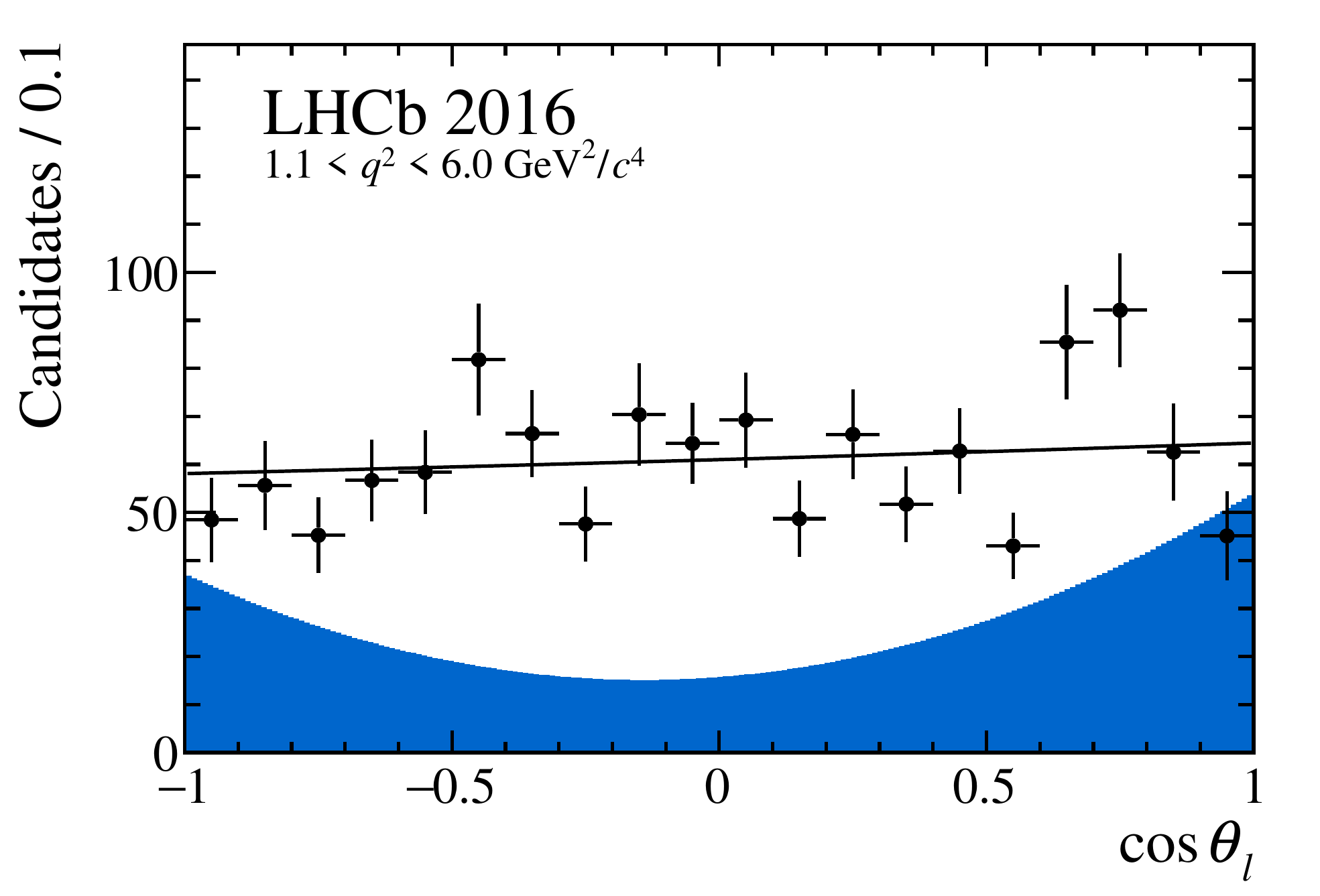}
 \includegraphics[width=0.32\textwidth]{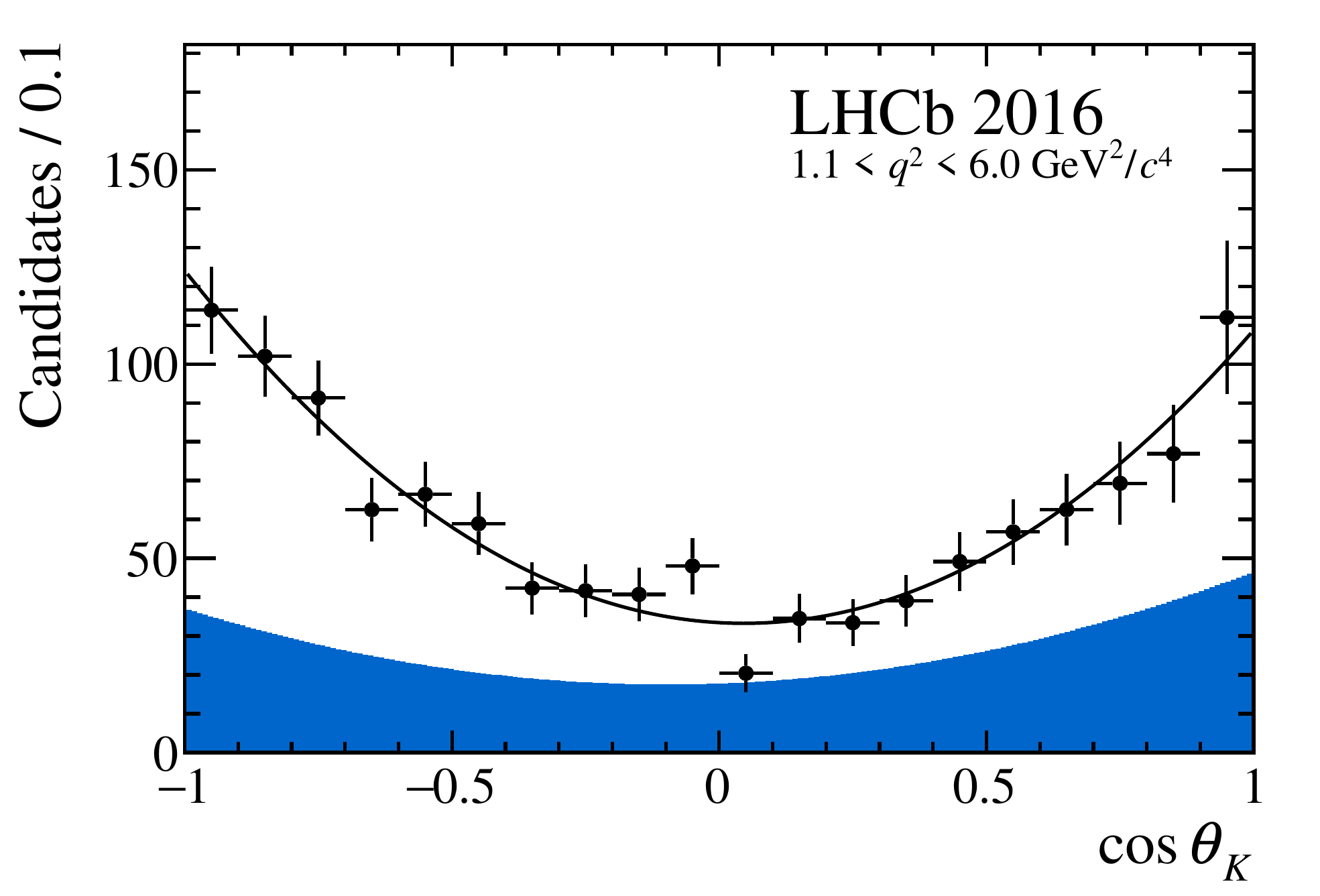}
 \includegraphics[width=0.32\textwidth]{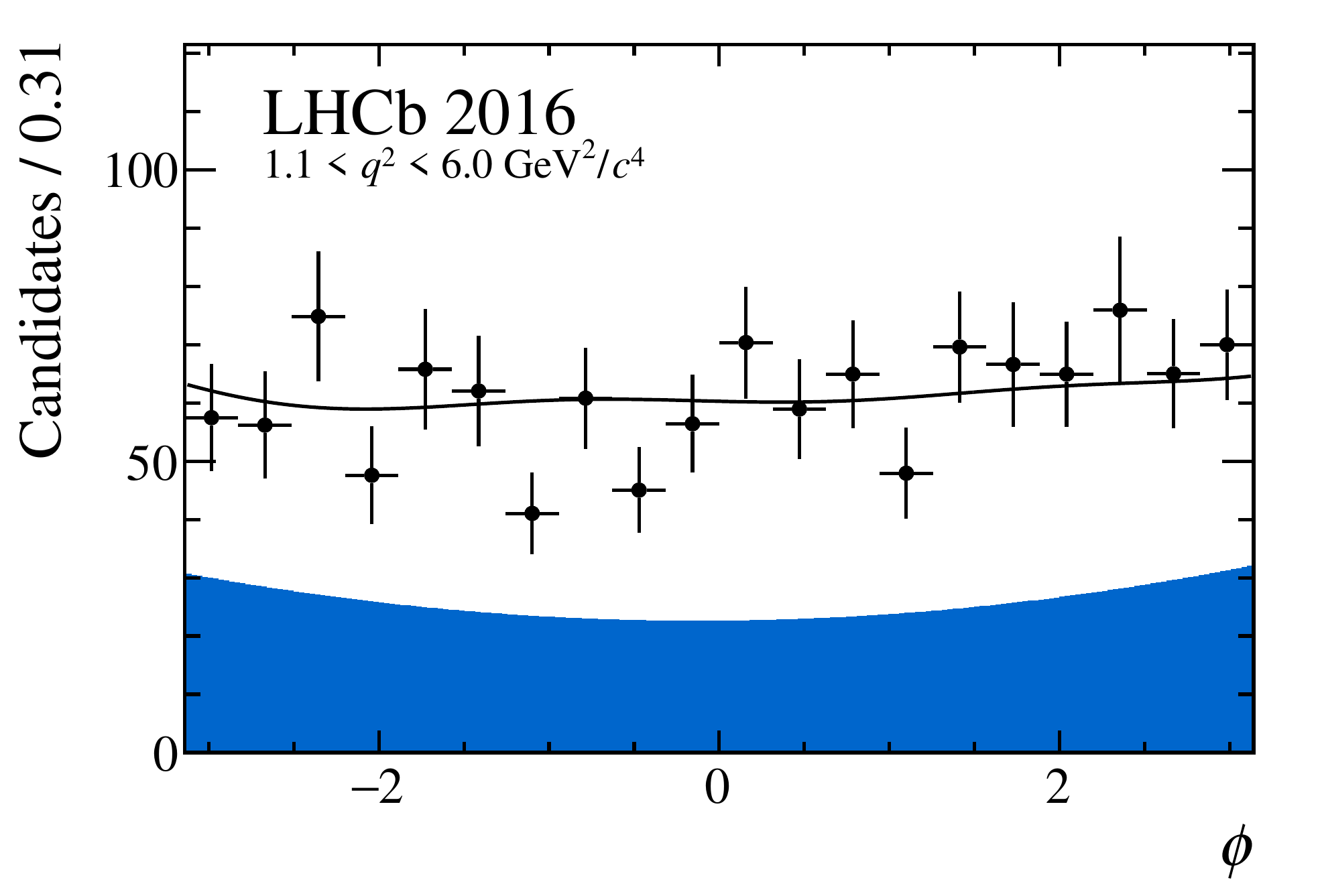}
 \includegraphics[width=0.32\textwidth]{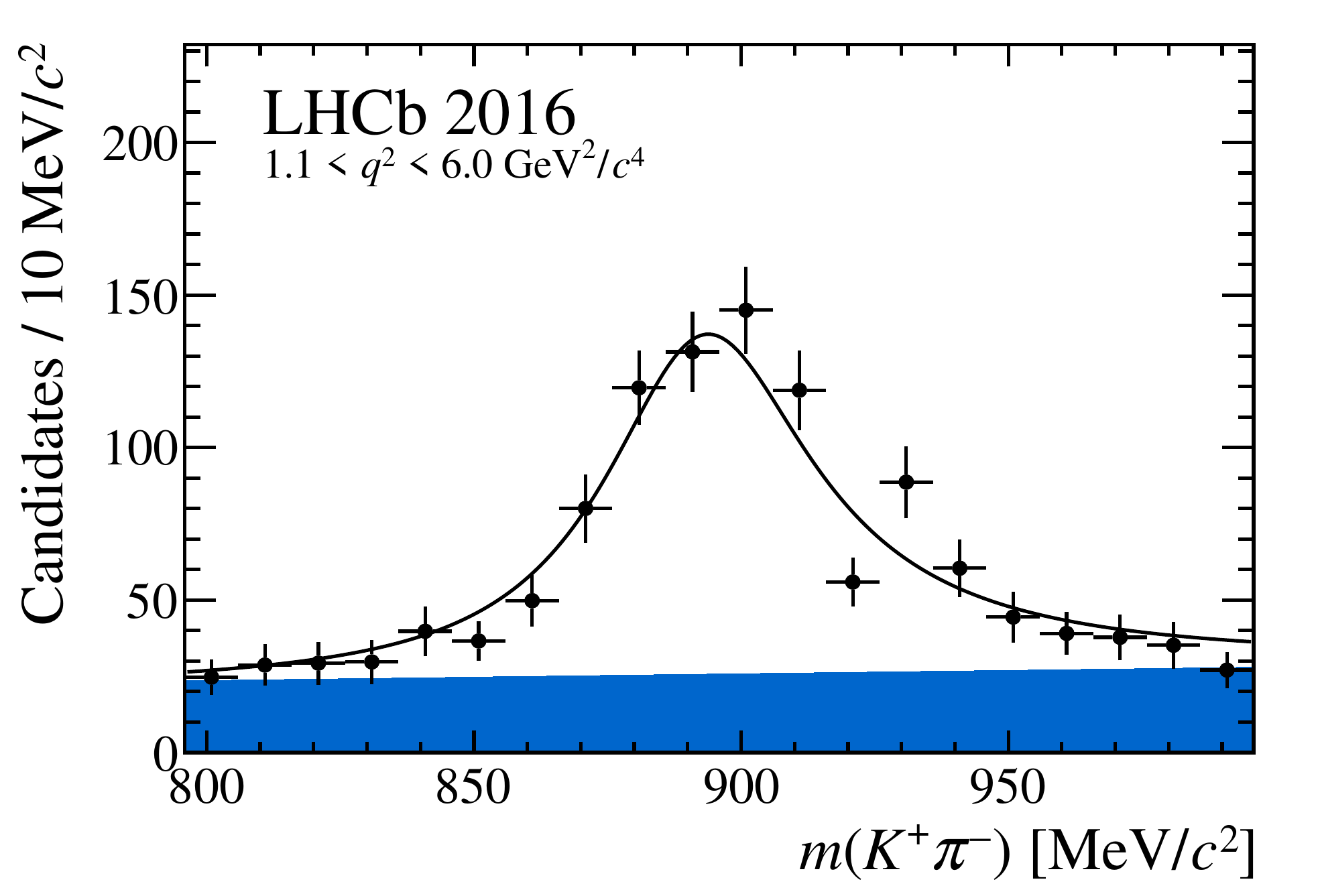}
 \includegraphics[width=0.32\textwidth]{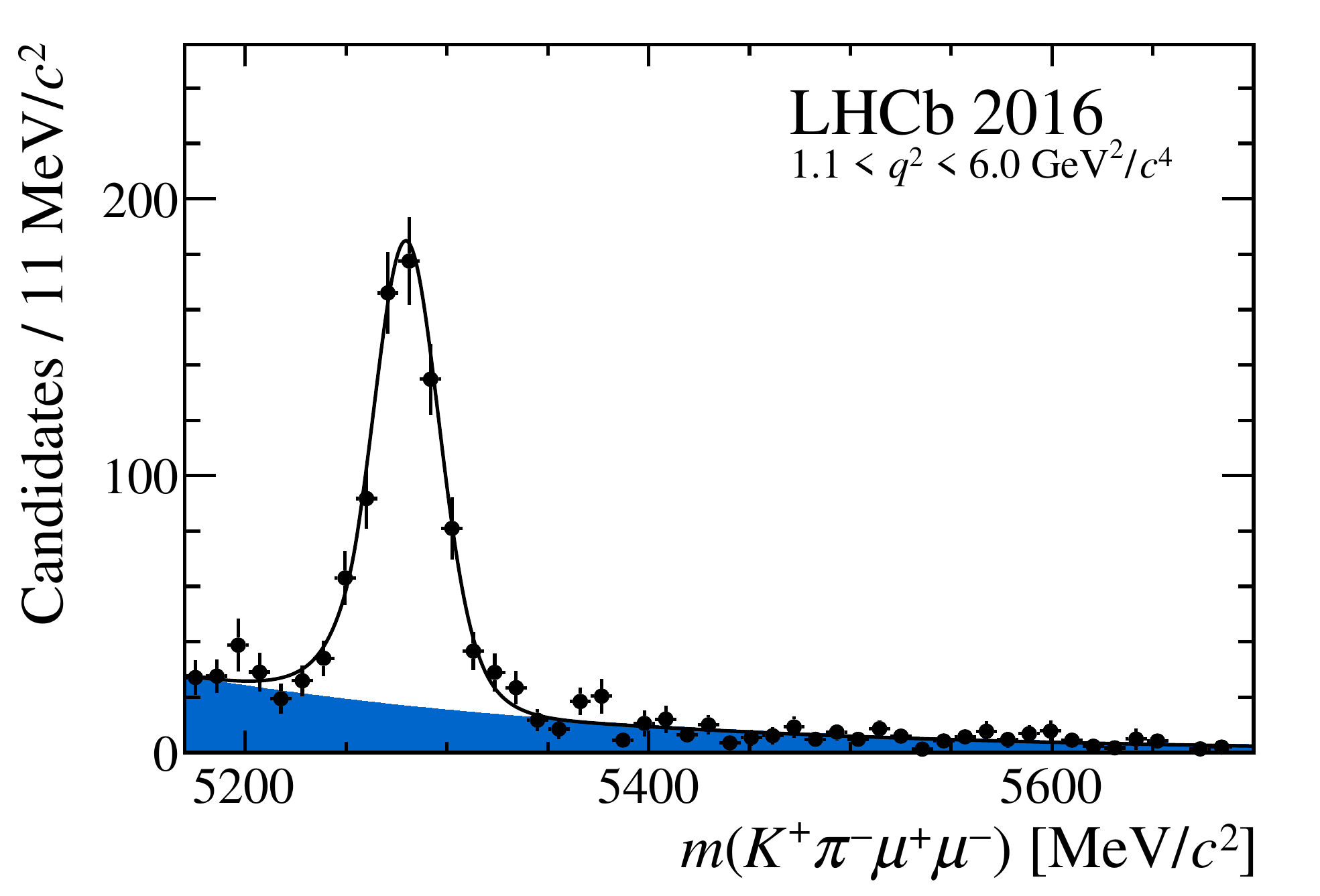}
 \caption{Projections of the fitted probability density function on the decay angles, \Mkpi and \Mkpimm for the bin $1.1<q^2<6.0\gevgevcccc$. The blue shaded region indicates background. \label{fig:projectionsi}}
 \end{figure}

\begin{figure}
   \centering
 \includegraphics[width=0.32\textwidth]{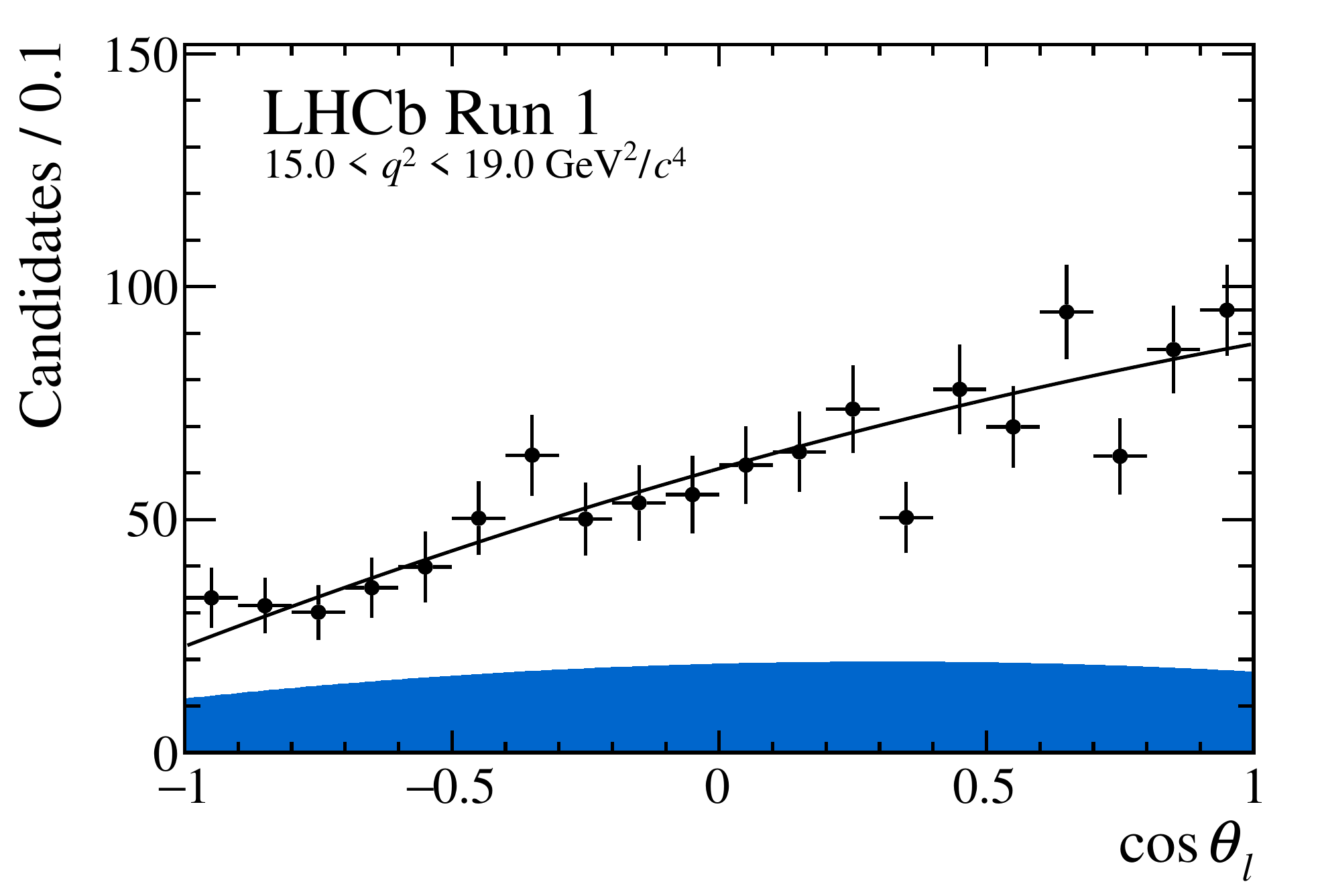}
 \includegraphics[width=0.32\textwidth]{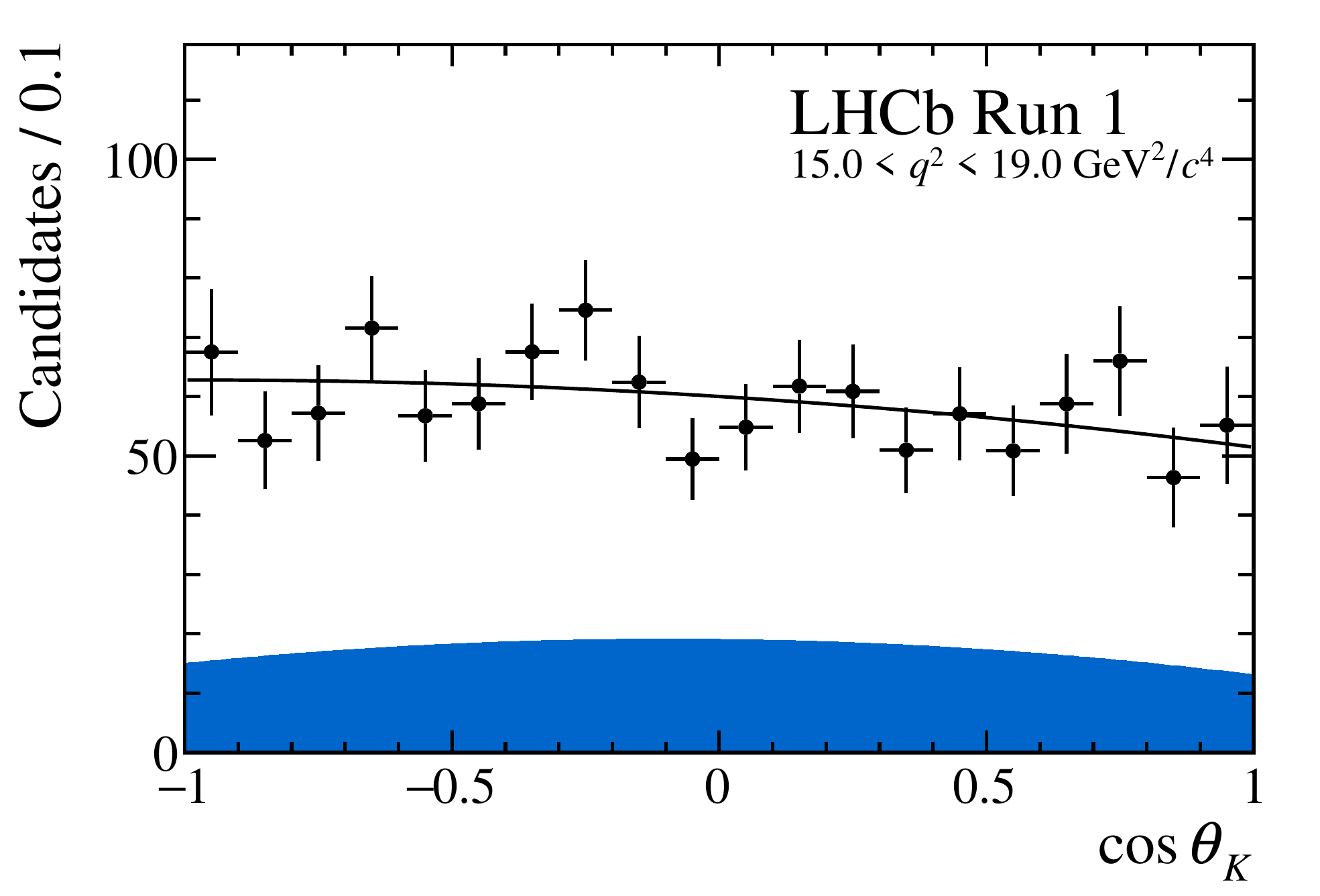}
 \includegraphics[width=0.32\textwidth]{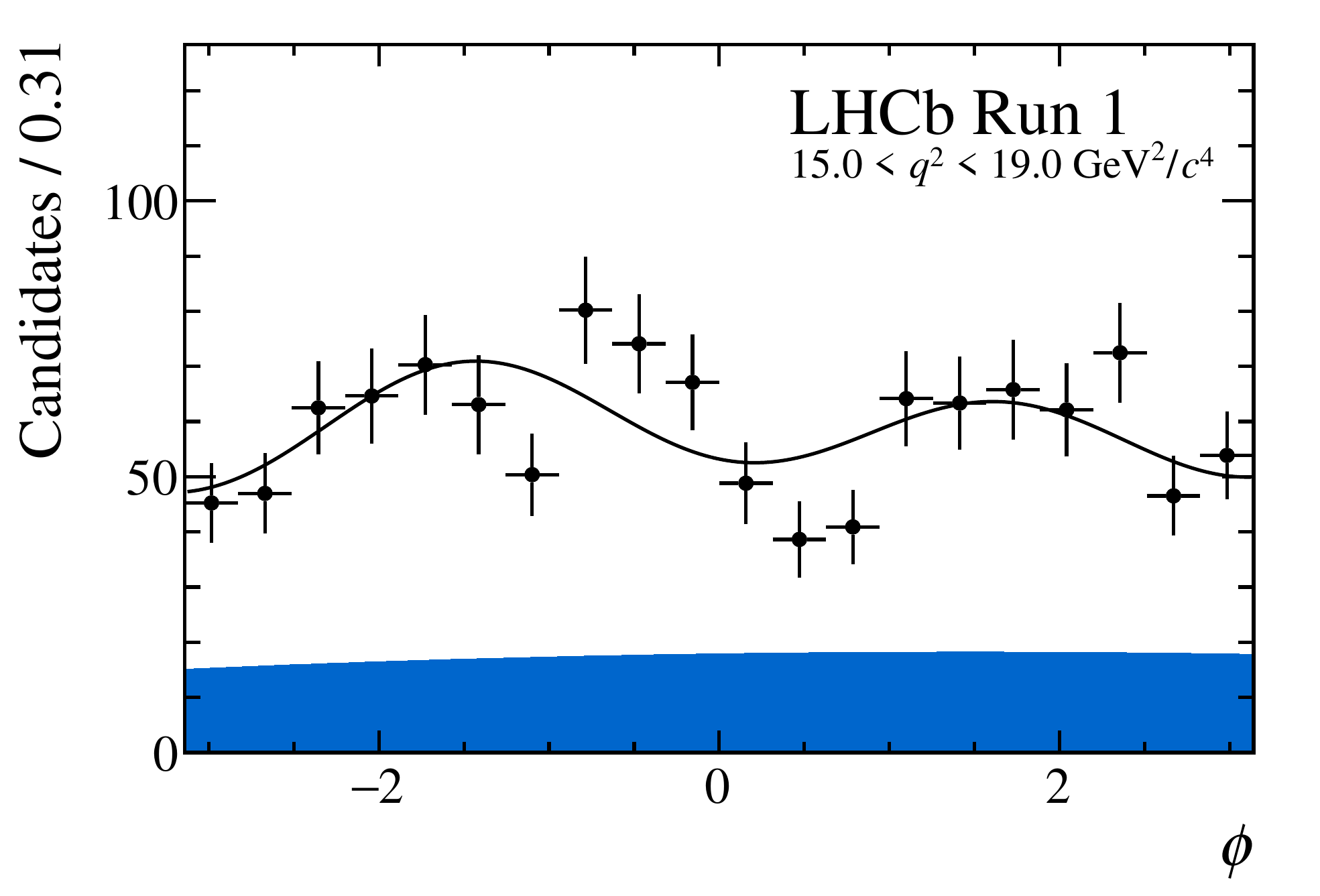}
 \includegraphics[width=0.32\textwidth]{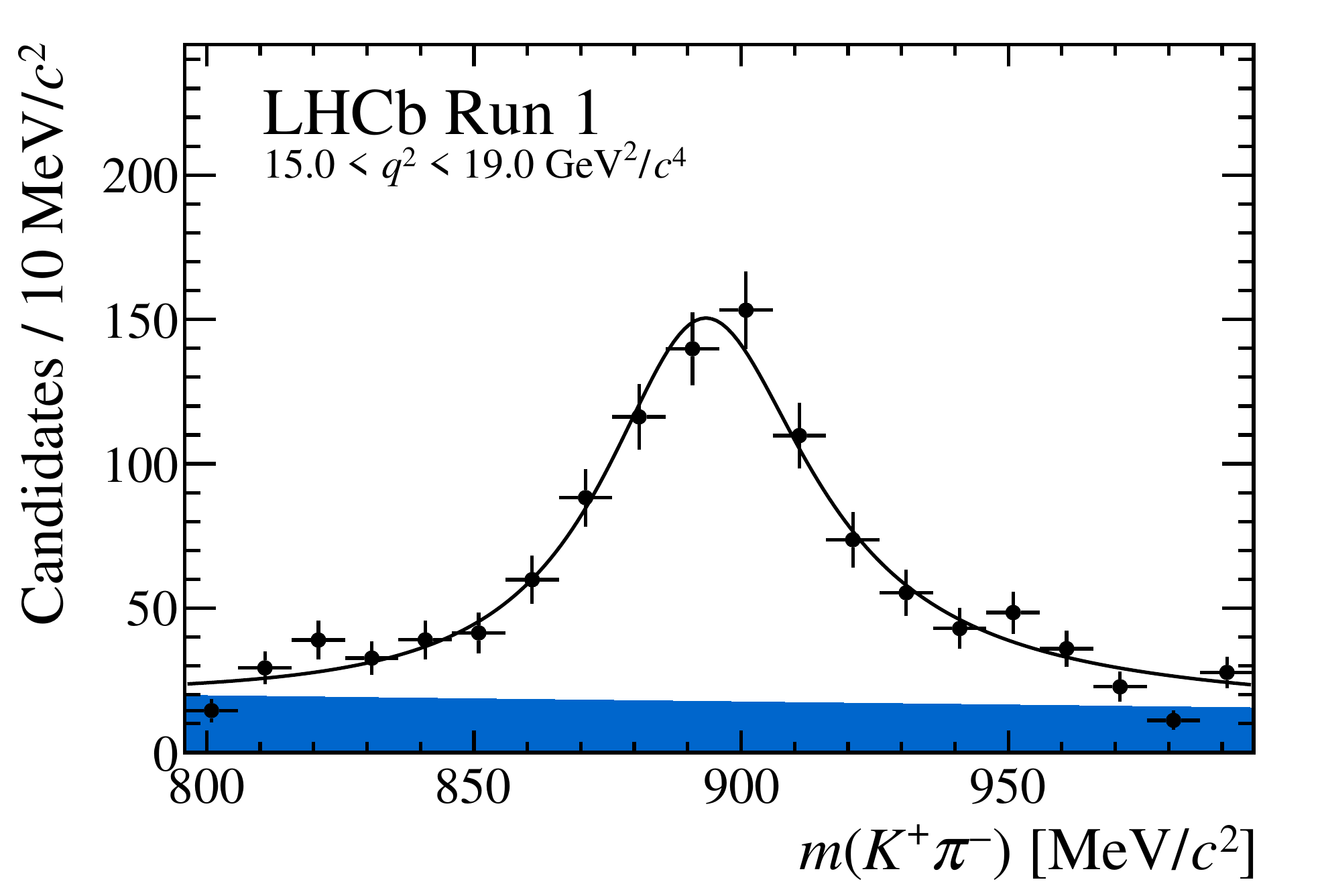}
 \includegraphics[width=0.32\textwidth]{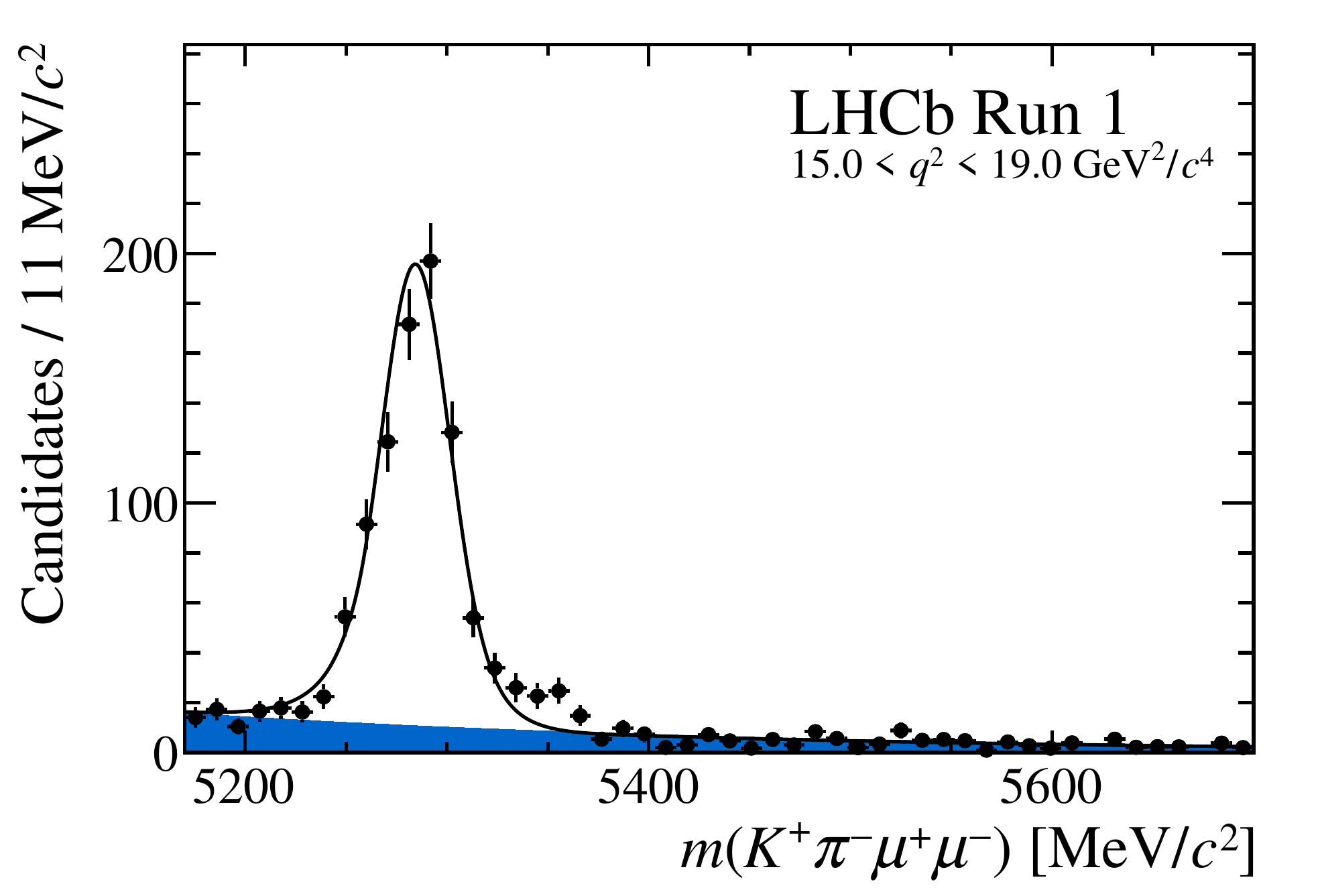}\\[0.5cm]
 \includegraphics[width=0.32\textwidth]{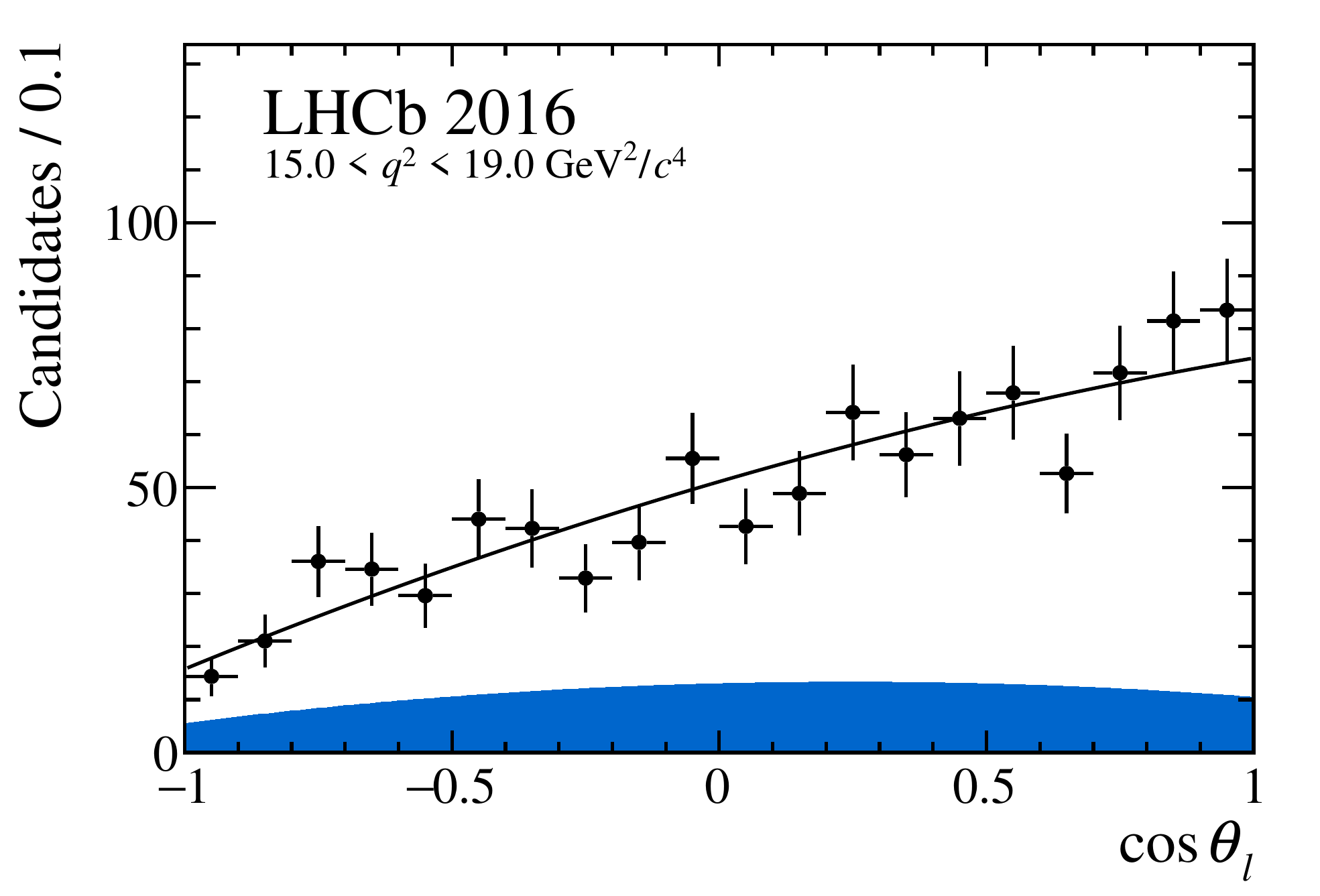}
 \includegraphics[width=0.32\textwidth]{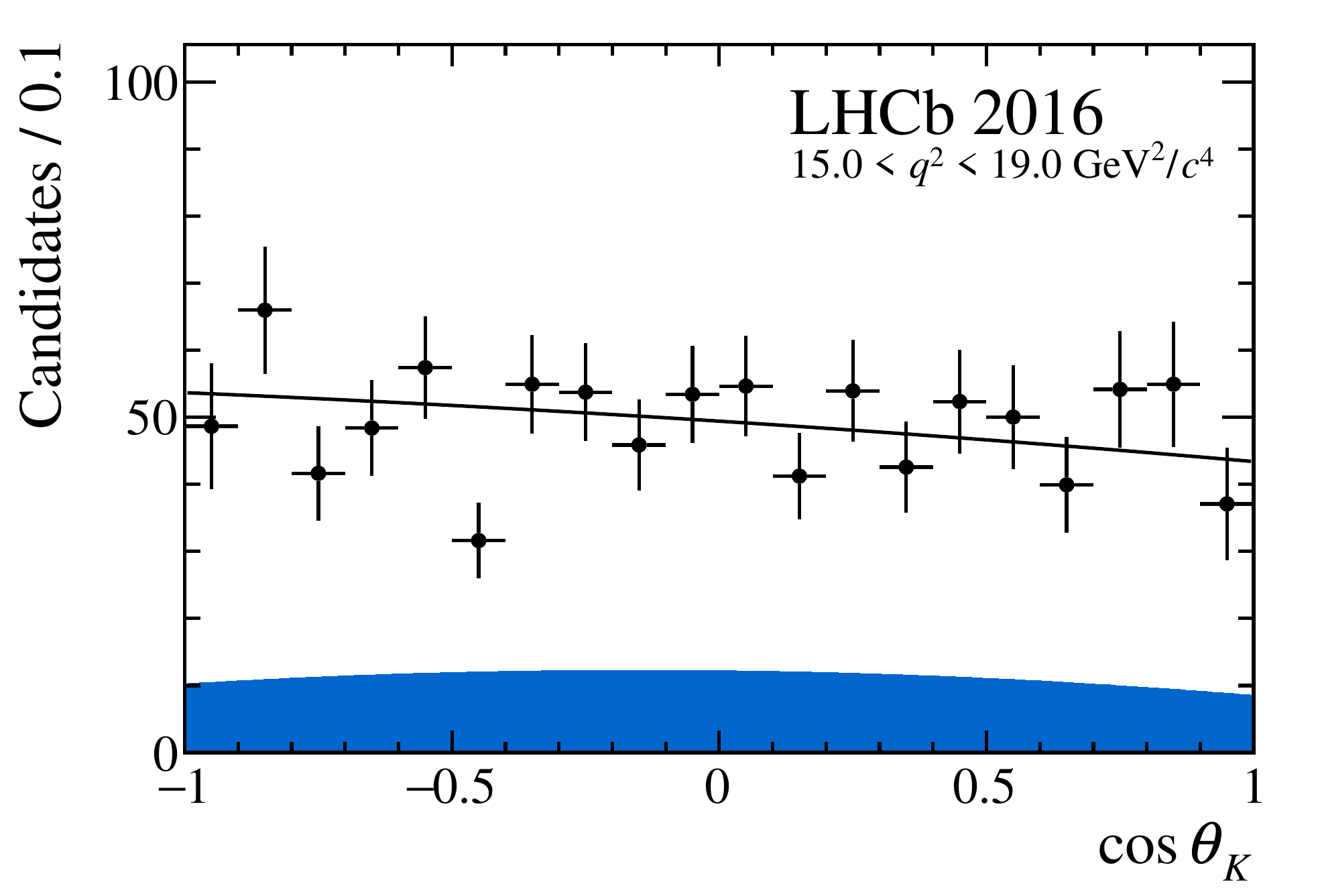}
 \includegraphics[width=0.32\textwidth]{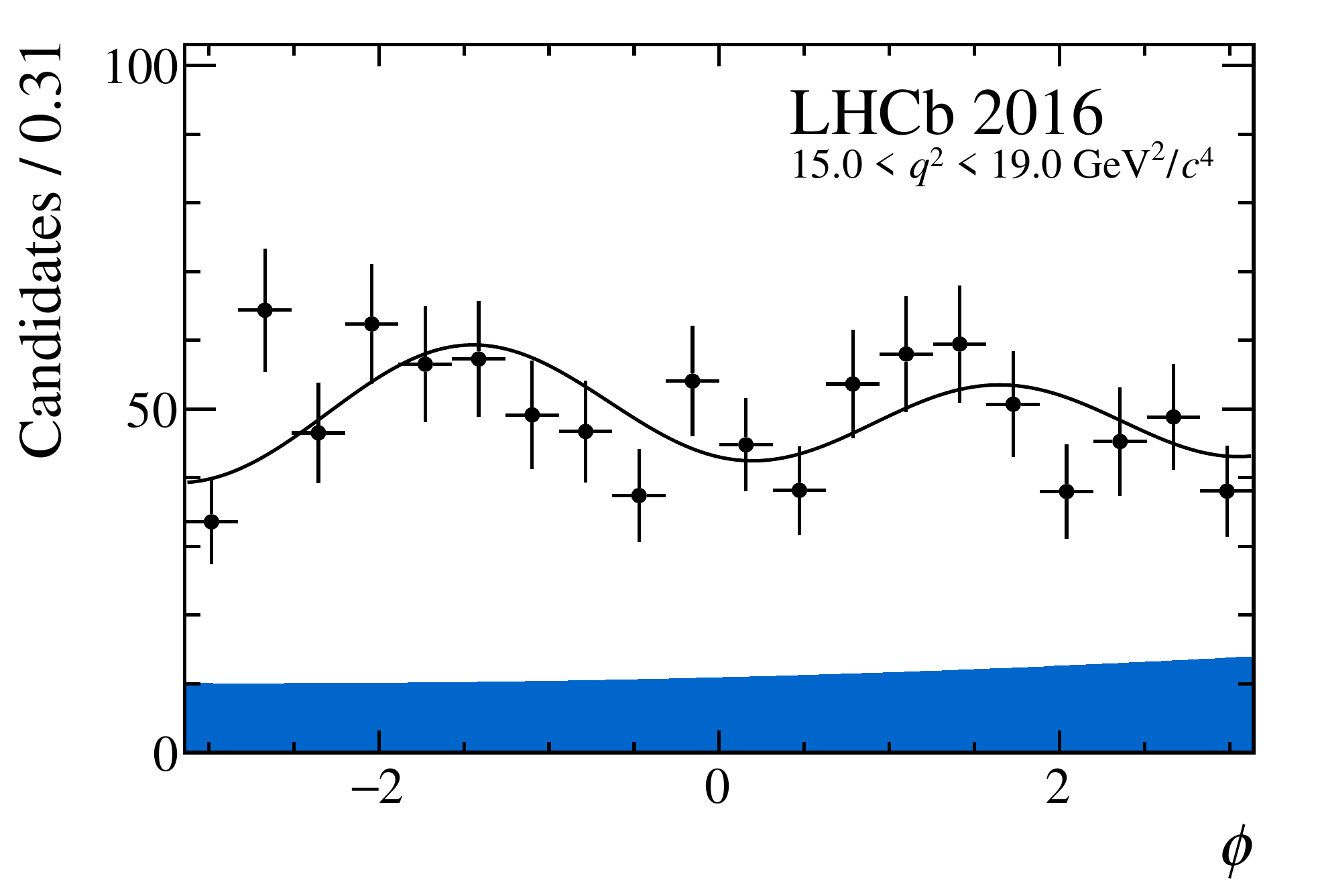}
 \includegraphics[width=0.32\textwidth]{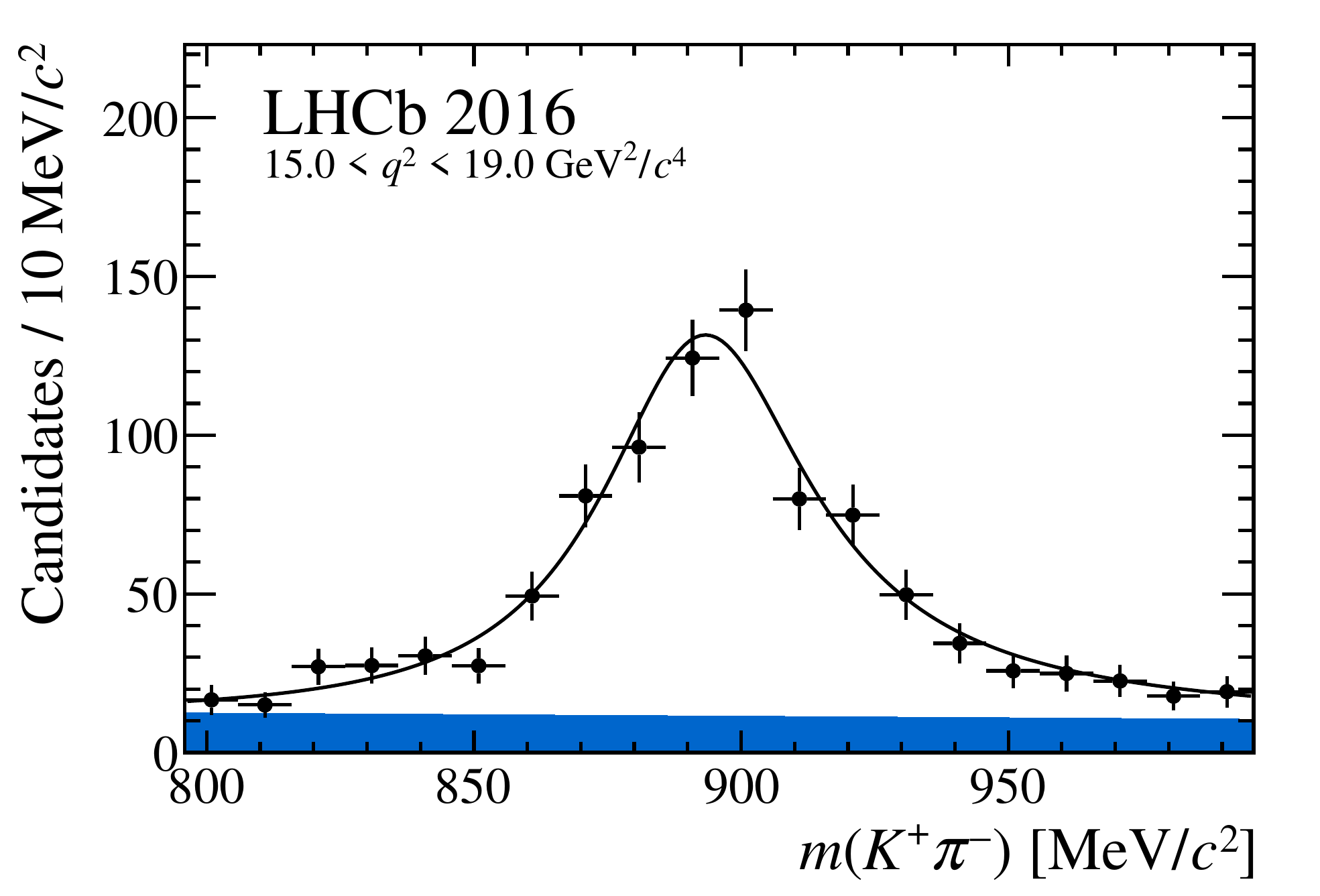}
 \includegraphics[width=0.32\textwidth]{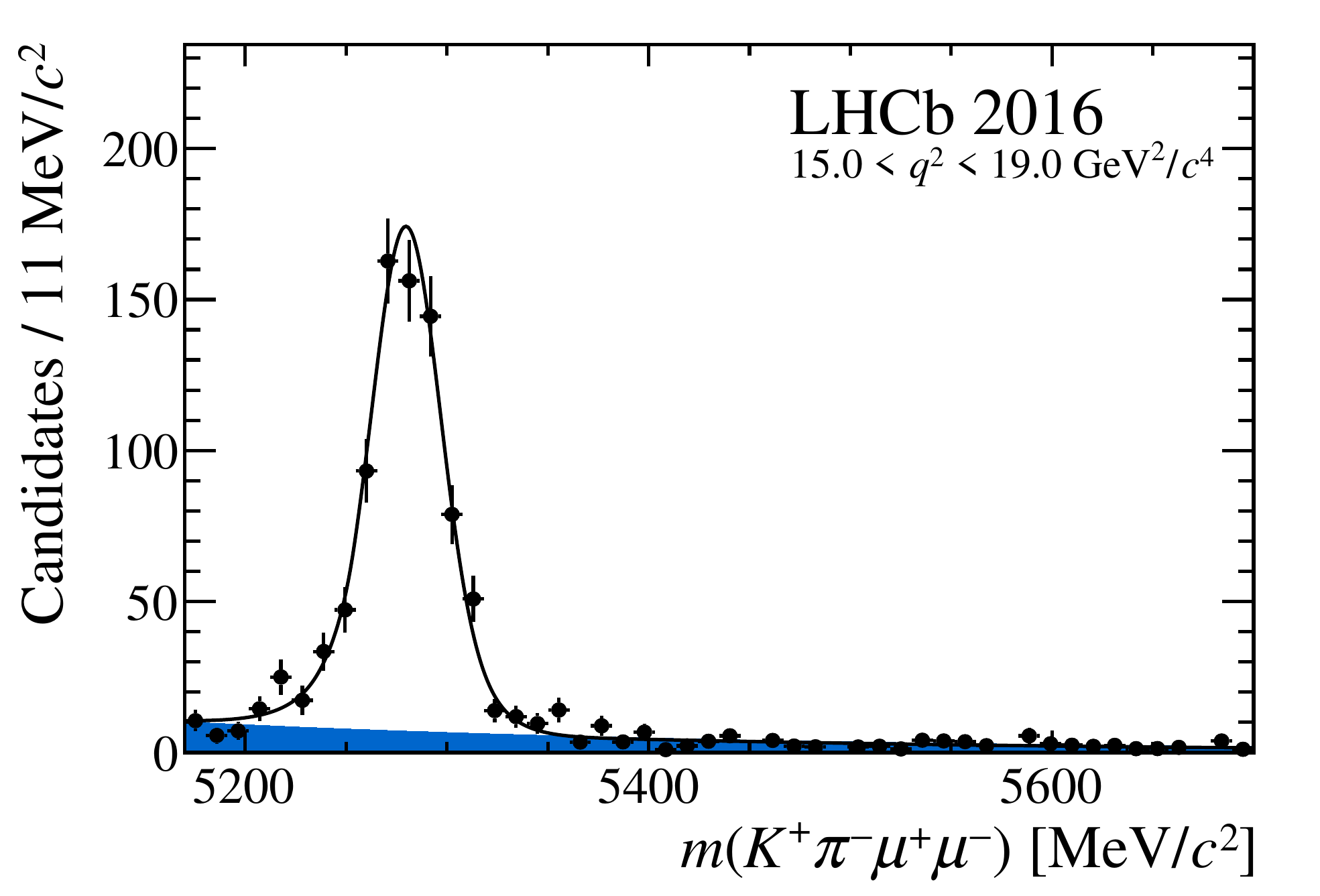}
 \caption{Projections of the fitted probability density function on the decay angles, \Mkpi and \Mkpimm for the bin $15.0<q^2<19.0\gevgevcccc$. The blue shaded region indicates background. \label{fig:projectionsj}}
 \end{figure}

The angular and mass distributions of the candidates in bins of \qsq for the Run~1 and the 2016 data, along with the projections of the simultaneous fit, are shown in Figs.~\ref{fig:projectionsa}--\ref{fig:projectionsj}.

\clearpage

\addcontentsline{toc}{section}{References}
\setboolean{inbibliography}{true}
\bibliographystyle{LHCb}
\bibliography{main,standard,LHCb-PAPER,LHCb-CONF,LHCb-DP,LHCb-TDR}

\newpage
 % LHCb collaboration author list
% Data extracted on March 4th, 2020 at 2:43pm for reference date 21-Jan-2020
\centerline
{\large\bf LHCb collaboration}
\begin
{flushleft}
\small
R.~Aaij$^{31}$,
C.~Abell{\'a}n~Beteta$^{49}$,
T.~Ackernley$^{59}$,
B.~Adeva$^{45}$,
M.~Adinolfi$^{53}$,
H.~Afsharnia$^{9}$,
C.A.~Aidala$^{81}$,
S.~Aiola$^{25}$,
Z.~Ajaltouni$^{9}$,
S.~Akar$^{66}$
% ...

{\footnotesize \it

% ...
$ ^{9}$Universit{\'e} Clermont Auvergne, CNRS/IN2P3, LPC, Clermont-Ferrand, France\\
% ...
$ ^{25}$INFN Sezione di Milano, Milano, Italy\\
% ...
$ ^{31}$Nikhef National Institute for Subatomic Physics, Amsterdam, Netherlands\\
% ...
$ ^{45}$Instituto Galego de F{\'\i}sica de Altas Enerx{\'\i}as (IGFAE), Universidade de Santiago de Compostela, Santiago de Compostela, Spain\\
% ...
$ ^{49}$Physik-Institut, Universit{\"a}t Z{\"u}rich, Z{\"u}rich, Switzerland\\
% ...
$ ^{53}$H.H. Wills Physics Laboratory, University of Bristol, Bristol, United Kingdom\\
% ...
$ ^{59}$Oliver Lodge Laboratory, University of Liverpool, Liverpool, United Kingdom\\
% ...
$ ^{66}$Los Alamos National Laboratory (LANL), Los Alamos, United States\\
$ ^{67}$Syracuse University, Syracuse, NY, United States\\
% ...
$ ^{81}$University of Michigan, Ann Arbor, United States, associated to $^{67}$\\
\bigskip
% ...
\medskip
}
\end{flushleft}

\end{document}